\documentstyle[11pt,epsfig,axodraw,amsmath]{article}
\def\msusy{\wt{m}}
\def\QB{Q^{\,\wt{\scriptscriptstyle B}}}
\def\QG{Q^{\,\wt{\scriptscriptstyle G}}}
\def\QW{Q^{\,\wt{\scriptscriptstyle W}}}

\def\CB{C^{\,\wt{\scriptscriptstyle B}}}
\def\CG{C^{\,\wt{\scriptscriptstyle G}}}
\def\CW{C^{\,\wt{\scriptscriptstyle W}}}
\def\CH{C^{\,\wt{\scriptscriptstyle H}}}
\def\QN{Q^{\,{\chi}^0_i}}

\def\CN{C^{\,{\chi}^0_i}}

\def\lle{_{\sss L}}
\def\rr{_{\sss R}}
\def\lr{_{\sss L,R}}

\def\ov{\overline}
\def\wt{\widetilde}
\def\sss{\scriptscriptstyle}
\catcode`@11
\def\seceqaa{\@addtoreset{equation}{section}
\def\theequation{A\arabic{equation}}}
\def\seceqbb{\@addtoreset{equation}{section}
\def\theequation{B\arabic{equation}}}
\def\seceqcc{\@addtoreset{equation}{section}
\def\theequation{C\arabic{equation}}}
\def\seceqdd{\@addtoreset{equation}{section}
\def\theequation{D\arabic{equation}}}
\catcode`@11
\def\beq{\begin{equation}}
\def\eeq{\end{equation}}
\def\bea{\begin{eqnarray}}
\def\eea{\end{eqnarray}}

\def\eps{\epsilon}
\def\non{\nonumber}

\def\mgss{M_{\tilde{g}}}
\def\mchi{M_{\tilde{\chi}}}

\begin{document}
\begin{titlepage}
\begin{center}
{\Large\bf Towards Large Volume Big Divisor $D3$/$D7$ ``$\mu$-Split Supersymmetry" and Ricci-Flat Swiss-Cheese Metrics, and Dimension-Six Neutrino Mass Operators}
\vskip 0.1in Mansi Dhuria$^{(a)}$ \footnote{email: mansidph@iitr.ernet.in}
and
 { Aalok Misra$^{(a),(b)}$\footnote{e-mail: aalokfph@iitr.ernet.in
}\\
(a) Department of Physics, Indian Institute of Technology,
Roorkee - 247 667, Uttaranchal, India\\
(b) Physics Department, Theory Unit, CERN, CH1211, Geneva 23, Switzerland } \vskip 0.5 true in
\date{\today}
\end{center}
\thispagestyle{empty}
\begin{abstract}
We show that it is possible to realize a ``$\mu$-split SUSY" scenario \cite{mu split susy} in the context of large volume limit of type IIB compactifications on Swiss-Cheese Calabi-Yau orientifolds in the presence of a mobile space-time filling $D3$-brane and a (stack of) $D7$-brane(s) wrapping the ``big" divisor. For this, we investigate the possibility of getting one Higgs to be light while other to be heavy in addition to a heavy Higgsino mass parameter. Further, we examine the existence of long lived gluino that manifests one of the major consequences of $\mu$-split SUSY scenario,  by computing its decay width as well as lifetime corresponding to the three-body decays of the gluino into either a quark, a squark and a neutralino or a quark, squark and Goldstino, as well as two-body decays of the gluino into either a neutralino and a gluon or a Goldstino and a gluon. Guided by the geometric K\"{a}hler potential for $\Sigma_B$ obtained in \cite{D3_D7_Misra_Shukla} based on GLSM techniques, and the Donaldson's algorithm  \cite{Donaldson_i} for obtaining numerically a Ricci-flat metric, we give details of our calculation in \cite{ferm_masses_MS} pertaining to our proposed metric for the full Swiss-Cheese Calabi-Yau  (the geometric K\"{a}hler potential being needed to be included in the full moduli space K\"{a}hler potential in the presence of the mobile space-time filling $D3$-brane), but for simplicity of calculation, close to the big divisor, which is Ricci-flat in the large volume limit. Also, as an application of the one-loop RG flow solution for the Higgsino mass parameter, we show that the contribution to the neutrino masses at the EW scale from dimension-six operators arising from the K\"{a}hler potential, is suppressed relative to the Weinberg-type dimension-five operators.

\end{abstract}
\end{titlepage}
\section{Introduction}
Despite the success of Standard Model in High Energy Physics, failure of naturalness and fine tuning requirements in the Higgs Sector remain basic motivations to construct theories beyond Standard Model. The supersymmetric extension of the Standard Model  can solve the fine tuning problem in the Higgs/scalar sector, however for this one  requires supersymmetric particles at TeV scale. Though it is possible to achieve gauge coupling unification and a good dark matter candidate, yet the existence of naturally large supersymmetric contribution to flavour changing neutral current, experimental value of electron dipole moment (EDM) for natural CP violating phase and dimension-five proton decays are serious issues that can not be solved elegantly in supersymmetric Standard Model. Also, lack of existence of light Higgs boson is one of the major tensions in MSSM. More recently, an alternative approach to SUSY has been adopted by Arkani-Hamed and Dimopoulos\cite{HamidSplitSUSY} in which they argued given that fine tuning anyway seems to be required to obtain a small and positive cosmological constant (which is one of the most serious issues), one is hence also allowed to assume fine tuning in other sectors of the theory (Higgs Sector) which is a less serious issue in the string theory landscape. Therefore in order to understand the right amount of cosmological constant (cc) in the `string theory landscape' in which different choices of string vacua are available depending on different SUSY breaking scales, one may prefer the high SUSY breaking scale region where one can obtain a small cc as well as finely tuned light Higgs boson. The realistic model based on high scale $(m_s\sim {10}^{10}$ GeV) SUSY breaking is named as split SUSY Model. In this scenario all scalar particles acquire heavy masses except one Higgs doublet which is finely tuned to be light while fermions (possibly also gaugino and Higgsino) are light. This  interesting class of model has attracted considerable attention though it abandons the primary reason for introducing supersymmetry. This scenario removes all unrealistic features of MSSM while preserves all good features ( possibly gauge coupling unification and dark matter candidate). As discussed in \cite{aspects_split susy}, dimension-five proton decay get naturally suppressed due to ultra heavy scalar masses and contribution to electron dipole moment (EDM) arising at two loop reach to the order of current experimental limits ignoring the effect of CP violating phase as compared to the case of low energy supersymmetric models. In \cite{sudhir kumar gupta} it is shown that the lightest neutralino can still be taken as a good dark matter candidate in split SUSY. Also gauge coupling unification remains inherent in split SUSY see \cite{haba and okada}. One of the other striking feature of this model based on heavy squark masses is non trivial gluino decay discussed in \cite{gluino decay pattern}. Kinematically favored three body gluino decays $\tilde{g}\rightarrow{\chi_i}^0{\bar{q}_J}q_J$  or $\tilde{g}\rightarrow{\chi_i}^\pm{\bar{q}_I}q_J$ (where ${\chi_i}^0$ ,${\chi_i}^\pm$ correspond to neutralinos and charginos, $q_{I,J}, {\bar{q}_{I,J}}$ correspond to quarks and antiquarks) occuring via virtual squarks get considerably suppressed due to heavy squark masses and hence gluino remains long lived. Therefore measuring life time of gluino can be adopted as indirect way to measure heavy squark mass i.e limit of SUSY breaking scale in split SUSY scenario.

 Despite explaining many unresolved issues of phenomenology in the context of split SUSY, the notorius ${\mu}$ problem still
 remains unsolved according to which the stable vaccum that spontaneously breaks electroweak symmetry requires ${\mu}$ to be of the order of supersymmetry breaking scale.  However in case of split SUSY scenario one is assuming ${\mu}$ to be light while
 supersymmetry breaking scale to be very high. The other alternative  to solve this serious $\mu$ problem has been discussed by authors in \cite{mu split susy} in which one discusses a further split in the split SUSY scenario by raising the
 ${\mu}$ parameter to a large value which could be about the same as the sfermion mass or the SUSY breaking scale; this scenario is dubbed as ${\mu}$-split SUSY scenario. In addition to solving the $\mu$ problem, all the nice features of split
 supersymmeric model like gauge coupling unification, dark matter candidate remain protected in this scenario.

With the promising approach of string theory to phenomenology as well as cosmology, it is quite interesting to realize the split SUSY scenario within a string theoretic framework. The signatures of the same in the context of type I and type IIA string theory were obtained respectively in \cite{type 1 string} and \cite{Kokorelis}. Recently, in the context of type IIB (``big divisor") LVS $D3/D7$ Swiss cheese phenomenology, the authors of \cite{ferm_masses_MS} explicitly showed the possibility of generating light fermion masses as well as heavy squark/sleptons masses    including a space-time filling mobile $D3$ brane and stack(s) of (fluxed) $D7$- branes wrapping the ``Big" divisor. Matter fields (quarks, leptons and their superpartners) are identified with the (fermionic superpartners of) Wilson line moduli whereas Higgses are identified with space-time filling mobile $D3$-brane position moduli. The plan of the rest of the paper is as follows. Building up on the same set up summarized in section {\bf 2}, in section {\bf 3}, we evaluate the masses of Higgs doublets at the Electroweak scale  up to one loop. Solving for eigenvalues of the Higgs(ino) mass matrix, we have explored the possibilty of realizing one eigenvalue of Higgs doublet to be light and other to be heavy in addition to a heavy Higgsino mass parameter which shows ${\mu}$-split SUSY scenario in the context of L(arge) V(olume) S(cenarios) coined as large volume ${\mu}$ split SUSY scenario. In order to seek the other striking evidence of ${\mu}$-split SUSY in the context of LVS, in section {\bf 4.1}, we calculate the tree-level three-body gluino decay into a quark, anti-quark and the lightest neutralino (which after diagonalizing the neutralino mass matrix, turns out to be largely a neutral gaugino with a small admixture of the Higgsinos) at SUSY Breaking scale; using  one-loop RG analysis of the effective dimension-six gluino decay operators, we show that the couplings of the effective theory are of the same order at the EW scale as at the squark mass scale (and therefore we conjecture at the string scale) and then using the approach discussed, e.g. in \cite{Manuel Toharia,Griffiths_particle}, we calculate the decay width and hence the lifetime of the gluino. In section {\bf 4.2}, we calculate the decay width and lifetime corresponding to the two-body decay of the gluino to a neutralino and gluon. Finally, in {\bf 4.3}, we calculate the decay widths and lifetimes corresponding to the gluino three-body decay into a quark, anti-quark and Goldstino and the gluino two-body decay into a Goldstino and gluon. Due to the presence of a mobile space-time filling $D3$-brane, one needs to include the geometrical K\"{a}hler potential in the moduli space K\"{a}hler potential. Guided by earlier estimates (See \cite{D3_D7_Misra_Shukla}) in the large volume limit of the geometric K\"{a}hler potential for the divisors of the Swiss-Cheese Calabi-Yau, as well as the Donaldson's algorithm for numerical construction of Ricci-flat metrics, in section {\bf 5} we construct a metric for the Swiss-Cheese Calabi-Yau in a coordinate patch, for simplicity, close to the big divisor, which in the large volume limit, is Ricci-flat.  Using the one-loop RG flow result for $\hat{\mu}$ of section {\bf 2}, we evaluate the contribution to neutrino masses of dimension-six operators from the K\"{a}hler potential in section {\bf 6} which naturally turns out to be extremely suppressed as compared to the dimension-five Weinberg-type operators. Section {\bf 7} has the concluding remarks. There are four appendices.
 \section{Setup}
In this section, we first describe our setup: type IIB compactification on the orientifold of a ``Swiss-Cheese Calabi-Yau" in the large volume limit including perturbative $\alpha^\prime$ and world sheet instanton corrections as well as one-loop corrections to the K\"{a}hler potential, and the instanton-generated superpotential written out respecting the (subgroup, under orientifolding, of) $SL(2,{\bf Z})$ symmetry of the underlying parent type IIB theory, in the presence of a mobile space-time filling $D3-$brane and stacks of $D7$-branes wrapping the ``big" divisor along with magnetic fluxes. This is followed by a summary of evaluation of soft supersymmetry breaking parameters, showing the possibility of getting light fermions and heavy scalar superpartners and generating (less than) $eV$ mass scales relevant to Majorana neutrino mass scales.

In  \cite{dSetal,largefNL_r_axionicswisscheese}, we  addressed some cosmological issues like $dS$ realization, embedding inflationary scenarios and realizing non-trivial non-Gaussianities in the context of type IIB Swiss-Cheese Calabi Yau orientifold in LVS. This has been done with the inclusion of (non-)perturbative $\alpha^{\prime}$-corrections to the K\"{a}hler potential and non-perturbative instanton contribution to the superpotential. The Swiss-Cheese Calabi Yau we are using, is a projective variety in ${\bf WCP}^4[1,1,1,6,9]$ given as
\begin{equation}
\label{eq:hyper}
x_1^{18} + x_2^{18} + x_3^{18} + x_4^3 + x_5^2 - 18\psi \prod_{i=1}^5x_i - 3\phi x_1^6x_2^6x_3^6 = 0,
\end{equation}
 which has two (big and small) divisors $\Sigma_B(x_5=0)$ and $\Sigma_S(x_4=0)$ . From Sen's orientifold-limit-of-F-theory point of view  corresponding to type IIB compactified on a Calabi-Yau three fold $Z$-orientifold with $O3/O7$ planes, one requires a Calabi-Yau four-fold $X_4$ elliptically fibered (with projection $\pi$) over a 3-fold $B_3(\equiv CY_3-$orientifold)  where $B_3$ could either be a Fano three-fold or an $n$-twisted ${\bf CP}^1$-fibration over ${\bf CP}^2$ such that pull-back of the divisors in $CY_3$ automatically satisfy Witten's unit-arithmetic genus condition  \cite{DDF,denef_LesHouches}. The toric data of $B_3$ consists of five divisors, three of which are pullbacks of three lines in ${\bf CP}^2$ and the other two are sections of the aforementioned fibration. From the point of view of M-theory compactified on $X_4$, the non-perturbative superpotential receives non-zero contributions from $M5$-brane instantons involving wrapping around uplifts {\bf V} to $X_4$ of ``vertical" divisors ($\pi({\rm\bf V})$ is a proper subset of $B_3$) in $B_3$. These vertical divisors are either components of singular fibers or are pull-backs of smooth divisors in $B_3$. There exists a Weierstrass model $\pi_0:{\cal W}\rightarrow B_3$ and its resolution
$\mu: X_4\rightarrow {\cal W}$. For $n=6$ \cite{DDF}, the $CY_4$ will be the resolution of a Weierstrass model with $D_4$ singularity along the first section and an $E_{6/7/8}$ singularity along the second section. The Calabi-Yau three-fold $Z$ then turns out to be a unique Swiss-Cheese Calabi Yau - an elliptic fibration over ${\bf CP}^2$ in ${\bf WCP}^4[1,1,1,6,9]$ given by (\ref{eq:hyper}). We would be assuming an $E_8$-singularity as this corresponds to
 $h^{1,1}_-(CY_3)=h^{2,1}(CY_4)\neq0$\cite{denef_LesHouches} which is what we will be needing and using. The required Calabi-Yau has $h^{1,1}=2, h^{2,1}=272$. The same has a large discrete symmetry group given by $\Gamma={\bf Z}_6\times{\bf Z}_{18}$ (as mentioned in \cite{D3_D7_Misra_Shukla}) relevant to construction of the mirror a la Greene-Plesser prescription. However, as is common in such calculations (See \cite{DDF,denef_LesHouches,Kachru_et_al}), one assumes that one is working with a subset of periods of $\Gamma$-invariant cycles - the six periods corresponding to the two complex structure deformations in (\ref{eq:hyper}) will coincide with the six periods of the mirror - the complex structure moduli absent in (\ref{eq:hyper}) will appear only at a higher order in the superpotential because of $\Gamma$-invariance and can be consistently set to zero (See \cite{Kachru_et_al}).

As shown in \cite{D3_D7_Misra_Shukla}, in order to support MSSM (-like) models and for resolving the tension between LVS cosmology and LVS phenemenology within a string theoretic setup, a mobile space-time filling $D3-$brane and stacks of $D7$-branes wrapping the ``big" divisor $\Sigma_B$ along with magnetic fluxes, are included. The appropriate ${\cal N}=1$ coordinates in the presence of a single $D3$-brane and a single $D7$-brane wrapping the big divisor $\Sigma^B$ along with $D7$-brane fluxes were obtained in \cite{jockersetal}; the same along with the details of the holomorphic isometric involution involved in orientifolding, as well as  expansion of the complete K\"{a}hler potential (including  the geometric K\"{a}hler potential) and the (non-perturbative) superpotential as a power series in fluctuations about Higgses' vevs and the corresponding  extremum values of the Wilson line moduli, have been summarized in \cite{ferm_masses_MS}.

Now, in the context of intersecting brane world scenarios \cite{int_brane_SM,Ibanez et al}, bifundamental leptons and quarks are obtained respectively from open strings stretched between $U(2)$ and $U(1)$ stacks, and $U(3)$ and $U(2)$ stacks of $D7$-branes; the adjoint gauge fields correspond to open strings starting and ending on the same $D7$-brane. In Large Volume Scenarios, however,  one considers four stacks of different numbers of multiple $D7$-branes  wrapping $\Sigma_B$ but with different  choices of magnetic $U(1)$ fluxes turned on, on the two-cycles which are non-trivial in the Homology of $\Sigma_B$ and not the ambient Swiss Cheese Calabi-Yau. The inverse gauge coupling constant squared for the $j$-th gauge group ($j:SU(3), SU(2),U(1)$), up to open string one-loop level, using \cite{Maldaetal_Wnp_pref,Wilson 1loop,Jockers_thesis}, will be given by
\begin{equation}
\label{eq:1overgsquared}
\frac{1}{g_{j{=SU(3)\ {\rm or}\ SU(2)}}^2} = Re(T_{S/B}) + ln\left(\left.P\left(\Sigma_S\right)\right|_{D3|_{\Sigma_B}}\right) + ln\left(\left.{\bar P}\left(\Sigma_S\right)\right|_{D3|_{\Sigma_B}}\right) + {\cal O}\left({\rm U(1)-Flux}_j^2\right),
\end{equation}
 where
${\rm U(1)-Flux}_j$ are abelian magnetic fluxes for the $j-$th stack. Also, $\left.P\left(\Sigma_s\right)\right|_{D3|_{\Sigma_B}}$ implies the defining hypersurface for the small divisor $\Sigma_S$ written out in terms of the position moduli of the mobile $D3$-brane, restricted to the big divisor $\Sigma_B$. In the dilute flux approximation, the ``ln" terms in the right hand side of (\ref{eq:1overgsquared})are of ${\cal O}\left( ln{\cal V}\right)$, which for ${\cal V}\sim10^6$ is taken to be of the same order as $\sigma^B+{\bar\sigma}^{\bar B} - C_{1{\bar 1}}|a_1|^2\sim{\cal V}^{\frac{1}{18}}$ appearing in $Re(T_B)$. For ${1}/{g_{U(1)}^2}$ there is a model-dependent numerical prefactor multiplying the right hand side of the $1/g_j^2$-relation. In the dilute flux approximation, $\alpha_i(M_s)/\alpha_i(M_{EW}), i=SU(3),SU(2),U(1)_Y$, are hence unified. By turning on different $U(1)$ fluxes on, e.g., the $3_{QCD}+2_{EW}$ $D7$-brane stacks in the LVS setup, $U(3_{QCD}+2_{EW})$ is broken down to $U(3_{QCD})\times U(2_{EW})$ and the four-dimensional Wilson line moduli $a_{I(=1,...,h^{0,1}_-(\Sigma_B))}$ and their fermionic superpartners $\chi^I$ that are valued, e.g., in the $adj(U(3_{QCD}+2_{EW}))$ to begin with, decompose into the bifundamentals $(3_{QCD},{\bar 2}_{EW})$ and its complex conjugate, corresponding to the bifundamental left-handed quarks of the Standard Model (See \cite{bifund_ferm}).  Further, the main idea then behind realizing $O(1)$ gauge coupling is the competing contribution to the gauge kinetic function (and hence to the gauge coupling) coming from the $D7$-brane Wilson line moduli as compared to the volume of the big divisor $\Sigma_B$, after constructing local (i.e. localized around the location of the mobile $D3$-brane in the Calabi-Yau) appropriate involutively-odd harmonic distribution one-form on the big divisor that lies in $coker\left(H^{(0,1)}_{{\bar\partial},-}(CY_3)\stackrel{i^*}{\rightarrow}
H^{(0,1)}_{{\bar\partial},-}(\Sigma_B)\right)$, the immersion map $i$ being defined as:
$i:\Sigma^B\hookrightarrow CY_3$. This will also entail stabilization of the Wilson line moduli at around $ {\cal V}^{-\frac{1}{4}}$ for vevs of around ${\cal V}^{\frac{1}{36}}$ of the $D3$-brane position moduli, the Higgses in our setup. Extremization of the ${\cal N}=1$ potential, as shown in \cite{D3_D7_Misra_Shukla} and mentioned earlier on, shows that this is indeed true.  This way the gauge couplings corresponding to the gauge theories living on stacks of $D7$ branes wrapping the ``big" divisor $\Sigma_B$ (with different $U(1)$ fluxes on the two-cycles inherited from $\Sigma_B$) will be given by:
$g_{YM}^{-2}\sim {\cal V}^{\frac{1}{18}}$, $T_B$ being the appropriate ${\cal N}=1$ K\"{a}hler coordinate  and the relevant text below the same) and $\mu_3$ related to the $D3$-brane tension,
implying a finite (${\cal O}(1)$) $g_{YM}$ for ${\cal V}\sim10^6$.

As discussed in \cite{Sparticles_Misra_Shukla}, for the type IIB Swiss-Cheese orientifold considered in our work, guided, e.g., by the vanishingly small Yukawa couplings $\hat{Y}_{\tilde{\cal A}_1^2{\cal Z}_i}$ obtained from an  $ED3$-instanton-generated superpotential (See Table 1), the spacetime filling mobile D3-brane position moduli $z_i$'s and the Wilson line moduli $a_{I}$'s could be respectively identified with Higgses and the first two generations of sparticles (squarks/sleptons) of some (MS)SM-like model.
With a (partial) cancelation between the volume of the ``big" divisor and the Wilson line
contribution (required for realizing $\sim O(1) g_{YM}$ in our setup), in \cite{D3_D7_Misra_Shukla}, we calculated in the large volume limit, several soft supersymmetry breaking parameters. The same relevant to this paper can be summarized in table 1.
 \begin{table}[htbp]
\centering
\begin{tabular}{|l|l|}
\hline
Gravitino mass &  $ m_{\frac{3}{2}}\sim{\cal V}^{-\frac{n^s}{2} - 1}$ \\
Gaugino mass & $ M_{\tilde g}\sim m_{\frac{3}{2}}$\\ \hline
$D3$-brane position moduli  & $ m_{{\cal Z}_i}\sim {\cal V}^{\frac{19}{36}}m_{\frac{3}{2}}$ \\
(Higgs) mass & \\
Wilson line moduli mass & $ m_{\tilde{\cal A}_1}\sim {\cal V}^{\frac{73}{72}}m_{\frac{3}{2}}$\\ \hline
A-terms & $A_{pqr}\sim n^s{\cal V}^{\frac{37}{36}}m_{\frac{3}{2}}$\\
& $\{p,q,r\} \in \{{{\tilde{\cal A}_1}},{{\cal Z}_i}\}$\\
\hline
Physical $\mu$-terms & $\hat{\mu}_{{\cal Z}_i{\cal Z}_j}\sim{\cal V}^{\frac{37}{36}}m_{\frac{3}{2}}$ \\
& $\hat{\mu}_{{\cal A}_1{\cal Z}_i}\sim{\cal V}^{-\frac{3}{4}}m_{\frac{3}{2}}$\\
& $\hat{\mu}_{{\cal A}_1{\cal A}_1}\sim{\cal V}^{-\frac{33}{36}}m_{\frac{3}{2}}$\\
\hline
Physical Yukawa couplings&
$\hat{Y}_{{\cal Z}_1{\cal Z}_2\tilde{\cal A}_1}\sim {\cal V}^{-\frac{17}{72}}m_{\frac{3}{2}}$\\
&$\hat{Y}_{\tilde{\cal A}_1^2{\cal Z}_i}\sim {\cal V}^{-\frac{127}{72}}m_{\frac{3}{2}}$\\
&$\hat{Y}_{\tilde{\cal A}_1\tilde{\cal A}_1\tilde{\cal A}_1}\sim {\cal V}^{-\frac{85}{24}}m_{\frac{3}{2}}$\\
\hline
Physical $\hat{\mu}B$-terms & $\left(\hat{\mu}B\right)_{{\cal Z}_1{\cal Z}_2}\sim{\cal V}^{\frac{37}{18}}m_{\frac{3}{2}}^2$\\
\hline
\end{tabular}
\caption{{\small Results on Soft SUSY Parameters Summarized}}
\end{table}

Fermion (Quark/Lepton) masses are generated by giving some VEVs to Higgses in $\int d^4 x \,e^{\hat{K}/2} Y_{ijk} {z}^{i} \psi^{j}\psi^{k}$. The (canonically normalized) fermionic mass matrix is generated by ${\hat {Y}_{ijk}<{z}_i }>$. The  mass of the fermionic superpartner of $\tilde{\cal A}_1$ in \cite{ferm_masses_MS} (which based on the near-vanishing value of the Yukawa coupling $\hat{Y}_{\tilde{\cal A}_1^2{\cal Z}_i}$ in Table 1, is conjectured to be a first/second generation quark/lepton) turns out to be given by:${\cal V}^{-\frac{199}{72}-\frac{n^s}{2}}$ in units of $M_p$, which  implies a range of fermion mass $m_{\rm ferm}\sim{\cal O}({\rm MeV-GeV})$ for Calabi Yau volume ${\cal V}\sim {\cal O}(6\times10^5-10^5)$. Interestingly, the mass-scale of $0.5$ MeV- the electronic mass scale- could be realized with ${\cal V}\sim 6.2\times 10^5, n^s=2$.

The non-zero neutrino masses are generated through the Weinberg(-type) dimension-five operators written out schematically as: $\int d^4x\int d^2\theta e^{\hat{K}/2}\times\left({\cal Z}^2{\cal A}_1^2\in\frac{\partial^2{\cal Z}^4}{\partial {\cal Z}^2}{\cal A}_1^2\right)$,  and is given as:
$m_{\nu}={v^2 sin^2\beta \hat{{\cal O}}_{{\cal Z}_i{\cal Z}_j{\cal Z}_k{\cal Z}_l}}/{2M_p}
$ where $\hat{{\cal O}}_{{\cal Z}_i{\cal Z}_i{\cal Z}_i{\cal Z}_i}\equiv$coefficient of the physical/normalized quartic in  ${\cal Z}_i$ in the superpotential,  and is given as
$\hat{{\cal O}}_{{\cal Z}_i{\cal Z}_i{\cal Z}_i{\cal Z}_i}={\frac{{e^\frac{\hat{K}}{2}}{\cal O}_{{\cal Z}_i{\cal Z}_j{\cal Z}_k{\cal Z}_l}}{{\sqrt{\hat{K}_{{\cal Z}_i{\bar{\cal Z}}_{\bar i}}\hat{K}_{{\cal Z}_j{\bar{\cal Z}}_{\bar j}}\hat{K}_{{\cal Z}_k{\bar{\cal Z}}_{\bar k}}\hat{K}_{{\cal Z}_l{\bar{\cal Z}}_{\bar l}}}}}}$ \cite{conlon_neutrino},
$v sin\beta\equiv\langle H_u\rangle$ and $sin\beta$ is defined via
$tan\beta={\langle H_u\rangle}/{\langle H_d\rangle}$; in our setup (See \cite{ferm_masses_MS}):
 \begin{eqnarray}
 {\cal O}_{{\cal Z}_i{\cal Z}_j{\cal Z}_k{\cal Z}_l}&\sim& \frac{2^{n^s}}{24}10^2\left(\mu_3 n^s (2\pi \alpha^\prime)^2\right)^4{\cal V}^{\frac{n^s}{2}+\frac{1}{9}}
e^{-n^s{\rm vol}(\Sigma_s)+in^s\mu_3(2\pi\alpha^\prime)^2{\cal V}^{\frac{1}{18}}(\alpha+i\beta)}
\end{eqnarray}
 Now, $z_i\sim\alpha_i{\cal V}^{\frac{1}{36}}, i=1,2; \beta\sim\alpha_1\alpha_2$ and
${\rm vol}(\Sigma_S)=\gamma_3 ln {\cal V}$ such that
$\gamma_3 ln{\cal V} + \mu_3l^2\beta {\cal V}^{\frac{1}{18}} = ln {\cal V}$, along with
$\hat{K}_{{\cal Z}_i{\bar{\cal Z}}_{\bar i}}\sim\frac{{\cal V}^{\frac{1}{72}}}
{\sqrt{\sum_\beta n^0_\beta}}$, and the assumption that the holomorphic isometric involution $\sigma$ as part of
the Swiss-Cheese orientifolding action $(-)^{F_L}\Omega\cdot \sigma$ is such that
$\sum_\beta n^0_\beta\sim\frac{{\cal V}}{{\cal O}(1)}$. By analying the RG running of coefficient $\kappa_{ij}$ of dimension-five operator $\kappa_{ij}L_iH.L_jH$ and $\langle H_u\rangle$, it was shown in \cite{ferm_masses_MS} that one can generate a neutrino mass of $\stackrel{<}{\sim}1eV$ in our setup.

\section{Realizing Large Volume $\mu$-Split SUSY}

In this section, we show that the eigenvalues of the Higgs mass matrix at the EW scale obtained from the solutions to the one-loop RG flow equations assuming non-universality in the open string moduli masses, results in an eigenvalue corresponding to the mass-squared of one of the Higgs doublet to be negative and small and the other to be large and positive with a heavy Higgsino (in addition to heavy squarks/sleptons and light quarks/leptons already demonstrated in \cite{D3_D7_Misra_Shukla,Sparticles_Misra_Shukla,ferm_masses_MS}) implying the existence of $D3/D7$ $\mu$-Split LVS.

The supersymmetric extension of SM constrains all soft terms (i.e ${\hat{\mu}B}$, ${m_{Z_1}^2}$, ${m_{Z_2}^2}$) appearing in the Higgs sector superpotential to be of the order of SUSY breaking scale. In case of split supersymmetry scenario, SUSY breaking scale is high. However, in order to get one light Higgs doublet at EW scale in this scenario, one needs these soft terms to be of $TeV$ order. Since fine tuning is allowed one can assume ${{\hat\mu}B}\sim m_{{\cal A}_I}^2$ (where $m_{{\cal A}_I}$'s correspond to squark/slepton masses scale which is of the order of high supersymmetry breaking scale as in case of split SUSY, and $\hat{\mu}_{{\cal Z}_1{\cal Z}_2}$ is the Higgsino mass parameter). As Higgsino mass contribution ($\hat{\mu}_{{\cal Z}_1{\cal Z}_2}$ parameter) is small in most of split SUSY models, one needs $B >>\hat{\mu}_{{\cal Z}_1{\cal Z}_2}$ in order to have $\hat\mu{B}\sim{m^2_{{\cal A}_I}}$. In an alternate approach to split SUSY scenario called ``$\mu$-split SUSY scenario"  \cite{mu split susy}, according to which one can assume even $\hat\mu\sim {m_{{\cal A}_I}}\sim B$ i.e large $\mu$ parameter to get ${{\hat\mu}B}\sim{m^2_{{\cal A}_I}}$, this choice appears more natural and also helps to alleviate the $``{\mu}$ problem"; see also \cite{Split SUSY mu}. In the LVS set up discussed earlier, values of $\hat\mu$ and B terms pertaining to SUSY breaking parameters has been summarized in results in\cite{D3_D7_Misra_Shukla}, which are of the order ${{\hat\mu}^2}\sim{\hat\mu{B}}\sim{m^2_{{\cal A}_I}}$ (scalar masses) i.e ${\hat\mu}\sim{B}\sim{m_{{\cal A}_I}}$ as in case of $\mu$ split SUSY. Henceforth we are seeking for $\mu$ split SUSY scenario in the context of LVS.

 The Higgs masses after soft supersymmetry breaking is given by $(m_{Z_i}^{2}+\hat{\mu}_{Z_i}^{2})^{1/2}$ (where $m_z$'s correspond to  mobile $D3$- Brane position moduli masses (to be identified with soft Higgs scalar mass parameter)) and the Higgsino mass is given by  $\hat{\mu}_{Z_i}$. Had the supersymmetry been unbroken, Higgs(sino) masses would have had been degenerate with cofficient $\hat{\mu}_{Z_i}$. Nevertheless we have defined SUSY breaking but we are still justified to use RG flow equation' solutions because $\hat{\mu}_{Z_i}>>m_{Z_i}$.
 However, due to lack of universality in moduli masses but universality in trilinear $A_{ijk}$ couplings, we need to use solution of RG flow equation for moduli masses as given in \cite{Nath+Arnowitt}.
\begin{equation}
\label{eq:heavy_H_I}
m_{{\cal Z}_{1}}^{2}(t)=m_{o}^{2}(1+\delta_1)+m_{1/2}^{2} g(t)+\frac{3}{5}S_0p,
\end{equation}
where
\begin{eqnarray}
\label{eq:S_0_def}
& & S_0=Tr(Ym^2)=m_{{\cal Z}_2}^2-m_{{\cal Z}_1}^2+\sum_{i=1}^{n_g}(m_{\tilde q_L}^2-2
m_{\tilde u_R}^2 +m_{\tilde d_R}^2 - m_{\tilde l_L}^2 + m_{\tilde e_R}^2)
\end{eqnarray}
in which all the masses are at the string scale and
 $n_g$ is the number of generations. $p$ is defined by
$p=\frac{5}{66}[1-(\frac{\tilde\alpha_1(t)}
{\tilde\alpha_1(M_s)})]$
where  $\tilde\alpha_1\equiv g_1^2/(4\pi)^2$ and  $g_1$ is the
$U(1)_Y$ gauge coupling constant. Further,
\begin{equation}
\label{eq:light_H_I}
 m_{{\cal Z}_{2}}^{2}(t) = m_0^2\Delta_{{\cal Z}_2} + m_{1/2}^{2}e(t) + A_{o}m_{1/2}f(t) + m_{o}^{2}h(t) -k(t)A_{o}^{2} - \frac{3}{5}S_0p
\end{equation}
where $\Delta_{{\cal Z}_2}$ is given by
\begin{eqnarray}
\label{eq:Delta_def}
\Delta_{{\cal Z}_2}=\frac{(D_0-1)}{2}(\delta_2+\delta_3+\delta_4)+ \delta_2;  D_0=1-6 {\cal Y}_t \frac{F(t)}{E(t)}
\end{eqnarray}
Here ${\cal Y}_t\equiv\hat{Y}_t^2(M_s)/(4\pi)^2$ where $\hat{Y}_t(M_s)$ is the physical top Yukawa coupling at the string scale which following \cite{Ibanez_et_al} will be set to 0.08, and
\begin{equation}
\label{eq:E_def}
E(t)=(1+\beta_3t)^{\frac{16}{3b_3}}(1+\beta_2t)^{\frac{3}{b_2}}
(1+\beta_1t)^{\frac{13}{9b_1}}
\end{equation}
In equation (\ref{eq:E_def})  $\beta_i\equiv\alpha_i(M_s)b_i/4\pi$
($\alpha_1=(5/3)\alpha_Y$), $b_i$ are
the one loop
beta function coefficients defined  by $(b_1,b_2,b_3)=
(33/5,1,-3)$,
and $F(t)=\int_0^t E(t)dt$.

The gauge couplings in 2HDM/(MS)SM, up to one-loop, obey the following equation:
\begin{equation}
\label{eq:g_i_1leqn}
16\pi^2\frac{dg_i}{dt}=b_ig_i^3,
\end{equation}
whose solution for $i\equiv U(1)$ is:
\begin{equation}
\label{eq:g_1_1lsoln}
\frac{1}{g_1^2(M_{EW})}=\frac{1}{g_1^2(M_s)}+\frac{33}{40\pi^2}ln\left(\frac{M_s^2}{M_{EW}^2}\right).
\end{equation}
For $M_s\sim 10^{15}GeV$ and $M_{EW}\sim 500 GeV$, the second term on the right hand side of of                                 (\ref{eq:g_1_1lsoln}) is $4.7$, and using $g_1^2(M_{EW})\sim\frac{4\pi}{100}$ yields:
\begin{equation}
\label{eq:g_1_squared_I}
1-\frac{g_1^2(M_{EW})}{g_1^2(M_S)}\sim\frac{19\pi}{100}.
\end{equation}
In other words $g_1^2(M_{S})\sim\frac{12\pi}{100}\approx0.4$. In the dilute flux approximation,
$g_1^2(M_S)=g_2^2(M_S)=g_3^2(M_S)$. To ensure $E(t)\in\bf{R}$, the $SU(3)$-valued $1+\beta_3t>0$, which for $t=57$ \emph{(justified in appendix {A})} implies that $g_3^2(M_s)<\frac{(4\pi)^2}{3\times57}\sim{\cal O}(1)$. Hence, the above choice of $g_3^2(M_s)=0.4$ is fine. However, this implies that the Wilson line modulus can not yield a neutrino mass scale of $\leq{\cal O}(eV)$ as argued in \cite{ferm_masses_MS} which requires an ${\cal O}(1)$ $g_i(M_s)$ (to ensure that a Planckian Higgs vev RG flows - to one loop - to $246GeV$).

From (\ref{eq:heavy_H_I}), the appendix and \cite{D3_D7_Misra_Shukla}, one sees that:
\begin{equation}
\label{eq:heavy_H_II}
\hskip-0.1in m_{{\cal Z}_1}^2(M_{EW})\sim m_{{\cal Z}_1}^2(M_s)+{(0.39)}m_{3/2}^2+\frac{1}{22}\times\frac{19\pi}{100}\times S_0,
\end{equation}
and
\begin{equation}
\label{eq:light_H_II}
m_{{\cal Z}_2}^2(M_{EW})\sim m_0^2\delta_2+{(0.32)}m_{3/2}^2+{(-0.03)}n^s\hat{\mu}_{{\cal Z}_1{\cal Z}_2}m_{3/2}+{(0.96)} m_0^2-{(0.01)}(n^s)^2\hat{\mu}_{{\cal Z}_1{\cal Z}_2}-\frac{19\pi}{2200}\times S_0,
\end{equation}
where we used $A_{{\cal Z}_i{\cal Z}_i{\cal Z}_i}\sim n^s\hat{\mu}_{{\cal Z}_1{\cal Z}_2}$ (See \cite{D3_D7_Misra_Shukla}).
The solution for RG flow equation for $\hat{\mu}^2$ to one loop order is given by \cite{Nath+Arnowitt}:
\begin{eqnarray}
\label{eq:muhat_I}
& & \hat{\mu}^2_{{\cal Z}_i{\cal Z}_i}=-\biggl[m_0^2 C_1+A_0^2 C_2 +m_{\frac{1}{2}}^2C_3+m_{\frac{1}{2}}
A_0C_4-\frac{1}{2}M_Z^2 +\frac{19\pi}{2200}\left(\frac{tan^2\beta+1}{tan^2\beta-1}\right)S_0\biggr],
\end{eqnarray}
wherein
\begin{eqnarray}
\label{eq:defs_muhat_1LRG}
& & C_1=\frac{1}{tan^2\beta-1}(1-\frac{3 D_0-1}{2}tan^2\beta) +
\frac{1}{tan^2\beta-1}\left(\delta_1-\delta_2tan^2\beta-\frac{ D_0-1}{2}(\delta_2 +
\delta_3+\delta_4)tan^2\beta\right);\nonumber\\
 & & C_2=-\frac{tan^2\beta}{tan^2\beta-1}k(t); C_3=-\frac{1}{tan^2\beta-1}\left(g(t)- tan^2\beta e(t)\right);~C_4=-\frac{tan^2\beta}{tan^2\beta-1}f(t),
\end{eqnarray}
and where the functions $e(t),f(t),g(t),k(t)$ are as defined
in the appendix. The overall minus sign on the right hand side of (\ref{eq:muhat_I}) indicates that our $\hat{\mu}_{{\cal Z}_1{\cal Z}_2}^2$ is negative of $\mu^2$ of \cite{Nath+Arnowitt}. In the large $tan\beta$ (but less than 50)-limit and assuming $\delta_1=\delta_2=0$, one sees that:
\begin{equation}
\label{eq:muhat_II}
\hskip -0.3in\hat{\mu}_{{\cal Z}_1{\cal Z}_2}^2\sim-\biggl[\left(\frac{1}{2}+\frac{{\cal O}({10}^3)}{2}\right)m_0^2-{(0.01)}(n^s)^2\hat{\mu}_{{\cal Z}_1{\cal Z}_2}^2+{(0.32)}m_{3/2}^2-1/2 M_{EW}^2+{(0.03)}n^s\hat{\mu}_{{\cal Z}_1{\cal Z}_2}m_{3/2}+\frac{19\pi}{2200}S_0\biggr].
\end{equation}
From (\ref{eq:light_H_II}) and (\ref{eq:muhat_II}) one therefore sees that the mass-squared of one of the two Higgs doublets, $m_{H_2}^2$, at the $EW$ scale is given by:
\begin{equation}
\label{eq:muhat_III}
\hskip -0.3in m^2_{H_2}=m^2_{{\cal Z}_2}+\hat{\mu}^2_{{\cal Z}_i{\cal Z}_i}= \left(\left(-\frac{1}{2}-\frac{{\cal O}({10}^3}{2}\right)m^2_{0}-{(0.06)}n^s\hat{\mu}_{{\cal Z}_1{\cal Z}_2}m_{3/2}\right)+\frac{1}{2}M_{EW}^2-\frac{19\pi}{1100}S_0.
\end{equation}
From \cite{D3_D7_Misra_Shukla}, we notice:
\begin{equation}
\label{eq:mu_m_Z_sq}
\hat{\mu}_{{\cal Z}_1{\cal Z}_2}m_{3/2}\sim m_{{\cal Z}_i}^2,
\end{equation}
using which in (\ref{eq:muhat_III}), one sees that for an ${\cal O}(1)\ n^s$,
\begin{equation}
\label{eq:muhat_IV}
m^2_{H_2}(M_{EW})\sim\frac{1}{2}M_{EW}^2-\frac{19\pi}{1100}S_0-\frac{{\cal O}({10}^3)}{2}{\cal V}m_{3/2}^2.
\end{equation}
We have assumed at $m_{{\cal Z}_1}(M_s)=m_{{\cal Z}_2}(M_s)$ (implying $\delta_1=\delta_2=0$ but $\delta_{3,4}\neq0$). So, $S_0\approx m_{\rm squark/slepton}^2$, which in our setup could be of ${\cal O}(\hat{\mu}^2)$. Further,
\begin{equation}
\label{eq:heavy_Higgs}
m^2_{H_1}(M_{EW})=\left(m^2_{{\cal Z}_1}+\hat{\mu}_{{\cal Z}_1{\cal Z}_2}^2\right)(M_{EW})
 \sim m^2_{{\cal Z}_1}(M_s)+\frac{1}{2}M_{EW}^2+(0.01)(n^s)^2{\cal V}^2m_{3/2}^2.
\end{equation}
In the results on Soft SUSY Parameters summarized in section {\bf 2}, one finds that $\hat{\mu}B\sim\hat{\mu}^2$ at the string scale. By assuming the same to be valid at the string and EW scales, the Higgs mass matrix at the $EW$-scale can thus be expressed as:
\begin{eqnarray}
\label{eq:Higss_mass_matrix}
& & \left(\begin{array}{cc}
m^2_{H_1} & \hat{\mu}B\\
\hat{\mu}B & m^2_{H_2}\end{array}\right) \sim\left(\begin{array}{cc}
m^2_{H_1} & \xi\hat{\mu}^2\\
\xi\hat{\mu}^2 & m^2_{H_2}
\end{array}\right).
\end{eqnarray}
The eigenvalues are given by:
\begin{eqnarray}
\label{eq:eigenvalues}
& & \frac{1}{2}\biggl(m^2_{H_1}+m^2_{H_2}\pm\sqrt{\left(m^2_{H_1}-m^2_{H_1}\right)^2+4\xi^2\hat{\mu}^4}\biggr).
\end{eqnarray}
As (for ${\cal O}(1)\ n^s$)
\begin{eqnarray}
\label{eq:evs_1}
& & m^2_{H_1}+m^2_{H_2}\sim0.01{\cal V}^2m^2_{3/2}-0.06S_0+...,\nonumber\\
& & m^2_{H_1}-m^2_{H_2}\sim0.01{\cal V}^2m^2_{3/2}+0.06S_0+...,\nonumber\\
& & \hat{\mu}_{{\cal Z}_1{\cal Z}_2}^2\sim0.01{\cal V}^2m^2_{3/2}-0.03S_0+...,
\end{eqnarray}
one sees that the eigenvalues are:
\begin{eqnarray}
\label{eq:evs_2}
& & 0.01{\cal V}^2m^2_{3/2}-0.06S_0+...\pm\sqrt{\left(0.01{\cal V}^2m^2_{3/2}+0.06S_0+...\right)^2+\xi^2\left(0.02{\cal V}^2m^2_{3/2}-0.06S_0\right)^2}.\nonumber\\
& &
\end{eqnarray}
Hence, assuming a universality w.r.t. to the $D3$-brane position moduli masses ($m_{Z_{1,2}}$) and lack of the same for the squark/slepton masses, if $S_0$ and $\xi$ are fine tuned as follows:
\begin{equation}
\label{eq:evs_3}
0.01{\cal V}^2m^2_{3/2}\sim-0.06S_0\ {\rm and}\ \xi\sim\frac{2}{3}+\frac{{\cal O}(10)}{{\cal V}^2}\left(\frac{m^2_{EW}}{m^2_{3/2}}\right),
\end{equation}
one sees that one obtains one light Higgs doublet (corresponding to the negative sign of the square root)
and one heavy Higgs doublet (corresponding to the positive sign of the square root). Note, however, the squared Higgsino mass parameter $\hat{\mu}_{{\cal Z}_1{\cal Z}_2}$ then turns out to be heavy with a value, at the EW scale of around $0.01{\cal V}m_{3/2}$  i.e to the order of squark/slepton mass squared scale which is possible in case of $\mu$ split SUSY scenario discussed above.  This shows the possibility of realizing  $\mu$ split SUSY scenario in the context of LVS phenomenology named as large volume ``$\mu$-split SUSY" scenario.
\section{Gluino Decay}

Another important phenomenological implication of $\mu$ split SUSY scenario is based on longevity of gluinos, since the squarks which mediate its decay are ultra-heavy. The absence of the scalars at the TeV scale affects both the production and decays of the gauginos, However in this paper, we will focus only on decay width as well as life time calculations of gluino decay. Since scalar superpartner masses are  heavier than gluino, tree-level two-body decays of gluino $\tilde{g}\rightarrow\tilde{q}q$ are forbidden and hence one considers kinematically allowed three body decays.

\subsection{$\tilde{g}\rightarrow q{\bar q}\chi_n$}

 We first discuss gluino three-body decays that involve the process like $\tilde{g}\rightarrow q{\bar q}\chi_n$ - $\tilde{g}$ being a gaugino, $q/{\bar q}$ being quark/anti-quark and $\chi_n$ being a neutralino. More specifically, e.g., the gluino decays into an anti-quark and an off-shell squark and the off-shell squark decays into a quark and neutralino.

 \begin{center}
\begin{picture}(500,180)(50,0)
\Text(90,130)[]{$\tilde{g}$}
\Line(60,120)(110,120)
\Gluon(60,120)(110,120){5}{4}
\ArrowLine (140,150)(110,120)
\Text(147,150)[]{$\bar{q}_I$}
\DashArrowLine (110,120)(130,90){4}
\Text(107,100)[]{$\tilde{q}_J$}
\ArrowLine(130,90)(160,120)
\Text(167,120)[]{$\tilde{\chi}_3^0$\hskip 1.5cm +}
\ArrowLine(130,90)(160,60)
\Text(167,65)[]{${q_K}$}
\Text(110,30)[]{(a)}
\Text(250,130)[]{$\tilde{g}$}
\Line(210,120)(260,120)
\Gluon(210,120)(260,120){5}{4}
\ArrowLine (290,150)(260,120)
\Text(297,150)[]{$q_K$}
\DashArrowLine(260,120)(280,90){4}{}
\Text(257,100)[]{$\tilde{q}_J$}
\ArrowLine(280,90)(310,120)
\Text(320,120)[]{$\bar{q}_I$}
\ArrowLine(280,90)(310,60)
\Text(317,65)[]{$\tilde{\chi}_3^0$}
\Text(280,30)[]{(b)}
\end{picture}
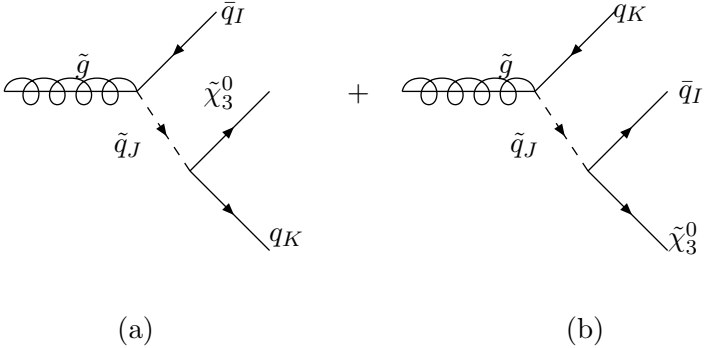
{\sl Figure 1: Three-body gluino decay diagrams}
\end{center}
 Using the two Wilson line moduli of section {\bf IV}, the superpotential can then be written as:
\begin{equation}
\label{eq:W}
W\sim \left(1 + z_1^{18} + z_2^{18} + z_3^2 - 3\phi_0z_1^6z_2^6\right)^{n_s}\frac{e^{in^s\left({\cal V}^{7/6}a_1^2+{\cal V}^{2/3}a_1a_2+{\cal V}^{1/6}a_2^2+\mu_3(2\pi\alpha^\prime)^2(z_1z_2+z_1^2+z_2^2)\right)}}{{\cal V}^{n^s}}.
\end{equation}
We need a gaugino-quark-squark vertex and a neutralino-quark-squark vertex. For the former, the basic idea is to generate a term of the type ${\bar Q}_L\tilde{q}_R\lambda_{\tilde{g}}H_L$ wherein $Q_L$ and $H_L$ are respectively the $SU(2)_L$ quark and Higgs doublets, $\tilde{q}_R$ is an $SU(2)_L$ singlet and $\lambda_{\tilde{g}}$ is the gluino. After spontaneous breaking of the EW symmetry when $H^0$ in $H_L$ acquires a non-zero vev $\langle H^0\rangle$, this term generates:
$\langle H^0\rangle{\bar Q}_L\tilde{q}_R\lambda_{\tilde{g}}$.  From \cite{Wess_Bagger}, the first vertex arises from the following term in the fermionic sector of the ${\cal N}=1$ gauged supergravity action:
\begin{equation}
\label{eq:gaugino+quark+squark}
g_{YM}g_{\alpha {\bar J}}X^\alpha{\bar\chi}^{\bar J}\lambda_{\tilde{g}},
\end{equation}
where $X^\alpha$ corresponds to the components of a killing isometry vector. From \cite{jockersetal}, one notes that $X^\alpha=-6i\kappa_4^2\mu_7Q_\alpha$, where $\alpha=S/B, Q_\alpha=2\pi\alpha^\prime\int_{\Sigma_B}i^*\omega_\alpha\wedge P_-\tilde{f}$ where $P_-$ is a harmonic zero-form on $\Sigma_B$ taking value +1 on $\Sigma_B$ and $-1$ on $\sigma(\Sigma_B)$ - $\sigma$ being a holomorphic isometric involution as part of the Calabi-Yau orientifold - and $\tilde{f}\in\tilde{H}^2_-(\Sigma^B)\equiv{\rm coker}\left(H^2_-(CY_3)\stackrel{i^*}{\rightarrow}H^2_-(\Sigma^B)\right)$. Writing the K\"{a}hler potential as:
\begin{eqnarray}
\label{eq:K}
& & \frac{K}{M_p^2}\sim-2 ln\biggl[\left(\frac{\sigma_B}{M_p}-{\cal V}^{2/3+1/2}\frac{|a_1|^2}{M_p^2}+{\cal V}^{2/3}\frac{\left(a_1{\bar a}_2+h.c.\right)}{M_p^2}+{\cal V}^{1/6}\frac{|a_2|^2}{M_p^2}+\mu_3{\cal V}^{\frac{1}{18}}\right)^{3/2}\nonumber\\
& & -\left(\frac{\sigma_S}{M_p}++\mu_3{\cal V}^{\frac{1}{18}}\right)^{3/2}+\sum n^0_\beta(...)\biggr],
\end{eqnarray}
and
\begin{eqnarray}
\label{eq:gYM_ReT}
& & g_{YM}\sim \left( Re(T_B)\right)^{-\frac{1}{2}}\nonumber\\
& & \sim\left(\frac{\sigma_B+{\bar\sigma}_B}{M_p}-{\cal V}^{\frac{7}{6}}\frac{|a_1|^2}{M_p^2}+{\cal V}^{2/3}\frac{\left(a_1{\bar a}_2+h.c.\right)}{M_p^2}+{\cal V}^{1/6}\frac{|a_2|^2}{M_p^2}+\mu_3{\cal V}^{\frac{1}{18}}+\mu_3(2\pi\alpha^\prime)^2\frac{\{|z_1|^2+|z_2|^2+z_1{\bar z}_2+z_2{\bar z}_1\}}{M_p^2}\right)^{-\frac{1}{2}}\nonumber\\
& &
\end{eqnarray}
where $\sigma_{B,S}$ are the $B/S$ divisors' volumes, and considering fluctuations:
$a_{1,2}\rightarrow a_{1,2}+{\cal V}^{-\frac{1}{4}}M_p$, the fluctuation in  $g_{YM}g_{\alpha {\bar a}_I}$
linear in (the fluctuation) $a_1$ (for concreteness; one can similarly work out fluctuations in $a_2$) as well as linear in $z_i$ can be shown to be given by (See appendix C for details):
\begin{equation}
\label{eq:gYMgBIbar_zi_aI}
g_{YM}g_{B {\bar a}_I}\rightarrow-{\cal V}^{\frac{13}{36}}z_ia_1\delta^1_I + {\cal V}^{-\frac{5}{36}}z_ia_1\delta^2_I,
\end{equation}
which for $z_i\rightarrow\langle z_i\rangle\sim V^{\frac{1}{36}}$ (in $M_p=1$ units) yields:
$-{\cal V}^{\frac{7}{18}}a_1\delta^1_I+{\cal V}^{-\frac{1}{9}}\delta^2_I$.

The dominant contribution to the physical gluino-quark-squark vertex
is proportional to
\begin{eqnarray}
\label{eq:gluino-quark-squark}
& & \frac{{\cal V}^{-1}{\cal V}^{\frac{1}{3}}f\left(-{\cal V}^{\frac{7}{18}}\ {\rm or}\ {\cal V}^{-\frac{1}{9}}\right)}{\left(\sqrt{\hat{K}_{{\cal A}_1{\bar {\cal A}}_1}}\right)^2} \sim \tilde{f}\left({\cal V}^{-\frac{37}{36}}\ {\rm or}\ {\cal V}^{-\frac{59}{36}}\right)
\end{eqnarray}
where $Q_B\sim{\cal V}^{\frac{1}{3}}f\left(2\pi\alpha^\prime\right)^2M_p$.\footnote{A small note on dimensional analysis: $\kappa_4^2\mu_7\left(2\pi\alpha^\prime\right)Q_Bg_{\sigma^B{\bar a}_{\bar j}}{\bar\chi}^{\bar j}\lambda$ has dimensions $M_p^4$ - utilizing $\kappa_4^2\mu_7\sim{\cal V}^{-1}\left(2\pi\alpha^\prime\right)^{-3}$, one sees that $Q_B$ has dimensions of $\left(2\pi\alpha^\prime\right)^2M_p$. Using the definition of $Q_B$, we will estimate $Q_B$ by
${\cal V}^{\frac{1}{3}}\left(2\pi\alpha^\prime\right)^2$, where the integral of $i^*\omega_B\wedge P_-\tilde{f}$ over $\Sigma_B$ is approximated by integrals of $i^*\omega_B$ and $P_-\tilde{f}$ over two-cycles non-trivial in the cohomology of $\Sigma_B$ and estimated/parametrized as ${\cal V}^{\frac{1}{3}}\left(2\pi\alpha^\prime\right)$ and $f$ respectively.}

The terms in the ${\cal N}=1$ supergravity action \cite{Wess_Bagger} relevant to the Neutralino/Higgsino-quark-squark vertex are:
\begin{equation}
\label{eq:higgsino_quark_squark-1}
i \sqrt{g}g_{I{\bar j}}{\bar\chi}^{\bar j}{\bar\sigma}^\mu\bigtriangledown_\mu\chi^I+\frac{e^{\frac{K}{2}}}{2}\left({\cal D}_iD_jW\right)\chi^i\chi^J+{\rm h.c.},
\end{equation}
where, disregarding the contribution from the gauge fields supported on the $D7$-brane( stacks) as the vertex we are interested in has no gauge fields associated with it,
\begin{eqnarray}
\label{eq:higgsino_quark_squark-2}
 \bigtriangledown_\mu\chi^{I,\alpha}&=&\partial_\mu\chi^{I,\alpha}+\chi^{I,\beta}\omega_\mu^\alpha\ _\beta
+ \Gamma^I_{JK}\partial_\mu a^J\chi^{K,\alpha} +\Gamma^I_{jK}\partial_\mu z^j\chi^K -\frac{1}{4}\left(\partial_JK\partial_\mu a^J-{\rm c.c.}\right)\chi^{I,\alpha},
\end{eqnarray}
wherein $\omega_\mu^\alpha\ _\beta=\delta^\alpha_\nu\delta^\rho_\beta\omega_\mu^\nu\ _\rho=\delta^\alpha_\nu \delta^\rho_\beta g^{\nu\lambda}\omega_{\mu\lambda\rho}$ and
\begin{eqnarray}
\label{eq:higgsino_quark_squark-3}
 \omega_{\mu\lambda\rho}& = & \frac{1}{2}\Biggl[-\frac{i}{2}e_{\rho a}\left(\psi_\lambda\sigma^a{\bar\psi}_\mu-\psi_\mu\sigma^a{\bar\psi}_\lambda\right)-\frac{i}{2}\left(
\psi_\mu\sigma^a{\bar\psi}_\rho-\psi_\rho\sigma^a{\bar\psi}_\mu\right) +\frac{i}{2}\left(\psi_\rho\sigma^a{\bar\psi}_\lambda-\psi_\lambda\sigma^a{\bar\psi}_\rho\right)
-e_{\rho a}\left(\partial_\mu e_\lambda\ ^a\right)\nonumber\\
& & -e_{\lambda a}\left(\partial_\rho e_\mu\ ^a-\partial_\mu e_\rho\ ^a\right)+e_{\mu a}\left(\partial_\lambda e_\rho\ ^a - \partial_\rho e_\lambda\ ^a\right)\Biggr].
\end{eqnarray}
Now, in generic type IIB orientifold compactifications with three-form fluxes \cite{GKP}, taking a warped metric ansatz:
\begin{equation}
\label{eq:warped_metric}
ds^2=e^{2A(y)}ds^2_{{\bf R}^{1,3}}+e^{-2A(y)}\tilde{g}_{mn}dy^mdy^n,
\end{equation}
the warp factor $e^{2A(y)}$ satisfies:
\begin{equation}
\label{eq:warp-eom}
\tilde{\bigtriangledown}^2\left(e^{4A(y)}\right)=\frac{e^{2A(y)}|G_{mnp}|^2}{12{\rm Im}(\tau)}
+2e^{-6A(y)}\partial_m e^{4A(y)}\partial^m e^{4A(y)}+2\kappa_{10}^2e^{2A(y)}\mu_3\rho_3^{\rm loc},
\end{equation}
$\rho_3^{\rm loc}$ corresponding to the localized $D3$-brane charge density corresponding to $D7$-branes wrapping $\Sigma_B$. Consider now a scaling of the unwarped metric: $\tilde{g}_{mn}\rightarrow\lambda^2\tilde{g}_{mn}$. The left hand side of (\ref{eq:warp-eom}) scales like
$\lambda^{-2}$, the first term on the right hand side scales like $\lambda^{-6}$ and the second term scales like $\lambda^{-2}$. Further, the five-form Bianchi identity becomes:
\begin{equation}
\label{eq:Bianchi_F_5}
d\tilde{F}_5=\left(2\pi\right)^4\left(\alpha^\prime\right)^2\rho^{\rm loc}_3dV_\perp+H_3\wedge F_3,
\end{equation}
where
\begin{eqnarray}
\label{eq:F_5_tilde}
& & \tilde{F}_5=F_5-\frac{1}{2}C_2\wedge H_3+\frac{1}{2}B_2\wedge F_3\nonumber\\
& & =\left(1+*_{10}\right)d\left(e^{4A(y)}\right)\wedge dx^0\wedge dx^1\wedge dx^2\wedge dx^3.
\end{eqnarray}
In (\ref{eq:Bianchi_F_5}), $dV_\perp$ is the volume form transverse to $\Sigma_B$ implying that the last term on the right hand side of (\ref{eq:warp-eom}) will scale like $\lambda^{-2}$. Hence,
\begin{equation}
\label{eq:warp_scale}
e^{2A(y)}\sim 1 + \frac{1}{\lambda^4}.
\end{equation}
We would assume that
\begin{equation}
\label{eq:warp}
e^{2A(y)} = 1 + \frac{(\alpha^\prime)^2}{\left(\sqrt{\tilde{g}_{mn}y^my^n}\right)^4},
\end{equation}
implying that $\partial_\mu e_\rho\ ^a=0$ (as derivatives in (\ref{eq:higgsino_quark_squark-3}) are with respect to ${\bf R}^{1,3}$ coordinates).

\vspace*{0.5cm}
 \begin{center}
\begin{picture}(500,150)(50,0)
\Text(100,120)[]{$\tilde{q}_{a_1}$}
\DashArrowLine(110,120)(160,120){5}
\Text(210,152)[]{$\tilde{\chi}_3^0$}
\ArrowLine(160,120)(210,150)
\Text(210,88)[]{$q_{a_2}$}
\ArrowLine(160,120)(210,90)
\Text(160,50)[]{$(a)$}
\Text(290,120)[]{$\tilde{q}_{a_1}$}
\DashArrowLine(300,120)(350,120){5}
\Text(400,152)[]{$\tilde{\chi}_3^0$}
\ArrowLine(350,120)(390,150)
\Text(400,88)[]{$ q_{a_1} $}
\ArrowLine(350,120)(390,90)
\Text(350,50)[]{$(b)$}
\end{picture}
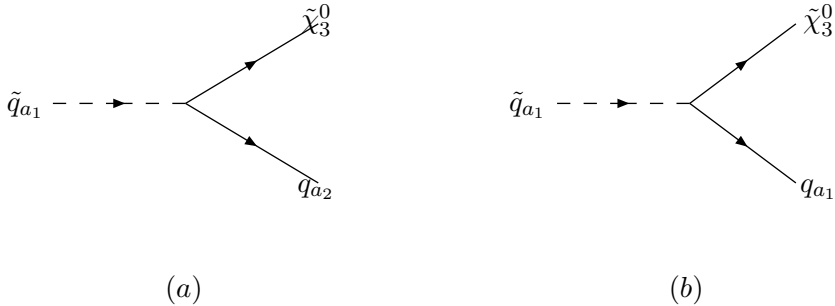
{\sl Figure 2: Two possibilities for Squark-Quark-Neutralino vertices considered}
\end{center}

For evaluation of the Higgsino-quark-squark vertex, consider:
\begin{eqnarray}
\label{eq:KaI_zi}
& & \hskip-0.5in\frac{K}{M_p^2}\sim-2 ln\biggl[\left({\cal V}^{\frac{2}{3}}-{\cal V}^{\frac{7}{6}}\frac{|a_1|^2}{M_p^2}+{\cal V}^{2/3}\frac{\left(a_1{\bar a}_2+h.c.\right)}{M_p^2}+{\cal V}^{1/6}\frac{|a_2|^2}{M_p^2}+\mu_3{\cal V}^{\frac{1}{18}}+\mu_3(2\pi\alpha^\prime)^2\frac{\{|z_1|^2+|z_2|^2+z_1{\bar z}_2+z_2{\bar z}_1\}}{M_p^2}\right)^{3/2}\nonumber\\
& & -\left({\cal V}^{\frac{1}{18}}+\mu_3(2\pi\alpha^\prime)^2\frac{\{|z_1|^2+|z_2|^2+z_1{\bar z}_2+z_2{\bar z}_1\}}{M_p^2}\right)^{3/2}+\sum n^0_\beta(...)\biggr],
\end{eqnarray}
and fluctuations of $g_{a_I{\bar z}_{\bar j}}=\partial_{a_I}{\bar\partial}_{{\bar z}_{{\bar j}}}K$ about $a_I\sim{\cal V}^{-\frac{1}{4}}$ at $z_1\rightarrow\langle z_1\rangle\sim{\cal V}^{\frac{1}{36}}$, $z_2\rightarrow\langle z_2\rangle\sim1.3{\cal V}^{\frac{1}{36}}$. A rigorous calculation would involve redoing the one-Wilson-line-modulus calculation of \cite{D3_D7_Misra_Shukla} for the two-Wilson-line-moduli case to diagonalize the matrix $\hat{K}_{C_i{\bar C}_{\bar j}}$; we will for this paper, be content with assuming that $\hat{K}_{{\cal A}_1{\bar {\cal A}}_1}\approx\hat{K}_{{\cal A}_2{\bar {\cal A}}_2}$ to get a rough estimate.

Now, to figure out the contribution to the squark-quark-neutralino vertex (Fig.2) from $i \sqrt{g}g_{i{\bar J}}{\bar\chi}^{\bar J}{\bar\sigma}^\mu\bigtriangledown_\mu\chi^i$, one needs to work the contribution from:
\begin{equation}
\label{eq:kin qqtildeneutralino}
ig_{i{\bar J}}{\bar\chi}^{\bar I}\left[{\bar\sigma}\cdot\partial\chi^i+\Gamma^i_{Lj}{\bar\sigma}\cdot\partial a^L\chi^j
+\frac{1}{4}\left(\partial_{a_1}K{\bar\sigma}\cdot a_1 - {\rm c.c.}\right)\chi^i\right].
\end{equation}
\begin{itemize}
\item
Utilizing that under fluctuations: $z_i\rightarrow z_i+{\cal V}^{\frac{1}{36}}, a_I\rightarrow a_I+{\cal V}^{-\frac{1}{4}}$ (See appendix C):
\begin{equation}
\label{eq:giJbar_fluctuation}
g_{i{\bar a}_1}\rightarrow {\cal V}^{\frac{11}{18}}a_1;\ g_{i{\bar a}_2}\rightarrow-{\cal V}^{\frac{1}{9}}a_1,
\end{equation}
one obtains a contribution $i\tilde{f}{\cal V}^{-\frac{7}{72}}{\bar\sigma}\cdot p_{\tilde{\chi}_3^0}$ for Fig.2(a) and $i\tilde{f}{\cal V}^{-\frac{43}{72}}{\bar\sigma}\cdot p_{\tilde{\chi}_3^0}$ for Fig.2(b).

\item
Using (\ref{eq:KaI_zi}) (in $M_p=1$ units), the fluctuation of the moduli space metric about $a_I\sim{\cal V}^{-\frac{1}{4}}$, i.e., $a_I\rightarrow {\cal V}^{-\frac{1}{4}} + a_I$, is given by:
\begin{eqnarray}
\label{eq:gAIzibar_expa1}
& & g_{A{\bar B}}=\left(\begin{array}{ccc} g_{a_1{\bar a}_{\bar 1}} & g_{a_1{\bar a}_{\bar 2}}
& g_{a_1{\bar z}_{\bar i}} \\
g_{a_2{\bar a}_{\bar 1}} & g_{a_2{\bar a}_{\bar 2}} & g_{a_2{\bar z}_{\bar i}}\\
g_{z_i{\bar a}_{\bar 1}} &g_{z_i{\bar a}_{\bar 2}} & g_{z_i{\bar z}_{\bar i}}\end{array}
\right)
\sim\left(\begin{array}{ccc}-{\cal V}^{\frac{3}{4}} - a_1{\cal V}^{\frac{3}{2}} & {\cal V}^{-\frac{7}{12}} + a_1{\cal V}& -{\cal V}^{-\frac{5}{36}} + a_1{\cal V}^{\frac{11}{18}}\\
{\cal V}^{-\frac{7}{12}} + a_1{\cal V} & {\cal V}^{-\frac{1}{4}} - a_1\sqrt{\cal V} &
-{\cal V}^{-\frac{23}{36}} - a_1{\cal V}^{\frac{1}{9}} \\
{\cal V}^{-\frac{5}{36}} + a_1{\cal V}^{\frac{11}{18}} & -{\cal V}^{-\frac{23}{36}} - a_1{\cal V}^{\frac{1}{9}} &
{\cal V}^{-\frac{11}{12}} + a_1{\cal V}^{-\frac{1}{6}}\end{array}\right),
\end{eqnarray}
implying:
\begin{eqnarray}
\label{eq:gAIzibar_inv}
& & g^{A{\bar B}}\sim\left(\begin{array}{ccc}-{\cal V}^{-\frac{3}{4}} + a_1 & {\cal V}^{-\frac{13}{36}} + a_1\sqrt{\cal V} &
{\cal V}^{\frac{1}{36}} + {\cal O}(a_1^2)\\
{\cal V}^{-\frac{13}{36}} + a_1\sqrt{\cal V} & {\cal V}^{\frac{1}{4}} + a_1{\cal V} & {\cal V}^{\frac{19}{36}} - a_1{\cal V}^{\frac{4}{9}}\\
{\cal V}^{\frac{1}{36}} + {\cal O}(a_1^2) & {\cal V}^{\frac{19}{36}} - a_1{\cal V}^{\frac{4}{9}} &
{\cal V}^{\frac{11}{12}} - a_1{\cal V}^{\frac{5}{3}} \end{array}\right).
\end{eqnarray}
Further:
\begin{eqnarray}
\label{eq:dzgaIzibar_expa1}
& & \partial_{z_i}g_{A{\bar B}}\sim\left(\begin{array}{ccc}
{\cal V}^{\frac{11}{6}} + a_1{\cal V}^{\frac{49}{36}} & -{\cal V}^{\frac{1}{9}} - a_1{\cal V}^{\frac{31}{36}}&
{\cal V}^{-\frac{1}{6}} + a_1{\cal V}^{\frac{7}{12}}\\
-{\cal V}^{\frac{1}{9}} - a_1{\cal V}^{\frac{31}{36}} & {\cal V}^{-\frac{7}{18}} + a_1{\cal V}^{\frac{13}{36}} &
-{\cal V}^{-\frac{2}{3}} + a_1{\cal V}^{\frac{1}{12}}\\
{\cal V}^{-\frac{1}{6}} + a_1{\cal V}^{\frac{7}{12}} & -{\cal V}^{-\frac{2}{3}} + a_1{\cal V}^{\frac{1}{12}} &
\frac{1}{\cal V} - a_1{\cal V}^{-\frac{11}{36}}
\end{array}
\right).
\end{eqnarray}
Using (\ref{eq:gAIzibar_expa1})-(\ref{eq:dzgaIzibar_expa1}),(\ref{eq:affine}), one sees that:
\begin{equation}
\label{eq:affine values}
\Gamma^{z_i}_{a_1z_j}\sim{\cal V}^{\frac{67}{36}};
\end{equation}
also one can show that:
\begin{equation}
\label{eq:giJbar}
g_{z_i{\bar a}_1}\sim{\cal V}^{-\frac{5}{36}},\ g_{z_i{\bar a}_2}\sim{\cal V}^{-\frac{23}{36}}.
\end{equation}
Utilizing (\ref{eq:affine values}) and (\ref{eq:giJbar}), one obtains the following contribution from $ig_{i{\bar J}}{\bar\chi}^{\bar J}\Gamma^i_{a_1j}{\bar\sigma}\cdot a_1\chi^i$ to the squark-quark-Neutralino vertex (Fig.2):
$i\tilde{f}{\cal V}^{\frac{73}{72}}\frac{{\bar\sigma}\cdot p_{\tilde{q}}}{M_p}$ for Fig.2(a) and $i\tilde{f}{\cal V}^{\frac{37}{72}}\frac{{\bar\sigma}\cdot p_{\tilde{q}}}{M_p}$ for Fig.2(b).

\item
Using (\ref{eq:giJbar}) and
\begin{equation}
\label{eq:dKa1}
\partial_{a_1}K\sim{\cal V}^0,
\end{equation}
one sees that the contribution from $\frac{1}{4}ig_{i{\bar J}}{\bar\chi}^{\bar I}\left(\partial_{a_1}K{\bar\sigma}\cdot a_1 - {\rm c.c.}\right)\chi^i$ is given by: $i\tilde{f}{\cal V}^{-\frac{61}{72}}\frac{{\bar\sigma}\cdot p_{\tilde{q}}}{M_p}$ for Fig.2(a)  and $i\tilde{f}{\cal V}^{-\frac{97}{72}}\frac{{\bar\sigma}\cdot p_{\tilde{q}}}{M_p}$ for Fig.2(b).
\end{itemize}

Putting everything together, one obtains the following contribution to Fig.2 from (\ref{eq:kin qqtildeneutralino}):
\begin{equation}
\label{eq:kinetic I}
\frac{i\tilde{f}{\bar\sigma}\cdot\left({\cal V}^{-\frac{7}{72}}\frac{p_{\tilde{\chi}_3^0}}{M_p}+{\cal V}^{\frac{73}{72}}\frac{p_{\tilde{q}}}{M_p}+{\cal V}^{-\frac{61}{72}}\frac{p_{\tilde{q}}}{M_p}\right)}{\sqrt{\hat{K}_{{\cal Z}_i{\bar{\cal Z}}_{\bar i}}}\left(\sqrt{\hat{K}_{{\cal A}_I{\bar{\cal A}}_{\bar I}}}\right)^2\sim {\cal V}^{\frac{1}{3}}}\sim i\tilde{f}{\bar\sigma}\cdot\left({\cal V}^{-\frac{31}{72}}\frac{p_{\tilde{\chi}_3^0}}{M_p}+{\cal V}^{\frac{2}{3}}\frac{p_{\tilde{q}}}{M_p}\right)
\end{equation}
 for Fig.2(a,) and
 \begin{equation}
 \label{eq:kinetic II}
 \frac{i\tilde{f}{\bar\sigma}\cdot\left({\cal V}^{-\frac{31}{72}}\frac{p_{\tilde{\chi}_3^0}}{M_p}+{\cal V}^{\frac{37}{72}}\frac{p_{\tilde{q}}}{M_p}+{\cal V}^{-\frac{97}{72}}\frac{p_{\tilde{q}}}{M_p}\right)}{\sqrt{\hat{K}_{{\cal Z}_i{\bar{\cal Z}}_{\bar i}}}\left(\sqrt{\hat{K}_{{\cal A}_I{\bar{\cal A}}_{\bar I}}}\right)^2\sim {\cal V}^{\frac{1}{3}}}\sim i\tilde{f}{\bar\sigma}\cdot\left({\cal V}^{-\frac{55}{72}}\frac{p_{\tilde{\chi}_3^0}}{M_p}+{\cal V}^{\frac{13}{72}}\frac{p_{\tilde{q}}}{M_p}\right)
 \end{equation}
  for Fig.2(b).

Also,
\begin{equation}
\label{eq:DdW_def}
{\cal D}_iD_JW=\partial_i\partial_JW + \left(\partial_i\partial_JK\right)W+\partial_iKD_JW +
\partial_JK\partial_iW - \left(\partial_iK\partial_JK\right)W - \Gamma^k_{iJ}D_kW - \Gamma^K_{iJ}D_KW;
\end{equation}
in our setup $\partial_JW=0$.
 As shown in appendix C, the fluctuation of $e^{\frac{K}{2}}{\cal D}_iD_JW$ with respect to $a_1$ about $a_I\sim{\cal V}^{-\frac{1}{4}}$ and at $z_1\rightarrow\langle z_1\rangle\sim{\cal V}^{\frac{1}{36}}$, $z_2\rightarrow\langle z_2\rangle\sim{\cal V}^{\frac{1}{36}}$ is given by:
\begin{itemize}
\item
\begin{eqnarray}
\label{eq:Higgsino_quark_sq_2}
& & e^{\frac{K}{2}}\left(\left(\partial_i\partial_{a_1}K\right)W+\partial_iKD_{a_1}W +
\partial_{a_1}K\partial_iW - \left(\partial_iK\partial_{a_1}K\right)W\right)\chi^i\chi^{a_1}\rightarrow {\cal V}^{-\frac{11}{9}}a_1\chi^i\chi^{a_1};\nonumber\\
& & e^{\frac{K}{2}}\left(\left(\partial_i\partial_{a_2}K\right)W+\partial_iKD_{a_2}W +
\partial_{a_2}K\partial_iW - \left(\partial_iK\partial_{a_2}K\right)W\right)\chi^i\chi^{a_2}\rightarrow -{\cal V}^{-\frac{31}{18}}a_1\chi^i\chi^{a_2}.
\end{eqnarray}

\item
using:
\begin{eqnarray}
\label{eq:exphalfKDW}
 e^{\frac{K}{2}}D_{z_i}W\sim-{\cal V}^{-\frac{71}{36}} - {\cal V}^{-\frac{71}{36}}a_1,  e^{\frac{K}{2}}D_{a_1}W\sim{\cal V}^{-2} - {\cal V}^{-\frac{5}{4}}a_1,  e^{\frac{K}{2}}D_{a_2}W\sim{\cal V}^{-\frac{5}{2}} - {\cal V}^{-\frac{7}{4}}a_1;
\end{eqnarray}
as well as:
\begin{eqnarray}
\label{eq:affine}
& & \Gamma^{z_i}_{z_ia_1}\sim{\cal V}^{\frac{67}{36}} + {\cal V}^{\frac{5}{9}}a_1, \Gamma^{z_i}_{z_ia_2}\sim-{\cal V}^{\frac{1}{4}}+{\cal V}a_1,  \Gamma^{a_1}_{z_ia_1}\sim-{\cal V}^{\frac{13}{12}} + {\cal V}^{\frac{11}{6}}a_1;\nonumber\\
& & \Gamma^{a_1}_{z_ia_2}\sim{\cal V}^{-\frac{3}{4}} + {\cal V}^{\frac{1}{9}}a_1, \Gamma^{a_2}_{z_ia_1}\sim {\cal V}^{\frac{53}{36}} + {\cal V}^{\frac{7}{3}}a_1, \Gamma^{a_2}_{z_ia_2}\sim - {\cal V}^{-\frac{1}{4}} + {\cal V}^{\frac{31}{36}}a_1;
\end{eqnarray}
one sees that:
\begin{equation}
\label{eq:affine_cont_DdW}
\frac{e^{\frac{K}{2}}}{2}\left(\Gamma^{z_i}_{z_ia_1}D_{z_i}W + \Gamma^{a_1}_{z_ia_1}D_{a_1}W + \Gamma^{a_2}_{z_ia_1}D_{a_2}W\right)\chi^i\chi^{a_1}\sim\left( - {\cal V}^{-\frac{1}{9}}a_1\right)
\chi^i\chi^{a_1},
\end{equation}
and
\begin{equation}
\label{eq:affine_cont_DdW}
\frac{e^{\frac{K}{2}}}{2}\left(\Gamma^{z_i}_{z_ia_2}D_{z_i}W + \Gamma^{a_1}_{z_ia_2}D_{a_1}W + \Gamma^{a_2}_{z_ia_2}D_{a_2}W\right)\chi^i\chi^{a_1}\sim\left( - {\cal V}^{-\frac{35}{36}}a_1\right)
\chi^i\chi^{a_2}.
\end{equation}
\end{itemize}
This implies that the contribution to the Higgsino-quark-squark vertex from $\frac{e^{\frac{K}{2}}}{2}\left({\cal D}_iD_JW\right)\chi^i\chi^J+{\rm h.c.}$ is:
\begin{eqnarray}
\label{eq:Higgsino-quark-squark-term_2}
& & \frac{\left({\cal V}^{-\frac{1}{9}}\ {\rm for}\ \chi^i\chi^{a_1},\ {\rm or}\ {\cal V}^{-\frac{35}{36}}\ {\rm for}\ \chi^i\chi^{a_2}\right)}{\sqrt{\hat{K}_{{\cal Z}_i{\bar{\cal Z}}_i}}\left(\sqrt{\hat{K}_{{\cal A}_1{\bar {\cal A}}_1}}\right)^2}\sim\frac{\left({\cal V}^{-\frac{1}{9}}\ {\rm or}\ {\cal V}^{-\frac{35}{36}}\right)}{{\cal V}^{\frac{1}{3}}}
\nonumber\\
& & \sim {\cal V}^{-\frac{4}{9}}\ {\rm or}\ {\cal V}^{-\frac{47}{36}}.
\end{eqnarray}
This implies that the contribution to the neutralino-squark-quark vertex from $\frac{e^{\frac{K}{2}}}{2}{\cal D}_iD_J\chi^i\chi^J$ will be given by:
\begin{equation}
\label{eq:chi3_q_sq}
\tilde{f}{\cal V}^{-\frac{51}{72}}\left({\cal V}^{-\frac{4}{9}}\ {\rm or}\ {\cal V}^{-\frac{47}{36}}\right)
\sim\tilde{f}\left({\cal V}^{-\frac{41}{36}}\ {\rm or}\ {\cal V}^{-\frac{145}{72}}\right).
\end{equation}

Hence, if $a_1$ is the required squark then the physical Higgsino-quark-squark vertex will be given by:
\begin{equation}
\label{eq:Higgsino_q-sq_a1}
 i\tilde{f}[{\bar\sigma}\cdot \left({\cal V}^{-\frac{31}{72}}\frac{p_{\tilde{\chi}_3^0}}{M_p}+{\cal V}^{\frac{2}{3}}\frac{p_{\tilde{q}}}{M_p})+ {\cal V}^{-\frac{41}{36}}\right]
\end{equation}
for Fig.2(a), and
\begin{equation}
\label{eq:Higgsino_q-sq_a2}
i\tilde{f}[\left[{\bar\sigma}\cdot \left({\cal V}^{-\frac{55}{72}}\frac{p_{\tilde{\chi}_3^0}}{M_p} + {\cal V}^{\frac{13}{72}}\frac{p_{\tilde{q}}}{M_p}\right)+ {\cal V}^{-\frac{145}{72}} \right ]
\end{equation}
for Fig.2(b). Including the contribution from $\lambda^0\rightarrow-\tilde{\chi}_3^0$ (essentially using (\ref{eq:gluino-quark-squark}) with the understanding $\lambda_{\tilde{g}}\rightarrow\lambda^0\rightarrow-\tilde{\chi}_3^0$)
and for subsequent use,
\begin{eqnarray}
\label{eq:G+X}
& & G^{{q/{\bar q}}_{a_1}}_{\tilde{q}_{a_1}}\sim\tilde{f}{\cal V}^{-\frac{37}{36}},\ G^{{q/{\bar q}}_{a_2}}_{\tilde{q}_{a_1}}\sim\tilde{f}{\cal V}^{-\frac{59}{36}};\nonumber\\
& & X^{{q/{\bar q}}_{a_1}}_{\tilde{q}_{a_1}}\sim i\tilde{f}\left[{\cal V}^{\frac{2}{3}}{\bar\sigma}\cdot\frac{p_{q/{\bar q}}}{M_p} + {\cal V}^{-\frac{37}{36}}\right],\ X^{{q/{\bar q}}_{a_2}}_{\tilde{q}_{a_1}}\sim i\tilde{f}\left[{\cal V}^{\frac{13}{72}}{\bar\sigma}\cdot\frac{p_{q/{\bar q}}}{M_p} + {\cal V}^{-\frac{59}{36}}\right].
\end{eqnarray}

To calculate the contribution of opertors at EW scale, one need to derive  the RGE from squark mass scale $\tilde m$ to EW scale. The non-renormalizable interactions  produced by integrating the heavy squarks and effective Lagrangian at
the matching scale $\tilde m$ is given by \cite{Guidice_et_al}
\beq
{\cal L} = \frac{1}{\msusy^2}\,\sum_{i=1}^7 \CB_i\,\QB_i + \sum_{i=1}^2 \CW_i\,\QW_i+\frac{1}{\tilde{m}^2}\sum_{i=1}^5C_i^{\widetilde{H}}Q_i^{\widetilde{H}} .
\eeq
where ${C_i}'s$ are Wilson coefficients inducing interaction of Gluino, quark and antiquark.
Integrating out the heavy squarks and sleptons,the $G$-parity odd dimension-six operators that figure in the effective Lagrangian  that can be written as \cite{Guidice_et_al}:
\begin{eqnarray}
\label{eq:effective operators}
 & & \QB_1 = \ov{\wt{B}} \,\gamma^{\mu}\,
\gamma_{5}\, {\tilde g}^a\; \otimes\;\sum_{k=1}^2 \;
\ov{q}\lle^{\,(k)} \,\gamma_{\mu} \, T^a \,q\lle^{\,(k)}, \QB_2  = \ov{\wt{B}}  \,\gamma^{\mu}\,
\gamma_{5}\, {\tilde g}^a\; \otimes\;
\sum_{k=1}^2\;\ov{u}\rr^{\,(k)} \,\gamma_{\mu} \, T^a \, u\rr^{\,(k)} ;\nonumber\\
 & & \QB_3  =  \ov{\wt{B}} \,\gamma^{\mu}\,
\gamma_{5}\, {\tilde g}^a\; \otimes\;
\sum_{k=1}^2\;\ov{d}\rr^{\,(k)} \,\gamma_{\mu} \, T^a\, d\rr^{\,(k)}, \QB_4  =  \ov{\wt{B}} \,\gamma^{\mu}\,
\gamma_{5}\, {\tilde g}^a\; \otimes\;
\ov{q}\lle^{\,(3)} \,\gamma_{\mu} \, T^a \,q\lle^{\,(3)};\nonumber\\
& & \QB_5  =   \ov{\wt{B}} \,\gamma^{\mu}\,
\gamma_{5}\, {\tilde g}^a\; \otimes\;
\ov{t}\rr \,\gamma_{\mu} \, T^a \, t\rr,  \QB_6 =  \ov{\wt{B}} \,\gamma^{\mu}\,
\gamma_{5}\, {\tilde g}^a\; \otimes\;
\ov{b}\rr \,\gamma_{\mu} \ T^a\, b\rr, \QB_7 = \ov{\wt{B}} \,\sigma^{\mu\nu} \,
\gamma_5 \, {\tilde g}^a\;G^a_{\mu\nu};
\end{eqnarray}
where $\ov{\wt{B}}$ are Bino component of neutral gaugino's, $k$ is a generation index, $T^a$ are the SU(3) generators and
$G^a_{\mu\nu}$ is the gluon field strength.
Assuming all Wilson coefficient to be almost equal in our set up, solution to  general RG expression becomes:
\begin{eqnarray}
\label{eq:CBmu}
\CB_i (\mu) &=& \eta_s^{-\frac {9}{10}}
\left(O(1)+ O(1)y +  O(1)z\right)\CB_i (\widetilde{m})
\end{eqnarray}
where $y=\eta_s^{\,4/5}-1$, $z=(\eta_s^{\,8/15}\,\eta_t^{-1/3}-1)/3$ and
\begin{eqnarray}
\label{eq:etas}
\eta_s \equiv \frac{\tilde{\alpha}_3(\widetilde{m} )}{\tilde{\alpha}_3(\mu)}&=&1+\frac{5}{2\pi}\tilde{\alpha}_3(\widetilde{m})
\ln \frac{\mu}{\widetilde{m}} \,,\nonumber\\
\eta_t \equiv \frac{{\cal Y}_t(\widetilde{m} )}{{\cal Y}_t(\mu )}&=&
\eta_s^\frac{8}{5} - \frac{3{\cal Y}_t(\widetilde{m})
}{2\tilde{\alpha}_3(\widetilde{m} )}
\left(  \eta_s^\frac{8}{5} -\eta_s \right),
\end{eqnarray}
Using results of \cite{Ibanez_Lopez},
\begin{equation}
\label{eq:RG_Y_t I}
{\cal Y}_t(\mu=\widetilde{m})=\frac{{\cal Y}_t(m_s)E(\mu=\widetilde{m})}{\left(1+6{\cal Y}_t(m_s)F(\mu=\widetilde{m})\right)}
\end{equation}
for $\widetilde{m}= 10^{12}GeV, t\sim 14, E(\widetilde{m})\sim 1.39, F(\widetilde{m})\sim 16.58$,
   putting values  one gets ${\cal Y}_t(\widetilde{m})\sim (7.4)\times (10)^{-4}$.  Therefore $\eta_s= 0.66,\eta_t= 0.52, y= -0.28,z= -8\times {10}^{-3} $.

Solving  equation (\ref{eq:CBmu}),
\begin{equation}
\label{eq:CB}
\CB_i (m_{EW})= 1.45~\CB_i (\widetilde{m}),
\end{equation}
and we will therefore assume that $\CB_i(m_{EW})\sim{\cal O}(1)~\CB_i(m_S)$. The use of MSSM-based RG flow equations above and later, in our setup, is justifiable by noting (from Table 1), e.g., (a) the R-parity conserving Yukawa couplings $\hat{Y}_{{\cal A}_1^2{\cal Z}_i}$ analogous to the first two-generation Yukawa couplings in MSSM are very small and the R-parity violating Yukawa couplings $\hat{Y}_{{\cal Z}_1{\cal Z}_2{\cal A}_1}, \hat{Y}_{{\cal A}_1{\cal A}_1{\cal A}_1}$ are extremely suppressed, as well as (b) the R-parity conserving $\hat{\mu}_{{\cal Z}_1{\cal Z}_2}$ - like MSSM - is non-trivial and the R-parity violating $\hat{\mu}_{{\cal Z}_i{\cal A}_1}$ as well as the R-parity conserving $\hat{\mu}_{{\cal A}_1{\cal A}_1}$, as in MSSM, are extremely suppressed. Of course, the $t=0$ values in MSSM and our setup are quite different.

The dimension six operators corresponding to neutral Wino component is
\begin{eqnarray}
\label{eq:qw1}
\QW_1= \ov{\wt{W}^{\sss A}} \,\gamma^{\mu}\,\gamma_5
\, {\tilde g}^a\; \otimes\;\sum_{k=1}^2 \;
\ov{q}\lle^{\,(k)} \,\gamma^{\mu} \,\tau^{\sss A}\,T^a\, q\lle^{\,(k)}; \QW_2= \ov{\wt{W}^{\sss A}} \,\gamma^{\mu}\,\gamma_5
\, {\tilde g}^a\; \otimes\;
\ov{q}\lle^{\,(3)} \,\gamma^{\mu} \,\tau^{\sss A}\,T^a\, q\lle^{\,(3)}.
\end{eqnarray}
The contribution of Wilson coefficient corresponding to this operator at EW scale can be solved by using analytic RG solution to the following equation:
\begin{eqnarray}
\label{CW1}
\CW_i(\mu)= \CW_i(\widetilde{m} )\eta_s^{\left
(\frac{\gamma_{s,i}}{10}+\frac{8
\gamma_{t,i}}{45}\right)} \
\eta_t ^{-\frac{\gamma_{t,i}}{9}}, i=1,2
\end{eqnarray}
where   $\gamma_{s,1}= \gamma_{s,2}= -3N_C$, $\gamma_{t,1}=0, \gamma_{t,2}=1 $, $ N_C=3,N_F=6$. Equation (\ref{CW1}) becomes:
 \begin{eqnarray}
\label{eq:CW2}
\CW_1(m_{EW})= \eta_s^{-\frac{9}{10}}\CW_i(\widetilde{m} )= 1.45~\CW_1(\widetilde{m} );\nonumber\\
\CW_2(m_{EW})= {\eta_s}^{-0.7}{\eta_t}^{-0.1}\CW_i(\widetilde{m} )=1.42~\CW_2(\widetilde{m} ).
 \end{eqnarray}
again we will therefore assume that $\CW_i(m_{EW})\sim{\cal O}(1)\CW_i(m_S)$.

 Now we are able to calculate the contribution of wilson coefficient corresponding to the operators involving neutralinos and quarks. The corresponding operator and wilson coefficients involving neutralino are given as:
\begin{eqnarray}
\QN_{1\,q\lle,q\rr} =
\ov{\chi^0_i}\,\gamma^{\mu}\,\gamma_5\,\,{\tilde g}^a\;\otimes\;
\sum_{k=1}^2\,\ov{q}^{\,(k)}\lr\,\gamma_{\mu}\,T^a\,q^{(k)}\lr,\nonumber\\
\CN_{1\,q_L} = \CB_1\,N_{i1} + \CW_1\,N_{i2}\,,\;\;\;\;\;
\CN_{1\,q_R} = \CB_2\,N_{i1}\,
\end{eqnarray}

Assuming $ N_{i1/2}\sim O(1)$ and using equation (\ref{eq:CB}) and (\ref{eq:CW2}), one gets
\begin{eqnarray}
\CN_{1\,q_L}(m_{EW})&=& 1.45~\CB_i (\widetilde{m}) +1.45~\CW_1(\widetilde{m} )= 1.45~\CN_{1\,q_L}(\widetilde m);\nonumber\\
\CN_{1\,q_R}(m_{EW})&=& 1.45~\CB_i (\widetilde{m})= 1.45~\CN_{1\,q_R}(\widetilde m).
\end{eqnarray}
From above, one can conclude that  results for Wilson coefficients do not change much upon RG-flow from squark mass scale $(\tilde m)$ down to EW scale, and therefore we will assume $\CN_{1\,q_{L/R}}(m_{EW})\sim{\cal O}(1) \CN_{1\,q_{L/R}}(m_S)$.

The analytical formulae to calculate decay width for three body tree level gluino decay channel as given in \cite{Manuel Toharia} is:-
\begin{eqnarray}
\label{eq:neutralinowidth}
\hspace{-.5cm}\Gamma(\tilde{g}\to\chi_{\rm n}^{o}q_{{}_I} \bar{q}_{{}_J} )
=  {g_s^2\over256 \pi^3 \mgss^3 } \sum_{i,j} \int ds_{13} ds_{23}\ {1\over2}
{\cal R}e\Big(A_{ij}(s_{23}) + B_{ij}(s_{13}) -  2 \eps_{n} \eps_{{g}}\
C_{ij}(s_{23},s_{13})\Big)
\end{eqnarray}
where the integrand is the square of the spin-averaged total amplitude
and $i,j=1,2,..,6$ are the indices of the squarks mediating the decay. The limits of integration in (\ref{eq:neutralinowidth}) are
\begin{eqnarray}
\label{eq:limits}
\hspace{-.3cm}s_{13}^{max}(s_{23})\!\! &=&\!\! m_{q_{{}_I}}^2 + \mchi^2 +{1\over2
  s_{23}}\left[(\mgss^2-m_{q_{{}_I}}^2-s_{23})(s_{23}-m_{q_{{}_J}}^2+\mchi^2)
  +\lambda^{1/2}(s_{23},\mgss^2,m_{q_{{}_I}}^2)\ \lambda^{1/2}(s_{23},m_{q_{{}_J}}^2,\mchi^2)\right]\nonumber \\
\hspace{-.3cm}s_{13}^{min}(s_{23})\!\! &=&\!\! m_{q_{{}_I}}^2 + \mchi^2 +{1\over2
  s_{23}}\left[(\mgss^2-m_{q_{{}_I}}^2-s_{23})(s_{23}-m_{{q_{{}_J}}}^2+\mchi^2)
  -\lambda^{1/2}(s_{23},\mgss^2,m_{q_{{}_I}}^2)\ \lambda^{1/2}(s_{23},m_{q_{{}_J}}^2,\mchi^2)\right] \nonumber\\
\hspace{-.3cm}s_{23}^{max}\ \ \  &=& (\mgss - m_{q_{{}_I}})^2\nonumber  \\
\hspace{-.3cm}s_{23}^{min}\ \ \  &=& (\mchi + m_{q_{{}_J}})^2
\end{eqnarray}
where
$$ \lambda(x,y,z)=x^2+y^2+z^2-2 xy-2xz -2yz $$ and the
kinematical variables are $s_{13}=(k_1+k_3)^2$ and $s_{23}=(k_2+k_3)^2$.
The $A_{ij}$ terms in (\ref{eq:neutralinowidth}) represent the contributions from the gluino decay channel involving gluino$\rightarrow$squark+quark and squark$\rightarrow$neutralino+anti-quark, whereas
 the $B_{ij}$ terms come from channel gluino$\rightarrow$squark+anti-quark and squark$\rightarrow$neutralino+quark. The same are defined as:
\begin{eqnarray*}
\label{eq:ABdefs}
& & A_{ij}\nonumber\\
 & &\hskip-0.8in \sim{\Big( {1\over2}(\mgss^2 +m_{{}_I}^2 - s_{23}) {\rm Tr}\left[{G_i^{q_{{}_I}}G_j^{{q_{{}_I}}{{}^\dagger}}}\right]+m_{{}_I} \mgss
{\rm Tr}\left[{G_i^{q_{{}_I}}\widetilde{G}_j^{{q_{{}_I}}{{}^\dagger}}}\right] \Big)
\Big({1\over2}(s_{23}-\mchi^2 -m_{{}_J}^2) {\rm Tr}\left[{X^{q_{{}_J}}_iX_j^{q_{{}_J}{}^\dagger}}\right]-m_{{}_J} \mchi
{\rm Tr}\left[{X^{q_{{}_J}}_i\widetilde{X}_j^{q_{{}_J}{}^\dagger}}\right] \Big)
 \over\left(s_{23}-M_{\tilde{{q}}_i}^2\right)\left(
 s_{23}-M_{\tilde{{q}}_j}^2     \right) }\ \ \ \ \ \ \ \ \non\\
& & B_{ij}\nonumber\\
 & &\hskip-0.8in \sim{\Big( {1\over2}(\mgss^2 +m_{{}_J}^2 - s_{13}) {\rm Tr}\left[{G_i^{q_{{}_J}}G_j^{{q_{{}_J}}{{}^\dagger}}}\right]+m_{{}_J} \mgss
{\rm Tr}\left[{G_i^{q_{{}_J}}\widetilde{G}_j^{{q_{{}_J}}{{}^\dagger}}}\right] \Big)
\Big({1\over2}(s_{13}-\mchi^2 -m_{{}_I}^2) {\rm Tr}\left[{X^{q_{{}_I}}_iX_j^{q_{{}_I}{}^\dagger}}\right]-m_{{}_I} \mchi
{\rm Tr}\left[{X^{q_{{}_I}}_i\widetilde{X}_j^{q_{{}_I}{}^\dagger}}\right] \Big)
 \over\left(s_{13}-M_{\tilde{{q}}_i}^2\right)\left(
 s_{13}-M_{\tilde{{q}}_j}^2     \right) },\nonumber\\
 & &
\end{eqnarray*}
where $X^{q_{{}_I}}_i$ represents the neutralino-$\tilde{q}_i$(squark)-$q_I$(quark) vertex and $G_i^{q_{{}_J}}$ represents the $\tilde{g}$(gluino)-$q_I-\tilde{q}_I$ vertex.
 The $C_{ij}$'s represent the interference terms:
\begin{eqnarray}
\label{eq:C_{ij}}
\hspace{-13.2cm} C_{ij}&=&{T_{ij}  \over  \left(s_{23}-M_{\tilde{{q}}_i}^2\right)
  \left(s_{13}-M_{\tilde{{q}}_j}^2  \right)}
\end{eqnarray}
with $T_{ij}$ defined by:
\begin{eqnarray}
\hspace{-0.9in}T_{ij}\!\!\!\!&=&\!\!\!\! K_1^{ij} \Big[(s_{13}\! -\!  \mchi^2\! -\! m_{q_{I}}^2)(\mgss^2\! +\!  m_{q_{J}}^2\!  -\! s_{13})\!
+\!  (s_{23}\!  -\! \mchi^2\! -\! m_{q_{J}}^2)(\mgss^2\! +\! m_{q_{I}}^2 \! -\!  s_{23})\!
-\! (\mgss^2\! +\! \mchi^2 \!  -\! s_{23}\! -\! s_{13} )(s_{23}\! +\! s_{13}\! -\! m_{q_{I}}^2-\! m_{q_{J}}^2 )\Big]\non\\
\hspace{-1.2cm}\ &&-\  4 \mchi \mgss m_{q_{J}} m_{q_{I}}\ K^{ij}_2
+\ 2\mgss m_{q_{J}} \left(s_{13}-\mchi^2 -m_{q_{I}}^2\right)\  K^{ij}_3\
+\ 2m_{q_{I}}m_{q_{J}}\left(s_{23}+s_{13}-m_{q_{I}}-m_{q_{J}} \right)\  K_4^{ij}\ \non\\
\hspace{-1.2cm}\ &&+\ 2\mgss m_{q_{I}}\left(s_{23}-\mchi^2-m_{q_{J}}^2 \right)\ K_5^{ij}\
-\ 2\mchi m_{q_{J}}\left(\mgss^2+m_{q_{I}}^2 -s_{23}\right)\ K_6^{ij}\ \non\\
\hspace{-1.2cm}\ && -\ 2\mchi\mgss \left(\mgss^2+\mchi^2 -s_{13}-s_{23}\right)\ K_7^{ij}\
-\ 2\mchi m_{q_{I}} \left(\mgss^2+m_{q_{J}}^2-s_{13}\right)\ K_8^{ij}.\non
\end{eqnarray}
where for our case, $$K_a^{ij} (a=1,..,8)\sim{\rm Tr}\left[X^{q_{{}_J}}_i X_j^{q_{{}_I}{}^\dagger} G_i^{q_{{}_I}} G_j^{q_{{}_I}{{}^\dagger}} \right]$$.

  Keeping neutralino ${\tilde\chi_3}^0$ mass to be around $\frac{1}{2}$ of  gaugino mass (see Appendix {\bf B}) and hence
  putting the value of gaugino and gluino mass to be of the order $m_{3/2}$ and squark mass term to be of the order
  ${\cal{V}}^{\frac{73}{72}}m_{3/2}$ respectively ($m_{3/2}\sim 10^3{TeV}$ being a gravitino mass, ${\cal{V}}\sim{10}^6$ being
    the Calabi-Yau Volume) as given in results in \cite{D3_D7_Misra_Shukla}, using above analytic expressions the limits of integration in our case becomes:-
$$s_{23 \max }= \left(m_{\frac{3}{2}}-m_q\right)^2,\ s_{23 \min }= \left(m_q+\frac{m_{\frac{3}{2}}}{2}\right)^2;$$
$$s_{13 \max }= \frac{\left(m_{\frac{3}{2}}^2-s_{23}\right) m_{\frac{3}{2}}^2}{4 s_{23}}+\frac{m_{\frac{3}{2}}^2}{4}, s_{13 \min }= \frac{5 m_{\frac{3}{2}}^2}{4}-s_{23}.$$

With the help of above expressions, we will calculate decay width of Gluino in four different cases discussed below:-
\begin{itemize}
\item Both quark and antiquark appearing in three body decay of gluino has been approximated  by wilson line moduli $a_1$ as shown in Feynman Graph 1:-
\begin{center}
\begin{picture}(500,180)(50,0)
\Text(90,130)[]{$\tilde{g}$}
\Line(60,120)(110,120)
\Gluon(60,120)(110,120){5}{4}
\ArrowLine (140,150)(110,120)
\Text(147,150)[]{$\bar{q}_{a_1}$}
\DashArrowLine (110,120)(130,90){4}
\Text(107,100)[]{$\tilde{q}_{a_1}$}
\ArrowLine(130,90)(160,120)
\Text(167,120)[]{$\tilde{\chi}_3^0$\hskip 1.5cm +}
\ArrowLine(130,90)(160,60)
\Text(167,65)[]{${q_{a_1}}$}
\Text(250,130)[]{$\tilde{g}$}
\Line(210,120)(260,120)
\Gluon(210,120)(260,120){5}{4}
\ArrowLine (290,150)(260,120)
\Text(297,150)[]{$q_{a_1}$}
\DashArrowLine(260,120)(280,90){4}{}
\Text(257,100)[]{$\tilde{q}_{a_1}$}
\ArrowLine(280,90)(310,120)
\Text(320,120)[]{$\bar{q}_{a_1}$}
\ArrowLine(280,90)(310,60)
\Text(317,65)[]{$\tilde{\chi}_3^0$}
\end{picture}
{\sl Figure 3: Both quark and anti-quark appearing in three body decay of gluino has been approximated  by the fermionic superpartner of the Wilson line modulus $a_1$}
\end{center}

\vskip -0.3in
\end{itemize}

For this particular case:-

\hskip -0.85in{\small $A_{ij}\left(Tr\left[G^{{\bar q}_{a_1}}_{\tilde{q}_{a_1}}G^{{\bar q}_{a_1}}_{\tilde{q}_{a_1}}\ ^\dagger\right]
Tr\left[X^{q_{a_1}}_{\tilde{q}_{a_1}}X^{q_{a_1}}_{\tilde{q}_{a_1}}\ ^\dagger\right]\sim\tilde{f}^4{\cal V}^{-\frac{37}{18}}\left\{Tr\left[{\cal V}^{\frac{2}{3}}{\bar\sigma}\cdot\frac{\left(p_{\tilde{\chi}_3^0+p_q}\right)}{M_p} + {\cal V}^{-\frac{37}{36}}{\bf 1}\right]^2\sim 2\left(\frac{{\cal V}^{-\frac{8}{3}}}{4}+{\cal V}^{\frac{4}{3}}\frac{\left({\bf p}_{\tilde{\chi}_3^0}+{\bf p}_{q}\right)^2}{M_p^2}+{\cal V}^{-\frac{37}{18}}+..\right)\right\}\sim\tilde{f}^4{\cal V}^{-\frac{85}{18}}\right)$},

 {\small \hskip-0.85in$B_{ij}\left(Tr\left[G^{q_{a_1}}_{\tilde{q}_{a_1}}G^{q_{a_1}}_{\tilde{q}_{a_1}}\ ^\dagger\right]
Tr\left[X^{{\bar q}_{a_1}}_{\tilde{q}_{a_1}}X^{{\bar q}_{a_1}}_{\tilde{q}_{a_1}}\ ^\dagger\right]\sim\tilde{f}^4{\cal V}^{-\frac{37}{18}}\left\{Tr\left[{\cal V}^{\frac{2}{3}}{\bar\sigma}\cdot\frac{\left(p_{\tilde{\chi}_3^0}+p_q\right)}{M_p} + {\cal V}^{-\frac{37}{36}}{\bf 1}\right]^2\sim 2\left(\frac{{\cal V}^{-\frac{8}{3}}}{4}+{\cal V}^{\frac{4}{3}}\frac{\left({\bf p}_{\tilde{\chi}_3^0}+{\bf p}_{q}\right)^2}{M_p^2}+{\cal V}^{-\frac{37}{18}}+..\right)\right\}\sim\tilde{f}^4{\cal V}^{-\frac{85}{18}}\right)$},

\hskip-0.8in{\small $C\Biggl(Tr\left[G^{{\bar q}_{a_1}}_{\tilde{q}_{a_1}}G^{q_{a_1}}_{\tilde{q}_{a_1}}\ ^\dagger X^{q_{a_1}}_{\tilde{q}_{a_1}}X^{{\bar q}_{a_1}}_{\tilde{q}_{a_1}}\ ^\dagger\right]\sim\tilde{f}^4{\cal V}^{-\frac{37}{18}}\Biggl\{Tr\left[\left({\cal V}^{\frac{2}{3}}{\bar\sigma}\cdot\frac{\left(p_{\tilde{\chi}_3^0}+p_q\right)}{M_p}+{\cal V}^{-\frac{37}{36}}{\bf 1}\right)\left({\cal V}^{\frac{2}{3}}{\bar\sigma}\cdot\frac{\left(p_{\tilde{\chi}_3^0}+p_{\bar q}\right)}{M_p}+{\cal V}^{-\frac{37}{36}}{\bf 1}\right)\right]$} \\
{\small $\sim2\left(\frac{{\cal V}^{-\frac{8}{3}}}{4}+{\cal V}^{\frac{4}{3}}\frac{\left({\bf p}_{\tilde{\chi}_3^0}+{\bf p}_{q}\right)\cdot\left({\bf p}_{\tilde{\chi}_3^0}+{\bf p}_{\bar q}\right)}{M_p^2}+{\cal V}^{-\frac{37}{18}}+..\right)\Biggr\}
\sim\tilde{f}^4{\cal V}^{-\frac{85}{18}}\Biggr)$}

Putting above values for various vertex elements and solving equation (\ref{eq:neutralinowidth}) $-$ (\ref{eq:C_{ij}}) , dominating contribution of decay width in given domain of integration is:
\begin{eqnarray}
 \label{eq: Decay width 1}
 \hspace{-.5cm}\Gamma(\tilde{g}\to\chi_{\rm n}^{o}q_{{}_I} \bar{q}_{{}_J} )
&\sim&{g_s^2 O(1) \over256 \pi^3 m_{\frac{3}{2}}^3 }\left[
{\tilde{f}^4{\cal V}^{-\frac{85}{18}} m_{\frac{3}{2}}^4 }{{\cal V}^4}+ \tilde{f}^4{\cal V}^{-\frac{85}{18}}{\cal V}^2 m_{\frac{3}{2}}^4 +\tilde{f}^4{\cal V}^{-\frac{85}{18}} {\cal V}^4  m_{\frac{3}{2}}^4\right] \nonumber\\
& & \sim {O(1) g_s^2\over256 \pi^3 m_{\frac{3}{2}}^3 }\left( \tilde{f}^4{\cal V}^{-\frac{85}{18}}{\cal V}^4 m_{\frac{3}{2}}^4  \right)\sim O(10^{-4})\tilde{f}^4{\cal V}^{-2.7}{m_{pl}} \nonumber\\
& & \sim O(10^{-2})\tilde{f}^4 GeV
 \end{eqnarray}
In case of N=1 Supergravity, modified D-term scalar potential in presence of background $D7$ fluxes  is defined as \cite{Jockers_thesis}
   \begin{equation}
   \label{eq:Dterm}
V_D = {\frac{108\kappa_4^2\mu_7}{{{\cal{K}}^2}{ReT_\Lambda}}}({{\cal{K}}_{Pa}}{\cal{B}}^a-{\cal{Q}_{\alpha}}{\emph{v}}^{\alpha})^2,
\end{equation}
The first term of $V_D$ can be minimized for ${\cal{B}}^a= 0$ and second term has extra contribution coming from additional D7 Brane fluxes.
Now, the corresponding F-term scalar potential in LVS limit has been calculated in \cite{D3_D7_Misra_Shukla} as
\begin{equation}
\label{eq:Fterm}
V_f\sim e^K G^{\sigma^{{\alpha}}{\bar{\sigma}^{\bar{\alpha}}}}D_{\sigma^{\alpha}}W^\alpha{\bar{D}_{\bar{\sigma}^{\bar{\alpha}}}}\bar{W}\sim{\cal{V}}^{19/18}m_{3/2}^2
\sim{\cal{V}}^{-3}m_{pl}^2
\end{equation}
 In case of dilute flux approximation $V_D < V_F$. Therefore from (\ref{eq:Dterm}) and (\ref{eq:Fterm})
 \begin{equation}
 {\frac{108\kappa_4^2\mu_7}{{{\cal{K}}^2}{ReT_\Lambda}}}({\cal{Q}_{\alpha}}{\emph{v}}^{\alpha})^2 < {\cal{V}}^{-3}m_{pl}^2
 \end{equation}
 where $ \kappa_4^2\mu_7\sim\frac{1}{\cal{V}}, {\cal{K}}=1/6Y$(Volume of physical Calabi Yau), $Q_{\alpha}\sim {\cal{V}}^{1/3}f,{ReT_\Lambda}\sim {\cal{V}}^{2/3}$
 (volume of "Big" Divisor) and ${\cal{V}}^{\alpha}\sim{\cal{V}}^{1/3}$(Internal volume of 2-cycle),
 Solving this
 \begin{equation}
 \frac{10^{4}f^2{\cal{V}}^{4/3}}{{\cal{V}}^3{\cal{V}}^{2/3}}<{\cal{V}}^{-3}
 \end{equation}
 i.e   ${f^2}< 10^{-8}$.

Utilizing this value of ${f^2}$, Decay width of gluino i.e equation (\ref{eq: Decay width 1}) becomes:
$ \Gamma(\tilde{g}\to\chi_{\rm n}^{o}q_{{}_I} \bar{q}_{{}_J} )\sim O(10^{-2})\tilde{f}^4 < O(10^{-18})GeV$.
 Further, Life time of gluino is given as
   \begin{eqnarray}
   \tau &=&\frac{\hbar}{\Gamma}\sim\frac{10^{-34} Jsec}{10^{-2}f^4 GeV}\sim\frac{10^{-22}}{f^4} > {10^{-6}}sec
   \end{eqnarray}

 \begin{itemize}
  \item quark appearing in three body decay of gluino has been approximated  by wilson line moduli $a_1$ and antiquark  has been approximated  by wilson line moduli $a_1$ as shown in Feynman Graph 2:-
 \end{itemize}
\begin{center}
\begin{picture}(500,180)(50,0)
\Text(90,130)[]{$\tilde{g}$}
\Line(60,120)(110,120)
\Gluon(60,120)(110,120){5}{4}
\ArrowLine (140,150)(110,120)
\Text(147,150)[]{$\bar{q}_{a_2}$}
\DashArrowLine (110,120)(130,90){4}
\Text(107,100)[]{$\tilde{q}_{a_1}$}
\ArrowLine(130,90)(160,120)
\Text(167,120)[]{$\tilde{\chi}_3^0$\hskip 1.5cm +}
\ArrowLine(130,90)(160,60)
\Text(167,65)[]{${q_{a_1}}$}
\Text(250,130)[]{$\tilde{g}$}
\Line(210,120)(260,120)
\Gluon(210,120)(260,120){5}{4}
\ArrowLine (290,150)(260,120)
\Text(297,150)[]{$q_{a_1}$}
\DashArrowLine(260,120)(280,90){4}{}
\Text(257,100)[]{$\tilde{q}_{a_1}$}
\ArrowLine(280,90)(310,120)
\Text(320,120)[]{$\bar{q}_{a_2}$}
\ArrowLine(280,90)(310,60)
\Text(317,65)[]{$\tilde{\chi}_3^0$}
\end{picture}
{\sl Figure 4: (Anti-)Quark appearing in three body decay of gluino has been approximated  by the superpartner of the Wilson line modulus $a_1$ and antiquark}
\end{center}
For Fig.4 diagrams:

 \hskip -0.89in{\small $A_{ij}\left(Tr\left[G^{{\bar q}_{a_2}}_{\tilde{q}_{a_1}}G^{{\bar q}_{a_2}}_{\tilde{q}_{a_1}}\ ^\dagger\right]
Tr\left[X^{q_{a_1}}_{\tilde{q}_{a_1}}X^{q_{a_1}}_{\tilde{q}_{a_1}}\ ^\dagger\right]\sim\tilde{f}^4{\cal V}^{-\frac{59}{18}}\left\{Tr\left[{\cal V}^{\frac{2}{3}}{\bar\sigma}\cdot\frac{\left(p_{\tilde{\chi}_3^0+p_q}\right)}{M_p}+{\cal V}^{-\frac{37}{36}}{\bf 1}\right]^2\sim 2\left(\frac{{\cal V}^{-\frac{8}{3}}}{4}+{\cal V}^{\frac{4}{3}}\frac{\left({\bf p}_{\tilde{\chi}_3^0}+{\bf p}_{q}\right)^2}{M_p^2}+{\cal V}^{-\frac{37}{18}}+..\right)\right\}\sim\tilde{f}^4{\cal V}^{-\frac{107}{18}}\right)$},

 {\small \hskip-0.89in$B_{ij}\left(Tr\left[G^{q_{a_1}}_{\tilde{q}_{a_1}}G^{q_{a_1}}_{\tilde{q}_{a_1}}\ ^\dagger\right]
Tr\left[X^{{\bar q}_{a_2}}_{\tilde{q}_{a_1}}X^{{\bar q}_{a_2}}_{\tilde{q}_{a_1}}\ ^\dagger\right]\sim\tilde{f}^4{\cal V}^{-\frac{37}{18}}\left\{Tr\left[{\cal V}^{\frac{13}{72}}{\bar\sigma}\cdot\frac{\left(p_{\tilde{\chi}_3^0}+p_q\right)}{M_p} + {\cal V}^{-\frac{59}{36}}{\bf 1}\right]^2\sim 2\left(\frac{{\cal V}^{-\frac{65}{18}}}{4}+{\cal V}^{\frac{13}{36}}\frac{\left({\bf p}_{\tilde{\chi}_3^0}+{\bf p}_{q}\right)^2}{M_p^2}+{\cal V}^{-\frac{59}{18}}+..\right)\right\}\sim\tilde{f}^4{\cal V}^{-\frac{102}{18}}\right)$},

\hskip-0.8in{\small $C\Biggl(Tr\left[G^{{\bar q}_{a_2}}_{\tilde{q}_{a_1}}G^{q_{a_1}}_{\tilde{q}_{a_1}}\ ^\dagger X^{q_{a_1}}_{\tilde{q}_{a_1}}X^{{\bar q}_{a_2}}_{\tilde{q}_{a_1}}\ ^\dagger\right]\sim\tilde{f}^4{\cal V}^{-\frac{8}{3}}\Biggl\{Tr\left[\left({\cal V}^{\frac{2}{3}}{\bar\sigma}\cdot\frac{\left(p_{\tilde{\chi}_3^0}+p_q\right)}{M_p}+{\cal V}^{-\frac{37}{36}}{\bf 1}\right)\left({\cal V}^{\frac{13}{72}}{\bar\sigma}\cdot\frac{\left(p_{\tilde{\chi}_3^0}+p_{\bar q}\right)}{M_p}+{\cal V}^{-\frac{59}{36}}{\bf 1}\right)\right]$} \\
{\small $\sim2\left(\frac{{\cal V}^{-\frac{19}{6}}}{4}+{\cal V}^{\frac{61}{72}}\frac{\left({\bf p}_{\tilde{\chi}_3^0}+{\bf p}_{q}\right)\cdot\left({\bf p}_{\tilde{\chi}_3^0}+{\bf p}_{\bar q}\right)}{M_p^2}+{\cal V}^{-\frac{96}{36}}+..\right)\Biggr\}
\sim\tilde{f}^4{\cal V}^{-\frac{35}{6}}\Biggr)$}

Three body gluino decay width for this case is:
\begin{eqnarray}
 \label{eq: Decay width 2}
 \hspace{-.5cm}\Gamma(\tilde{g}\to\chi_{\rm n}^{o}q_{{}_I} \bar{q}_{{}_J} )
&\sim&{g_s^2 O(1) \over256 \pi^3 m_{\frac{3}{2}}^3 }\left[
{\tilde{f}^4{\cal V}^{-\frac{107}{18}} m_{\frac{3}{2}}^4 }{{\cal V}^4}+ \tilde{f}^4{\cal V}^{-\frac{102}{18}}{\cal V}^2 m_{\frac{3}{2}}^4 +\tilde{f}^4{\cal V}^{-\frac{35}{6}} {\cal V}^4 m_{\frac{3}{2}}^4\right] \nonumber\\
& & \sim { O(1) g_s^2\over256 \pi^3 m_{\frac{3}{2}}^3 }\left( \tilde{f}^4{\cal V}^{-5.8}{\cal V}^4 m_{\frac{3}{2}}^4  \right)\sim O(10^{-4})\tilde{f}^4{\cal V}^{-3.8}m_{pl} \nonumber\\
& & \sim O(10^{-9})\tilde{f}^4 GeV < O(10^{-25})GeV
 \end{eqnarray}
Life time of gluino is given as
   \begin{eqnarray}
   \tau &=&\frac{\hbar}{\Gamma}\sim\frac{10^{-34} J sec}{10^{-9}f^4 GeV}\sim\frac{10^{-15}}{f^4}> {10^{1}}sec
   \end{eqnarray}

\begin{itemize}
\item Both quark and antiquark appearing in three body decay of gluino has been approximated  by wilson line moduli $a_2$ as shown in Feynman Graph 3:-

    \end{itemize}
\begin{center}
\begin{picture}(500,180)(50,0)
\Text(90,130)[]{$\tilde{g}$}
\Line(60,120)(110,120)
\Gluon(60,120)(110,120){5}{4}
\ArrowLine (140,150)(110,120)
\Text(147,150)[]{$\bar{q}_{a_2}$}
\DashArrowLine (110,120)(130,90){4}
\Text(107,100)[]{$\tilde{q}_{a_1}$}
\ArrowLine(130,90)(160,120)
\Text(167,120)[]{$\tilde{\chi}_3^0$\hskip 1.5cm +}
\ArrowLine(130,90)(160,60)
\Text(167,65)[]{${q_{a_2}}$}
\Text(250,130)[]{$\tilde{g}$}
\Line(210,120)(260,120)
\Gluon(210,120)(260,120){5}{4}
\ArrowLine (290,150)(260,120)
\Text(297,150)[]{$q_{a_2}$}
\DashArrowLine(260,120)(280,90){4}{}
\Text(257,100)[]{$\tilde{q}_{a_1}$}
\ArrowLine(280,90)(310,120)
\Text(320,120)[]{$\bar{q}_{a_2}$}
\ArrowLine(280,90)(310,60)
\Text(317,65)[]{$\tilde{\chi}_3^0$}
\end{picture}
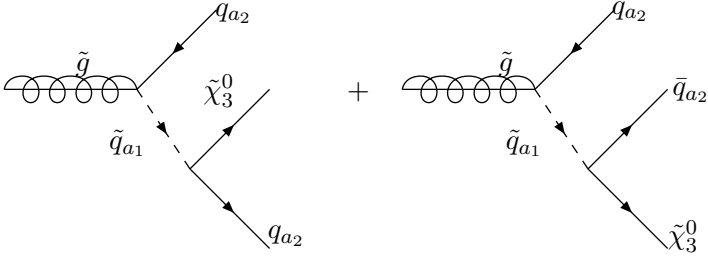
{\sl Figure 5: Both quark and antiquark appearing in three body decay of gluino have been approximated  by the superpartner of the Wilson line modulus $a_2$}
\end{center}
For Fig.5 diagrams:

\hskip -0.8in{\small $A_{ij}\left(Tr\left[G^{{\bar q}_{a_2}}_{\tilde{q}_{a_1}}G^{{\bar q}_{a_2}}_{\tilde{q}_{a_1}}\ ^\dagger\right]
Tr\left[X^{q_{a_1}}_{\tilde{q}_{a_1}}X^{q_{a_1}}_{\tilde{q}_{a_1}}\ ^\dagger\right]\sim\tilde{f}^4{\cal V}^{-\frac{59}{18}}\left\{Tr\left[{\cal V}^{\frac{13}{72}}{\bar\sigma}\cdot\frac{\left(p_{\tilde{\chi}_3^0+p_q}\right)}{M_p} + {\cal V}^{-\frac{59}{36}}{\bf 1}\right]^2\sim 2\left(\frac{{\cal V}^{-\frac{65}{18}}}{4}+{\cal V}^{\frac{13}{36}}\frac{\left({\bf p}_{\tilde{\chi}_3^0}+{\bf p}_{q}\right)^2}{M_p^2}+{\cal V}^{-\frac{59}{18}}+..\right)\right\}\sim\tilde{f}^4{\cal V}^{-\frac{124}{18}}\right)$},

 {\small \hskip-0.84in$B_{ij}\left(Tr\left[G^{q_{a_1}}_{\tilde{q}_{a_1}}G^{q_{a_1}}_{\tilde{q}_{a_1}}\ ^\dagger\right]
Tr\left[X^{{\bar q}_{a_2}}_{\tilde{q}_{a_1}}X^{{\bar q}_{a_2}}_{\tilde{q}_{a_1}}\ ^\dagger\right]\tilde{f}^4{\cal V}^{-\frac{59}{18}}\left\{Tr\left[{\cal V}^{\frac{13}{72}}{\bar\sigma}\cdot\frac{\left(p_{\tilde{\chi}_3^0+p_{\bar q}}\right)}{M_p} + {\cal V}^{-\frac{59}{36}}{\bf 1}\right]^2\sim 2\left(\frac{{\cal V}^{-\frac{65}{18}}}{4}+{\cal V}^{\frac{13}{36}}\frac{\left({\bf p}_{\tilde{\chi}_3^0}+{\bf p}_{\bar q}\right)^2}{M_p^2}+{\cal V}^{-\frac{59}{18}}+..\right)\right\}\sim\tilde{f}^4{\cal V}^{-\frac{124}{18}}\right)$},

\hskip-0.8in{\small $C\Biggl(Tr\left[G^{{\bar q}_{a_2}}_{\tilde{q}_{a_1}}G^{q_{a_1}}_{\tilde{q}_{a_1}}\ ^\dagger X^{q_{a_1}}_{\tilde{q}_{a_1}}X^{{\bar q}_{a_2}}_{\tilde{q}_{a_1}}\ ^\dagger\right]\sim\tilde{f}^4{\cal V}^{-\frac{59}{18}}\Biggl\{Tr\left[\left({\cal V}^{\frac{13}{72}}{\bar\sigma}\cdot\frac{\left(p_{\tilde{\chi}_3^0+p_q}\right)}{M_p} + {\cal V}^{-\frac{59}{36}}{\bf 1}\right)\left({\cal V}^{\frac{13}{72}}{\bar\sigma}\cdot\frac{\left(p_{\tilde{\chi}_3^0+p_{\bar q}}\right)}{M_p} + {\cal V}^{-\frac{59}{36}}{\bf 1}\right)\right]$} \\
{\small $\sim2\left(\frac{{\cal V}^{-\frac{65}{18}}}{4}+{\cal V}^{\frac{13}{36}}\frac{\left({\bf p}_{\tilde{\chi}_3^0}+{\bf p}_{q}\right)\cdot\left({\bf p}_{\tilde{\chi}_3^0}+{\bf p}_{\bar q}\right)}{M_p^2}+{\cal V}^{-\frac{59}{18}}+..\right)\Biggr\}
\sim\tilde{f}^4{\cal V}^{-\frac{124}{18}}\Biggr)$}

Decay width of Gluino for this particular case is:
\begin{eqnarray}
 \label{eq: Decay width 2}
 \hspace{-.5cm}\Gamma(\tilde{g}\to\chi_{\rm n}^{o}q_{{}_I} \bar{q}_{{}_J} )
&\sim&{g_s^2 O(1)\over256 \pi^3 m_{\frac{3}{2}}^3 }\left[
{\tilde{f}^4{\cal V}^{-\frac{124}{18}} m_{\frac{3}{2}}^4 }{{\cal V}^4}+ \tilde{f}^4{\cal V}^{-\frac{124}{18}}{\cal V}^2 m_{\frac{3}{2}}^4  +\tilde{f}^4{\cal V}^{-\frac{124}{18}} {\cal V}^4 m_{\frac{3}{2}}^4\right] \nonumber\\
& & \sim {O(1) g_s^2\over256 \pi^3 m_{\frac{3}{2}}^3 }\left( \tilde{f}^4{\cal V}^{-\frac{124}{18}}{\cal V}^4 m_{\frac{3}{2}}^4  \right)\sim O(10^{-4})\tilde{f}^4{\cal V}^{-5}m_{pl} \nonumber\\
& & \sim O(10^{-16})\tilde{f}^4 GeV < O(10^{-32})GeV
 \end{eqnarray}
Life time of gluino is given as
   \begin{eqnarray}
   \tau &=&\frac{\hbar}{\Gamma}\sim\frac{10^{-34} Jsec}{10^{-16}f^4 GeV}\sim\frac{10^{-8}}{f^4}> {10^{8}}sec
   \end{eqnarray}

\begin{itemize}
\item quark appearing in three body decay of gluino has been approximated  by wilson line moduli $a_2$ and antiquark  has been approximated  by wilson line moduli $a_1$ as shown in Feynman Graph 4:-
\end{itemize}
\begin{center}
\begin{picture}(500,180)(50,0)
\Text(90,130)[]{$\tilde{g}$}
\Line(60,120)(110,120)
\Gluon(60,120)(110,120){5}{4}
\ArrowLine (140,150)(110,120)
\Text(147,150)[]{$\bar{q}_{a_1}$}
\DashArrowLine (110,120)(130,90){4}
\Text(107,100)[]{$\tilde{q}_{a_1}$}
\ArrowLine(130,90)(160,120)
\Text(167,120)[]{$\tilde{\chi}_3^0$\hskip 1.5cm +}
\ArrowLine(130,90)(160,60)
\Text(167,65)[]{${q_{a_2}}$}
\Text(250,130)[]{$\tilde{g}$}
\Line(210,120)(260,120)
\Gluon(210,120)(260,120){5}{4}
\ArrowLine (290,150)(260,120)
\Text(297,150)[]{$q_{a_2}$}
\DashArrowLine(260,120)(280,90){4}{}
\Text(257,100)[]{$\tilde{q}_{a_1}$}
\ArrowLine(280,90)(310,120)
\Text(320,120)[]{$\bar{q}_{a_1}$}
\ArrowLine(280,90)(310,60)
\Text(317,65)[]{$\tilde{\chi}_3^0$}
\end{picture}
{\sl Figure 6: (Anti-)Quark appearing in three-body decay of gluino has been approximated  by the superpartner of the Wilson line modulus $(a_1)a_2$ }
\end{center}
For Fig.6 diagrams:

\hskip -0.88in{\small $A_{ij}\left(Tr\left[G^{{\bar q}_{a_1}}_{\tilde{q}_{a_1}}G^{{\bar q}_{a_1}}_{\tilde{q}_{a_1}}\ ^\dagger\right]
Tr\left[X^{q_{a_2}}_{\tilde{q}_{a_1}}X^{q_{a_2}}_{\tilde{q}_{a_1}}\ ^\dagger\right]\sim\tilde{f}^4{\cal V}^{-\frac{37}{18}}\left\{Tr\left[{\cal V}^{\frac{13}{72}}{\bar\sigma}\cdot\frac{\left(p_{\tilde{\chi}_3^0+p_q}\right)}{M_p} + {\cal V}^{-\frac{59}{36}}{\bf 1}\right]^2\sim 2\left(\frac{{\cal V}^{-\frac{65}{18}}}{4}+{\cal V}^{\frac{13}{36}}\frac{\left({\bf p}_{\tilde{\chi}_3^0}+{\bf p}_{q}\right)^2}{M_p^2}+{\cal V}^{-\frac{59}{18}}+..\right)\right\}\sim\tilde{f}^4{\cal V}^{-\frac{102}{18}}\right)$},

 {\small \hskip-0.88in$B_{ij}\left(Tr\left[G^{q_{a_2}}_{\tilde{q}_{a_2}}G^{q_{a_1}}_{\tilde{q}_{a_1}}\ ^\dagger\right]
Tr\left[X^{{\bar q}_{a_1}}_{\tilde{q}_{a_1}}X^{{\bar q}_{a_1}}_{\tilde{q}_{a_1}}\ ^\dagger\right]\sim\tilde{f}^4{\cal V}^{-\frac{59}{18}}\left\{Tr\left[{\cal V}^{\frac{2}{3}}{\bar\sigma}\cdot\frac{\left(p_{\tilde{\chi}_3^0}+p_q\right)}{M_p} + {\cal V}^{-\frac{37}{36}}{\bf 1}\right]^2\sim 2(\frac{{\cal V}^{-\frac{8}{3}}}{4}+{\cal V}^{\frac{4}{3}}\frac{\left({\bf p}_{\tilde{\chi}_3^0}+{\bf p}_{q}\right)^2}{M_p^2}+{\cal V}^{-\frac{37}{18}}+..)\right\}\sim\tilde{f}^4{\cal V}^{-\frac{107}{18}}\right)$},

\hskip-0.8in{\small $C\Biggl(Tr\left[G^{{\bar q}_{a_2}}_{\tilde{q}_{a_1}}G^{q_{a_1}}_{\tilde{q}_{a_1}}\ ^\dagger X^{q_{a_1}}_{\tilde{q}_{a_1}}X^{{\bar q}_{a_2}}_{\tilde{q}_{a_1}}\ ^\dagger\right]\sim\tilde{f}^4{\cal V}^{-\frac{8}{3}}\Biggl\{Tr\left[\left({\cal V}^{\frac{2}{3}}{\bar\sigma}\cdot\frac{\left(p_{\tilde{\chi}_3^0}+p_q\right)}{M_p}+{\cal V}^{-\frac{37}{36}}{\bf 1}\right)\left({\cal V}^{\frac{13}{72}}{\bar\sigma}\cdot\frac{\left(p_{\tilde{\chi}_3^0}+p_{\bar q}\right)}{M_p}+{\cal V}^{-\frac{59}{36}}{\bf 1}\right)\right]$} \\
{\small $\sim2\left(\frac{{\cal V}^{-\frac{19}{6}}}{4}+{\cal V}^{\frac{61}{72}}\frac{\left({\bf p}_{\tilde{\chi}_3^0}+{\bf p}_{q}\right)\cdot\left({\bf p}_{\tilde{\chi}_3^0}+{\bf p}_{\bar q}\right)}{M_p^2}+{\cal V}^{-\frac{96}{36}}+..\right)\Biggr\}
\sim\tilde{f}^4{\cal V}^{-\frac{35}{6}}\Biggr)$}

Decay width of Gluino for this  case is:
\begin{eqnarray}
 \label{eq: Decay width 2}
 \hspace{-.5cm}\Gamma(\tilde{g}\to\chi_{\rm n}^{o}q_{{}_I} \bar{q}_{{}_J} )
&\sim&{g_s^2 O(1) \over256 \pi^3 m_{\frac{3}{2}}^3 }\left[
{\tilde{f}^4{\cal V}^{-\frac{102}{18}} m_{\frac{3}{2}}^4 }{{\cal V}^4}+ \tilde{f}^4{\cal V}^{-\frac{169}{36}}{\cal V}^2 m_{\frac{3}{2}}^4  + \tilde{f}^4{\cal V}^{-\frac{107}{18}} {\cal V}^4 m_{\frac{3}{2}}^4\right] \nonumber\\
& & \sim {O(1) g_s^2\over256 \pi^3 m_{\frac{3}{2}}^3 }\left( \tilde{f}^4{\cal V}^{-5.8}{\cal V}^4 m_{\frac{3}{2}}^4  \right)\sim O(10^{-4})\tilde{f}^4{\cal V}^{-3.8}m_{pl} \nonumber\\
& & \sim O(10^{-9})\tilde{f}^4 GeV < O(10^{-25})GeV
 \end{eqnarray}
Life time of gluino is given as
   \begin{eqnarray}
   \tau &=&\frac{\hbar}{\Gamma}\sim\frac{10^{-34} Jsec}{10^{-9}f^4 GeV}\sim\frac{10^{-15}}{f^4}> {10^{1}}sec
   \end{eqnarray}
   From above four cases discussed, one can approximate life time of Gluino as $\tau>{10^{-6}}, {10^{1}}, {10^{8}}, {10^{1}}$ sec (depending $a_1, a_2$ to be relevant quark and antiquark)  i.e one can enhance life time of Gluino, thus proving existence of long lived Gluino in the context of LVS $\mu$ split SUSY Scenario.
   \subsection{$\tilde{g}\rightarrow\tilde{\chi}_3^0+g$}

\begin{center}
\begin{picture}(1000,200)(50,0)
\Text(140,130)[]{$\tilde{g}$}
\Line(110,120)(160,120)
\Gluon(110,120)(160,120){5}{4}
\Text(140,130)[]{$\tilde{g}$}
\ArrowLine (160,120)(190,150)
\Text(180,152)[]{$q_I$}
\DashArrowLine (190,90)(160,120){4}
\Text(170,100)[]{$\tilde{q}_R$}
\DashArrowLine (190,150)(190,90){4}
\Gluon(190,90)(230,90){5}{4}
\Text(240,90)[]{$g_\mu$}
\ArrowLine(190,150)(230,150)
\Text(240,150)[]{$\tilde{\chi}_3^0$}
\Text(180,30)[]{(a)}
\Line(310,120)(360,120)
\Gluon(310,120)(360,120){5}{4}
\Text(340,130)[]{$\tilde{g}$}
\DashArrowLine (360,120)(390,150){4}
\Text(380,152)[]{$\tilde{q}_R$}
\ArrowLine (390,90)(360,120)
\Text(370,100)[]{$q_I$}
\ArrowLine (390,150)(390,90)
\Gluon(390,90)(430,90){5}{4}
\Text(440,90)[]{$g_\mu$}
\ArrowLine(390,150)(430,150)
\Text(440,150)[]{$\tilde{\chi}_3^0$}
\Text(380,30)[]{(b)}
\end{picture}
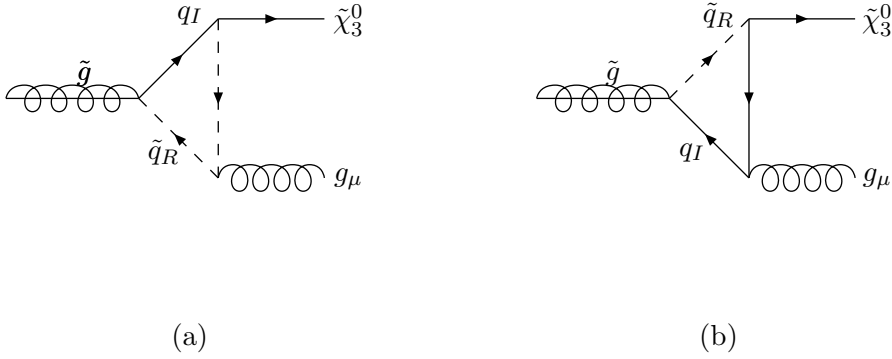
{\sl Figure 7: Diagrams contributing to one-loop two-body gluino decay}
\end{center}

Relevant to Figs.(a) and (b), the gluino-quark-squark vertex and the neutralino-quark-squark vertex will be given by (\ref{eq:G+X}). To figure out the quark-quark-gluon vertex, one notes that after elimination of the RR two-form axion $D_2^\alpha$ in favor of its dual axion $\rho_\alpha$, the Green-Schwarz term $\mu_7(2\pi\alpha^\prime)Q_\alpha\int_{{\bf R}^{3,1}}dD^\alpha_2\wedge A$ - $A$ being the $D7$ gauge field and $Q_\alpha=2\pi\alpha^\prime\int_{\Sigma_B\cup\sigma(\Sigma_B)}i^*\omega_\alpha\wedge P_-\tilde{f}$ - induced by the flux $\tilde{f}$ modifies the gauged isometry of the bulk. In particular, $T_B$ becomes charged with a covariant derivative given by:
\begin{equation}
\label{eq:cov_derT}
\bigtriangledown_\mu T_B=\partial_\mu T_B + 12i\pi\alpha^\prime\kappa_4^2\mu_7Q_BA_\mu,
\end{equation}
and consequently the $D$ term generated corresponding to the killing isometry vector: $$X=X^B\partial_B=-12i\pi\alpha^\prime\kappa_4^2\mu_7Q_B\partial_{T_B}$$
 is given by:
\begin{equation}
\label{eq:D_term}
D^B=\frac{4\pi\alpha^\prime\kappa_4^2\mu_7Q_Bv^B}{\cal V}
\end{equation}
 The quark-quark-gluon vertex relevant to Fig.(a), from \cite{Wess_Bagger} is given by:
  \begin{eqnarray}
  \label{eq:qqgl 1}
& &  g_{YM}g_{I{\bar J}}{\bar\chi}^{\bar J}{\bar\sigma}\cdot A\ {\rm Im}\left(X^BK + i D^B\right)\chi^I,\nonumber\\
& & \sim g_{YM}g_{I{\bar J}}{\bar\chi}^{\bar J}{\bar\sigma}\cdot A\left\{6\kappa_4^2\mu_72\pi\alpha^\prime Q_BK+\frac{12\kappa_4^2\mu_72\pi\alpha^\prime Q_Bv^B}{\cal V}\right\}\chi^I
  \end{eqnarray}
  which utilizing the fact:
  \begin{equation}
  \label{eq:gaabar}
  g_{a_1{\bar a}_{\bar 1}}\sim{\cal V}^{\frac{3}{4}},\ g_{a_1{\bar a}_{\bar 2}}\sim{\cal V}^{\frac{1}{4}},\ g_{a_2{\bar a}_{\bar 2}}\sim{\cal V}^{-\frac{1}{4}},
  \end{equation}
  as well as $g_{YM}\sim{\cal V}^{-\frac{1}{36}}, v^B\sim{\cal V}^{\frac{1}{3}}, Q_B\sim{\cal V}^{\frac{1}{3}}(2\pi\alpha^\prime)^2\tilde{f}$, yields for the quark-quark-gluon vertex:
 \begin{eqnarray}
 \label{eq:qqgl 2}
& &  \frac{\left({\cal V}^{\frac{1}{9}}\delta^I_{a_1}\delta^J_{a_1}+{\cal V}^{-\frac{7}{18}}\delta^I_{a_{1/2}}\delta^J_{a_{2/1}}+{\cal V}^{-\frac{8}{9}}\delta^I_{a_2}\delta^J_{a_2}\right)\tilde{f}{\bar\sigma}\cdot\epsilon}{\left(\sqrt{\hat{K}_{{\cal A}_1{\bar {\cal A}}_1}}\right)^2\sim{\cal V}^{\frac{31}{36}}}\nonumber\\
& & \sim\left({\cal V}^{-\frac{3}{4}}\delta^I_{a_1}\delta^J_{a_1}+{\cal V}^{-\frac{5}{4}}\delta^I_{a_{1/2}}\delta^J_{a_{2/1}}+{\cal V}^{-\frac{7}{4}}\delta^I_{a_2}\delta^J_{a_2}\right)\tilde{f}{\bar\sigma}\cdot\epsilon.
 \end{eqnarray}

  The gauge kinetic term for squark-squark-gluon vertex, relevant to Fig.(b) will be given by $ {\frac{1}{{\cal V}^2}}G^{\sigma_B{\bar\sigma}_B}\tilde{\bigtriangledown}_\mu T_B\tilde{\bigtriangledown}^\mu {\bar T}_{\bar B}$. This implies that the following term generates the required squark-squark-gluon vertex:
\begin{equation}
\label{eq:sq sq gl}
\hskip-0.63in {\frac{6i\kappa_4^2\mu_72\pi\alpha^\prime Q_BG^{\sigma_B{\bar\sigma}_B}}{{\cal V}^2}}A^\mu\partial_\mu\left(\kappa_4^2\mu_7(2\pi\alpha^\prime)^2C_{1{\bar 1}}a_1{\bar a}_{\bar 1}\right)\xrightarrow[{\small \kappa_4^2\mu_7(2\pi\alpha^\prime)^2C_{1{\bar 1}}\sim{\cal V}^{\frac{7}{6}}}]{\small G^{\sigma_B{\bar\sigma}_B}\sim{\cal V}^{\frac{37}{36}},}\frac{{\cal V}^{\frac{55}{36}}\epsilon\cdot\left(2k-(p_{\tilde{\chi}_3^0}+p_{\tilde{g}})\right)}{\left(\sqrt{\hat{K}_{{\cal A}_1{\bar {\cal A}}_1}}\right)^2}\sim{\cal V}^{-\frac{4}{3}}\tilde{f}\left[ 2\epsilon\cdot k-\epsilon\cdot\left(p_{\tilde{\chi}_3^0}+p_{\tilde{g}}\right)\right].
\end{equation}

\begin{center}
\begin{picture}(500,180)(-100,0)
\Text(280,120)[]{${\cal V}^{-\frac{4}{3}}{\tilde f}\left[ 2\epsilon\cdot k-\epsilon\cdot\left(p_{\tilde{\chi}_3^0}+p_{\tilde{g}}\right)\right]$}
\DashArrowLine(0,120)(-50,120){5}
\Text(-80,120)[]{$\tilde{q}_R(k-p_{\tilde{g}})$}
\Text(65,153)[]{$\tilde{q}_R(k-p_{\tilde{\chi}_3^0})$}
\DashArrowLine(40,150)(0,120){5}
\Text(54,88)[]{$ g_\mu (\epsilon_\mu)$}
\Gluon(0,120)(40,90){5}{4}
\Text(0,81)[]{(\bf a)}
\end{picture}
\vskip-1.3in
\begin{picture}(500,200)(-100,0)
\Text(-57,120)[]{$\tilde{g}$}
\Text(280,120)[]{$\tilde{f}\left({\cal V}^{-\frac{37}{36}}\delta^I_{a_1}+{\cal V}^{-\frac{59}{36}}\delta^I_{a_2}\right)$}
\Line(-50,120)(0,120)
\Gluon(-50,120)(0,120){5}{4}
\ArrowLine(0,120)(40,150)
\Text(47,150)[]{$\tilde{q}_R$}
\DashArrowLine(40,90)(0,120){4}
\Text(47,90)[]{$\bar{q}_I$}
\Text(0,81)[]{(\bf b)}
\end{picture}
\vskip-1.6in
\begin{picture}(500,220)(-100,0)
\Text(253,120)[]{$i\tilde{f}\left[\left({\cal V}^{\frac{2}{3}}{\bar\sigma}\cdot\frac{\left(p_{\tilde{\chi}_3^0}+p_{q/{\bar q}}\right)}{M_p} + {\cal V}^{-\frac{37}{36}}\right)\delta^I_{a_1}+\left({\cal V}^{\frac{13}{72}}{\bar\sigma}\cdot\frac{\left(p_{\tilde{\chi}_3^0}+p_{q/{\bar q}}\right)}{M_p} + {\cal V}^{-\frac{59}{36}}\right)\delta^I_{a_2}\right]$}
\ArrowLine(-50,120)(0,120)
\Text(-70,120)[]{$q_I(k)$}
\Text(57,150)[]{$\tilde{\chi}_3^0(p_{\tilde{\chi}_3^0})$}
\ArrowLine(0,120)(40,150)
\Text(63,83)[]{$ \tilde{q}_R(k-p_{\tilde{\chi}_3^0}) $}
\DashArrowLine(0,120)(40,90){5}
\Text(0,81)[]{(\bf c)}
\end{picture}
\vskip-1in
\begin{picture}(500,180)(-100,0)
\Text(280,120)[]{$\left({\cal V}^{-\frac{3}{4}}\delta^I_{a_1}\delta^J_{a_1}+{\cal V}^{-\frac{5}{4}}\delta^I_{a_{1/2}}\delta^J_{a_{2/1}}+{\cal V}^{-\frac{7}{4}}\delta^I_{a_2}\delta^J_{a_2}\right)\tilde{f}{\bar\sigma}\cdot\epsilon$}
\ArrowLine(0,120)(-50,120)
\Text(-55,120)[]{$q_J$}
\Text(47,158)[]{$g_\mu(\epsilon_\mu)$}
\Gluon(40,150)(0,120){5}{4}
\Text(47,88)[]{$ q_I $}
\ArrowLine(40,90)(0,120)
\Text(0,81)[]{(\bf d)}
\end{picture}
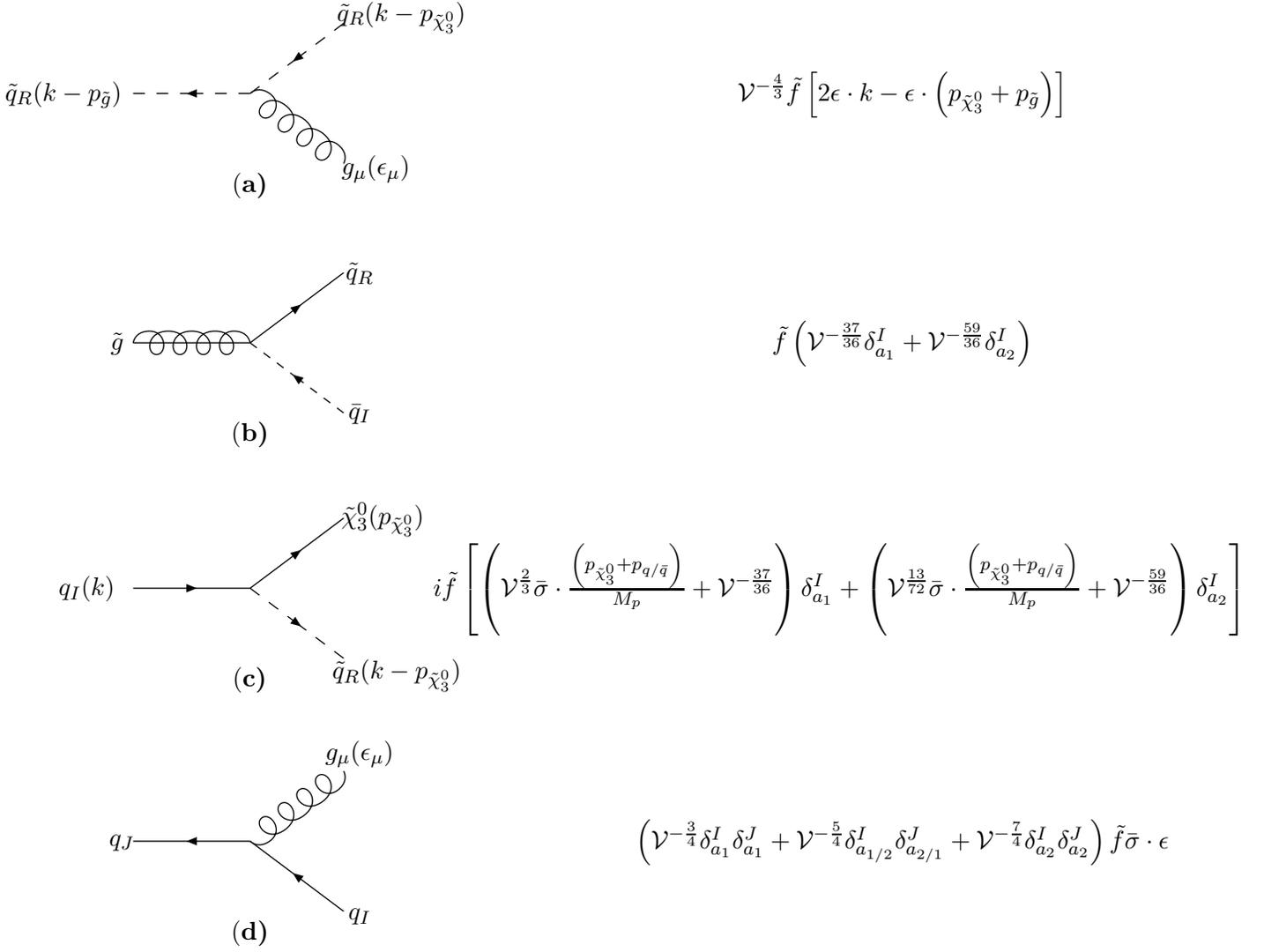
{\sl Figure 8: Diagrams corresponding  to contribution of different vertex elements in  one loop  gluino decay}
\end{center}

To calculate decay width of one loop two body Gluino, one needs to calculate coupling vertices at EW scale first. The effective operator involving a neutralino and a gluon is given as \cite{Guidice_et_al}
\begin{equation}
\QN_g =
\ov{\chi^0_i}\,\sigma^{\mu\nu}\,\gamma_5\,{\tilde g}^a\,G_{\mu\nu}^a\,.
\end{equation}
where
\begin{equation}
\label{coeffg}
\left.C^{\,{\chi}^0_i}_g\right._{\rm \!\!\!eff}
(\mu) =\CB_7(\mu)\,N_{i1}+\CH_5(\mu)\,N_{i4}\,v +
\frac{g_s\,h_t}{8\pi^2}\,\CH_2(\mu)\,\,N_{i4}\,v\,\ln\frac{m_t^2}{\mu^2}\,,
\end{equation}
$\CB_7(\mu)$ corresponds to effective Bino's coupling while $\CH_5(\mu)$ and $\CH_2(\mu)$ corresponds to effective Higgsino coupling.
Since we assume non-universalty of squarks in our set up, $\CB_7(\widetilde m)$ defined in \cite{Guidice_et_al} will be non-zero and RG solution to same is given as:
\begin{equation}
\CB_7(\mu)= \CB_7(\widetilde{m} )\eta_s^{\left
(\frac{\gamma_{s,i}}{10}+\frac{8
\gamma_{t,i}}{45}\right)} \
\eta_t ^{-\frac{\gamma_{t,i}}{9}},
\end{equation}
for this particular case $$\gamma_{s,7}=1/3(2N_F-18N_C),N_C=3,N_F=6; \gamma_{t,7}=0, $$  $N_{i1}\sim O(1)$, putting above values:
\begin{equation}
\label{eq:CBEW}
\CB_7(m_{EW})= \eta_s^{\frac{9}{10}}\CB_7(\widetilde{m} )= 0.69\CB_i(\widetilde{m} )
\end{equation}  where $\eta_s =0.66 $.
Using results  $\CH_{2,5}(m_{EW})\sim O(1)\CH_{2,5}(\widetilde m)$ as given in \cite{Guidice_et_al} and equation (\ref{eq:CBEW}), from (\ref{coeffg}), one gets:
\begin{equation}
\label{coeffg1}
\left.C^{\,{\chi}^0_i}_g\right._{\rm \!\!\!eff}
(m_{EW})\sim O(1) \left.C^{\,{\chi}^0_i}_g\right._{\rm \!\!\!eff}(\widetilde m)
\end{equation}
i.e.  behavior of Wilson coefficients corresponding to  two body gluino decay does not change much upon RG evolution to EW scale. Now, for simplicity of calculations, we will assume that it is only the $a_1$ squark and its fermionic superpartner  which are circulating in the loop. Using the vertices calculated above relevant to Figs. 7(a) and 7(b), and the Feynman rules of \cite{2comp} one obtains for the scattering amplitude:
\begin{eqnarray}
\label{eq:M_I}
& & {\cal M}\sim\tilde{f}^3M_p\int\frac{d^4k}{\left(2\pi\right)^4}\nonumber\\
& & \times{\cal V}^{-\frac{37}{36}}\left(\frac{i{\bar\sigma}\cdot k}{k^2-m_q^2+i\epsilon}\right)\left({\cal V}^{\frac{-4}{3}}+{\cal V}^{\frac{2}{3}}\frac{{\bf\sigma}\cdot({\bf p_{\tilde{\chi}_3^0}+k})}{M_p}+{\cal V}^{-\frac{37}{36}}\right)\left(\frac{i}{\left[\left(k-p_{\tilde{\chi}_3^0}\right)^2-m^2_{\tilde{q}}+i\epsilon\right]}
\right)\nonumber\\
& & \times\left({\cal V}^{-\frac{4}{3}}\epsilon\cdot\left(2k-p_{\tilde{\chi}_3^0}-p_{\tilde{g}}\right)\right)\left(\frac{i}
{\left[\left(k-p_{\tilde{g}}\right)^2-m^2_{\tilde{q}}+i\epsilon\right]}\right)\nonumber\\
& & + \tilde{f}^3M_p\int\frac{d^4k}{\left(2\pi\right)^4}{\cal V}^{-\frac{37}{36}}\left(\frac{i}
{\left[\left(k+p_{\tilde{\chi}_3^0}\right)^2-m^2_{\tilde{q}}+i\epsilon\right]}\right)\left({\cal V}^{\frac{-4}{3}}+{\cal V}^{\frac{2}{3}}\frac{{\bf\sigma}\cdot({\bf p_{\tilde{\chi}_3^0}+k})}{M_p}+{\cal V}^{-\frac{37}{36}}\right)\left(\frac{i{\bar\sigma}\cdot k}{k^2-m_q^2+i\epsilon}\right)\nonumber\\
& & \times\left({\cal V}^{-\frac{3}{4}}{\bar\sigma}\cdot\epsilon\right)\left(\frac{i{\bar\sigma}\cdot\left(k-p_{g_\mu}\right)}{\left[
\left(k-p_{g_\mu}\right)^2-m^2_q+i\epsilon\right]}\right)
\end{eqnarray}
Using the 1-loop integrals of \cite{Pass+Velt}:
\begin{eqnarray}
\label{eq:passvelt}
& & \frac{1}{i}\int\frac{d^4k}{\left(2\pi\right)^4}\frac{\left(k_\mu,\ k_\mu k_\nu\right)}{\left(k^2-m_1^2+i\epsilon\right)
\left[\left(k+p_1\right)^2-m_2^2+i\epsilon\right]\left[\left(k+p_1+p_2\right)^2-m_3^2+i\epsilon\right]}\nonumber\\
& & =4\pi^2\Biggl[p_{1\mu}C_{11}+p_{2\mu}C_{12}, p_{1\mu}p_{1\nu}C_{21}+p_{2\mu}p_{2\nu}C_{22}+\left(p_{1\mu}p_{2\nu}+p_{1\nu}p_{2\mu}\right)C_{23}+\eta_{\mu\nu}C_{24}\Biggr];
\nonumber\\
& & (a) m_1=m_q, m_2=m_3=m_{\tilde{q}};\ p_1=-p_{\tilde{\chi}_3^0}, p_2=-p_{g_\mu};\nonumber\\
& & (b) m_1=m_3=m_q, m_2=m_{\tilde{q}};\ p_1=p_{\tilde{\chi}_3^0}, p_2=-p_{\tilde{g}}.
\end{eqnarray}

First we will calculate different one loop three point functions $C_{ij}$'s. The formulae's used to calculate these functions for cases (a) and (b) are :-
\begin{eqnarray}
\label{Cdef}
& & f_1=m_1^2-m_2^2-p_1^2;\nonumber\\
& & f_2=m_2^2-m_3^2+p_1^2-p_5^2;\nonumber\\
& & R_1=
\begin{array}{l}
 {\frac{1}{2} \left(B_0(1,3)-B_0(2,3)+C_0 f_1\right)}
\end{array}
;\nonumber\\
& & R_2=\frac{1}{2} \left(B_0(1,2)-B_0(1,3)+C_0 f_1\right);\nonumber\\
& & X= \left(
\begin{array}{ll}
 p_1^2 & p_1 p_2 \\
 p_1 p_2 & p_2^2
\end{array}
\right);\nonumber\\
& & \left(
\begin{array}{l}
 C_{11} \\
 C_{12}
\end{array}
\right)= X^{-1}\left(
\begin{array}{l}
 R_1 \\
 R_2
\end{array}
\right);\nonumber\\
& & C_{24}= -\frac{1}{2} C_0 m_1^2+\frac{1}{4} \left(B_0(1,3)-C_{11} f_1-C_{12}
   f_2\right)+\frac{1}{4};\nonumber\\
& & R_3= \left(
\begin{array}{l}
 \frac{1}{2} \left(B_0(2,3)+B_1(1,3)+C_{11} f_1\right)-C_{24}
\end{array}
\right);\nonumber\\
& & R_4= \left(
\begin{array}{l}
 \frac{1}{2} \left(B_1(1,3)-B_1(2,3)+C_{12} f_1\right)
\end{array}
\right);\nonumber\\
& & R_5= \left(
\begin{array}{l}
 \frac{1}{2} \left(B_1(1,2)-B_1(1,3)+C_{11} f_2\right)
\end{array}
\right);\nonumber\\
& & R_6= \left(
\begin{array}{l}
 \frac{1}{2} \left(C_{12} f_2-B_1(1,3)\right)-C_{24}
\end{array}
\right);\nonumber\\
& & \left(
\begin{array}{l}
 C_{21} \\
 C_{23}
\end{array}
\right)= X^{-1}\left(
\begin{array}{l}
 R_3 \\
 R_5
\end{array}
\right);\nonumber\\
& & \left(
\begin{array}{l}
 C_{22} \\
 C_{23}
\end{array}
\right)=X^{-1}\left(
\begin{array}{l}
 R_4 \\
 R_6
\end{array}
\right);\nonumber\\
\end{eqnarray}

Various C-functions are related to $C_0$ and two point functions $B_0, B_1$. Therefore It becomes necessary to compute these functions first.
\begin{itemize}
\item $C_0$ has been evaluated using expression given in \cite{appell}.
\begin{eqnarray}
\label{eq:C0}
{\hskip-2.5 cm}& & C_0(p_1^2,p_2^2,p_3^3;m_1^2,m_2^2,m_3^2) =  -\frac{i}{(4\pi)^2}
\frac{1}{\lambda^{1/2}(p_1^2,p_2^2,p_3^2)}  {\times} \sum_{i=1}^{2} \biggl( [ R(x_i,y)-R(x_i',y) ] -
[\alpha{\leftrightarrow}(1-\alpha), 1{\leftrightarrow}3 ] \biggr)
\end{eqnarray}
where $R$ has been written in terms of an Appell's function $R(x,y) = \frac{1}{2}\ xy\ F_3[1,1,1,1;3;x,y]$ and definitions of various variables used in above expression have been given in \cite{appell}.

\item The two point functions appearing in results are defined by omitting one of the factor in the denominator from three point one loop function function defined in (\ref{eq:passvelt}), for e.g $B_0(1,2)$ is defined as:
\begin{eqnarray}
& & \frac{1}{i}\int\frac{d^4k}{\left(2\pi\right)^4}\frac{1}{\left(k^2-m_1^2+i\epsilon\right)
\left[\left(k+p_1\right)^2-m_2^2+i\epsilon\right]}
\end{eqnarray}
The equations used to explicitly evaluate B-functions are:-
\begin{eqnarray}
\label{B def}
& &B_0= \Delta-\left[ln(p^2 -i\epsilon)+\sum{ln(1-x_j)+F(1,x_j)}\right],\nonumber\\
& & B_1= -\frac{1}{2}\Delta+\frac{1}{2}\left[ln(p^2 -i\epsilon)+\sum{ln(1-x_j)+F(2,x_j)}\right].\nonumber\\
\end{eqnarray}
where $\Delta$ is the divergent pole in $D$(dimensionality)-4, and $x_i$ are roots of equation $-p^2 x^2+\left(p^2+m_2^2-m_1^2\right)x + m_1^2=0$,
and:
\begin{equation}
\label{F function}
\int_0^1 x^n \log \left(x-x_1\right) \, dx =\frac{1}{n+1}\left[{ln(1-x_1)+F(n+1,x_1)}\right].
\end{equation}

\end{itemize}

 Using (\ref{Cdef}-\ref{F function}), values of relevant two point, three point and other functions in our case are given as:- \\
\underline{R functions}
\begin{eqnarray}
\label{eq: R functions}
& & \begin{array}{l}
R^{(a)}_1=0.75, R^{(a)}_2= 0.7; \nonumber\\
R^{(b)}_1= -24.3, R^{(b)}_2= -29.4; \nonumber\\
\end{array} \nonumber\\
& & \begin{array}{l}
R^{(b)}_3= R^{(b)}_4 = R^{(b)}_5 = R^{(b)}_6= O(10){\cal V}^2. \nonumber\\
\end{array} \nonumber\\
\end{eqnarray}
\underline{Two point functions}:- The functions evaluated using (\ref{B def}) and (\ref{F function}) also  contains UV divergent piece as given below:
\begin{eqnarray}
\label{eq: two point functions}
& & \begin{array}{l}
\label{array:11}
B^{(a)}_0(1,2)= \Delta - 53.2,  B^{(a)}_0(1,3)= \Delta - 54.6;\nonumber\\
B^{(a)}_0(2,3)= \Delta - 55.2,  B^{(b)}_0(1,2)= \Delta - 54.6;\nonumber\\
B^{(b)}_0(1,3)= \Delta + 4.6,   B^{(b)}_0(2,3)=\Delta + 53.2 ;\nonumber\\
\end{array}\nonumber\\
& & \begin{array}{l}
B^{(b)}_1(1,2)=-\frac{1}{2}\Delta + 2{\cal V}^2 \sim  -\frac{1}{2}\Delta + 2\times 10^{12};\nonumber\\
B^{(b)}_1(1,3)=-\frac{1}{2}\Delta - 2.3;\nonumber\\
B^{(b)}_1(2,3)= -\frac{1}{2}\Delta -0.5{\cal V}^2 \sim -\frac{1}{2}\Delta - 0.5\times 10^{12}.\nonumber\\
\end{array} \nonumber\\
\end{eqnarray}
\underline{Three point one loop functions}: The functions have been evaluated using formulae given in \ref{Cdef}. UV  divergent piece get cancelled for all C's functions except $C_{24}$. As from (\ref{Cdef}),\\

$ C_{24}= -\frac{1}{2} C_0 m_1^2+\frac{1}{4} \left(B_0(1,3)-C_{11} f_1-C_{12}
   f_2\right)+\frac{1}{4}$\\
here, $B_0(1,3)$ is UV divergent while all other quantities are finite, putting values, one gets,
$C^{(a)}_{24}= \Delta + O(1){\cal V}^2$. considering finite piece of $C_{24}$ and calculating all other C's functions using (\ref{Cdef}-\ref{eq:C0}), results are given below:
\begin{eqnarray}
\label{eq: three point functions}
& & \begin{array}{l}
C^{(a)}_{24}= O(1){\cal V}^2, C^{(b)}_{24}=  O(1){\cal V}^2 ;\nonumber\\
C^{(a)}_0= \frac{{0.3}{\cal V}^2}{M^2_{P}}{GeV}^{-2}\sim {0.3}\times {10}^{-24}{GeV}^{-2};\nonumber\\
C^{(b)}_{11}= \frac{{-7.8}{\cal V}^4}{M^2_{P}}{GeV}^{-2}\sim {-7.8}\times {10}^{-12} {GeV}^{-2};\nonumber\\
C^{(b)}_{12}= \frac{{-8.1}{\cal V}^4}{M^2_{P}}{GeV}^{-2}\sim {-8.1}\times {10}^{-12} {GeV}^{-2};\nonumber\\
\end{array}\nonumber\\
& & \begin{array}{l}
C^{(b)}_{21}= C^{(b)}_{22}= C^{(b)}_{23}\sim O(10){GeV}^{-2};\nonumber\\
C^{(b)}_0= \frac{{O({10}^{-3})}{\cal V}^2}{M^2_{P}}{GeV}^{-2}\sim  {10}^{-27}{GeV}^{-2}.\nonumber\\
\end{array} \nonumber\\
\end{eqnarray}
Now, equation(\ref{eq:M_I}) can be evaluated to yield:
\begin{eqnarray}
\label{eq:M_II}
& & {\bar u}(p_{\tilde{\chi}_3^0})\Biggl(\tilde{f}^3{\cal V}^{-\frac{61}{18}}\Biggl[\left\{{\bar\sigma}\cdot p_{\tilde{\chi}_3^0}C^{(a)}_{11}+{\bar\sigma}\cdot p_{g_\mu}C^{(a)}_{12}\right\}(2\epsilon\cdot p_{\tilde{\chi}_3^0})+\left\{{\bar\sigma}\cdot p_{\tilde{\chi}_3^0}\epsilon\cdot p_{\tilde{\chi}_3^0}C^{(a)}_{21}+{\bar\sigma}\cdot p_{g_\mu}\epsilon\cdot p_{\tilde{\chi}_3^0}C^{(a)}_{23}
+{\bar\sigma}\cdot\epsilon C^{(a)}_{24}\right\}\Biggr]\nonumber\\
& & +\tilde{f}^3{\cal V}^{-\frac{50}{18}}\Biggl[-\left\{{\bar\sigma}\cdot p_{\tilde{\chi}_3^0}C^{(b)}_{11}
+{\bar\sigma}\cdot p_{\tilde{g}}C^{(b)}_{12}\right\}{\bar\sigma}\cdot\epsilon{\bar\sigma}\cdot p_{g_\mu}+
{\bar\sigma}\cdot p_{\tilde{\chi}_3^0}{\bar\sigma}\cdot\epsilon {\bar\sigma}\cdot p_{\tilde{\chi}_3^0}C^{(b)}_{21} + {\bar\sigma}\cdot p_{\tilde{g}}{\bar\sigma}\cdot\epsilon {\bar\sigma}\cdot p_{\tilde{g}}C^{(b)}_{22}\nonumber\\
& & -\biggl({\bar\sigma}\cdot p_{\tilde{\chi}_3^0}{\bar\sigma}\cdot\epsilon{\bar\sigma}\cdot p_{\tilde{g}} + {\bar\sigma}\cdot p_{\tilde{g}}{\bar\sigma}\cdot\epsilon {\bar\sigma}\cdot p_{\tilde{\chi}_3^0}\biggr)C^{(b)}_{23}+ {\bar\sigma}_\mu{\bar\sigma}\cdot\epsilon{\bar\sigma}^\mu C^{(b)}_{24}\Biggr]\Biggr)
u(p_{\tilde{g}}),
\end{eqnarray}
which equivalently could be rewritten as:
\begin{equation}
\label{eq:M_III}
\tilde{f}^3{\bar u}(p_{\tilde{\chi}_3^0})\Biggl[{\bar\sigma}\cdot{\cal A}+{\bar\sigma}\cdot p_{\tilde{\chi}_3^0}{\bar\sigma}\cdot\epsilon{\bar\sigma}\cdot{\cal B}_1 + {\bar\sigma}\cdot p_{g_\mu}
{\bar\sigma}\cdot\epsilon{\bar\sigma}\cdot{\cal B}_2 + D_3{\bar\sigma}_\mu{\bar\sigma}\cdot\epsilon{\bar\sigma}^\mu
C^{(b)}_{24} \Biggr]u(p_{\tilde{g}}),
\end{equation}
where
\begin{eqnarray}
\label{eq:AB1B2}
& & {\cal A}^\mu\equiv {\cal V}^{-\frac{61}{18}}\left[p_{\tilde{\chi}_3^0}^\mu\epsilon\cdot p_{\tilde{\chi}_3^0}\left(2C^{(a)}_{11} + C^{(a)}_{21}\right) + p_{g_\mu}^\mu\epsilon\cdot p_{\tilde{\chi}_3^0}\left(C^{(a)}_{12} + C^{(a)}_{23}\right) + \epsilon^\mu C^{(a)}_{24}\right];\nonumber\\
& & {\cal B}_1^\mu\equiv {\cal V}^{-\frac{50}{18}}\left[-p_{g_\mu}^\mu\left(C^{(b)}_{11} + C^{(b)}_{12} + C^{(b)}_{23}-C^{(b)}_{22}\right) + p_{\tilde{\chi}_3^0}^\mu\left(C^{(b)}_{21} + C^{(b)}_{22} - 2C^{(b)}_{23}\right)\right];\nonumber\\
& & {\cal B}_2^\mu\equiv {\cal V}^{-\frac{50}{18}}\left[p_{g_\mu}^\mu\left(C^{(b)}_{12}+C^{(b)}_{22}\right) + p_{\tilde{\chi}_3^0}^\mu\left( C^{(b)}_{22} - C^{(b)}_{23}\right)\right];\nonumber\\
& & {D_3}\equiv {\cal V}^{-\frac{50}{18}}
\end{eqnarray}
Replacing ${\bar u}(p_{\tilde{\chi}_3^0}){\bar\sigma}\cdot p_{\tilde{\chi}_3^0}$ by $m_{\tilde{\chi}_3^0}{\bar u}(p_{\tilde{\chi}_3^0})$ and ${\bar\sigma}\cdot p_{\tilde{g}}u(p_{\tilde{g}})$ by $m_{\tilde {g}}$, and using $\epsilon\cdot p_{\tilde{\chi}_3^0}=0$,(\ref{eq:M_III}) be simplified to:
\begin{equation}
\label{eq:M_IV}
{\cal M}\sim\tilde{f}^3{\bar u}(p_{\tilde{\chi}_3^0})\left[A{\bar\sigma}\cdot\epsilon + B_1{\bar\sigma}\cdot\epsilon
{\bar\sigma}\cdot p_{\tilde{\chi}_3^0} + B_2{\bar\sigma}\cdot p_{\tilde{g}}{\bar\sigma}\cdot\epsilon +
D_1{\bar\sigma}\cdot p_{\tilde{g}}{\bar\sigma}\cdot\epsilon{\bar\sigma}\cdot p_{\tilde{\chi}_3^0} + {D_3} {\bar\sigma}_\mu{\bar\sigma}\cdot\epsilon{\bar\sigma}^\mu C^{(b)}_{24}\right]u(p_{\tilde{g}}),
\end{equation}
where
\begin{eqnarray}
\label{eq:AB1B2D1 defs}
& & A\equiv {\cal V}^{-\frac{61}{18}}C^{(a)}_{24} - m_{\tilde{g}}m_{\tilde{\chi}_3^0} {\cal V}^{-\frac{50}{18}}\Biggl\{
 C^{(b)}_{11} + 2 C^{(b)}_{12} + C^{(b)}_{23}-C^{(b)}_{22}\Biggr\},\nonumber\\
 & & B_1\equiv {\cal V}^{-\frac{50}{18}}\left(C^{(b)}_{11} + 2 C^{(b)}_{12} +  C^{(b)}_{21}\right)m_{\tilde{\chi}_3^0},\nonumber\\
 & & B_2\equiv {\cal V}^{-\frac{50}{18}}\left(C^{(b)}_{12}+C^{(b)}_{22}\right)m_{\tilde{g}},\nonumber\\
 & & D_1\equiv {\cal V}^{-\frac{50}{18}}\left(- C^{(b)}_{12}- C^{(b)}_{23}\right).
\end{eqnarray}
Strictly speaking, one also needs to add to (\ref{eq:M_II}) the contribution of one-loop graphs wherein the direction of arrows is opposite to the one considered in the above calculation. This will have the effect of $p_{\tilde{g},\tilde{\chi}_3^0,g_\mu}\rightarrow -p_{\tilde{g},\tilde{\chi}_3^0,g_\mu}$ to which the one-loop integrals are insensitive, as well as $\left\{{\bar\sigma}\cdot p_{\tilde{\chi}_3^0}C^{(b)}_{11}
+{\bar\sigma}\cdot p_{\tilde{g}}C^{(b)}_{12}\right\}{\bar\sigma}\cdot\epsilon{\bar\sigma}\cdot p_{g_\mu}\rightarrow
{\bar\sigma}\cdot p_{g_\mu}{\bar\sigma}\cdot\epsilon\left\{{\bar\sigma}\cdot p_{\tilde{\chi}_3^0}C^{(b)}_{11}
+{\bar\sigma}\cdot p_{\tilde{g}}C^{(b)}_{12}\right\}$ in the second line of (\ref{eq:M_II}). However, this does not change the estimate of the decay width in what follows in which we do not worry about adding these contributions. Utilizing values of C's calculated in (\ref{eq: three point functions}),
 \begin{eqnarray}
 \label{eq: A, B and D functions}
& & \begin{array}{l}
 A \sim O(1){\cal V}^{-0.8}, B_1\sim O(1){\cal V}^{-1.8}{GeV}^{-1}, B_2\sim O(10){\cal V}^{-1.8}{GeV}^{-1},  D_1\sim O(10){\cal V}^{-2.8}{GeV}^{-2};{D_3}\sim {\cal V}^{-\frac{50}{18}};\nonumber\\
 \end{array}\nonumber\\
\end{eqnarray}
\begin{eqnarray}
\label{eq:spinavGamma}
& &  \sum_{\tilde{g}\ {\rm and}\ \tilde{\chi}_3^0\ {\rm spins}}|{\cal M}|^2
\sim \tilde{f}^6 Tr\Biggl(\sigma\cdot p_{\tilde{\chi}_3^0}\left[A{\bar\sigma}\cdot\epsilon + B_1{\bar\sigma}\cdot\epsilon
{\bar\sigma}\cdot p_{\tilde{\chi}_3^0} + B_2{\bar\sigma}\cdot p_{\tilde{g}}{\bar\sigma}\cdot\epsilon +
D_1{\bar\sigma}\cdot p_{\tilde{g}}{\bar\sigma}\cdot\epsilon{\bar\sigma}\cdot p_{\tilde{\chi}_3^0} + D_3 {\bar\sigma}_\mu{\bar\sigma}\cdot\epsilon{\bar\sigma}^\mu C^{(b)}_{24}\right]\nonumber\\
& & \times\sigma\cdot p_{\tilde{g}}\left[A{\bar\sigma}\cdot\epsilon + B_1{\bar\sigma}\cdot\epsilon
{\bar\sigma}\cdot p_{\tilde{\chi}_3^0} + B_2{\bar\sigma}\cdot p_{\tilde{g}}{\bar\sigma}\cdot\epsilon +
D_1{\bar\sigma}\cdot p_{\tilde{g}}{\bar\sigma}\cdot\epsilon{\bar\sigma}\cdot p_{\tilde{\chi}_3^0} + D_3 {\bar\sigma}_\mu{\bar\sigma}\cdot\epsilon{\bar\sigma}^\mu C^{(b)}_{24}\right]^\dagger\Biggr),
\end{eqnarray}
which at:
\begin{equation}
\label{eq:kinematic_point}
p^0_{\tilde{\chi}_3^0}=\sqrt{m_{\tilde{\chi}_3^0}^2c^4+\rho^2},
p^1_{\tilde{\chi}_3^0}=p^2_{\tilde{\chi}_3^0}=p^3_{\tilde{\chi}_3^0}=\frac{\rho}{\sqrt{3}}
=\frac{1}{\sqrt{3}}\frac{c\left(m_{\tilde{g}}^2 - m_{\tilde{\chi}_3^0}^2\right)}{2m_{\tilde{g}}},
\end{equation}
yields:
\begin{eqnarray}
\label{eq:trace}
& &  \hskip -0.6in\frac{\tilde{f}^6}{256}{m_{\tilde{g}}}^2 \left[6 {m_{\tilde{g}}}^2 ({B_1}+{D_1} {m_{\tilde{g}}})^2+\left\{8
   {A_1}+16 D_3{C_{24}}+{m_{\tilde{g}}} \left(\left(5+\sqrt{3}\right) {B_1}+8 {B_2}+\left(5+\sqrt{3}\right) {D_1}
   {m_{\tilde{g}}}\right)\right\}^2\right],\nonumber\\
   & &
\end{eqnarray}
in the rest frame of the gluino.

Incorporating results of (\ref{eq:AB1B2D1 defs})in equation (\ref{eq:trace}), one gets
\begin{eqnarray}
\label{eq:spinavGamma result}
& &  \sum_{\tilde{g}\ {\rm and}\ \tilde{\chi}_3^0\ {\rm spins}}|{\cal M}|^2 \sim O(100){\tilde{f}^6}{D_3}^2 {\cal V}^4 m_{\tilde{g}}^2
\end{eqnarray}
Now, using standard two-body decay results (See \cite{Griffiths_particle}), the decay width $\Gamma$ is given by the following expression:
\begin{equation}
\label{eq:Gamma}
\Gamma=\frac{\sum_{\tilde{g}\ {\rm and}\ \tilde{\chi}_3^0\ {\rm spins}}|{\cal M}|^2\left(m_{\tilde{g}}^2-m_{\tilde{\chi}_3^0}^2\right)}{16\pi\hbar m_{\tilde{g}}^3}
\end{equation}

Using result of (\ref{eq:spinavGamma}) and $m_{\tilde{g}}\sim {10}^6$ GeV, $m_{\tilde{\chi_3^0}}\sim {\frac{1}{2}}m_{\tilde{g}}\sim \frac{1}{2}\times {10}^6$ GeV, two body decay width is given as:
\begin{equation}
\label{eq:Gamma1}
\Gamma= \frac{3}{4} \frac{O(100) \tilde{f}^6{D_3^2}{{\cal V}^4}{m_{\tilde {g}}}^4} {16\pi m_{\tilde{g}}^3}\sim \tilde{f}^6 {10}^{-4} GeV
\end{equation}
since $\tilde{f}^2 < {10}^{-8}$ as calculated above, $\Gamma < {10}^{5}$ GeV.  Life time of gluino is given as:
   \begin{eqnarray}
   \tau &=&\frac{\hbar}{\Gamma}\sim\frac{10^{-34} Jsec}{10^{-4}f^6 GeV}\sim\frac{10^{-20}}{f^6} sec > {10}^{4}sec
   \end{eqnarray}
   \subsection{Gluino($\tilde{g}$) decays into Goldstino($\tilde{G}$)}

\begin{itemize}
\item
We first consider the three-body decay of the gluino into Goldstino and a quark and anti-quark: $\tilde{g}\rightarrow\tilde{G}+q+{\bar q}$. For simplicity, we will
consider only the Wilson line modulus $a_1$ and its ferimonic superpartner. So, as in the three-body of the gluino into a neutralino and a quark and anti-quark, the following tree-level diagrams are relevant:
\begin{center}
\begin{picture}(500,180)(50,0)
\Text(90,130)[]{$\tilde{g}$}
\Line(60,120)(110,120)
\Gluon(60,120)(110,120){5}{4}
\ArrowLine (140,150)(110,120)
\Text(147,150)[]{$\bar{q}_{a_1}$}
\DashArrowLine (110,120)(130,90){4}
\Text(107,100)[]{$\tilde{q}_{a_1}$}
\ArrowLine(130,90)(160,120)
\Text(167,120)[]{$\tilde{G}$\hskip 1.5cm +}
\ArrowLine(130,90)(160,60)
\Text(167,65)[]{${q_{a_1}}$}
\Text(250,130)[]{$\tilde{g}$}
\Line(210,120)(260,120)
\Gluon(210,120)(260,120){5}{4}
\ArrowLine (290,150)(260,120)
\Text(297,150)[]{$q_{a_1}$}
\DashArrowLine(260,120)(280,90){4}{}
\Text(257,100)[]{$\tilde{q}_{a_1}$}
\ArrowLine(280,90)(310,120)
\Text(320,120)[]{$\bar{q}_{a_1}$}
\ArrowLine(280,90)(310,60)
\Text(317,65)[]{$\tilde{G}$}
\end{picture}
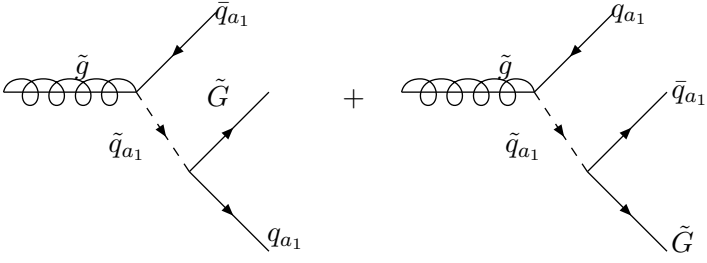
{\sl Figure 9: Three body Gluino decay into Gravitino with the assumption that both quarks and squarks have been approximated by (the superpartner of the) Wilson line modulus $a_1$}
\end{center}
The gluino-(anti-)quark-squark vertex will again be given by (\ref{eq:gluino-quark-squark}). The gravitino-quark-squark vertex would come from a term of the type ${\bar\psi}_\mu\tilde{q}_Rq_LH_L$ where $\psi_\mu$ is the gravitino field. After spontaneous breaking of the EW symmetry by a non-zero vev to $H^0$, the above yields: $\langle H^0\rangle{\bar\psi}_\mu
\tilde{q}_Rq_L$, which in ${\cal N}=1$ gauged supergravity lagrangian of \cite{Wess_Bagger} is given by:
\begin{equation}
\label{eq:gravitino-q-sq}
-g_{I{\bar J}}\left(\partial_\mu {\bar a}^{\bar J}\right)\chi^I\sigma^\nu{\bar\sigma}_\mu\psi_\nu - \frac{i}{2}e^{\frac{K}{2}}\left(D_IW\right)\chi^I\sigma^\mu{\bar\psi}_\mu + {\rm h.c.}.
\end{equation}
From \cite{gravitinomodexp}, the gravitino field can be decomposed into the spin-$\frac{1}{2}$ Goldstino field $\tilde{G}$
via:
\begin{equation}
\label{eq:Goldstino_decomp}
\psi_\nu=\rho_\nu + \sigma_\nu\tilde{G},\ \tilde{G}=-\frac{1}{3}\sigma^\mu\psi_\mu,
\end{equation}
$\rho_\nu$ being a spin-$\frac{3}{2}$ field.
Hence, the Goldstino-content of (\ref{eq:gravitino-q-sq}), using $\sigma^\nu{\bar\sigma}^\mu\sigma_\nu=-2{\bar\sigma}^\nu$,
is given by:
\begin{equation}
\label{eq:Goldstino-s-sq_I}
2g_{I{\bar J}}\left(\partial_\mu {\bar a}^{\bar J}\right)\chi^I{\bar\sigma}^\mu\tilde{G} + \frac{3i}{2}e^{\frac{K}{2}}\left(D_IW\right)\chi^I\tilde{G} + {\rm h.c.}
\end{equation}
Now,
utilizing:
\begin{eqnarray}
\label{eq:Goldstino-s-sq_II}
& & g_{a_1{\bar a}_{\bar 1}}\sim z_i{\cal V}^{\frac{11}{18}}\Biggr|_{z_i\rightarrow\langle z_i\rangle\sim{\cal V}^{\frac{1}{36}}}\sim{\cal V}^{\frac{23}{36}}\nonumber\\
& & e^{\frac{K}{2}}D_{a_1}W\Biggr|_{a_1\rightarrow a_1+{\cal V}^{-\frac{1}{4}}}\sim z_i{\cal V}^{-\frac{11}{9}}a_1|_{z_i\rightarrow\langle z_i\rangle\sim{\cal V}^{\frac{1}{36}}}\sim{\cal V}^{-\frac{43}{36}}a_1,
\end{eqnarray}
one obtains:
\begin{center}
\begin{picture}(500,180)(-100,0)
\Text(180,120)[]{$\frac{\left({\cal V}^{\frac{23}{36}}{\bar\sigma}\cdot\frac{p_{\tilde{q}}}{M_p} + {\cal V}^{-\frac{43}{36}}\right)}{\left(\sqrt{\hat{K}_{{\cal A}_1{\bar {\cal A}}_1}}\right)^2\sim{\cal V}^{\frac{31}{36}}}
\sim {\cal V}^{-\frac{8}{36}}{\bar\sigma}\cdot\frac{p_{\tilde{q}}}{M_p} + {\cal V}^{-\frac{37}{18}}$}
\ArrowLine(-50,120)(0,120)
\Text(-58,120)[]{$q_I$}
\Text(47,158)[]{$\tilde{G}$}
\ArrowLine(0,120)(40,150)
\Text(47,83)[]{$ \tilde{q}_R $}
\DashArrowLine(0,120)(40,90){5}
\end{picture}
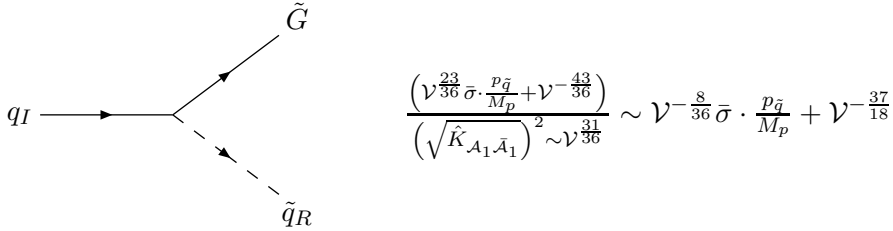
{\sl Figure 10: The Goldstino-quark-squark vertex}
\end{center}

For this particular case:-

\hskip -0.4in{\small $A_{ij}\left(Tr\left[G^{{\bar q}_{a_1}}_{\tilde{q}_{a_1}}G^{{\bar q}_{a_1}}_{\tilde{q}_{a_1}}\ ^\dagger\right]
Tr\left[\tilde{G}^{q_{a_1}}_{\tilde{q}_{a_1}}\tilde{G}^{q_{a_1}}_{\tilde{q}_{a_1}}\ ^\dagger\right]\sim\tilde{f}^2{\cal V}^{-\frac{37}{18}}\left\{Tr\left[{\cal V}^{-\frac{8}{36}}{\bar\sigma}\cdot\frac{\left(p_{\tilde{G}}+p_q\right)}{M_p} + {\cal V}^{-\frac{37}{18}}{\bf 1}\right]^2\sim 2\left({\cal V}^{-\frac{37}{9}}+{\cal V}^{-\frac{8}{18}}\frac{\left({\bf p}_{\tilde{G}}+{\bf p}_{q}\right)^2}{M_p^2}\right)\right\}\sim\tilde{f}^2{\cal V}^{-\frac{37}{6}}\right)$},

 {\small \hskip-0.64in$B_{ij}\left(Tr\left[G^{q_{a_1}}_{\tilde{q}_{a_1}}G^{q_{a_1}}_{\tilde{q}_{a_1}}\ ^\dagger\right]
Tr\left[\tilde{G}^{{\bar q}_{a_1}}_{\tilde{q}_{a_1}}\tilde{G}^{{\bar q}_{a_1}}_{\tilde{q}_{a_1}}\ ^\dagger\right]\sim\tilde{f}^2{\cal V}^{-\frac{37}{18}}\left\{Tr\left[{\cal V}^{-\frac{8}{36}}{\bar\sigma}\cdot\frac{\left(p_{\tilde{G}}+p_q\right)}{M_p} + {\cal V}^{-\frac{37}{18}}{\bf 1}\right]^2\sim 2\left({\cal V}^{-\frac{37}{9}}+{\cal V}^{-\frac{8}{18}}\frac{\left({\bf p}_{\tilde{G}}+{\bf p}_{q}\right)^2}{M_p^2}\right)\right\}\sim\tilde{f}^2{\cal V}^{-\frac{37}{6}}\right)$},

\hskip-0.6in{\small $C\Biggl(Tr\left[G^{{\bar q}_{a_1}}_{\tilde{q}_{a_1}}G^{q_{a_1}}_{\tilde{q}_{a_1}}\ ^\dagger \tilde{G}^{q_{a_1}}_{\tilde{q}_{a_1}}\tilde{G}^{{\bar q}_{a_1}}_{\tilde{q}_{a_1}}\ ^\dagger\right]\sim\tilde{f}^2{\cal V}^{-\frac{37}{18}}\Biggl\{Tr\left[\left({\cal V}^{-\frac{8}{36}}{\bar\sigma}\cdot\frac{\left(p_{\tilde{G}}+p_q\right)}{M_p} + {\cal V}^{-\frac{37}{18}}{\bf 1}\right)\left({\cal V}^{-\frac{8}{36}}{\bar\sigma}\cdot\frac{\left(p_{\tilde{G}}+p_{\bar q}\right)}{M_p} + {\cal V}^{-\frac{37}{18}}{\bf 1}\right)\right]$} \\
{\small $\sim2\left({\cal V}^{-\frac{37}{9}}+{\cal V}^{-\frac{8}{18}}\frac{\left({\bf p}_{\tilde{G}}+{\bf p}_{q}\right)\cdot\left({\bf p}_{\tilde{G}}+{\bf p}_{\bar q}\right)}{M_p^2}\right)\Biggr\}
\sim\tilde{f}^2{\cal V}^{-\frac{37}{6}}\Biggr)$}

Utilizing the values of vertex elements calculated above and from (\ref{eq:neutralinowidth}), we can calculate decay width for Gluino in this particular case. Using (\ref{eq:limits}), limits of integration in this case are:
$$s_{23 \max }= \left(m_{\frac{3}{2}}-m_q\right){}^2,\ s_{23 \min }= m_q^2;$$
$$s_{13 \max }= m_{\frac{3}{2}}^2-s_{23}, s_{13 \min }= 0$$
where $m_{\tilde{g}}= m_{\frac{3}{2}} \sim {10}^6 GeV, m_{\tilde{G}}= 0$. The decay width of Gluino  is given as:
\begin{eqnarray}
 \label{eq: Decay width goldstino}
 \hspace{-.5cm}\Gamma(\tilde{g}\to\chi_{\rm n}^{o}q_{{}_I} \bar{q}_{{}_J} )
&\sim&{g_s^2\over256 \pi^3 m_{\frac{3}{2}}^3 }\left[-
\tilde{f}^2{\cal V}^{-\frac{37}{6}} m_{\frac{3}{2}}^4 {18{\cal V}^4}+ 9 \tilde{f}^2{\cal V}^{-\frac{37}{6}}{\cal V}^4 m_{\frac{3}{2}}^4 - 8\tilde{f}^2{\cal V}^{-\frac{37}{6}} {\cal V}^2 m_q m_{\frac{3}{2}}^3\right] \nonumber\\
& & \sim { g_s^2\over256 \pi^3 m_{\frac{3}{2}}^3 }( \tilde{f}^2 O(10){\cal V}^{-6}{\cal V}^4 m_{\frac{3}{2}}^4 )\sim O(10^{-3}){\cal V}^{-2}\tilde{f}^2 m_{\frac{3}{2}} \nonumber\\
& & \sim O(10^{3}){\cal V}^{-2}\tilde{f}^2 GeV < O(10^{-{17}})GeV
 \end{eqnarray}
The life time of gluino is given as:
   \begin{eqnarray}
   \tau &=&\frac{\hbar}{\Gamma}\sim\frac{10^{-34} Jsec}{10^{-9}f^2 GeV}\sim\frac{10^{-15}}{f^2}> 10^{-{7}}sec
   \end{eqnarray}

\item
We now consider the two-body decay of the gluino into a Goldstino and a gluon:
\begin{center}
\begin{picture}(1000,200)(50,0)
\Text(140,130)[]{$\tilde{g}$}
\Line(110,120)(160,120)
\Gluon(110,120)(160,120){5}{4}
\DashArrowLine (160,120)(190,150){4}
\Text(180,152)[]{$\tilde{q}_R$}
\ArrowLine(190,90)(160,120)
\Text(170,100)[]{${q_I}$}
\ArrowLine(190,150)(190,90)
\Gluon(190,90)(230,90){5}{4}
\Text(240,90)[]{$g_\mu$}
\ArrowLine(190,150)(230,150)
\Text(235,150)[]{$\tilde{G}$}
\Text(180,30)[]{(b)}
\Line(310,120)(360,120)
\Gluon(310,120)(360,120){5}{4}
\Text(340,130)[]{$\tilde{g}$}
\ArrowLine (360,120)(390,150)
\Text(380,152)[]{$q_I$}
\DashArrowLine (390,90)(360,120){4}
\Text(370,100)[]{$\tilde{q}_R$}
\DashArrowLine (390,150)(390,90){4}
\Gluon(390,90)(430,90){5}{4}
\Text(440,90)[]{$g_\mu$}
\ArrowLine(390,150)(430,150)
\Text(440,150)[]{$\tilde{G}$}
\Text(380,30)[]{(a)}
\end{picture}
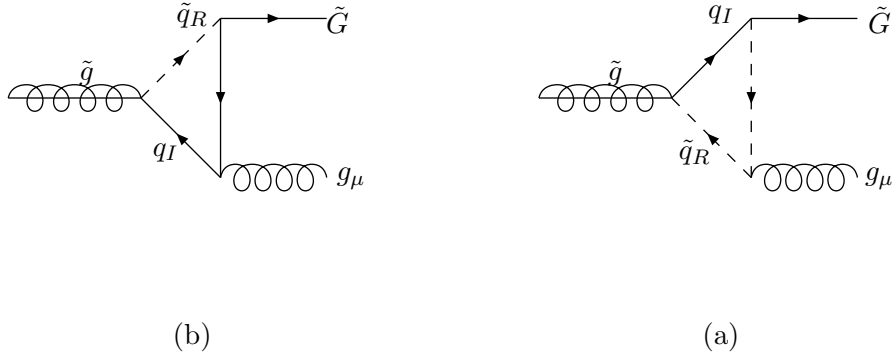
{\sl Figure 11:  Diagrams contributing to one-loop Gluino decay into Goldstino and gluon}
\end{center}
\end{itemize}
The matrix element for the above will be given by:
\begin{eqnarray}
\label{eq:MGold_I}
& & {\cal M}\sim{\tilde{f}}^2\int\frac{d^4k}{\left(2\pi\right)^4}\nonumber\\
& & \times{\cal V}^{-\frac{37}{36}}\left(\frac{i{\bar\sigma}\cdot k}{k^2-m_q^2+i\epsilon}\right)\left({\cal V}^{-\frac{37}{18}}+{\cal V}^{-\frac{8}{36}}\frac{{\bf\sigma}\cdot({\bf -p_{\tilde{G}}+k})}{M_p}\right)\left(\frac{i}{\left[\left(k-p_{\tilde{G}}\right)^2-m^2_{\tilde{q}}+i\epsilon\right]}
\right)\left({\cal V}^{-\frac{4}{3}}\epsilon\cdot\left(2k-p_{\tilde{G}}-p_{\tilde{g}}\right)\right)\nonumber\\
& & \times\left(\frac{i}
{\left[\left(k-p_{\tilde{g}}\right)^2-m^2_{\tilde{q}}+i\epsilon\right]}\right)\nonumber\\
& & + {\tilde{f}}^2\int\frac{d^4k}{\left(2\pi\right)^4}{\cal V}^{-\frac{37}{36}}\left(\frac{i}
{\left[\left(k+p_{\tilde{G}}\right)^2-m^2_{\tilde{q}}+i\epsilon\right]}\right)\left({\cal V}^{-\frac{37}{18}}+{\cal V}^{-\frac{8}{36}}\frac{{\bf\sigma}\cdot({\bf p_{\tilde{G}}+k})}{M_p}\right)\left(\frac{i{\bar\sigma}\cdot k}{k^2-m_q^2+i\epsilon}\right)\left({\cal V}^{-\frac{3}{4}}{\bar\sigma}\cdot\epsilon\right)\nonumber\\
& & \times\left(\frac{i{\bar\sigma}\cdot\left(k-p_{g_\mu}\right)}{\left[
\left(k-p_{g_\mu}\right)^2-m^2_q+i\epsilon\right]}\right)
\end{eqnarray}

Using the same approach as used in {\bf 4.2}, first we need to calculate relevant coupling at EW scale, Lagrangian and effective operators for Gluino-gluon-Goldstino coupling are given as \cite{Guidice_et_al}:
\begin{equation}
\label{eq:Laggold}
{\cal L} = \frac{1}{{\widetilde m}^2}\,\sum_{i=1}^5 \CG_i\,\QG_i\; \nonumber\\
\end{equation} and
\begin{eqnarray}
\QG_1= \ov{\wt{G}} \,\gamma^{\mu}\,
\gamma_{5}\, {\tilde g}^a\; \otimes\;\sum_{k=1,2 \atop q=u,d}\;
\ov{q}^{\,(k)} \,\gamma_{\mu} \, T^a \,q^{\,(k)},
\QG_2= \ov{\wt{G}} \,\gamma^{\mu}\,
\gamma_{5}\, {\tilde g}^a\; \otimes\;
\ov{q}\lle^{\,(3)} \,\gamma_{\mu} \, T^a \,q\lle^{\,(3)};\nonumber\\
\QG_3= \ov{\wt{G}} \,\gamma^{\mu}\,
\gamma_{5}\, {\tilde g}^a\; \otimes\;
\ov{t}\rr \,\gamma_{\mu} \, T^a \, t\rr,
\QG_4= \ov{\wt{G}} \,\gamma^{\mu}\,
\gamma_{5}\, {\tilde g}^a\; \otimes\;
\ov{b}\rr \,\gamma_{\mu}\,  T^a\, b\rr,
\QG_5= \ov{\wt{G}} \,\sigma^{\mu\nu} \,\gamma_5\,
 {\tilde g}^a\;G^a_{\mu\nu}
\end{eqnarray}
($\widetilde m$ is the squark mass scale)
Assuming all $ \CG_i$ (Wilson coefficient corresponding to aforementioned coupling) to be equal for $i=1,2,3,4$ in our set up, the RG evolution of same for the goldstino
operators has the simple analytic form given below:
\begin{eqnarray}
\label{eq:CG}
& & \CG_i (m_{EW})= \eta_s^{-\frac {9}{10}} [1 +O(1)y]\CG_i (\widetilde m), \CG_5(m_{EW})= \eta_s^{-\frac {7}{5}}\,\CG_5 (\widetilde m )\,
\end{eqnarray}
Using value of $\eta_s$= 0.66 and $y= -0.3$ calculated in section {\bf 4.1},
\begin{eqnarray}
\label{eq:CG1}
& & \CG_i (m_{EW})= 1.45~\CG_i (\widetilde m),  \CG_5(m_{EW})=1.8~\CG_i (\widetilde m),
\end{eqnarray}
and we therefore assume $\CG_i(m_{EW})\sim{\cal O}(1)~\CG_i(m_S)$, i.e., the Wilson coefficients corresponding to Gluino-Goldstino- Gluon coupling do not change much upon RG evolution to EW scale.

Analogous to {\bf 4.2}, (\ref{eq:MGold_I}) can be written as:
\begin{eqnarray}
\label{eq:MGold_II}
& & {\tilde{f}}^2{\cal V}^{-\frac{159}{36}}\Biggl[\left\{{\bar\sigma}\cdot p_{\tilde{G}}C^{(a)}_{11}+{\bar\sigma}\cdot p_{g_\mu}C^{(a)}_{12}\right\}(2\epsilon\cdot p_{\tilde{G}})+\left\{{\bar\sigma}\cdot p_{\tilde{G}}\epsilon\cdot p_{\tilde{G}}C^{(a)}_{21}+{\bar\sigma}\cdot p_{g_\mu}\epsilon\cdot p_{\tilde{G}}C^{(a)}_{23}
+{\bar\sigma}\cdot\epsilon C^{(a)}_{24}\right\}\Biggr]\nonumber\\
& & +\tilde{f}^2{\cal V}^{-\frac{23}{6}}\Biggl[-\left\{{\bar\sigma}\cdot p_{\tilde{G}}C^{(b)}_{11}
+{\bar\sigma}\cdot p_{\tilde{g}}C^{(b)}_{12}\right\}{\bar\sigma}\cdot\epsilon{\bar\sigma}\cdot p_{g_\mu}+
{\bar\sigma}\cdot p_{\tilde{G}}{\bar\sigma}\cdot\epsilon {\bar\sigma}\cdot p_{\tilde{G}}C^{(b)}_{21} + {\bar\sigma}\cdot p_{\tilde{g}}{\bar\sigma}\cdot\epsilon {\bar\sigma}\cdot p_{\tilde{g}}C^{(b)}_{22}\nonumber\\
& & -\biggl({\bar\sigma}\cdot p_{\tilde{G}}{\bar\sigma}\cdot\epsilon{\bar\sigma}\cdot p_{\tilde{g}} + {\bar\sigma}\cdot p_{\tilde{g}}{\bar\sigma}\cdot\epsilon {\bar\sigma}\cdot p_{\tilde{G}}\biggr)C^{(b)}_{23}+ {\bar\sigma}_\mu{\bar\sigma}\cdot\epsilon{\bar\sigma}^\mu C^{(b)}_{24}\Biggr],
\end{eqnarray}
which equivalently could be rewritten as:
\begin{equation}
\label{eq:MGold_III}
\tilde{f}^2{\bar u}(p_{\tilde{G}})\Biggl[{\bar\sigma}\cdot{\cal A}+{\bar\sigma}\cdot p_{\tilde{G}}{\bar\sigma}\cdot\epsilon{\bar\sigma}\cdot{\cal B}_1 + {\bar\sigma}\cdot p_{g_\mu}
{\bar\sigma}\cdot\epsilon{\bar\sigma}\cdot{\cal B}_2 + {D_4}{\bar\sigma}_\mu{\bar\sigma}\cdot\epsilon{\bar\sigma}^\mu
C^{(b)}_{24} \Biggr]u(p_{\tilde{g}}),
\end{equation}
where
\begin{eqnarray}
\label{eq:AB1B2}
& & {\cal A}^\mu\equiv {\cal V}^{-\frac{159}{36}}\left[p_{\tilde{G}}^\mu\epsilon\cdot p_{\tilde{G}}\left(2C^{(a)}_{11} + C^{(a)}_{21}\right) + p_{g_\mu}^\mu\epsilon\cdot p_{\tilde{G}}\left(C^{(a)}_{12} + C^{(a)}_{23}-C^{(b)}_{22}\right) + \epsilon^\mu C^{(a)}_{24}\right];\nonumber\\
& & {\cal B}_1^\mu\equiv {\cal V}^{-\frac{23}{6}}\left[-p_{g_\mu}^\mu\left(C^{(b)}_{11} + C^{(b)}_{12} + C^{(b)}_{23}-C^{(b)}_{22}\right) + p_{\tilde{G}}^\mu\left(C^{(b)}_{21} + C^{(b)}_{22} - 2C^{(b)}_{23}\right)\right];\nonumber\\
& & {\cal B}_2^\mu\equiv {\cal V}^{-\frac{23}{6}}\left[p_{g_\mu}^\mu\left(C^{(b)}_{12}+C^{(b)}_{22}\right) + p_{\tilde{G}}^\mu\left( C^{(b)}_{22} - C^{(b)}_{23}\right)\right];\nonumber\\
& & D_4 \equiv {\cal V}^{-\frac{23}{6}}.
\end{eqnarray}
This time around replacing ${\bar u}(p_{\tilde{G}}){\bar\sigma}\cdot p_{\tilde{G}}$ by  0 and ${\bar\sigma}\cdot p_{\tilde{g}}u(p_{\tilde{g}})$
by $m_{\tilde{g}}u(p_{\tilde{g}})$, (\ref{eq:MGold_III}) can be rewritten as:
\begin{equation}
\label{eq:MGold_IV}
{\tilde{f}}^2{\bar u}(p_{\tilde{G}})\left(A_2{\bar\sigma}\cdot\epsilon + B_3{\bar\sigma}\cdot p_{\tilde{g}}{\bar\sigma}\cdot\epsilon + D_2{\bar\sigma}\cdot p_{\tilde{g}}{\bar\sigma}\cdot\epsilon{\bar\sigma}\cdot p_{\tilde{G}} + D_4 {\bar\sigma}_\mu{\bar\sigma}\cdot\epsilon{\bar\sigma}^\mu C^{(b)}_{24}\right)u(p_{\tilde{g}}),
\end{equation}
where
\begin{eqnarray}
\label{eq:A2B3D2+defs}
& & A_2\equiv {\cal V}^{-\frac{159}{36}}C^{(a)}_{24}; B_3\equiv {\cal V}^{-\frac{23}{6}}M_{\tilde{g}}(C^{(b)}_{12}++C^{(b)}_{22}); D_2\equiv {\cal V}^{-\frac{23}{6}}\left(-C^{(b)}_{12} + C^{(b)}_{23}\right); D_4 \equiv {\cal V}^{-\frac{23}{6}}.
\end{eqnarray}

 Using equation (\ref{Cdef}-\ref{eq:C0}), Results of various C's functions required for this particular  case are:-
\begin{eqnarray}
\label{eq: three point goldstino functions}
& & \begin{array}{l}
C^{(a)}_{24}= -(0.25){\cal V}^2,C^{(b)}_{24}=  O(1){\cal V}^2;\nonumber\\
C^{(a)}_0= \frac{{0.2}{\cal V}^2}{M_p^2}{GeV}^{-2}\sim {0.2}\times {10}^{-24}{GeV}^{-2};\nonumber\\
C^{(b)}_{12}= \frac{{58}{\cal V}^4}{M_p^2}{GeV}^{-2}\sim {58}\times {10}^{-12} {GeV}^{-2};\nonumber\\
C^{(b)}_{11}= \frac{{56}{\cal V}^4}{M_p^2}{GeV}^{-2}\sim {56}\times {10}^{-12} {GeV}^{-2};\nonumber\\
\end{array}\nonumber\\
& & \begin{array}{l}
 C^{(b)}_{22}= C^{(b)}_{23}\sim -60 {GeV}^{-2};\nonumber\\
C^{(b)}_0= \frac{{0.2}{\cal V}^2}{M_p^2}{GeV}^{-2}\sim -{0.3}\times {10}^{-24}{GeV}^{-2}.\nonumber\\
\end{array} \nonumber\\
\end{eqnarray}

Utilizing (\ref{eq: three point goldstino functions}), one gets: $A_2\equiv O(1) {\cal V}^{-1.5}$, $B_3\equiv O(100){\cal V}^{-5}$, $D_2\equiv {\cal V}^{-4}, D_4 \equiv {\cal V}^{-\frac{23}{6}}$
\begin{eqnarray}
\label{eq:spinavGammaGold}
& &  \sum_{\tilde{g}\ {\rm and}\ \tilde{G}\ {\rm spins}}|{\cal M}|^2
\sim \tilde{f}^4 Tr\Biggl(\sigma\cdot p_{\tilde{G}}\left[A_2{\bar\sigma}\cdot\epsilon + B_3{\bar\sigma}\cdot p_{\tilde{g}}{\bar\sigma}\cdot\epsilon + D_2{\bar\sigma}\cdot p_{\tilde{g}}{\bar\sigma}\cdot\epsilon{\bar\sigma}\cdot p_{\tilde{G}} + D_4{\bar\sigma}_\mu{\bar\sigma}\cdot\epsilon{\bar\sigma}^\mu C^{(b)}_{24}\right]\nonumber\\
& & \times\sigma\cdot p_{\tilde{g}}\left[A_2{\bar\sigma}\cdot\epsilon + B_3{\bar\sigma}\cdot p_{\tilde{g}}{\bar\sigma}\cdot\epsilon + D_2{\bar\sigma}\cdot p_{\tilde{g}}{\bar\sigma}\cdot\epsilon{\bar\sigma}\cdot p_{\tilde{G}} + D_4{\bar\sigma}_\mu{\bar\sigma}\cdot\epsilon{\bar\sigma}^\mu C^{(b)}_{24}\right]^\dagger\Biggr),
\end{eqnarray}
which at:
\begin{equation}
\label{eq:kinematic_point}
p^0_{\tilde{G}}=m_{\tilde{g}}/2,p^1_{\tilde{G}}=p^2_{\tilde{G}}=p^3_{\tilde{G}}=\frac{m_{\tilde{g}}}{2\sqrt{3}},
\end{equation}
yields:
\begin{equation}
\label{eq:spinavgammaGold II}
\tilde{f}^4 {m_{\tilde{g}}}^2 \left[{D_2}^2 {m_{\tilde{g}}}^4+\frac{1}{6} \left(6 {A_2}+12{D_4}
   {C_{24}^{(b)}}+{m_{\tilde{g}}} \left(6 {B_3}+\left(3+\sqrt{3}\right) {D_2} {m_{\tilde{g}}}\right)\right)^2\right]\sim 144\tilde{f}^4 {m_{\tilde{g}}}^2 ({D_4}{C_{24}^{(b)}})^2.
\end{equation}
So, using results from \cite{Griffiths_particle}, the decay width comes out to be equal to:
\begin{equation}
\label{eq:GammaGold}
\Gamma=\frac{ \sum_{\tilde{g}\ {\rm and}\ \tilde{G}\ {\rm spins}}|{\cal M}|^2}{16\pi\hbar m_{\tilde{g}}}\sim (0.3)m_{\tilde{g}}\tilde{f}^4  ({D_4}{C_{24}^{(b)}})^2 \sim {10}^{-18}\tilde{f}^4  GeV.
\end{equation}
since $\tilde{f}^2 < {10}^{-8}$ as calculated above, $\Gamma < {10}^{14}$ GeV.  Life time of gluino is given as:
   \begin{eqnarray}
   \tau &=&\frac{\hbar}{\Gamma}\sim\frac{10^{-34} Jsec}{10^{-18}f^4 GeV}\sim\frac{10^{-6}}{f^4} sec > {10^{10}}sec
   \end{eqnarray}
\section{Geometric K\"{a}hler Potential for the Swiss-Cheese Calabi-Yau}

In principle, due to the presence of a mobile $D3$-brane, one must also include the geometric K\"{a}hler potential $K_{\rm geom}$ of the Swiss-Cheese Calabi-Yau in the moduli space K\"{a}hler potential. In \cite{D3_D7_Misra_Shukla}, given that we had restricted the mobile $D3$-brane to $\Sigma_B$, one had estimated (in the large volume limit) $K_{\rm geom}\sim
\frac{{\cal V}^{-\frac{1}{3}}}{\sqrt{ln {\cal V}}}$ summarized as follows. Using GLSM techniques and the toric data for the given Swiss-Cheese Calabi-Yau, the geometric K\"{a}hler potential for the divisor ${\Sigma_B}$ (and ${\Sigma_S}$) in the LVS limit was evaluated in \cite{D3_D7_Misra_Shukla} in terms of derivatives of genus-two Siegel theta functions as well as
two Fayet-Iliopoulos parameters corresponding to the two $C^*$ actions in the  two-dimensional ${\cal N}=2$ supersymmetric gauge theory whose target space is our toric variety Calabi-Yau, and a parameter $\zeta$ encoding the information about the $D3-$brane position moduli-independent (in the LVS limit) period matrix of the hyperelliptic curve $w^2=P(z)$, $P(z)$ being the sextic in the exponential of the vector superfields eliminated as auxiliary fields, corresponding to $\Sigma_B$. To be a bit more specific, one can show that upon elimination of the vector superfield (in the IR limit of the GLSM), one obtains an octic in $e^{2V_2}$, $V_2$ being one of the two real gauge superfields. Using Umemura's result \cite{Umemura} on expressing the roots of an algebraic polynomial of  degree $n$ in terms of Siegel theta functions
of genus $g(>1)=[(n+2)/2]$ :  $\theta\left[\begin{array}{c} \mu\\
\nu
\end{array}\right](z,\Omega)$ for $\mu,\nu\in{\bf R}^g, z\in {\bf C}^g$ and $\Omega$ being a complex symmetric
$g\times g$ period matrix with $Im(\Omega)>0$ defined as follows:
$$
\theta\left[\begin{array}{c} \mu\\
\nu
\end{array}\right](z,\Omega)=\sum_{n\in{\bf Z}^g}e^{i\pi(n+\mu)^T\Omega(n+\mu)+2i\pi(n+\mu)^T(z+\nu)}.$$
Hence for an octic, one needs to use Siegel theta functions of genus five. The period matrix $\Omega$ will be defined as follows:
$$\Omega_{ij}=\left(\sigma\right)^{-1}_{ik}\rho_{kj}$$
where $$\sigma_{ij}\equiv\oint_{A_j}dz \frac{z^{i-1}}{\sqrt{z(z-1)(z-2)P(z)}}$$ and
$$\rho_{ij}\equiv\oint_{B_j}\frac{z^{i-1}}{\sqrt{z(z-1)(z-2)P(z)}},$$
$\{A_i\}$ and $\{B_i\}$ being a canonical basis of cycles satisfying: $A_i\cdot A_j=B_i\cdot B_j=0$ and
$A_i\cdot B_j=\delta_{ij}$. Umemura's result then is that a root:
$$\hskip-0.3in\frac{1}{2\left(\theta\left[\begin{array}{ccccc}
\frac{1}{2} & 0 & 0 & 0 & 0 \\
0 & 0 & 0 & 0 & 0  \end{array}\right](0,\Omega)\right)^4
\left(\theta\left[\begin{array}{ccccc}
\frac{1}{2} & \frac{1}{2} & 0 & 0 & 0 \\
0 & 0 & 0 & 0 & 0  \end{array}\right](0,\Omega)\right)^4}$$
$$\hskip-0.3in\times\Biggl[\left(\theta\left[\begin{array}{ccccc}
\frac{1}{2} & 0 & 0 & 0 & 0 \\
0 & 0 & 0 & 0 & 0  \end{array}\right](0,\Omega)\right)^4\left(\theta\left[\begin{array}{ccccc}
\frac{1}{2} & \frac{1}{2} & 0 & 0 & 0 \\
0 & 0 & 0 & 0 & 0  \end{array}\right](0,\Omega)\right)^4$$
$$\hskip-0.3in+ \left(\theta\left[\begin{array}{ccccc}
0 & 0 & 0 & 0 & 0 \\
0 & 0 & 0 & 0 & 0  \end{array}\right](0,\Omega)\right)^4
\left(\theta\left[\begin{array}{ccccc}
0 & \frac{1}{2} &  0 & 0 & 0 \\
0 & 0 & 0 & 0 & 0  \end{array}\right](0,\Omega)\right)^4$$
$$\hskip-0.3in- \left(\theta\left[\begin{array}{ccccc}
0 & 0 & 0 & 0 & 0 \\
\frac{1}{2} & 0 & 0 & 0 & 0  \end{array}\right](0,\Omega)\right)^4
\left(\theta\left[\begin{array}{ccccc}
0 & \frac{1}{2} & 0 & 0 & 0 \\
\frac{1}{2} & 0 & 0 & 0 & 0 \end{array} \right](0,\Omega)\right)^4\Biggr].$$
In the LVS limit, the octic reduces to a sextic. Umemura's result would require the use of genus-four Siegel theta functions. However, using the results of
\cite{Zhivkov}, one can express the roots of a sextic in terms of derivatives of genus-two Siegel theta functions as follows:
$$\hskip-0.6in\left[\frac{\sigma_{22}\frac{d}{dz_1}\theta\left[\begin{array}{cc}
\frac{1}{2}&\frac{1}{2} \\
0&\frac{1}{2}
\end{array}\right]\left((z_1,z_2),\Omega\right)
- \sigma_{21}\frac{d}{dz_2}\theta\left[\begin{array}{cc}
\frac{1}{2}&\frac{1}{2} \\
0&\frac{1}{2}
\end{array}\right]\left((z_1,z_2),\Omega\right) }
{\sigma_{12}\frac{d}{dz_1}\theta\left[\begin{array}{cc}
\frac{1}{2}&\frac{1}{2} \\
0&\frac{1}{2}
\end{array}\right]\left((z_1,z_2),\Omega\right)
- \sigma_{12}\frac{d}{dz_2}\theta\left[\begin{array}{cc}
\frac{1}{2}&\frac{1}{2} \\
0&\frac{1}{2}
\end{array}\right]\left((z_1,z_2),\Omega\right)}\right]_{z_1=z_2=0},$$
etc.

 The symmetric period matrix corresponding to the hyperelliptic
curve $w^2=P(z)$ is given by:
$$\hskip-0.3in\left(\begin{array}{cc}
\Omega_{11} & \Omega_{12} \\
\Omega_{12} & \Omega_{22}
\end{array}\right)=\frac{1}{\sigma_{11}\sigma_{22}-\sigma_{12}\sigma_{21}}\left(\begin{array}{cc}
\sigma_{22} & -\sigma_{12} \\
-\sigma_{21} & \sigma_{11}
\end{array}\right)\left(\begin{array}{cc}
\rho_{11} & \rho_{12} \\
\rho_{21} & \rho_{22}
\end{array}\right),$$
where $\sigma_{ij}=\int_{z_*{A_j}}\frac{z^{i-1}dz}{\sqrt{P(z)}}$ and
$\rho_{ij}=\int_{z_*{B_j}}\frac{z^{i-1}dz}{\sqrt{P(z)}}$ where $z$ maps the $A_i$ and $B_j$ cycles to the
$z-$plane.

As mentioned earlier, if the space-time filling mobile $D3$-brane is free to explore the full Calabi-Yau, one would require the knowledge of the geometric K\"{a}hler potential of the full Calabi-Yau. We will now estimate $K_{\rm geom}$ using the Donaldson's algorithm \cite{Donaldson_i} and obtain a metric for the Swiss-Cheese Calabi-Yau in a coordinate patch and for simplicity, close to $\Sigma_B$, that is Ricci-flat in the large volume limit.

For simplicity, working near $x_5=0$ - setting $x_5=\epsilon$ - the no-where vanishing holomorphic three-form
$$\Omega=\oint \frac{dz_1\wedge dz_2\wedge dz_3\wedge dz_4}{P(\{z_i\})},$$
in the $x_2\neq0$-patch with $z_1=\frac{x_1}{x_2}, z_2=\frac{x_3}{x_2}, z_3=\frac{x_4}{x_2^6}, z_4=\frac{x_5}{x_2^6}$
and
$$P(\{z_i\})=1+z_1^{18}+z_2^{18}+z_3^3-\psi\prod_{i=1}^4z_i -\phi z_1^6z_2^6.$$
By the Griffiths residue formula, one obtains:
$$\Omega=\frac{dz_1\wedge dz_2\wedge dz_4}{\frac{\partial P}{\partial z_3}}=\frac{dz_1\wedge dz_2\wedge dz_4}{3z_3^2-\psi z_1z_2z_3},$$
which near $z_4\sim\epsilon$ gives
$$\frac{dz_1\wedge dz_2\wedge dz_4}{3(\phi z_1^6z_2^6-z_1^{18}-z_2^{18}-1)^{\frac{2}{3}}-\psi\epsilon z_1z_2}.$$
The crux of the Donaldson's algorithm is that the sequence
$$\frac{1}{k\pi}\partial_i{\bar\partial}_{\bar j}\left(ln\sum_{\alpha,\beta}h^{\alpha{\bar\beta}}s_\alpha{\bar s}_{\bar\beta}\right)$$
on $P(\{z_i\})$, in the $k\rightarrow\infty$-limit - which in practice implies $k\sim10$ - converges to a unique Calabi-Yau metric for the given K\"{a}hler class and complex structure; $h_{\alpha{\bar\beta}}$ is a balanced metric on the line bundle ${\cal O}_{P(\{z_i\})}(k)$ (with sections $s_\alpha$) for any $k\geq1$, i.e., $$T(h)_{\alpha{\bar\beta}}\equiv \frac{N_k}{\sum_{j=1}w_j}\sum_{i}\frac{s_\alpha(p_i)\overline{s_\beta(p_i)}w_i}
{h^{\gamma{\bar\delta}}
s_\gamma(p_i)\overline{s_\delta(p_i)}}=h_{\alpha{\bar\beta}},$$ where the weight at point $p_i$, $w_i\sim\frac{i^*(J_{GLSM}^3)}{\Omega\wedge{\bar\Omega}}$ with the embedding map $i:P(\{z_i\})\hookrightarrow{\bf WCP}^4$ and the number of sections is denoted by $N_k$. The above corresponds to a K\"{a}hler potential $$K=\frac{1}{k\pi} \, \, ln\sum_{\begin{array}{c}
i_1,...,i_k\\
{\bar j}_1,...,{\bar j}_{\bar k}
\end{array}}h^{(i_1...i_k),({\bar j}_{\bar 1}...{\bar j}_{\bar k})}z_{i_1}...z_{i_k}{\bar z}_{{\bar j}_{\bar 1}...{\bar j}_{{\bar k}}}$$ - the argument of the logarithm being of holomorphic, anti-holomorphic bidegree $(k,k)$. For simplicity, consider $k=2$ for which the sections $s_\alpha$ are given by monomials $z_1^{n_1}z_2^{n_2}z_3^{n_3}$ with $n_1+n_2+n_3\leq2$. Based on our earlier estimate of the geometric K\"{a}hler potential for $\Sigma_B$, we take the following ansatz for the geometric K\"{a}hler potential for the $CY_3$:
\begin{eqnarray}
\label{eq:K}
& & K=- r_1 ln\Biggl[\frac{1}{3\sqrt{r_1|z_1^{18}+z_2^{18}-\phi z_1^6z_2^6|^{\frac{2}{3}}}}\Biggl(r_2 - {\cal V}^{\frac{1}{18}}h^{z_1^2{\bar z}_1^2}\left(|z_1|^2+|z_2|^2+z_1{\bar z}_2+{\bar z}_1z_2\right)\nonumber\\
& & -\frac{{\cal V}^{\frac{1}{12}}h^{z_1^2{\bar z}_1^2}}{\epsilon}\left(z_1{\bar z}_4+{\bar z}_1z_4+z_2{\bar z}_4+{\bar z}_2z_4\right)+h^{z_1^2{\bar z}_1^2}\left(|z_1|^4+|z_2|^4+z_1^2{\bar z}_2^2+z_2^2{\bar z}_1^2+|z_1|^2(z_1{\bar z}_2+{\bar z}_1z_2)\right)\nonumber\\
& & +|z_1|^2\left(z_1{\bar z}_2+{\bar z}_1z_2+|z_1|^2|z_2|^2\right)+\frac{{\cal V}^{\frac{1}{36}}h^{z_1^2{\bar z}_1^2}}{\epsilon}\biggl(z_1^2{\bar z}_2{\bar z}_4+{\bar z}_1^2z_2z_4+z_2^2{\bar z}_1{\bar z}_4+{\bar z}_2^2z_1z_4\nonumber\\
& & +|z_1|^2\left(z_1{\bar z}_4+{\bar z}_1{\bar z}_4\right)
+ |z_2|^2\left(z_2{\bar z}_4+{\bar z}_2{\bar z}_4\right)+|z_1|^2\left(z_2{\bar z}_4+{\bar z}_2{\bar z}_4\right)
+ |z_2|^2\left(z_1{\bar z}_4+{\bar z}_1{\bar z}_4\right)\biggr)\Biggr)\sqrt{\zeta}
\Biggr]\nonumber\\
& & -r_2 ln\left[\left(\frac{\zeta}{r_1|z_1^{18}+z_2^{18}-\phi z_1^6z_2^6|^{\frac{2}{3}}}\right)^{\frac{1}{6}}\right],
\end{eqnarray}
where the balanced-metric and Ricci-flatness conditions are used to determine the unknown $h^{z_1^2,{\bar z}_1^2}$. Now, with $$w_i\sim{g_{z_1{\bar z}_{\bar 1}}g_{z_2{\bar z}_{\bar 2}}g_{z_4{\bar z}_{\bar 4}}}{|3(\phi z_1^6z_2^6-z_1^{18}-z_2^{18}-1)^{\frac{2}{3}}-\psi\epsilon z_1z_2|^2},$$ we will approximate $\frac{N_kw_i}{\sum_jw_j}\sim{\cal O}(1)$ localizing around the position of $D3$-brane, and in obvious notations and around $z_4\sim\epsilon$, the following is utilized in writing out the above ansatz for the K\"{a}hler potential:
 \begin{eqnarray}
 \label{eq:bal_relations}
& & \sum_{\alpha{\bar\beta}}h^{\alpha{\bar\beta}}s_\alpha{\bar s}_{\bar\beta}\sim h^{z_1^2{\bar z}_{\bar 1}^2}z_1^2{\bar z}_{\bar 1}^2\sim h^{z_1^2{\bar z}_{\bar 1}^2}{\cal V}^{\frac{1}{9}};\nonumber\\
& & T(h)_{z_iz_j{\bar z}_{\bar l}{\bar z}_{\bar 4}}\sim\frac{{\cal V}^{\frac{1}{12}}\epsilon}{h^{z_1^2{\bar z}_{\bar 1}^2}{\cal V}^{\frac{1}{9}}}\sim T(h)_{z_iz_4{\bar z}_{\bar k}{\bar z}_{\bar l}}\sim h_{z_iz_4{\bar z}_{\bar k}{\bar z}_{\bar l}}; \nonumber\\
& & T(h)_{z_iz_j{\bar z}_{\bar 4}^2}\sim T(h)_{z_iz_4{\bar z}_{\bar k}{\bar z}_{\bar 4}}
\sim\frac{{\cal V}^{\frac{1}{18}}\epsilon^2}{h^{z_1^2{\bar z}_{\bar 1}^2}{\cal V}^{\frac{1}{9}}}
\sim h_{z_iz_j{\bar z}_{\bar 4}};\nonumber\\
& & T(h)_{z_iz_4{\bar z}_{\bar 4}^2} \sim\frac{{\cal V}^{\frac{1}{36}}\epsilon^3}{h^{z_1^2{\bar z}_{\bar 1}^2}{\cal V}^{\frac{1}{9}}} \sim h_{z_iz_4{\bar z}_{\bar 4}}\sim0;\nonumber\\
& & T(h)_{z_i{\bar z}_{\bar j}}\sim\frac{z_i{\bar z}_{\bar j}}{h^{z_1^2{\bar z}_{\bar 1}^2}{\cal V}^{\frac{1}{9}}}\sim\frac{{\cal V}^{-\frac{1}{18}}}{h^{z_1^2{\bar z}_{\bar 1}^2}}\sim h_{z_i{\bar z}_{\bar j}};\nonumber\\
& & T(h)_{z_i{\bar z}_{\bar 4}}\sim\frac{z_i\epsilon}{h^{z_1^2{\bar z}_{\bar 1}^2}{\cal V}^{\frac{1}{9}}}\sim\frac{\epsilon{\cal V}^{-\frac{1}{12}}}{h^{z_1^2{\bar z}_{\bar 1}^2}}\sim h_{z_i{\bar z}_{\bar 4}}.
 \end{eqnarray}
Further, for a K\"{a}hler manifold, utilizing $\Gamma^l_{jk}=g^{l{\bar m}}\partial_j g_{k{\bar m}}$
and hence $R_{i{\bar j}}=-{\bar\partial}_{\bar j}\Gamma^k_{ik}$. Using the results of appendix D, one sees that for $z_1\sim {\cal V}^{\frac{1}{36}}$ and $z_2\sim 1.3{\cal V}^{\frac{1}{36}}$ and $r_2\sim{\cal V}^{\frac{1}{3}}$ and $r_1\sim\sqrt{ln {\cal V}}$, (\ref{eq:K}) yields:
\begin{eqnarray}
\label{eq:Ricci}
& & R_{z_1{\bar z}_1}\sim R_{z_2{\bar z}_2}\sim R_{z_1{\bar z}_{\bar 2}}\nonumber\\
& & \propto {\cal V}^{-\frac{1}{18}} - {\cal V}^{-\frac{5}{18}}h^{z_1^2{\bar z}_{\bar 1}^2} + {\cal V}^{-\frac{1}{2}}\left(h^{z_1^2{\bar z}_{\bar 1}^2}\right)^2 + {\cal V}^{-\frac{17}{18}}\left(h^{z_1^2{\bar z}_{\bar 1}^2}\right)^3 + {\cal O}\left({\cal V}^{-\frac{19}{18}}\right),
\end{eqnarray}
implying that the Ricci tensor, in the LVS limit, vanishes  for
 \begin{equation}
 \label{eq:h sol}
 h^{z_1^2{\bar z}_{\bar 1}^2}\sim{\cal V}^{\frac{2}{9}}.
 \end{equation}
 Hence, the LVS Ricci-flat metric's components near $(z_1,z_2,z_4)\sim({\cal V}^{\frac{1}{36}},{\cal V}^{\frac{1}{36}},\epsilon)$ are estimated to be:
 \begin{eqnarray}
 \label{eq:Ricci_flat_metric}
 & & g_{i{\bar j}}\sim\left(\begin{array}{ccc}
\frac{h^{z_1^2{\bar z}_{\bar 1}^2}\left(-{\cal V}^{\frac{2}{9}}\times {\cal O}(10) + h^{z_1^2{\bar z}_{\bar 1}^2}\times {\cal O}(10)\right)}{{\cal V}^{\frac{1}{36}}\left({\cal V}^{\frac{2}{9}} + h^{z_1^2{\bar z}_{\bar 1}^2}\times{\cal O}(10)\right)^2} &
\frac{h^{z_1^2{\bar z}_{\bar 1}^2}\left(-{\cal V}^{\frac{2}{9}}\times {\cal O}^\prime(10) + h^{z_1^2{\bar z}_{\bar 1}^2}\times {\cal O}^\prime(10)\right)}{{\cal V}^{\frac{1}{36}}\left({\cal V}^{\frac{2}{9}} + h^{z_1^2{\bar z}_{\bar 1}^2}\times{\cal O}(10)\right)^2} &
\frac{h^{z_1^2{\bar z}_{\bar 1}^2}\left(-{\cal V}^{\frac{2}{9}}\times {\cal O}(1) + h^{z_1^2{\bar z}_{\bar 1}^2}\times {\cal O}(1)\right)}{\epsilon\left({\cal V}^{\frac{2}{9}} + h^{z_1^2{\bar z}_{\bar 1}^2}\times{\cal O}(10)\right)^2}\\
\frac{h^{z_1^2{\bar z}_{\bar 1}^2}\left(-{\cal V}^{\frac{2}{9}}\times {\cal O}^\prime(10) + h^{z_1^2{\bar z}_{\bar 1}^2}\times {\cal O}^\prime(10)\right)}{{\cal V}^{\frac{1}{36}}\left({\cal V}^{\frac{2}{9}} + h^{z_1^2{\bar z}_{\bar 1}^2}\times{\cal O}(10)\right)^2} & \frac{h^{z_1^2{\bar z}_{\bar 1}^2}\left(-{\cal V}^{\frac{2}{9}}\times {\cal O}^{\prime\prime}(10) + h^{z_1^2{\bar z}_{\bar 1}^2}\times {\cal O}^{\prime\prime}(10)\right)}{{\cal V}^{\frac{1}{36}}\left({\cal V}^{\frac{2}{9}} + h^{z_1^2{\bar z}_{\bar 1}^2}\times{\cal O}(10)\right)^2} & \frac{h^{z_1^2{\bar z}_{\bar 1}^2}\left(-{\cal V}^{\frac{2}{9}}\times {\cal O}(1) + h^{z_1^2{\bar z}_{\bar 1}^2}\times {\cal O}(1)\right)}{{\cal V}^{\frac{1}{36}}\left({\cal V}^{\frac{2}{9}} + h^{z_1^2{\bar z}_{\bar 1}^2}\times{\cal O}(10)\right)^2} \\
\frac{h^{z_1^2{\bar z}_{\bar 1}^2}\left(-{\cal V}^{\frac{2}{9}}\times {\cal O}(1) + h^{z_1^2{\bar z}_{\bar 1}^2}\times {\cal O}(1)\right)}{\epsilon\left({\cal V}^{\frac{2}{9}} + h^{z_1^2{\bar z}_{\bar 1}^2}\times{\cal O}(10)\right)^2}&
\frac{h^{z_1^2{\bar z}_{\bar 1}^2}\left(-{\cal V}^{\frac{2}{9}}\times {\cal O}(1) + h^{z_1^2{\bar z}_{\bar 1}^2}\times {\cal O}(1)\right)}{{\cal V}^{\frac{1}{36}}\left({\cal V}^{\frac{2}{9}} + h^{z_1^2{\bar z}_{\bar 1}^2}\times{\cal O}(10)\right)^2} &
 \frac{{\cal O}(100)\left(h^{z_1^2{\bar z}_{\bar 1}^2}\right)^2{\cal V}^{\frac{1}{36}}}{2\epsilon^2\left({\cal V}^{\frac{2}{9}} + h^{z_1^2{\bar z}_{\bar 1}^2}\times{\cal O}(10)\right)^2}\\
\end{array}\right)\nonumber\\
& & {\rm which\ near}\ h^{z_1^2{\bar z}_{\bar 1}^2}={\cal V}^{\frac{2}{9}}\ {\rm yields}:\nonumber\\
& & \sim\left(\begin{array}{ccc}
{\cal V}^{-\frac{1}{36}} & {\cal V}^{-\frac{1}{36}} & 0\\
{\cal V}^{-\frac{1}{36}}& {\cal V}^{-\frac{1}{36}} & 0\\
0 & 0 & \frac{{\cal V}^{\frac{1}{36}}}{\epsilon^2} \end{array}\right).
\end{eqnarray}
This corrects the numerical values of $h^{z_1^2{\bar z}_{\bar 1}^2}$ and therefore $g_{1{\bar 1},2{\bar 2},3{\bar 3}}$
of \cite{ferm_masses_MS}.

\section{Dimension-Six Neutrino Masses}

As per \cite{d=6_m_nu}, neutrino masses can also be obtained from the K\"{a}hler potential via dimension-six operators:
\begin{equation}
\label{eq:dim_6}
\frac{1}{M_p^2}\int d^2\theta d^2{\bar\theta}\left(\kappa_{\alpha\beta}{{\bar{\cal Z}}_2\cdot{\cal A}_{I\alpha}{\cal Z}_1\cdot{\cal A}_{I\beta}}+\kappa_{\alpha\beta}^\prime{{\bar{\cal Z}}_1\cdot{\cal A}_{I\alpha} {\bar{\cal Z}}_1\cdot{\cal A}_{I\beta}}+{\rm h.c.}\right),
\end{equation}
$\alpha,\beta$ denoting the $SU(2)_{EW}$ indices.
 The understanding is that one considers the ${\bar\theta}^2$ component - $F^{{\cal Z}_{1,2}}$ - of one of the two ${\cal Z}_{1,2}$ superfields. Using Appendix A of \cite{D3_D7_Misra_Shukla}, e.g., ${\bar F}^{{\cal Z}_2}=e^{K/2}
 G^{{\bar{\cal Z}}_{\bar 2}{\cal Z}_i}D_iW\ni {\cal V}^{\frac{17}{18}}e^{K/2}\mu_{{\cal Z}_1{\cal Z}_2}{\cal Z}_2$. Working with canonically normalized fields, this implies that the neutrino mass will be given by:
 \begin{equation}
 \label{eq:nu_mass}
 \frac{{\cal V}^{\frac{17}{18}}\hat{\mu}_{{\cal Z}_1{\cal Z}_2}v^2}{M_p^2}\left(\frac{\kappa_S}{\left(\sqrt {K_{{\cal A}_I{\bar{\cal A}}_I}}\right)^2}sin^2\beta+\frac{\kappa^\prime}{\left(\sqrt {K_{{\cal A}_I {\bar {\cal A}}_I}}\right)^2}sin\beta cos\beta\right).
 \end{equation}
 In the large $tan\beta$-regime, the second term in (\ref{eq:nu_mass}) is dropped.

 The K\"{a}hler potential is given by:
{\small \begin{eqnarray}
\label{eq:K}
& &\hskip-0.5cm \frac{K}{M_p^2} = - ln\left(-i(\tau-{\bar\tau})\right) - ln\bigl(i\int_{CY_3}\Omega\wedge{\bar\Omega}\bigr) - 2 ln\Bigl[a\bigl(\frac{T_B + {\bar T}_B}{M_p} - \gamma K_{\rm geom}\bigr)^{\frac{3}{2}}-a\left(\frac{T_S + {\bar T}_S}{M_p} - \gamma K_{\rm geom}\right)^{\frac{3}{2}} \nonumber\\
 & & + \sum_{m,n\in{\bf Z}^2/(0,0)}
\frac{({\bar\tau}-\tau)^{\frac{3}{2}}}{(2i)^{\frac{3}{2}}} \frac{1}{|m+n\tau|^3}\Bigl\{\frac{\chi}{2}- 4\sum_{\beta\in H_2^-(CY_3,{\bf Z})} n^0_\beta cos(\frac{mk.{\cal B} + nk.c}{M_p})\Bigr\}\Bigr]
\end{eqnarray}}
In (\ref{eq:K}), apart from the usual tree-level contribution, the second  and third lines include the geometric K\"{a}hler potential (due to the presence of the mobile $D3$-brane) as well as perturbative and non-perturbative $\alpha^\prime$ corrections; $\{n^0_\beta\}$ are the genus-zero Gopakumar-Vafa invariants for the holomorphic curve $\beta\in H_2^-(CY_3,{\bf Z})$ that count the number of genus-zero rational curves. Further, $\gamma K_{\rm geom}$($\gamma\sim\kappa_4^2\mu_3\sim1/{\cal V}$), was estimated in \cite{D3_D7_Misra_Shukla} using GLSM techniques and was shown to be subdominant. Now, using the notations and technique of \cite{D3_D7_Misra_Shukla}, consider a holomorphic one-form
\begin{equation}
\label{eq:harm_1form_II}
A_2=\omega_2(z_1,z_2)dz_1+\tilde{\omega}_2(z_1,z_2)dz_2,
\end{equation}
where $\omega_2(-z_1,z_2)=\omega_2(z_1,z_2), \tilde{\omega}_2(-z_1,z_2)=-\tilde{\omega}_2(z_1,z_2)$ (under $z_1\rightarrow-z_1,z_{2,3}\rightarrow z_{2,3}$) and
$$\partial A_2=(1+z_1^{18}+z_2^{18}+z_3^3-\phi_0z_1^6z_2^6)dz_1\wedge dz_2$$
(implying $dA_2|_{\Sigma_B}=0$). Assuming $\partial_1\tilde{\omega}_2=-\partial_2\omega_2$, then
around $|z_3|\sim{\cal V}^{1/6}, |z_{1,2}|\sim{\cal V}^{1/36}$ - localized around the mobile $D3$-brane - one estimates the component of the distribution one-form (\ref{eq:harm_1form_II}):
 $$\tilde{\omega}_2(z_1,z_2)\sim z_1^{19}/19+z_2^{18}z_1+\sqrt{\cal V}z_1-\phi_0/7z_1^7z_2^6$$
 with $\omega_2(z_1,z_2)=-\tilde{\omega}_2(z_2,z_1)$ in the LVS limit, and utilizing the result of \cite{D3_D7_Misra_Shukla} pertaining to the $I=J=1$-term, one hence obtains: $$i\kappa_4^2\mu_7C_{I{\bar J}}a_I{\bar a}_{\bar J}\sim{\cal V}^{7/6}|a_1|^2+{\cal V}^{2/3}(a_1{\bar a}_{\bar 2}+c.c.)+{\cal V}^{1/6}|a_2|^2,$$ $a_2$ being another Wilson line modulus. The K\"{a}hler potential, in the LVS limit will then be of the form
\begin{eqnarray*}
& & K\sim-2 ln\biggl[\left({\cal V}^{\frac{2}{3}}-{\cal V}^{\frac{2}{3}+\frac{1}{2}}{\cal A}_1^\dagger{\cal A}_1+{\cal V}^{\frac{2}{3}}\left({\cal A}_1{\cal A}_2^\dagger+h.c.\right)+{\cal V}^{\frac{1}{6}}{\cal A}_2^\dagger{\cal A}_2+\beta_1{\cal Z}_i^\dagger{\cal Z}_i\right)^{\frac{3}{2}}\nonumber\\
& & -\left(\alpha_2{\cal V}^{\frac{1}{18}}+\beta_2{\cal Z}_i^\dagger{\cal Z}_i\right)^{\frac{3}{2}}+\sum n^0_\beta(...)\biggr]
\end{eqnarray*}
 - $a_{1,2}$ promoted to the Wilson line moduli superfields ${\cal A}_{1,2}$ and the $D3$-brane position moduli $z_i$ being promoted to the superfields ${\cal Z}_i$. The $a_I$'s can be stabilized at around ${\cal V}^{-1/4}$ (See \cite{D3_D7_Misra_Shukla}); consider fluctuations in $a_I$ about ${\cal V}^{-1/4}$: $a_I\rightarrow{\cal V}^{-1/4}+a_I$. Given that the genus-0 Gopakumar-Vafa invariants are very large for compact projective varieties, taking $\sum_\beta n^0_\beta(...)\sim{\cal V}$ one obtains the following:
 \begin{equation}
 \label{eq:mass_matrix}
 \left(\begin{array}{cc} \kappa_{{\cal A}_1{\cal A}_1{\cal Z}_i{\cal Z}_j}&\kappa_{{\cal A}_1{\cal A}_2{\cal Z}_i{\cal Z}_j}\\
 \kappa_{{\cal A}_1{\cal A}_2{\cal Z}_i{\cal Z}_j}&\kappa_{{\cal A}_2{\cal A}_2{\cal Z}_i{\cal Z}_j}
 \end{array}\right)\sim\left(\begin{array}{cc}{\cal V}^{3/4}&{\cal V}^{1/4}\\ {\cal V}^{1/4}&{\cal V}^{-1/4}\end{array}\right),
 \end{equation}
 which in the LVS limit has two eigenvalues: $0,{\cal V}^{3/4}$.

 From the first reference in \cite{d=6_m_nu}, one sees that the solution to the one-loop RG-flow equations for $\kappa_S$ is given by:
 $$\kappa_S(M_{EW})\sim\kappa_S(M_s)\left(1-\frac{u}{16\pi^2} ln\left(\frac{M_s}{M_{EW}}\right)\right),$$ where $u\equiv Tr(3\hat{Y}^\dagger_U \hat{Y}_U+3\hat{Y}^\dagger_D\hat{Y}_D+\hat{Y}^\dagger_L\hat{Y}_L)-3g_{SU(2)}^2-g_{U(1)}^2$ (evaluated at $M_s$), where $Y_{U,D,L}$ are the Yukawa couplings corresponding to the up-type quarks, down-type quarks and leptons. Neglecting the Yukawa couplings of the first two generations (the same was verified for the single Wilson-line modulus setup of \cite{D3_D7_Misra_Shukla} as discussed above) \cite{Ibanez_Lopez}
 $${\cal Y}_\tau(t)={\cal Y}_\tau(M_s)(1+\beta_2 t)^{3/b_2}(1+\beta_1t)^{3/b_1};$$ using $m_\tau(M_{EW})\sim1.8GeV,{\cal Y}_\tau(M_{EW})\sim7\times10^{-5}$ one hence obtains $${\cal Y}_\tau(M_S)\sim3\times10^{-5}.$$
 Therefore one will estimate:
 $$Tr(3\hat{Y}^\dagger_U \hat{Y}_U+3\hat{Y}^\dagger_D\hat{Y}_D+\hat{Y}^\dagger_L\hat{Y}_L)\sim(4\pi)^2(3{\cal\ Y}_t+{\cal Y}_\tau)
 \sim10^{-2}.$$ Given that $g_2^2(M_s)=0.4$,
 $$\kappa_S(M_{EW})\sim\kappa_S(M_s)(1-10^{-2}),$$ up to one loop.
 Now, using the third equation of (\ref{eq:evs_1}) pertaining to
 $$\hat{\mu}_{{\cal Z}_1{\cal Z}_2}(M_{EW})\sim10^{-1}{\cal V}m_{3/2}\sim10^5TeV$$ (for ${\cal V}=10^6$), the non-zero eigenvalue in the large $tan\beta$-limit, using (\ref{eq:nu_mass}) and $m_{3/2}\sim10^3 TeV$, correspond to:
 \begin{eqnarray}
 \label{eq:nu_mass_II}
 & & m_\nu(M_{EW})\sim\frac{{\cal V}^{\frac{17}{18}}\hat{\mu}_{{\cal Z}_1{\cal Z}_2}(M_{EW})v^2\times {\cal V}^{\frac{3}{4}}}{M_p^2\left(\sqrt {K_{{\cal A}_I{\bar{\cal A}}_I}}\right)^2}\sim\frac{{\cal V}^{\frac{17}{18}+1+\frac{3}{4}}(246)^2\times10^{6-1}}{{\cal V}^{\frac{29}{36}}(10^{18})^2}GeV,
 \end{eqnarray}
(using $m_{3/2}\sim10^6GeV$) which for ${\cal V}\sim10^6$ yields $m_\nu\sim10^{-7}eV$, clearly negligible as compared to the dimension-five operators' contribution worked out in \cite{ferm_masses_MS}.

\section{Summary and Discussion}

Split SUSY Models were proposed in which soft scalar masses (squarks/sleptons) are heavy while fermions (including the gauginos and Higgsino) remain light. Dealing  with all phenomenological issues successfully in the context of split SUSY, the ``$\mu$ - problem" still remained unresolved. The same could be addressed in a variant of split SUSY scenario according to which one can assume further splitting in split SUSY by raising the Higgsino mass term ($\mu$ parameter) to a large value, i.e, to the order of high supersymmetry breaking scale. The scenario based on this has been named as ``$\mu$-split SUSY" scenario.

 In this paper, we have studied in detail the possibility of generating $\mu$-split SUSY scenario in the context of type IIB Swiss-Cheese orientifold (involving isometric holomorphic involution) compactifications in the L(arge) V(olume) S(cenarios). Generation of very heavy scalar masses and light(superpartner) fermion masses that has already been realized in the context of L(arge) V(olume) S(cenario), has been adopted as one of the signatures of split supersymmetric behavior. To see it more clearly, we have tried to generate one light Higgs boson with the assumption  that fine tuning is allowed in case of split SUSY models. For this, using solution of  RG flow equation for the mobile $D3$-brane position moduli and Higgsino mass terms and further assuming gauge coupling up to one loop order and non-universality in squark/slepton masses (in addition to the non-universality between the Higgs' and squark/slepton masses), by diagonalizing the mass matrix for the Higgs doublet, we get one light Higgs (about $150 GeV$)   and one heavy Higgs, about a tenth of the squark masses. The  Higgsino also turns out to about a tenth of the squark mass. Since in our set up, $\mu$ value comes out to be of the order of squark/slepton mass scale i.e high scale, therefore we are in a position to define our scenario close to $\mu$-split SUSY scenario which we could refer to as "LVS $\mu$ split SUSY Scenario".

The most distinctive feature of split SUSY is based on longevity of gluino. Therefore, in order to seek striking evidence
of split SUSY in the context of LVS, in section {\bf 4.1},  we first estimated the decay width for tree-level
three-body gluino decay into a quark, squark and neutralino. By constructing the neutralino mass matrix and diagonalizing the
same, we identified the neutralino with a mass less than that of the gluino (this neutralino in the dilute flux approximation
is roughly half the mass of the gluino).  This neutralino turns out to be largely a neutral gaugino with a small admixture of
the Higgsinos. Using one-loop RG analysis of coefficients of the effective dimension-six gluino decay operators as given
in \cite{Guidice_et_al}, we showed that these coefficients at the EW scale are of the same order as that at the squark mass
scale; we assume that these coefficients at the EW scale will be of the same order as that at the string scale.
The lower bound on the gluino lifetime via this three-body decay channel was estimated to lie in the range: $10^{-6}-10^{8}$ seconds. Adopting the same approach as in section {\bf 4.1}, in section {\bf 4.2},  we calculated the decay width of one-loop
two-body gluino decay into gluon and neutralino, results of which, similar to the tree level gluino decay, yield large life time(s) of gluino  for this case. The high squark mass, helps to suppress the tree-level as well as one loop gluino decay width. The fact that we have obtained  suppressed  Gluino decay width  for squark masses of the order of $10^{12}GeV$, is in agreement with the previous theoretical studies based on gluino decays in split SUSY in literature (\cite{Guidice_et_al,Manuel Toharia}) and results based on collider phenomenology for stable gluino. As per collider point of view, production of  jets
 and missing energy will no longer remain the  signature to search  for indirect experimental evidence of gluino.
 Now, the probability of  production of ``R-Hadrons" formed by gluino pairs will give indirect experimental evidence of
 gluino, signatures of which have been studied recently in $pp$ collisions at LHC \cite{exp_Gluino pair} where cross section
 of gluino pair production
increases as  gluino life time increases, thus providing  indirect support to LVS $\mu$ split SUSY model discussed above. Further, in section {\bf 4.3}, going through the same analysis, we have estimated the tree-level as well as one-loop  decay width  of gluino decaying into Goldstino. This time around, the lifetime of the gluino in the three-body decay channel involving quark, squark and Goldstino, is estimated to have a lower bound of $10^{-7}$ seconds; the gluino lifetime in the two-body decay channel involving a Goldstino and gluon, like the neutralino two-body decay, is quite enhanced.

We should also keep in mind that the above analysis involves fine tuning at two levels - one, the stabilizing value of the Wilson line moduli to ensure a partial cancelation between the big divisor volume modulus and the quadratic Wilson line moduli contribution in an appropriate chiral ${\cal N}=1$ coordinate so as to obtain an ${\cal O}(1)$ gauge coupling constant despite wrapping stacks of $D7$-branes around the big divisor; second,
in the hypercharge weighted sum of soft scalar masses as well as the ${\cal O}(1)$ proportionality constant between the Higgsino-mass parameter squared and the soft SUSY parameter $\hat{\mu}B$, in order to obtain one light and one heavy Higgs at the EW scale.

 In section {\bf 5}, based on GLSM techniques and the Donaldson's algorithm for obtaining numerically a Ricci-flat metric, we have proposed a metric for the Swiss-Cheese Calabi-Yau used in a coordinate patch and for simplicity, close to the big divisor, which is Ricci-flat in the large volume limit. This, with some effort, could also be generalized for points finitely away from the big divisor.

 Further, as an  application to  one-loop RG-flow equation for Higgsino mass parameter, in section {\bf 6}, we have calculated the (first two generations') neutrino mass that  can  be obtained from the  dimension-six operators arising from the K\"{a}hler potential. The reason for doing so was the large value of the Higgsino mass parameter in our $\mu$-split SUSY LVS setup (as opposed to conventional split SUSY) and the fact that the dimension-six neutrino mass is proportional to the same. It turns out that the same is not sufficient to compensate the Planckian suppression of dimension-six contribution and one obtains $m_\nu\sim10^{-7}eV$ i.e extremely suppressed relative to the Weinberg-type dimension-five operators.

Inspired by the fact that neutralino seems to  appear as lightest supersymmetric particle in our set up and squark still appears  as propagator in tree level decay, study of neutralino decay properties and  exploring the possibility of the neutralino to be a dark matter candidate as well as reproducing the observable value of dark energy density in LVS $\mu$ split SUSY model, seem to be interesting aspects of our $\mu$-split SUSY to look into.
\section*{Acknowledgement}

MD is supported by a CSIR Junior Research Fellowship. AM would like to thank Albert Einstein Institute for Gravitation (Max Planck Institute) at Berlin, Ohio State University (specially S.Raby and C.Bobkov), University of California at Berkeley (specially O.Ganor), Northeastern University (specially P.Nath), the division of theoretical physics in the department of Mathematical Sciences at the University of Liverpool (specially T.Mohaupt), Centre for Research in String Theory at the Queen Mary College University of London (specially R.Russo), ETH Zurich (specially M.Gaberdiel) and the CERN theory group, for hospitality, financial support and discussions, where part of this work was done.

\appendix
\setcounter{equation}{0} \seceqaa
\section{Values of $t$-dependent Functions}
  In this appendix, we have collected the expressions for all parameters required for computation of Higgs masses \cite{Ibanez_et_al}. The parameters are generally evolved in solution of evolution equations for all mass parameters as well as coupling constants involved in supersymmetric theory. We have calculated their numerical values for our set up using
  $t= 2ln(M_s/M_{EW})=57$ where $M_s= \frac{M_{pl}}{\sqrt{\cal V}}={10}^{15} GeV$(string scale) for ${\cal V}={10}^6$ (Calabi-Yau volume), $M_{EW} = 500 GeV$ (Electroweak scale). Form factors
  $$e(t), h(t), f(t), k(t), q(t), l(t), r(t), s(t), k(t), D(t)$$ depend on Yukawa coupling while other parameters:
   $$ H_{1,...,8}(t), F_{2,...,4}(t), G_{1,2}(t)$$
   depend just on gauge coupling constant. The same are summarized below:

\begin{eqnarray*}
& &  l(t)=\left(q(t)\right)^2;\
l(57)=1.59
\nonumber\\
& &  q(t)=\frac{(t {\beta_1}+1)^{\frac{1}{2 {b_1}}} (t
   {\beta_2}+1)^{\frac{3}{2 {b_2}}}}{\sqrt[4]{6 {{\cal Y}_t}
   F(t)+1}},\
   q(57)=1.26\nonumber\\
   & &    {\rm where}\ \beta_i=b_i\frac{\alpha(M_s)}{4\pi};
\nonumber\\
& &  g(t)=\frac{{f_1}(t) \tilde\alpha(M_s)}{2}+\frac{3 {f_2}(t) \tilde\alpha(M_s)}{2};\
g(57)=0.39 \nonumber\\
& & r(t)=\frac{3 {{\cal Y}_t} F(t) q(t)}{E(t)};\
r(57)=-0.53\nonumber\\
& &
s(t)=\frac{3 {{\cal Y}_t} F(t) q(t)}{M(t)};\
s(57)=0.02
\nonumber\\
& & h(t)=\frac{1}{2} \left(\frac{3}{D(t)}-1\right);\
h(57)=0.96
\nonumber\\
\end{eqnarray*}

\begin{eqnarray*}
& & k(t)=\frac{3 {{\cal Y}_t} F(t)}{D(t)^2};\
k(57)=0.01
\nonumber\\
& & f(t)=-\frac{6 {{\cal Y}_t} {H3}(t)}{D(t)^2};\
f(57)=-0.03
\nonumber\\
& & D(t)=6 {{\cal Y}_t} F(t)+1;\
D(57)=1.02
\nonumber\\
& & e(t)=\frac{3}{2}\left[\frac{({H_2}(t)+6 {{\cal Y}_t}
   {H_4}(t))^2}{3 D(t)^2}+{H_8}(t)+\frac{{G_1}(t)+{{\cal Y}_t}
   {G_2}(t)}{D(t)}\right];\
e(57)=0.32
\nonumber\\
& & f_i(t)=\frac{1-\frac{1}{(t {\beta_i}+1)^2}}{{\beta_i}};\
 f_1(57)=45.85,\ f_2(57)=94.34,\ f_3(57)=259.09
\nonumber\\
& & h_i(t)=\frac{t}{t {\beta_i}+1};\
h_1(57)=30.03,\ h_2(57)=50.17,\ h_3(57)=96.32
\nonumber\\
& & F(t)=\int_0^tE(t^\prime)dt^\prime;\
F(57)=125.02\nonumber\\
& & H_1(t)=\frac{\left(\frac{3}{(t {\beta
   2}+1)^2}+\frac{16}{3 (t {\beta_3}+1)^2}+\frac{13}{15 (t
   {\beta_1}+1)^2}\right) \alpha (M_s)}{4 \pi };\
   H_1(57)=0.04
\nonumber\\
& & H_2(t)=\frac{\left(\frac{13 {h_1}(t)}{15}+3
   {h_2}(t)+\frac{16 {h_3}(t)}{3}\right) \alpha (M_s)}{4 \pi };\
   H_2(57)=1.65
\nonumber\\
& & H_3(t)=t E(t)-F(t);\
H_3(57)=119.37
\nonumber\\
& & H_4(t)=F(t) {H_2}(t)-{H_3}(t);\
H_4(57)=86.64\nonumber\\
& & H_5(t)=\frac{\left(-\frac{22 {f_1}(t)}{15}+6
   {f_2}(t)-\frac{16 {f_3}(t)}{3}\right) \alpha (0)}{4 \pi };\
   H_5(57)=-2.11
\nonumber\\
& & H_6(t)=\int_0^t {H_2}(t^\prime)^2 E(t^\prime) \, dt^\prime;\
H_6(57)=133.62
\nonumber\\
& &  H_7(t)=\frac{{h_1}(t) \alpha (M_s)}{4 \pi }+\frac{3
   {h_2}(t) \alpha (M_s)}{4 \pi };\
   H_7(57)=0.43
\nonumber\\
& & H_8(t)=\frac{\left(-\frac{{f_1}(t)}{3}+{f_2}(t
   )-\frac{8 {f_3}(t)}{3}\right) \alpha (M_s)}{4 \pi };\
   H_8(57)=-1.46\nonumber\\
& &  F_2(t)=\frac{\left(\frac{8 {f_1}(t)}{15}+\frac{8
   {f_3}(t)}{3}\right) \alpha (M_s)}{4 \pi };\
   F_2(57)=1.71
\nonumber\\
& & F_3(t)=F(t) {F_2}(t)-\int_0^t {F_2}(t^\prime) E(t^\prime) \,
   dt^\prime;\
   F_3(57)=107.29
   \end{eqnarray*}

   \begin{eqnarray*}
& & F_4(t)=\int_0^t {H_5}(t^\prime) P(t^\prime) \, dt^\prime;\
F_4(57)=-107.12
\nonumber\\
& & G_1(t)={F_2}(t)-\frac{{H_2}(t)^2}{3};\
G_1(57)=0.80
\nonumber\\
& & G_2(t)=2 F(t) {H_2}(t)^2-4 {H_4}(t)
   {H_2}(t)+6 {F_3}(t)-{F_4}(t)-2 {H_6}(t);\
   G_2(57)=591.47
   \end{eqnarray*}
   \section{The Neutralinos}
\setcounter{equation}{0} \seceqbb

In this appendix we work out the eigenalues and eigenvectors of the neutralino mass matrix and conclude that the neutralino ($\tilde{\chi}_3^0$) with a mass lighter than the gluino is largely a gaugino (such as the Bino) with a small admixture of the two Higgsinos. For this purpose, apart from the gaugino and Higgsino mass terms, we will need to look at the gaugino-Higgsino coupling term: $g_{YM}g_{\sigma^B{\cal Z}^i}X^B\tilde{H}^i\lambda^0$ where $\lambda^0$ is a neutral gaugino (such as the Bino). Analogous to ${\bf 4.1}$, we need to look for $\tilde{H}_LH_L\lambda^0$ where $\tilde{H}_L$ is an $SU(2)_L$ Higgsino doublet; after $H^0$ acquires a non-zero vev $\langle H^0\rangle$, this generates the term $\langle H^0\rangle\tilde{H}_L\lambda^0$.

Using the following form of the K\"{a}hler potential:
\begin{eqnarray}
\label{eq:K_neutr}
& & \hskip-0.3in\frac{K}{M_p^2}\sim-2ln\Biggl(\sum_\beta n^0_\beta ... +\left[\frac{\sigma_B+{\bar\sigma}_B}{M_p} - \frac{|a_1|^2}{M_p^2}{\cal V}^{\frac{7}{6}}+\frac{|a_2|^2}{M_p^2}{\cal V}^{\frac{1}{6}}+{\cal V}^{\frac{2}{3}}\frac{\left(a_2{\bar a}_1+a_1{\bar a}_2\right)}{M_p^2}+\mu_3\frac{\left(|z_1|^2+|z_2|^2+z_1{\bar z}_2+z_2{\bar z}_1\right)}{M_p^2}\right]^{\frac{3}{2}}\nonumber\\
& & -\left[\frac{\sigma_S+{\bar\sigma}_S}{M_p}+\mu_3\frac{\left(|z_1|^2+|z_2|^2+z_1{\bar z}_2+z_2{\bar z}_1\right)}{M_p^2}\right]^{\frac{3}{2}}\Biggr),
\end{eqnarray}
and utilizing (\ref{eq:gYM_ReT}) assuming $\sigma_B+{\bar\sigma}_B-{\cal V}^{\frac{7}{6}}|a_1|^2\sim{\cal V}^{\frac{1}{18}}$ [$\sigma_{B}$ (Big Divisor' Volume) was shown in \cite{ferm_masses_MS,D3_D7_Misra_Shukla} to be stabilized at ${\cal V}^{\frac{2}{3}}$  corresponding to $a_I$ being stabilized at ${\cal V}^{-\frac{1}{4}}$; to obtain ${\cal V}^{\frac{1}{18}}$ (the value at which the small divisor volume is stablized (See \cite{ferm_masses_MS})) to effect $\frac{1}{g^2_{YM}}\sim Re(T_S)$ analogous to \cite{conloncal}, one assumes a fine tuning that $|a_1|^2$ is stabilized at: ${\cal V}^{-\frac{1}{2}}- {\cal V}^{(-\frac{7}{6}+\frac{1}{18}=)-\frac{10}{9}}$], one obtains:
\begin{equation}
\label{eq:gYM_ReT_II}
g_{YM}g_{\sigma^B{\cal Z}^i}\sim \frac{z_i{\cal V}^{-\frac{55}{36}}}{M_p}\Biggr|_{z_i\sim{\cal V}^{\frac{1}{36}}M_p}\sim{\cal V}^{-\frac{3}{2}}.
\end{equation}
Hence, the gluino-Higgsino interaction strength using canonically noramlized fields, is given by: $\frac{\kappa_4^2\mu_7Q_B{\mu_3}{{\cal V}^{-\frac{3}{2}}}}{\left(\sqrt{K_{{\cal Z}_i {\bar {\cal Z}}_{\bar i}}}\right)}$, which taking $Q_B\sim{\cal V}^{\frac{1}{3}}\tilde{f}$ and $n^s=2$ yields a neutralino mass matrix:
\begin{eqnarray}
\label{eq:neut_mass}
& & \left(\begin{array}{ccc}
\frac{1}{{\cal V}^2} & -\frac{\tilde{f}}{{\cal V}^{\frac{121}{72}}} & -\frac{\tilde{f}}{{\cal V}^{\frac{121}{72}}}\\
-\frac{\tilde{f}}{{\cal V}^{\frac{121}{72}}} & 0 & {\cal V}^{-\frac{35}{36}}\\
-\frac{\tilde{f}}{{\cal V}^{\frac{121}{72}}}& {\cal V}^{-\frac{35}{36}} & 0
\end{array}\right)M_p,
\end{eqnarray}
with eigenvalues:
\begin{eqnarray}
\label{eq:neut evalues}
& & \Biggl\{\frac{1}{V^{35/36}},\frac{1}{2} \left(-\frac{\sqrt{V^{37/18}+2 V^{37/36}+8\tilde{f}^2
   V^{23/36}+1}}{V^2}-\frac{1}{V^{35/36}}+\frac{1}{V^2}\right),\nonumber\\
& & \left(\frac{-V^{43/18}+\sqrt{V^{37/18}+2 V^{37/36}+8 V^{23/36}\tilde{f}^2+1} V^{49/36}+V^{49/36}}{2
   V^{121/36}}\right)\Biggr\}M_p
\end{eqnarray}
and normalized eigenvectors:
\begin{eqnarray}
\label{eq:neut_evectors}
& & \left(\begin{array}{c}0 \\ -\frac{1}{\sqrt{2}} \\\frac{1}{\sqrt{2}} \end{array} \right),
\left(\begin{array}{c}
 -\frac{V^{37/36}-\sqrt{8 V^{23/36} \tilde{f}^2+\left(V^{37/36}+1\right)^2}+1}{2 \tilde{f} V^{23/72} \sqrt{\frac{\left(V^{37/36}-\sqrt{8 V^{23/36}
   \tilde{f}^2+\left(V^{37/36}+1\right)^2}+1\right)^2}{4 \tilde{f}^2 V^{23/36}}+2}} \\
 \frac{1}{\sqrt{\frac{\left(V^{37/36}-\sqrt{8 V^{23/36} \tilde{f}^2+\left(V^{37/36}+1\right)^2}+1\right)^2}{4 \tilde{f}^2 V^{23/36}}+2}} \\
 \frac{1}{\sqrt{\frac{\left(V^{37/36}-\sqrt{8 V^{23/36} \tilde{f}^2+\left(V^{37/36}+1\right)^2}+1\right)^2}{4 \tilde{f}^2 V^{23/36}}+2}}
\end{array}
\right),\nonumber\\
   & &
\left(
\begin{array}{c}
 -\frac{V^{37/36}+\sqrt{8 V^{23/36} \tilde{f}^2+\left(V^{37/36}+1\right)^2}+1}{2 \tilde{f} V^{23/72} \sqrt{\frac{\left(V^{37/36}+\sqrt{8 V^{23/36}
   \tilde{f}^2+\left(V^{37/36}+1\right)^2}+1\right)^2}{4 \tilde{f}^2 V^{23/36}}+2}} \\
 \frac{1}{\sqrt{\frac{\left(V^{37/36}+\sqrt{8 V^{23/36} \tilde{f}^2+\left(V^{37/36}+1\right)^2}+1\right)^2}{4 \tilde{f}^2 V^{23/36}}+2}} \\
 \frac{1}{\sqrt{\frac{\left(V^{37/36}+\sqrt{8 V^{23/36} \tilde{f}^2+\left(V^{37/36}+1\right)^2}+1\right)^2}{4 \tilde{f}^2 V^{23/36}}+2}}
\end{array}
\right)
\end{eqnarray}
We hence obtain the following neutralinos:
\begin{eqnarray}
\label{eq:neutralinos_I}
& & \tilde{\chi}_1^0\sim\frac{-\tilde{H}_1^0+\tilde{H}_2^0}{\sqrt{2}};\ {\rm mass}\sim{\cal V}^{-\frac{35}{36}}M_p>m_{\frac{3}{2}},\nonumber\\
& & \tilde{\chi}_2^0\sim \frac{\tilde{f}{\cal V}^{-\frac{51}{72}}}{\sqrt{2}}\lambda^0+\frac{\tilde{H}_1^0+\tilde{H}_2^0}{\sqrt{2}};\ {\rm mass}\sim{\cal V}^{-\frac{35}{36}}M_p>m_{\frac{3}{2}};\ CP:-,\nonumber\\.
& & \tilde{\chi}_3^0\sim-\lambda^0+\tilde{f}{\cal V}^{-\frac{51}{72}}\left(\tilde{H}_1^0+\tilde{H}_2^0\right);\ {\rm mass}\sim
\frac{1}{2}{\cal V}^{-2}M_p<m_{\frac{3}{2}}.
\end{eqnarray}
These could be inverted to read:
\begin{eqnarray}
\label{eq:neutralinos_II}
& & \lambda^0\sim-\tilde{\chi}_3^0-\tilde{f}{\cal V}^{-\frac{51}{72}}\tilde{\chi}_2^0,\nonumber\\
& & \tilde{H}_1^0\sim\frac{\tilde{\chi}_1^0-\tilde{\chi}_2^0}{\sqrt{2}}+\frac{\tilde{f}{\cal V}^{-\frac{51}{72}}}{2}\tilde{\chi}_3^0,\nonumber\\
& & \tilde{H}_2^0\sim\frac{\tilde{\chi}_1^0+\tilde{\chi}_2^0}{\sqrt{2}}+\frac{\tilde{f}{\cal V}^{-\frac{51}{72}}}{2}\tilde{\chi}_3^0.
\end{eqnarray}
Hence, for gluino-decay studies, it is $\tilde{\chi}_3^0$ - largely a gaugino $\lambda^0$ with a small admixture of the Higgsinos - which will be relevant. For squark-quark-neutralino vertex, we will work out squark-quark-gaugino vertex replacing the gaugino by $-\tilde{\chi}_3^0$ with mass half of that of the gluino and also the squark-quark-Higgsino vertex replacing the Higgsino by $\frac{\tilde{f}{\cal V}^{-\frac{51}{72}}}{2}\tilde{\chi}_3^0$, and then add these contributions.

\section{Moduli Space Metric Miscellania}
\setcounter{equation}{0} \seceqcc

The results of this appendix are used in various places in section {\bf 4}. In $M_p=1$ units,
\begin{eqnarray}
\label{eq:gBa2bar}
& & g_{\sigma^B{\bar a}_{\bar 2}}\sim\nonumber\\
& & \frac{9 \left(\sqrt{\cal V} {a_1}+{a_2}\right) \sqrt[6]{\cal V} \left(-{a_1} {{\bar a}_{\bar 1}} {\cal V}^{7/6}+({{\bar a}_{\bar 1}} {a_2}+{a_1} {{\bar a}_{\bar 2}}) {\cal V}^{2/3}+{a_2}
   {{\bar a}_{\bar 2}} \sqrt[6]{\cal V}+{\sigma^B}+{\bar\sigma}^{\bar B}+  |{z_1}+{z_2}|^2\right)}{{\cal X}^2}\nonumber\\
   & & -\frac{3 \left(\sqrt{\cal V} {a_1}+{a_2}\right) \sqrt[6]{\cal V}}{2 \sqrt{-|{a_1}|^2 {\cal V}^{7/6}+({{\bar a}_{\bar 1}}
   {a_2}+{a_1} {{\bar a}_{\bar 2}}) {\cal V}^{2/3}+|{a_2}|^2 \sqrt[6]{\cal V}+{\sigma^B}+{\bar\sigma}^{\bar B}+ |{z_1}+{z_2}|^2} {\cal X}},
\end{eqnarray}
where
\begin{equation}
\hskip-0.5in{\cal X}\equiv \left(-\left( |{z_1}+{z_2}|^2+\sqrt[18]{\cal V}\right)^{3/2}+\left(-|{a_1}|^2
   {\cal V}^{7/6}+({{\bar a}_{\bar 1}} {a_2}+{a_1} {{\bar a}_{\bar 2}}) {\cal V}^{2/3}+|{a_2}|^2 \sqrt[6]{\cal V}+{\sigma^B}+{\bar\sigma}^{\bar B}+ |{z_1}+{z_2}|^2\right)^{3/2}+\sum_{\beta\in H_2^-}n^0_\beta\right).
   \end{equation}
The coefficient of $z_i$ in $g_{YM}g_{\sigma^B{\bar a}_{\bar 2}}$ around ${\bar z}_{\bar i}\sim {\cal V}^{\frac{1}{36}},
{\bar a}_{\bar I}\sim{\cal V}^{-\frac{1}{4}}, a_2\sim{\cal V}^{-\frac{1}{4}}$ is given by:
\begin{eqnarray}
\label{eq:gYMgBa2barexpzi}
& & \frac{1}{\sqrt{-{\cal V}^{11/12} {a_1}+{\cal V}^{5/12} {a_1}+{\sigma^B}+{\bar\sigma}^{\bar B}+\sqrt[6]{\cal V}+2
    \sqrt[18]{\cal V}+\frac{1}{\sqrt[3]{\cal V}}}}\times\nonumber\\
    & &\hskip-0.93in \Biggl\{\frac{9}{2} \left(\sqrt{\cal V} {a_1}+\frac{1}{\sqrt[4]{\cal V}}\right) \sqrt[6]{\cal V} \Biggl(\frac{2  \sqrt[36]{\cal V}}{{\cal Y}^2}  -\left(3  \sqrt{-{\cal V}^{11/12} {a_1}+{\cal V}^{5/12}
   {a_1}+{\sigma^B}+{\bar\sigma}^{\bar B}+\sqrt[6]{\cal V}+2  \sqrt[18]{\cal V}+\frac{1}{\sqrt[3]{\cal V}}} \sqrt[36]{\cal V}-3  \sqrt{2 \sqrt[18]{\cal V}
   +\sqrt[18]{\cal V}} \sqrt[36]{\cal V}\right)\nonumber\\
   & & \times\frac{2  \left(-{\cal V}^{11/12} {a_1}+{\cal V}^{5/12} {a_1}+{\sigma^B}+{\bar\sigma}^{\bar B}+\sqrt[6]{\cal V}+2
   \sqrt[18]{\cal V}+\frac{1}{\sqrt[3]{\cal V}}\right)}{{\cal Y}^3}\Biggr) -\frac{3}{2} \left(\sqrt{\cal V}
   {a_1}+\frac{1}{\sqrt[4]{\cal V}}\right) \sqrt[6]{\cal V}\nonumber\\
    & & \times\Biggl[\frac{1}{\sqrt{-{\cal V}^{11/12} {a_1}+{\cal V}^{5/12}
   {a_1}+{\sigma^B}+{\bar\sigma}^{\bar B}+\sqrt[6]{\cal V}+2  \sqrt[18]{\cal V}+\frac{1}{\sqrt[3]{\cal V}}}{\cal Y}}\Biggl\{\frac{3  \sqrt{2 \sqrt[18]{\cal V} +\sqrt[18]{\cal V}} \sqrt[36]{\cal V}}{\cal Y}\nonumber\\
   & & -\frac{3  \sqrt{-{\cal V}^{11/12} {a_1}+{\cal V}^{5/12} {a_1}+{\sigma^B}+{\bar\sigma}^{\bar B}+\sqrt[6]{\cal V}+2
    \sqrt[18]{\cal V}+\frac{1}{\sqrt[3]{\cal V}}} \sqrt[36]{\cal V}}{\cal Y}\Biggr\}\nonumber\\
   & & -\frac{ \sqrt[36]{\cal V}}{\left(-{\cal V}^{11/12} {a_1}+{\cal V}^{5/12}
   {a_1}+{\sigma^B}+{\bar\sigma}^{\bar B}+\sqrt[6]{\cal V}+2  \sqrt[18]{\cal V}+\frac{1}{\sqrt[3]{\cal V}}\right)^{3/2} {\cal Y}}\Biggr]\Biggr\}\nonumber\\
    & &\hskip-0.8in -\frac{ \sqrt[36]{\cal V} \left(\frac{9 \left(\sqrt{\cal V} {a_1}+\frac{1}{\sqrt[4]{\cal V}}\right) \left(-{\cal V}^{11/12}
   {a_1}+{\cal V}^{5/12} {a_1}+{\sigma^B}+{\bar\sigma}^{\bar B}+\sqrt[6]{\cal V}+2  \sqrt[18]{\cal V}+\frac{1}{\sqrt[3]{\cal V}}\right) \sqrt[6]{\cal V}}{2 {\cal Y}^2}-\frac{3 \left(\sqrt{\cal V} {a_1}+\frac{1}{\sqrt[4]{\cal V}}\right) \sqrt[6]{\cal V}}{2 \sqrt{-{\cal V}^{11/12}
   {a_1}+{\cal V}^{5/12} {a_1}+{\sigma^B}+{\bar\sigma}^{\bar B}+\sqrt[6]{\cal V}+2  \sqrt[18]{\cal V}+\frac{1}{\sqrt[3]{\cal V}}} {\cal Y}}\right)}{\left(-{\cal V}^{11/12} {a_1}+{\cal V}^{5/12} {a_1}+{\sigma^B}+{\bar\sigma}^{\bar B}+\sqrt[6]{\cal V}+2
    \sqrt[18]{\cal V}+\frac{1}{\sqrt[3]{\cal V}}\right)^{3/2}},
   \end{eqnarray}
   where
   \begin{equation}
   {\cal Y}\equiv -\left(2 \sqrt[18]{\cal V}
   +\sqrt[18]{\cal V}\right)^{3/2}+\left(-{\cal V}^{11/12} {a_1}+{\cal V}^{5/12} {a_1}+{\sigma^B}+{\bar\sigma}^{\bar B}+\sqrt[6]{\cal V}+2
   \sqrt[18]{\cal V}+\frac{1}{\sqrt[3]{\cal V}}\right)^{3/2}+\sum_{\beta\in H_2^-}n^0_\beta.
   \end{equation}

The coefficient of $\left(a_1-{\cal V}^{-\frac{1}{4}}\right)$ in (\ref{eq:gYMgBa2barexpzi}) is given by:
\begin{eqnarray*}
\label{eq:gYMgBa2barexpzia_1}
& & -\frac{1}{2 \left(2 \sqrt[18]{\cal V}
   +{\sigma^B}+{\bar\sigma}^{\bar B}-{\cal V}^{2/3}+2 \sqrt[6]{\cal V}+\frac{1}{\sqrt[3]{\cal V}}\right)^{3/2}}\Biggr\{\left({\cal V}^{5/12}-{\cal V}^{11/12}\right) \Biggl[\frac{9}{2} \left(\sqrt[4]{\cal V}+\frac{1}{\sqrt[4]{\cal V}}\right) \sqrt[6]{\cal V} \Biggl(\frac{2
   \sqrt[36]{\cal V}}{{\cal Y}^2}\nonumber\\
   & & \hskip-0.8in-\frac{2 \left(3  \sqrt{2 \sqrt[18]{\cal V} +{\sigma^B}+{\bar\sigma}^{\bar B}-{\cal V}^{2/3}+2
   \sqrt[6]{\cal V}+\frac{1}{\sqrt[3]{\cal V}}} \sqrt[36]{\cal V}-3  \sqrt{2 \sqrt[18]{\cal V} +\sqrt[18]{\cal V}} \sqrt[36]{\cal V}\right) \left(2 \sqrt[18]{\cal V}
   +{\sigma^B}+{\bar\sigma}^{\bar B}-{\cal V}^{2/3}+2 \sqrt[6]{\cal V}+\frac{1}{\sqrt[3]{\cal V}}\right)}{{\cal Y}^3}\Biggr)\nonumber\\
   & & -\frac{3}{2} \left(\sqrt[4]{\cal V}+\frac{1}{\sqrt[4]{\cal V}}\right) \sqrt[6]{\cal V} \Biggl(\frac{\frac{3
    \sqrt{2 \sqrt[18]{\cal V} +\sqrt[18]{\cal V}} \sqrt[36]{\cal V}}{\cal Y}-\frac{3  \sqrt{2 \sqrt[18]{\cal V}
   +{\sigma^B}+{\bar\sigma}^{\bar B}-{\cal V}^{2/3}+2 \sqrt[6]{\cal V}+\frac{1}{\sqrt[3]{\cal V}}} \sqrt[36]{\cal V}}{\cal Y}}{\sqrt{2 \sqrt[18]{\cal V} +{\sigma^B}+{\bar\sigma}^{\bar B}-{\cal V}^{2/3}+2 \sqrt[6]{\cal V}+\frac{1}{\sqrt[3]{\cal V}}}
  {\cal Y}}\nonumber\\
  & & -\frac{ \sqrt[36]{\cal V}}{\left(2 \sqrt[18]{\cal V} +{\sigma^B}+{\bar\sigma}^{\bar B}-{\cal V}^{2/3}+2
   \sqrt[6]{\cal V}+\frac{1}{\sqrt[3]{\cal V}}\right)^{3/2}{\cal Y}}\Biggr)\Biggr]\Biggr\}+\nonumber\\
    & & \sqrt[36]{\cal V} \Biggl[\frac{3
   \left({\cal V}^{5/12}-{\cal V}^{11/12}\right) \left(\frac{9 \left(\sqrt[4]{\cal V}+\frac{1}{\sqrt[4]{\cal V}}\right) \left(2 \sqrt[18]{\cal V}
   +{\sigma^B}+{\bar\sigma}^{\bar B}-{\cal V}^{2/3}+2 \sqrt[6]{\cal V}+\frac{1}{\sqrt[3]{\cal V}}\right) \sqrt[6]{\cal V}}{2 {\cal Y}^2}-\frac{3 \left(\sqrt[4]{\cal V}+\frac{1}{\sqrt[4]{\cal V}}\right) \sqrt[6]{\cal V}}{2 \sqrt{2 \sqrt[18]{\cal V}
   +{\sigma^B}+{\bar\sigma}^{\bar B}-{\cal V}^{2/3}+2 \sqrt[6]{\cal V}+\frac{1}{\sqrt[3]{\cal V}}}{\cal Y}}\right)}{2 \left(2 \sqrt[18]{\cal V}
   +{\sigma^B}+{\bar\sigma}^{\bar B}-{\cal V}^{2/3}+2 \sqrt[6]{\cal V}+\frac{1}{\sqrt[3]{\cal V}}\right)^{5/2}}\nonumber\\
   & & -\frac{1}{\left(2 \sqrt[18]{\cal V}
   +{\sigma^B}+{\bar\sigma}^{\bar B}-{\cal V}^{2/3}+2 \sqrt[6]{\cal V}+\frac{1}{\sqrt[3]{\cal V}}\right)^{3/2}}\nonumber\\
   & & \times\Biggl\{\frac{9}{2} \sqrt[6]{\cal V} \Biggl(\frac{\sqrt{\cal V} \left(2
   \sqrt[18]{\cal V} +{\sigma^B}+{\bar\sigma}^{\bar B}-{\cal V}^{2/3}+2 \sqrt[6]{\cal V}+\frac{1}{\sqrt[3]{\cal V}}\right)+\left(\sqrt[4]{\cal V}+\frac{1}{\sqrt[4]{\cal V}}\right)
   \left({\cal V}^{5/12}-{\cal V}^{11/12}\right)}{{\cal Y}^2}\nonumber\\
   & & -\frac{3
   \left(\sqrt[4]{\cal V}+\frac{1}{\sqrt[4]{\cal V}}\right) \left(2 \sqrt[18]{\cal V} +{\sigma^B}+{\bar\sigma}^{\bar B}-{\cal V}^{2/3}+2
   \sqrt[6]{\cal V}+\frac{1}{\sqrt[3]{\cal V}}\right)^{3/2} \left({\cal V}^{5/12}-{\cal V}^{11/12}\right)}{{\cal Y}^3}\Biggr)\nonumber\\
   & & -\frac{3}{2} \sqrt[6]{\cal V}
   \Biggl(\frac{\frac{\sqrt{\cal V}}{\sqrt{2 \sqrt[18]{\cal V} +{\sigma^B}+{\bar\sigma}^{\bar B}-{\cal V}^{2/3}+2
   \sqrt[6]{\cal V}+\frac{1}{\sqrt[3]{\cal V}}}}-\frac{\left(\sqrt[4]{\cal V}+\frac{1}{\sqrt[4]{\cal V}}\right) \left({\cal V}^{5/12}-{\cal V}^{11/12}\right)}{2 \left(2 \sqrt[18]{\cal V}
   +{\sigma^B}+{\bar\sigma}^{\bar B}-{\cal V}^{2/3}+2 \sqrt[6]{\cal V}+\frac{1}{\sqrt[3]{\cal V}}\right)^{3/2}}}{\cal Y}
   \nonumber\\
   & & +\frac{\left(\sqrt[4]{\cal V}+\frac{1}{\sqrt[4]{\cal V}}\right) \left(\frac{3 \sqrt{2 \sqrt[18]{\cal V}
   +{\sigma^B}+{\bar\sigma}^{\bar B}-{\cal V}^{2/3}+2 \sqrt[6]{\cal V}+\frac{1}{\sqrt[3]{\cal V}}} {\cal V}^{11/12}}{2 {\cal Y}}-\frac{3 \sqrt{2 \sqrt[18]{\cal V} +{\sigma^B}+{\bar\sigma}^{\bar B}-{\cal V}^{2/3}+2
   \sqrt[6]{\cal V}+\frac{1}{\sqrt[3]{\cal V}}} {\cal V}^{5/12}}{2 {\cal Y}}\right)}{\sqrt{2 \sqrt[18]{\cal V}
   +{\sigma^B}+{\bar\sigma}^{\bar B}-{\cal V}^{2/3}+2 \sqrt[6]{\cal V}+\frac{1}{\sqrt[3]{\cal V}}}{\cal Y}}\Biggr)\Biggr\}\Biggr]\nonumber\\
    & & +\frac{1}{\sqrt{2 \sqrt[18]{\cal V}
   +{\sigma^B}+{\bar\sigma}^{\bar B}-{\cal V}^{2/3}+2 \sqrt[6]{\cal V}+\frac{1}{\sqrt[3]{\cal V}}}}\Biggl\{\frac{9}{2} \sqrt[6]{\cal V} \Biggl[\sqrt{\cal V}
   \Biggl(\frac{2  \sqrt[36]{\cal V}}{{\cal Y}^2}\end{eqnarray*}

   \begin{eqnarray*}
   & & \hskip-0.8in-\frac{2 \left(3 \sqrt{2 \sqrt[18]{\cal V}
   +{\sigma^B}+{\bar\sigma}^{\bar B}-{\cal V}^{2/3}+2 \sqrt[6]{\cal V}+\frac{1}{\sqrt[3]{\cal V}}} \sqrt[36]{\cal V}-3  \sqrt{2 \sqrt[18]{\cal V} +\sqrt[18]{\cal V}}
   \sqrt[36]{\cal V}\right) \left(2 \sqrt[18]{\cal V} +{\sigma^B}+{\bar\sigma}^{\bar B}-{\cal V}^{2/3}+2 \sqrt[6]{\cal V}+\frac{1}{\sqrt[3]{\cal V}}\right)}{{\cal Y}^3}\Biggr)\nonumber\\
   & & +\left(\sqrt[4]{\cal V}+\frac{1}{\sqrt[4]{\cal V}}\right) \Biggl[-\frac{6  \sqrt{2 \sqrt[18]{\cal V}
   +{\sigma^B}+{\bar\sigma}^{\bar B}-{\cal V}^{2/3}+2 \sqrt[6]{\cal V}+\frac{1}{\sqrt[3]{\cal V}}} \sqrt[36]{\cal V} \left({\cal V}^{5/12}-{\cal V}^{11/12}\right)}{{\cal Y}^3}\nonumber\\
   & &\hskip-0.8in -2 \Biggl(\frac{1}{{\cal Y}^3}\Biggl\{\left(3  \sqrt{2 \sqrt[18]{\cal V}
   +{\sigma^B}+{\bar\sigma}^{\bar B}-{\cal V}^{2/3}+2 \sqrt[6]{\cal V}+\frac{1}{\sqrt[3]{\cal V}}} \sqrt[36]{\cal V}-3  \sqrt{2 \sqrt[18]{\cal V} +\sqrt[18]{\cal V}}
   \sqrt[36]{\cal V}\right) \left({\cal V}^{5/12}-{\cal V}^{11/12}\right)\nonumber\\
   & & +\frac{3}{2}  \sqrt{2 \sqrt[18]{\cal V} +{\sigma^B}+{\bar\sigma}^{\bar B}-{\cal V}^{2/3}+2
   \sqrt[6]{\cal V}+\frac{1}{\sqrt[3]{\cal V}}} \sqrt[36]{\cal V} \left({\cal V}^{5/12}-{\cal V}^{11/12}\right)\Biggr\}
  \nonumber\\
  & &  -\frac{1}{2 {\cal Y}^4}\Biggl\{9 \left(3  \sqrt{2
   \sqrt[18]{\cal V} +{\sigma^B}+{\bar\sigma}^{\bar B}-{\cal V}^{2/3}+2 \sqrt[6]{\cal V}+\frac{1}{\sqrt[3]{\cal V}}} \sqrt[36]{\cal V}-3  \sqrt{2 \sqrt[18]{\cal V}
   +\sqrt[18]{\cal V}} \sqrt[36]{\cal V}\right)\nonumber\\
    & & \times\left(2 \sqrt[18]{\cal V} +{\sigma^B}+{\bar\sigma}^{\bar B}-{\cal V}^{2/3}+2
   \sqrt[6]{\cal V}+\frac{1}{\sqrt[3]{\cal V}}\right)^{3/2} \left({\cal V}^{5/12}-{\cal V}^{11/12}\right)\Biggr\}\Biggr)\Biggr]\Biggr]\nonumber\\
   & &\hskip-0.9in -\frac{3}{2}
   \sqrt[6]{\cal V} \Biggl[\sqrt{\cal V} \left(\frac{\frac{3  \sqrt{2 \sqrt[18]{\cal V} +\sqrt[18]{\cal V}} \sqrt[36]{\cal V}}{\cal Y}-\frac{3  \sqrt{2 \sqrt[18]{\cal V} +{\sigma^B}+{\bar\sigma}^{\bar B}-{\cal V}^{2/3}+2
   \sqrt[6]{\cal V}+\frac{1}{\sqrt[3]{\cal V}}} \sqrt[36]{\cal V}}{\cal Y}}{\sqrt{2 \sqrt[18]{\cal V}
   +{\sigma^B}+{\bar\sigma}^{\bar B}-{\cal V}^{2/3}+2 \sqrt[6]{\cal V}+\frac{1}{\sqrt[3]{\cal V}}} {\cal Y}}-\frac{ \sqrt[36]{\cal V}}{\left(2
   \sqrt[18]{\cal V} +{\sigma^B}+{\bar\sigma}^{\bar B}-{\cal V}^{2/3}+2 \sqrt[6]{\cal V}+\frac{1}{\sqrt[3]{\cal V}}\right)^{3/2} {\cal Y}}\right)\nonumber\\
   & &+\left(\sqrt[4]{\cal V}+\frac{1}{\sqrt[4]{\cal V}}\right) \Biggl[ \sqrt[36]{\cal V} \Biggl(\frac{3
   \left({\cal V}^{5/12}-{\cal V}^{11/12}\right)}{2 \left(2 \sqrt[18]{\cal V} +{\sigma^B}+{\bar\sigma}^{\bar B}-{\cal V}^{2/3}+2 \sqrt[6]{\cal V}+\frac{1}{\sqrt[3]{\cal V}}\right)^{5/2}
  {\cal Y}}\nonumber\\
  & & -\frac{\frac{3 \sqrt{2 \sqrt[18]{\cal V} +{\sigma^B}+{\bar\sigma}^{\bar B}-{\cal V}^{2/3}+2
   \sqrt[6]{\cal V}+\frac{1}{\sqrt[3]{\cal V}}} {\cal V}^{11/12}}{2 {\cal Y}}-\frac{3 \sqrt{2 \sqrt[18]{\cal V}
   +{\sigma^B}+{\bar\sigma}^{\bar B}-{\cal V}^{2/3}+2 \sqrt[6]{\cal V}+\frac{1}{\sqrt[3]{\cal V}}} {\cal V}^{5/12}}{2 {\cal Y}}}{\left(2 \sqrt[18]{\cal V} +{\sigma^B}+{\bar\sigma}^{\bar B}-{\cal V}^{2/3}+2
   \sqrt[6]{\cal V}+\frac{1}{\sqrt[3]{\cal V}}\right)^{3/2}{\cal Y}}\Biggr)\nonumber\\
   & & +\left(\frac{3 \sqrt{2 \sqrt[18]{\cal V}
   +\sqrt[18]{\cal V}} \sqrt[36]{\cal V}}{\cal Y}-\frac{3  \sqrt{2 \sqrt[18]{\cal V}
   +{\sigma^B}+{\bar\sigma}^{\bar B}-{\cal V}^{2/3}+2 \sqrt[6]{\cal V}+\frac{1}{\sqrt[3]{\cal V}}} \sqrt[36]{\cal V}}{\cal Y}\right)\nonumber\\
   & & \times \Biggl(\frac{\frac{3 \sqrt{2 \sqrt[18]{\cal V} +{\sigma^B}+{\bar\sigma}^{\bar B}-{\cal V}^{2/3}+2
   \sqrt[6]{\cal V}+\frac{1}{\sqrt[3]{\cal V}}} {\cal V}^{11/12}}{2 {\cal Y}}-\frac{3 \sqrt{2 \sqrt[18]{\cal V}
   +{\sigma^B}+{\bar\sigma}^{\bar B}-{\cal V}^{2/3}+2 \sqrt[6]{\cal V}+\frac{1}{\sqrt[3]{\cal V}}} {\cal V}^{5/12}}{2 {\cal Y}}}{\sqrt{2 \sqrt[18]{\cal V} +{\sigma^B}+{\bar\sigma}^{\bar B}-{\cal V}^{2/3}+2
   \sqrt[6]{\cal V}+\frac{1}{\sqrt[3]{\cal V}}} {\cal Y}}\nonumber\\
   & & -\frac{{\cal V}^{5/12}-{\cal V}^{11/12}}{2 \left(2 \sqrt[18]{\cal V}
   +{\sigma^B}+{\bar\sigma}^{\bar B}-{\cal V}^{2/3}+2 \sqrt[6]{\cal V}+\frac{1}{\sqrt[3]{\cal V}}\right)^{3/2}{\cal Y}}\Biggr)\nonumber\\
   & & +\frac{1}{\sqrt{2 \sqrt[18]{\cal V} +{\sigma^B}+{\bar\sigma}^{\bar B}-{\cal V}^{2/3}+2
   \sqrt[6]{\cal V}+\frac{1}{\sqrt[3]{\cal V}}} {\cal Y}}\end{eqnarray*}

   \begin{eqnarray*}
   & & \times\Biggl\{\frac{3  \sqrt{2 \sqrt[18]{\cal V} +\sqrt[18]{\cal V}} \sqrt[36]{\cal V} \left(\frac{3
   \sqrt{2 \sqrt[18]{\cal V} +{\sigma^B}+{\bar\sigma}^{\bar B}-{\cal V}^{2/3}+2 \sqrt[6]{\cal V}+\frac{1}{\sqrt[3]{\cal V}}} {\cal V}^{11/12}}{2 {\cal Y}}-\frac{3 \sqrt{2 \sqrt[18]{\cal V} +{\sigma^B}+{\bar\sigma}^{\bar B}-{\cal V}^{2/3}+2
   \sqrt[6]{\cal V}+\frac{1}{\sqrt[3]{\cal V}}} {\cal V}^{5/12}}{2{\cal Y}}\right)}{\cal Y}\nonumber\\
   & &-3
    \sqrt[36]{\cal V} \Biggl(\frac{{\cal V}^{5/12}-{\cal V}^{11/12}}{2 \sqrt{2 \sqrt[18]{\cal V} +{\sigma^B}+{\bar\sigma}^{\bar B}-{\cal V}^{2/3}+2
   \sqrt[6]{\cal V}+\frac{1}{\sqrt[3]{\cal V}}} {\cal Y}} +  \frac{\sqrt{2 \sqrt[18]{\cal V}
   +{\sigma^B}+{\bar\sigma}^{\bar B}-{\cal V}^{2/3}+2 \sqrt[6]{\cal V}+\frac{1}{\sqrt[3]{\cal V}}}}{\cal Y}\nonumber\\
   & &\hskip-0.8in \times \left(\frac{3 \sqrt{2 \sqrt[18]{\cal V}
   +{\sigma^B}+{\bar\sigma}^{\bar B}-{\cal V}^{2/3}+2 \sqrt[6]{\cal V}+\frac{1}{\sqrt[3]{\cal V}}} {\cal V}^{11/12}}{2 {\cal Y}}-\frac{3 \sqrt{2 \sqrt[18]{\cal V} +{\sigma^B}+{\bar\sigma}^{\bar B}-{\cal V}^{2/3}+2
   \sqrt[6]{\cal V}+\frac{1}{\sqrt[3]{\cal V}}} {\cal V}^{5/12}}{2 {\cal Y}}\right)\Biggr)\Biggr\}\Biggr]\Biggr]\Biggr\},
   \end{eqnarray*}
This needs to be simplified with the understanding that $\sigma^B+{\bar\sigma}^{\bar B} - |a_1|^2{\cal V}^{\frac{7}{6}}\sim{\cal V}^{\frac{1}{18}}$. After doing so, one obtains ${\cal V}^{-\frac{5}{36}} $ as in (\ref{eq:gYMgBIbar_zi_aI}).

Similarly, the coefficient of $z_i$ in $g_{YM}g_{\sigma^B{\bar a}_{\bar 1}}$ when expanded around ${\bar z}_{\bar i}\sim {\cal V}^{\frac{1}{36}},
{\bar a}_{\bar I}\sim{\cal V}^{-\frac{1}{4}}, a_2\sim{\cal V}^{-\frac{1}{4}}$ is given by:
\begin{eqnarray}
\label{eq:gYMgBa1barexpzi}
& & \frac{1}{\sqrt{-{\cal V}^{11/12} {a_1}+{\cal V}^{5/12} {a_1}+{\sigma^B}+{\bar\sigma}^{\bar B}+\sqrt[6]{\cal V}+2
    \sqrt[18]{\cal V}+\frac{1}{\sqrt[3]{\cal V}}}}\Biggl\{\frac{9}{2} \left(\frac{1}{\sqrt[4]{\cal V}}-{a_1} \sqrt{\cal V}\right) {\cal V}^{2/3} \Biggl[\frac{2  \sqrt[36]{\cal V}}{{\cal Y}^2}\nonumber\\
   & & -\frac{2 \left(3  \sqrt{-{\cal V}^{11/12} {a_1}+{\cal V}^{5/12}
   {a_1}+{\sigma^B}+{\bar\sigma}^{\bar B}+\sqrt[6]{\cal V}+2  \sqrt[18]{\cal V}+\frac{1}{\sqrt[3]{\cal V}}} \sqrt[36]{\cal V}-3  \sqrt{2 \sqrt[18]{\cal V}
   +\sqrt[18]{\cal V}} \sqrt[36]{\cal V}\right) }{{\cal Y}^3}\nonumber\\
   & & \times \left(-{\cal V}^{11/12} {a_1}+{\cal V}^{5/12} {a_1}+{\sigma^B}+{\bar\sigma}^{\bar B}+\sqrt[6]{\cal V}+2
   \sqrt[18]{\cal V}+\frac{1}{\sqrt[3]{\cal V}}\right)\Biggr]\nonumber\\
   & & -\frac{3}{2}
   \left(\frac{1}{\sqrt[4]{\cal V}}-{a_1} \sqrt{\cal V}\right) {\cal V}^{2/3} \Biggl[\frac{\frac{3  \sqrt{2 \sqrt[18]{\cal V} +\sqrt[18]{\cal V}} \sqrt[36]{\cal V}}{\cal Y}-\frac{3  \sqrt{-{\cal V}^{11/12} {a_1}+{\cal V}^{5/12} {a_1}+{\sigma^B}+{\bar\sigma}^{\bar B}+\sqrt[6]{\cal V}+2
    \sqrt[18]{\cal V}+\frac{1}{\sqrt[3]{\cal V}}} \sqrt[36]{\cal V}}{\cal Y}}{\sqrt{-{\cal V}^{11/12} {a_1}+{\cal V}^{5/12}
   {a_1}+{\sigma^B}+{\bar\sigma}^{\bar B}+\sqrt[6]{\cal V}+2  \sqrt[18]{\cal V}+\frac{1}{\sqrt[3]{\cal V}}}{\cal Y}}\nonumber\\
   & & -\frac{ \sqrt[36]{\cal V}}{\left(-{\cal V}^{11/12} {a_1}+{\cal V}^{5/12}
   {a_1}+{\sigma^B}+{\bar\sigma}^{\bar B}+\sqrt[6]{\cal V}+2  \sqrt[18]{\cal V}+\frac{1}{\sqrt[3]{\cal V}}\right)^{3/2} {\cal Y}}\Biggr]\Biggr\}\nonumber\\
   & & \hskip-0.8in-\frac{ \sqrt[36]{\cal V} \left(\frac{9 \left(\frac{1}{\sqrt[4]{\cal V}}-{a_1} \sqrt{\cal V}\right) \left(-{\cal V}^{11/12}
   {a_1}+{\cal V}^{5/12} {a_1}+{\sigma^B}+{\bar\sigma}^{\bar B}+\sqrt[6]{\cal V}+2  \sqrt[18]{\cal V}+\frac{1}{\sqrt[3]{\cal V}}\right) {\cal V}^{2/3}}{2 {\cal Y}^2}-\frac{3 \left(\frac{1}{\sqrt[4]{\cal V}}-{a_1} \sqrt{\cal V}\right) {\cal V}^{2/3}}{2 \sqrt{-{\cal V}^{11/12}
   {a_1}+{\cal V}^{5/12} {a_1}+{\sigma^B}+{\bar\sigma}^{\bar B}+\sqrt[6]{\cal V}+2  \sqrt[18]{\cal V}+\frac{1}{\sqrt[3]{\cal V}}}{\cal Y}}\right)}{\left(-{\cal V}^{11/12} {a_1}+{\cal V}^{5/12} {a_1}+{\sigma^B}+{\bar\sigma}^{\bar B}+\sqrt[6]{\cal V}+2
    \sqrt[18]{\cal V}+\frac{1}{\sqrt[3]{\cal V}}\right)^{3/2}},\nonumber\\
    &&
\end{eqnarray}
the coefficient of $\left(a_1-{\cal V}^{-\frac{1}{4}}\right)$ wherein is given by:
\begin{eqnarray*}
& & -\frac{1}{2 \left(2 \sqrt[18]{\cal V}
   +{\sigma^B}+{\bar\sigma}^{\bar B}-{\cal V}^{2/3}+2 \sqrt[6]{\cal V}+\frac{1}{\sqrt[3]{\cal V}}\right)^{3/2}}\Biggl\{\left({\cal V}^{5/12}-{\cal V}^{11/12}\right) \Biggl[\frac{9}{2} \left(\frac{1}{\sqrt[4]{\cal V}}-\sqrt[4]{\cal V}\right) {\cal V}^{2/3}\times\nonumber\\
    & &\biggl(\frac{2  \sqrt[36]{\cal V}}{{\cal Y}^2}-\frac{2 \left(3  \sqrt{2 \sqrt[18]{\cal V} +{\sigma^B}+{\bar\sigma}^{\bar B}-{\cal V}^{2/3}+2
   \sqrt[6]{\cal V}+\frac{1}{\sqrt[3]{\cal V}}} \sqrt[36]{\cal V}-3  \sqrt{2 \sqrt[18]{\cal V} +\sqrt[18]{\cal V}} \sqrt[36]{\cal V}\right) }{{\cal Y}^3}\nonumber\\
   & & \times \left(2 \sqrt[18]{\cal V}
   +{\sigma^B}+{\bar\sigma}^{\bar B}-{\cal V}^{2/3}+2 \sqrt[6]{\cal V}+\frac{1}{\sqrt[3]{\cal V}}\right)\biggr)\nonumber\\
   & & -\frac{3}{2} \left(\frac{1}{\sqrt[4]{\cal V}}-\sqrt[4]{\cal V}\right) {\cal V}^{2/3} \Biggl(\frac{\frac{3
   \sqrt{2 \sqrt[18]{\cal V} +\sqrt[18]{\cal V}} \sqrt[36]{\cal V}}{\cal Y}-\frac{3  \sqrt{2 \sqrt[18]{\cal V}
   +{\sigma^B}+{\bar\sigma}^{\bar B}-{\cal V}^{2/3}+2 \sqrt[6]{\cal V}+\frac{1}{\sqrt[3]{\cal V}}} \sqrt[36]{\cal V}}{\cal Y}}{\sqrt{2 \sqrt[18]{\cal V} +{\sigma^B}+{\bar\sigma}^{\bar B}-{\cal V}^{2/3}+2 \sqrt[6]{\cal V}+\frac{1}{\sqrt[3]{\cal V}}}
  {\cal Y}}\nonumber\\
   & & -\frac{ \sqrt[36]{\cal V}}{\left(2 \sqrt[18]{\cal V} +{\sigma^B}+{\bar\sigma}^{\bar B}-{\cal V}^{2/3}+2
   \sqrt[6]{\cal V}+\frac{1}{\sqrt[3]{\cal V}}\right)^{3/2} {\cal Y}}\Biggr)\Biggr]\Biggr\}+
   \nonumber\\
   & &  \sqrt[36]{\cal V} \Biggl[\frac{3
   \left({\cal V}^{5/12}-{\cal V}^{11/12}\right) \left(\frac{9 \left(\frac{1}{\sqrt[4]{\cal V}}-\sqrt[4]{\cal V}\right) \left(2 \sqrt[18]{\cal V}
   +{\sigma^B}+{\bar\sigma}^{\bar B}-{\cal V}^{2/3}+2 \sqrt[6]{\cal V}+\frac{1}{\sqrt[3]{\cal V}}\right) {\cal V}^{2/3}}{2 {\cal Y}^2}-\frac{3 \left(\frac{1}{\sqrt[4]{\cal V}}-\sqrt[4]{\cal V}\right) {\cal V}^{2/3}}{2 \sqrt{2 \sqrt[18]{\cal V}
   +{\sigma^B}+{\bar\sigma}^{\bar B}-{\cal V}^{2/3}+2 \sqrt[6]{\cal V}+\frac{1}{\sqrt[3]{\cal V}}}{\cal Y}}\right)}{2 \left(2 \sqrt[18]{\cal V}
   +{\sigma^B}+{\bar\sigma}^{\bar B}-{\cal V}^{2/3}+2 \sqrt[6]{\cal V}+\frac{1}{\sqrt[3]{\cal V}}\right)^{5/2}}\nonumber\\
   & & -\frac{1}{\left(2 \sqrt[18]{\cal V}
   +{\sigma^B}+{\bar\sigma}^{\bar B}-{\cal V}^{2/3}+2 \sqrt[6]{\cal V}+\frac{1}{\sqrt[3]{\cal V}}\right)^{3/2}}\nonumber\\
   & & \times\Biggl\{\frac{9}{2} {\cal V}^{2/3}
   \Biggl(\frac{\left(\frac{1}{\sqrt[4]{\cal V}}-\sqrt[4]{\cal V}\right) \left({\cal V}^{5/12}-{\cal V}^{11/12}\right)-\left(2 \sqrt[18]{\cal V} +{\sigma^B}+{\bar\sigma}^{\bar B}-{\cal V}^{2/3}+2
   \sqrt[6]{\cal V}+\frac{1}{\sqrt[3]{\cal V}}\right) \sqrt{\cal V}}{{\cal Y}^2}\nonumber\\
   & & -\frac{3
   \left(\frac{1}{\sqrt[4]{\cal V}}-\sqrt[4]{\cal V}\right) \left(2 \sqrt[18]{\cal V} +{\sigma^B}+{\bar\sigma}^{\bar B}-{\cal V}^{2/3}+2
   \sqrt[6]{\cal V}+\frac{1}{\sqrt[3]{\cal V}}\right)^{3/2} \left({\cal V}^{5/12}-{\cal V}^{11/12}\right)}{{\cal Y}^3}\Biggr)\nonumber\\
   & & -\frac{3}{2} {\cal V}^{2/3}
   \Biggl[\frac{-\frac{\left(\frac{1}{\sqrt[4]{\cal V}}-\sqrt[4]{\cal V}\right) \left({\cal V}^{5/12}-{\cal V}^{11/12}\right)}{2 \left(2 \sqrt[18]{\cal V}
   +{\sigma^B}+{\bar\sigma}^{\bar B}-{\cal V}^{2/3}+2 \sqrt[6]{\cal V}+\frac{1}{\sqrt[3]{\cal V}}\right)^{3/2}}-\frac{\sqrt{\cal V}}{\sqrt{2 \sqrt[18]{\cal V}
   +{\sigma^B}+{\bar\sigma}^{\bar B}-{\cal V}^{2/3}+2 \sqrt[6]{\cal V}+\frac{1}{\sqrt[3]{\cal V}}}}}{\cal Y}\nonumber\\
   & & +\frac{\left(\frac{1}{\sqrt[4]{\cal V}}-\sqrt[4]{\cal V}\right) \left(\frac{3 \sqrt{2 \sqrt[18]{\cal V}
   +{\sigma^B}+{\bar\sigma}^{\bar B}-{\cal V}^{2/3}+2 \sqrt[6]{\cal V}+\frac{1}{\sqrt[3]{\cal V}}} {\cal V}^{11/12}}{2 {\cal Y}}-\frac{3 \sqrt{2 \sqrt[18]{\cal V} +{\sigma^B}+{\bar\sigma}^{\bar B}-{\cal V}^{2/3}+2
   \sqrt[6]{\cal V}+\frac{1}{\sqrt[3]{\cal V}}} {\cal V}^{5/12}}{2{\cal Y}}\right)}{\sqrt{2 \sqrt[18]{\cal V}
   +{\sigma^B}+{\bar\sigma}^{\bar B}-{\cal V}^{2/3}+2 \sqrt[6]{\cal V}+\frac{1}{\sqrt[3]{\cal V}}} {\cal Y}}\Biggr]\Biggr\}\Biggr]\nonumber\\
   & & +\frac{1}{\sqrt{2 \sqrt[18]{\cal V}
   +{\sigma^B}+{\bar\sigma}^{\bar B}-{\cal V}^{2/3}+2 \sqrt[6]{\cal V}+\frac{1}{\sqrt[3]{\cal V}}}}\Biggl\{\frac{9}{2} {\cal V}^{2/3}
   \Biggl(\left(\frac{1}{\sqrt[4]{\cal V}}-\sqrt[4]{\cal V}\right)\nonumber\\
    & & \times\Biggl[-\frac{6  \sqrt{2 \sqrt[18]{\cal V} +{\sigma^B}+{\bar\sigma}^{\bar B}-{\cal V}^{2/3}+2
   \sqrt[6]{\cal V}+\frac{1}{\sqrt[3]{\cal V}}} \sqrt[36]{\cal V} \left({\cal V}^{5/12}-{\cal V}^{11/12}\right)}{{\cal Y}^3}-
   \end{eqnarray*}

   \begin{eqnarray*}
   & & 2 \Biggl(\frac{1}{{\cal Y}^3}\Biggl\{\left(3
   \sqrt{2 \sqrt[18]{\cal V} +{\sigma^B}+{\bar\sigma}^{\bar B}-{\cal V}^{2/3}+2 \sqrt[6]{\cal V}+\frac{1}{\sqrt[3]{\cal V}}} \sqrt[36]{\cal V}-3  \sqrt{2 \sqrt[18]{\cal V}
   +\sqrt[18]{\cal V}} \sqrt[36]{\cal V}\right)\nonumber\\
    & & \times\left({\cal V}^{5/12}-{\cal V}^{11/12}\right)+\frac{3}{2}  \sqrt{2 \sqrt[18]{\cal V}
   +{\sigma^B}+{\bar\sigma}^{\bar B}-{\cal V}^{2/3}+2 \sqrt[6]{\cal V}+\frac{1}{\sqrt[3]{\cal V}}} \sqrt[36]{\cal V} \left({\cal V}^{5/12}-{\cal V}^{11/12}\right)\Biggr\}\nonumber\\
   & & -\frac{1}{2 {\cal Y}^4}\Biggl\{9 \left(3  \sqrt{2 \sqrt[18]{\cal V} +{\sigma^B}+{\bar\sigma}^{\bar B}-{\cal V}^{2/3}+2
   \sqrt[6]{\cal V}+\frac{1}{\sqrt[3]{\cal V}}} \sqrt[36]{\cal V}-3  \sqrt{2 \sqrt[18]{\cal V} +\sqrt[18]{\cal V}} \sqrt[36]{\cal V}\right)\nonumber\\
   & & \times \left(2 \sqrt[18]{\cal V}
   +{\sigma^B}+{\bar\sigma}^{\bar B}-{\cal V}^{2/3}+2 \sqrt[6]{\cal V}+\frac{1}{\sqrt[3]{\cal V}}\right)^{3/2} \left({\cal V}^{5/12}-{\cal V}^{11/12}\right)\Biggr\}\Biggr)\Biggr] -\sqrt{\cal V} \Biggl(\frac{2  \sqrt[36]{\cal V}}{{\cal Y}^2}-\nonumber\\
   & & \hskip-0.5in\frac{2 \left(3  \sqrt{2 \sqrt[18]{\cal V} +{\sigma^B}+{\bar\sigma}^{\bar B}-{\cal V}^{2/3}+2
   \sqrt[6]{\cal V}+\frac{1}{\sqrt[3]{\cal V}}} \sqrt[36]{\cal V}-3  \sqrt{2 \sqrt[18]{\cal V} +\sqrt[18]{\cal V}} \sqrt[36]{\cal V}\right) \left(2 \sqrt[18]{\cal V}
   +{\sigma^B}+{\bar\sigma}^{\bar B}-{\cal V}^{2/3}+2 \sqrt[6]{\cal V}+\frac{1}{\sqrt[3]{\cal V}}\right)}{{\cal Y}^3}\Biggr)\Biggr)\nonumber\\
   & & -\frac{3}{2} {\cal V}^{2/3} \Biggl(\left(\frac{1}{\sqrt[4]{\cal V}}-\sqrt[4]{\cal V}\right) \Biggl[
   \sqrt[36]{\cal V} \Biggl(\frac{3 \left({\cal V}^{5/12}-{\cal V}^{11/12}\right)}{2 \left(2 \sqrt[18]{\cal V} +{\sigma^B}+{\bar\sigma}^{\bar B}-{\cal V}^{2/3}+2
   \sqrt[6]{\cal V}+\frac{1}{\sqrt[3]{\cal V}}\right)^{5/2}{\cal Y}}\nonumber\\
   & & -\frac{\frac{3 \sqrt{2 \sqrt[18]{\cal V}
   +{\sigma^B}+{\bar\sigma}^{\bar B}-{\cal V}^{2/3}+2 \sqrt[6]{\cal V}+\frac{1}{\sqrt[3]{\cal V}}} {\cal V}^{11/12}}{2 {\cal Y}}-\frac{3 \sqrt{2 \sqrt[18]{\cal V} +{\sigma^B}+{\bar\sigma}^{\bar B}-{\cal V}^{2/3}+2
   \sqrt[6]{\cal V}+\frac{1}{\sqrt[3]{\cal V}}} {\cal V}^{5/12}}{2{\cal Y}}}{\left(2 \sqrt[18]{\cal V}
   +{\sigma^B}+{\bar\sigma}^{\bar B}-{\cal V}^{2/3}+2 \sqrt[6]{\cal V}+\frac{1}{\sqrt[3]{\cal V}}\right)^{3/2} {\cal Y}}\Biggr)\nonumber\\
   & & +\left(\frac{3  \sqrt{2 \sqrt[18]{\cal V} +\sqrt[18]{\cal V}} \sqrt[36]{\cal V}}{\cal Y}-\frac{3  \sqrt{2 \sqrt[18]{\cal V} +{\sigma^B}+{\bar\sigma}^{\bar B}-{\cal V}^{2/3}+2
   \sqrt[6]{\cal V}+\frac{1}{\sqrt[3]{\cal V}}} \sqrt[36]{\cal V}}{\cal Y}\right)\nonumber\\
   & & \times \Biggl(\frac{\frac{3 \sqrt{2 \sqrt[18]{\cal V}
   +{\sigma^B}+{\bar\sigma}^{\bar B}-{\cal V}^{2/3}+2 \sqrt[6]{\cal V}+\frac{1}{\sqrt[3]{\cal V}}} {\cal V}^{11/12}}{2 {\cal Y}}-\frac{3 \sqrt{2 \sqrt[18]{\cal V} +{\sigma^B}+{\bar\sigma}^{\bar B}-{\cal V}^{2/3}+2
   \sqrt[6]{\cal V}+\frac{1}{\sqrt[3]{\cal V}}} {\cal V}^{5/12}}{2 {\cal Y}}}{\sqrt{2 \sqrt[18]{\cal V}
   +{\sigma^B}+{\bar\sigma}^{\bar B}-{\cal V}^{2/3}+2 \sqrt[6]{\cal V}+\frac{1}{\sqrt[3]{\cal V}}} {\cal Y}}\nonumber\\
   & & -\frac{{\cal V}^{5/12}-{\cal V}^{11/12}}{2 \left(2
   \sqrt[18]{\cal V} +{\sigma^B}+{\bar\sigma}^{\bar B}-{\cal V}^{2/3}+2 \sqrt[6]{\cal V}+\frac{1}{\sqrt[3]{\cal V}}\right)^{3/2} {\cal Y}}\Biggr)\nonumber\\
   & & +\frac{1}{\sqrt{2 \sqrt[18]{\cal V} +{\sigma^B}+{\bar\sigma}^{\bar B}-{\cal V}^{2/3}+2
   \sqrt[6]{\cal V}+\frac{1}{\sqrt[3]{\cal V}}} {\cal Y}}\nonumber\\
   & & \times\Biggl\{\frac{3  \sqrt{2 \sqrt[18]{\cal V} +\sqrt[18]{\cal V}} \sqrt[36]{\cal V} \left(\frac{3
   \sqrt{2 \sqrt[18]{\cal V} +{\sigma^B}+{\bar\sigma}^{\bar B}-{\cal V}^{2/3}+2 \sqrt[6]{\cal V}+\frac{1}{\sqrt[3]{\cal V}}} {\cal V}^{11/12}}{2 {\cal Y}}-\frac{3 \sqrt{2 \sqrt[18]{\cal V} +{\sigma^B}+{\bar\sigma}^{\bar B}-{\cal V}^{2/3}+2
   \sqrt[6]{\cal V}+\frac{1}{\sqrt[3]{\cal V}}} {\cal V}^{5/12}}{2{\cal Y}}\right)}{\cal Y}\nonumber\\
   & & -3
    \sqrt[36]{\cal V} \Biggl(\frac{{\cal V}^{5/12}-{\cal V}^{11/12}}{2 \sqrt{2 \sqrt[18]{\cal V} +{\sigma^B}+{\bar\sigma}^{\bar B}-{\cal V}^{2/3}+2
   \sqrt[6]{\cal V}+\frac{1}{\sqrt[3]{\cal V}}} {\cal Y}}+
   \end{eqnarray*}

   \begin{eqnarray*}
   & & \hskip-0.8in\frac{\sqrt{2 \sqrt[18]{\cal V}
   +{\sigma^B}+{\bar\sigma}^{\bar B}-{\cal V}^{2/3}+2 \sqrt[6]{\cal V}+\frac{1}{\sqrt[3]{\cal V}}} \left(\frac{3 \sqrt{2 \sqrt[18]{\cal V}
   +{\sigma^B}+{\bar\sigma}^{\bar B}-{\cal V}^{2/3}+2 \sqrt[6]{\cal V}+\frac{1}{\sqrt[3]{\cal V}}} {\cal V}^{11/12}}{2 {\cal Y}}-\frac{3 \sqrt{2 \sqrt[18]{\cal V} +{\sigma^B}+{\bar\sigma}^{\bar B}-{\cal V}^{2/3}+2
   \sqrt[6]{\cal V}+\frac{1}{\sqrt[3]{\cal V}}} {\cal V}^{5/12}}{2{\cal Y}}\right)}{\cal Y}\Biggr)\Biggr\}\Biggr]\nonumber\\
   & &\hskip-0.8in -\sqrt{\cal V} \Biggl[\frac{\frac{3  \sqrt{2
   \sqrt[18]{\cal V} +\sqrt[18]{\cal V}} \sqrt[36]{\cal V}}{\cal Y}-\frac{3  \sqrt{2 \sqrt[18]{\cal V}
   +{\sigma^B}+{\bar\sigma}^{\bar B}-{\cal V}^{2/3}+2 \sqrt[6]{\cal V}+\frac{1}{\sqrt[3]{\cal V}}} \sqrt[36]{\cal V}}{\cal Y}}{\sqrt{2 \sqrt[18]{\cal V} +{\sigma^B}+{\bar\sigma}^{\bar B}-{\cal V}^{2/3}+2 \sqrt[6]{\cal V}+\frac{1}{\sqrt[3]{\cal V}}}
   {\cal Y}}-\frac{ \sqrt[36]{\cal V}}{\left(2 \sqrt[18]{\cal V} +{\sigma^B}+{\bar\sigma}^{\bar B}-{\cal V}^{2/3}+2
   \sqrt[6]{\cal V}+\frac{1}{\sqrt[3]{\cal V}}\right)^{3/2}{\cal Y}}\Biggr]\Biggr)\Biggr\}.
\end{eqnarray*}
This needs to be simplified with the understanding that $\sigma^B+{\bar\sigma}^{\bar B} - |a_1|^2{\cal V}^{\frac{7}{6}}\sim{\cal V}^{\frac{1}{18}}$. After doing so, one obtains $-{\cal V}^{\frac{13}{36}} $ as in (\ref{eq:gYMgBIbar_zi_aI}).

Now,
\begin{eqnarray}
\label{eq:gzia1bar}
&& g_{z_i{\bar a}_{\bar 1}}\sim\nonumber\\
& & \hskip-0.8in-\frac{1}{{\cal W}_2}\Biggl\{3  \left({a_1} \sqrt{\cal V}-{a_2}\right) {\cal V}^{2/3} ({{\bar z}_{\bar 1}}+{{\bar z}_{\bar 2}}) \Biggl[2 \left({{\bar a}_{\bar 1}} {\cal V}^{2/3} \left({a_2}-{a_1}
   \sqrt{\cal V}\right)+ {z_1} {{\bar z}_{\bar 1}}+ {{\bar z}_{\bar 1}} {z_2}+ {z_1} {{\bar z}_{\bar 2}}+ {z_2}
   {{\bar z}_{\bar 2}}+{\cal V}^{2/3}+{{\bar a}_{\bar 2}} \left(\sqrt{\cal V} {a_1}+{a_2}\right) \sqrt[6]{\cal V}\right)^{3/2}\nonumber\\
   & & -3 \sqrt{ |z_1+z_2|^2+\sqrt[18]{\cal V}} \left({{\bar a}_{\bar 1}} {\cal V}^{2/3} \left({a_2}-{a_1} \sqrt{\cal V}\right)+ |z_1|^2+
   {{\bar z}_{\bar 1}} {z_2}+ {z_1} {{\bar z}_{\bar 2}}+ |z_2|^2+{\cal V}^{2/3}+{{\bar a}_{\bar 2}} \left(\sqrt{\cal V} {a_1}+{a_2}\right)
   \sqrt[6]{\cal V}\right)\nonumber\\
   & & +\left( |z_1+z_2|^2+\sqrt[18]{\cal V}\right)^{3/2}-V\Biggr]\Biggr\}
   \end{eqnarray}
   Around $z_i,{\bar z}_{\bar i}\sim {\cal V}^{\frac{1}{36}},
{\bar a}_{\bar I}\sim{\cal V}^{-\frac{1}{4}}, a_2\sim{\cal V}^{-\frac{1}{4}}$, the coefficient of
$\left(a_1-{\cal V}^{-\frac{1}{4}}\right)$ is given by:
   \begin{eqnarray}
   \label{eq:gzia1barexpa_1}
   & & -3  {\cal V}^{4/9} \Biggl[\frac{3 \left(\sqrt{(4 +1) \sqrt[18]{\cal V}}-\sqrt{4 \sqrt[18]{\cal V} +2 \sqrt[6]{\cal V}+\frac{1}{\sqrt[3]{\cal V}}}\right) {\cal V}^{5/12}
   \left(\sqrt{\cal V}-1\right)^2}{\sqrt{4 \sqrt[18]{\cal V} +2 \sqrt[6]{\cal V}+\frac{1}{\sqrt[3]{\cal V}}} {\cal Z}^2}\nonumber\\
   & & +\left(2 \left(4 \sqrt[18]{\cal V} +2
   \sqrt[6]{\cal V}+\frac{1}{\sqrt[3]{\cal V}}\right)^{3/2}+\left((4 +1) \sqrt[18]{\cal V}\right)^{3/2}-V-\frac{3 \left(4 {\cal V}^{7/18} +2 \sqrt{\cal V}+1\right) \sqrt{(4
   +1) \sqrt[18]{\cal V}}}{\sqrt[3]{\cal V}}\right)\nonumber\\
    & & \times\left(\frac{3 {\cal V}^{5/12} \left(\sqrt{\cal V}-1\right)^2}{{\cal Z}^3}+\frac{{\cal V}^{3/4} \left(8 {\cal V}^{7/18} +\sum_{\beta\in H_2^-}n^0_\beta+2
   \sqrt{\cal V}+3\right)}{2 \sqrt{4 \sqrt[18]{\cal V} +2 \sqrt[6]{\cal V}+\frac{1}{\sqrt[3]{\cal V}}} \left(4 {\cal V}^{7/18} +2 \sqrt{\cal V}+1\right) {\cal Z}^2}\right)\Biggr],
   \end{eqnarray}
   where
   \begin{equation}
   {\cal Z}\equiv\left({\cal V}^{-\frac{1}{3}} + \sqrt[18]{\cal V} + 2\sqrt[6]{\cal V}\right)^{\frac{3}{2}} - \sqrt[12]{\cal V} + \sum_{\beta\in H_2^-}n^0_\beta.
   \end{equation}
   This yields ${\cal V}^{\frac{11}{18}}$ as in (\ref{eq:giJbar_fluctuation}).

   Similarly,
   \begin{eqnarray}
   \label{eq:gzia2bar}
   & & g_{z_i{\bar a}_{\bar 2}}\sim\nonumber\\
    & & \hskip-0.9in\frac{1}{{\cal W}_2}\Biggl\{3  \left(\sqrt{\cal V} {a_1}+{a_2}\right) \sqrt[6]{\cal V} ({{\bar z}_{\bar 1}}+{{\bar z}_{\bar 2}}) \Biggl[2 \left({{\bar a}_{\bar 1}} {\cal V}^{2/3} \left({a_2}-{a_1}
   \sqrt{\cal V}\right)+ {z_1} {{\bar z}_{\bar 1}}+ {{\bar z}_{\bar 1}} {z_2}+ {z_1} {{\bar z}_{\bar 2}}+ {z_2}
   {{\bar z}_{\bar 2}}+{\cal V}^{2/3}+{{\bar a}_{\bar 2}} \left(\sqrt{\cal V} {a_1}+{a_2}\right) \sqrt[6]{\cal V}\right)^{3/2}\nonumber\\
   & & -3 \sqrt{ ({z_1}+{z_2})
   ({{\bar z}_{\bar 1}}+{{\bar z}_{\bar 2}})+\sqrt[18]{\cal V}} \left({{\bar a}_{\bar 1}} {\cal V}^{2/3} \left({a_2}-{a_1} \sqrt{\cal V}\right)+ {z_1} {{\bar z}_{\bar 1}}+
   {{\bar z}_{\bar 1}} {z_2}+ {z_1} {{\bar z}_{\bar 2}}+ {z_2} {{\bar z}_{\bar 2}}+{\cal V}^{2/3}+{{\bar a}_{\bar 2}} \left(\sqrt{\cal V} {a_1}+{a_2}\right)
   \sqrt[6]{\cal V}\right)\nonumber\\
   & & +\left( ({z_1}+{z_2}) ({{\bar z}_{\bar 1}}+{{\bar z}_{\bar 2}})+\sqrt[18]{\cal V}\right)^{3/2}-V\Biggr]\Biggr\},
   \end{eqnarray}
   where
   \begin{eqnarray}
  & &  {\cal W}_2\equiv 2 \sqrt{{{\bar a}_{\bar 1}} {\cal V}^{2/3}
   \left({a_2}-{a_1} \sqrt{\cal V}\right)+ {z_1} {{\bar z}_{\bar 1}}+ {{\bar z}_{\bar 1}} {z_2}+ {z_1} {{\bar z}_{\bar 2}}+
   {z_2} {{\bar z}_{\bar 2}}+{\cal V}^{2/3}+{{\bar a}_{\bar 2}} \left(\sqrt{\cal V} {a_1}+{a_2}\right) \sqrt[6]{\cal V}}
   \nonumber\\
   & &\hskip-0.8in \times\Biggl(\left({{\bar a}_{\bar 1}} {\cal V}^{2/3} \left({a_2}-{a_1}
   \sqrt{\cal V}\right)+ {z_1} {{\bar z}_{\bar 1}}+ {{\bar z}_{\bar 1}} {z_2}+ {z_1} {{\bar z}_{\bar 2}}+ {z_2}
   {{\bar z}_{\bar 2}}+{\cal V}^{2/3}+{{\bar a}_{\bar 2}} \left(\sqrt{\cal V} {a_1}+{a_2}\right) \sqrt[6]{\cal V}\right)^{3/2}\nonumber\\
   & & -\left( ({z_1}+{z_2})
   ({{\bar z}_{\bar 1}}+{{\bar z}_{\bar 2}})+\sqrt[18]{\cal V}\right)^{3/2}+\sum_{\beta\in H_2^-}n^0_\beta\Biggr)^2.
   \end{eqnarray}
   which when expanded around  around $z_i\sim{\cal V}^{\frac{1}{36}},{\bar z}_{\bar i}\sim {\cal V}^{\frac{1}{36}},
{\bar a}_{\bar I}\sim{\cal V}^{-\frac{1}{4}}, a_2\sim{\cal V}^{-\frac{1}{4}}$, yields as the coefficient of
$\left(a_1-{\cal V}^{-\frac{1}{4}}\right)$:
\begin{eqnarray}
\label{eq:gzia2barexpa_1}
& & 3  {\cal V}^{7/36} \Biggl[\left(2 \left(4 \sqrt[18]{\cal V} +2 \sqrt[6]{\cal V}+\frac{1}{\sqrt[3]{\cal V}}\right)^{3/2}+\left((4 +1)
   \sqrt[18]{\cal V}\right)^{3/2}-V+\frac{3 \left(-4 {\cal V}^{7/18} -2 \sqrt{\cal V}-1\right) \sqrt{(4 +1) \sqrt[18]{\cal V}}}{\sqrt[3]{\cal V}}\right)\nonumber\\
& & \times   \left(\frac{\frac{\sqrt{\cal V}}{\sqrt{4 \sqrt[18]{\cal V} +2 \sqrt[6]{\cal V}+\frac{1}{\sqrt[3]{\cal V}}}}-\frac{\left(\sqrt[4]{\cal V}+\frac{1}{\sqrt[4]{\cal V}}\right)
   \left({\cal V}^{5/12}-{\cal V}^{11/12}\right)}{2 \left(4 \sqrt[18]{\cal V} +2 \sqrt[6]{\cal V}+\frac{1}{\sqrt[3]{\cal V}}\right)^{3/2}}}{{\cal Z}^2}-\frac{3 \left(\sqrt[4]{\cal V}+\frac{1}{\sqrt[4]{\cal V}}\right)
   \left({\cal V}^{5/12}-{\cal V}^{11/12}\right)}{{\cal Z}^3}\right)\nonumber\\
   & & +\frac{\left(\sqrt[4]{\cal V}+\frac{1}{\sqrt[4]{\cal V}}\right) \left(3 \sqrt{4 \sqrt[18]{\cal V} +2
   \sqrt[6]{\cal V}+\frac{1}{\sqrt[3]{\cal V}}} \left({\cal V}^{5/12}-{\cal V}^{11/12}\right)+\frac{3 \sqrt{(4 +1) \sqrt[18]{\cal V}}
   \left({\cal V}^{5/4}-{\cal V}^{3/4}\right)}{\sqrt[3]{\cal V}}\right)}{\sqrt{4 \sqrt[18]{\cal V} +2 \sqrt[6]{\cal V}+\frac{1}{\sqrt[3]{\cal V}}} {\cal Z}^2}\Biggr]
   \end{eqnarray}
   This yields $-{\cal V}^{\frac{1}{9}}$ as in (\ref{eq:giJbar_fluctuation}).

Now,
   \begin{eqnarray*}
   & & e^{\frac{K}{2}}\left(\left(\partial_i\partial_{a_2}K\right)W+\partial_iKD_{a_2}W +
\partial_{a_2}K\partial_iW - \left(\partial_iK\partial_{a_2}K\right)W\right)\nonumber\\
& & \sim\frac{1}{{\cal W}_1}\Biggl\{ \left(\sqrt{\cal V}
   {{\bar a}_{\bar 1}}+{{\bar a}_{\bar 2}}\right) \left({z_1}^{18}-3 {\phi_0} {z_2}^6
   {z_1}^6+{z_2}^{18}+\left(-{z_1}^{18}+3 {\phi_0} {z_2}^6
   {z_1}^6-{z_2}^{18}\right)^{2/3}\right)\nonumber\\
   & & \times \Biggl[ 6\left({z_1}^{18}-3
   {\phi_0} {z_2}^6 {z_1}^6+{z_2}^{18}+\left(-{z_1}^{18}+3 {\phi_0} {z_2}^6
   {z_1}^6-{z_2}^{18}\right)^{2/3}\right) ({{\bar z}_{\bar 1}}+{{\bar z}_{\bar 2}})\nonumber\\
   & & \times\left({{\bar a}_{\bar 1}}
   {\cal V}^{2/3} \left({a_2}-{a_1} \sqrt{\cal V}\right)+ {z_1} {{\bar z}_{\bar 1}}+
   {{\bar z}_{\bar 1}} {z_2}+ {z_1} {{\bar z}_{\bar 2}}+ {z_2}
   {{\bar z}_{\bar 2}}+{\cal V}^{2/3}+{{\bar a}_{\bar 2}} \left(\sqrt{\cal V} {a_1}+{a_2}\right) \sqrt[6]{\cal V}\right)
   \nonumber\\
& & \times   \Biggl(\sqrt{{{\bar a}_{\bar 1}} {\cal V}^{2/3} \left({a_2}-{a_1} \sqrt{\cal V}\right)+ {z_1}
   {{\bar z}_{\bar 1}}+ {{\bar z}_{\bar 1}} {z_2}+ {z_1} {{\bar z}_{\bar 2}}+
   {z_2} {{\bar z}_{\bar 2}}+{\cal V}^{2/3}+{{\bar a}_{\bar 2}} \left(\sqrt{\cal V} {a_1}+{a_2}\right)
   \sqrt[6]{\cal V}}\nonumber\\
   & & -\sqrt{ ({z_1}+{z_2})
   ({{\bar z}_{\bar 1}}+{{\bar z}_{\bar 2}})+\sqrt[18]{\cal V}}\Biggr)\end{eqnarray*}

   \begin{eqnarray*}
   & & -2 \Biggl[12 \left(3 {z_1}^{17}-3 {\phi_0}
   {z_2}^6 {z_1}^5-\frac{2 \left({z_1}^{17}-{\phi_0} {z_1}^5
   {z_2}^6\right)}{\sqrt[3]{-{z_1}^{18}+3 {\phi_0} {z_2}^6
   {z_1}^6-{z_2}^{18}}}\right)\nonumber\\
   & & - (2 {z_1}+{z_2}) \left({z_1}^{18}-3
   {\phi_0} {z_2}^6 {z_1}^6+{z_2}^{18}+\left(-{z_1}^{18}+3 {\phi_0} {z_2}^6
   {z_1}^6-{z_2}^{18}\right)^{2/3}\right)\Biggr]\nonumber\\
   & & \times \left({{\bar a}_{\bar 1}} {\cal V}^{2/3}
   \left({a_2}-{a_1} \sqrt{\cal V}\right)+ {z_1} {{\bar z}_{\bar 1}}+
   {{\bar z}_{\bar 1}} {z_2}+ {z_1} {{\bar z}_{\bar 2}}+ {z_2}
   {{\bar z}_{\bar 2}}+{\cal V}^{2/3}+{{\bar a}_{\bar 2}} \left(\sqrt{\cal V} {a_1}+{a_2}\right) \sqrt[6]{\cal V}\right)\nonumber\\
   & & \times
   \Biggl[\left({{\bar a}_{\bar 1}} {\cal V}^{2/3} \left({a_2}-{a_1} \sqrt{\cal V}\right)+ {z_1}
   {{\bar z}_{\bar 1}}+ {{\bar z}_{\bar 1}} {z_2}+ {z_1} {{\bar z}_{\bar 2}}+
   {z_2} {{\bar z}_{\bar 2}}+{\cal V}^{2/3}+{{\bar a}_{\bar 2}} \left(\sqrt{\cal V} {a_1}+{a_2}\right)
   \sqrt[6]{\cal V}\right)^{3/2}\nonumber\\
   & & -\left( |z_1+z_2|^2+\sqrt[18]{\cal V}\right)^{3/2}+\sum_{\beta\in H_2^-}n^0_\beta\Biggr]\nonumber\\
   & & + \left({z_1}^{18}-3
   {\phi_0} {z_2}^6 {z_1}^6+{z_2}^{18}+\left(-{z_1}^{18}+3 {\phi_0} {z_2}^6
   {z_1}^6-{z_2}^{18}\right)^{2/3}\right) ({{\bar z}_{\bar 1}}+{{\bar z}_{\bar 2}})\nonumber\\
   & &  \Biggl(2
   \left({{\bar a}_{\bar 1}} {\cal V}^{2/3} \left({a_2}-{a_1} \sqrt{\cal V}\right)+ {z_1}
   {{\bar z}_{\bar 1}}+ {{\bar z}_{\bar 1}} {z_2}+ {z_1} {{\bar z}_{\bar 2}}+
   {z_2} {{\bar z}_{\bar 2}}+{\cal V}^{2/3}+{{\bar a}_{\bar 2}} \left(\sqrt{\cal V} {a_1}+{a_2}\right)
   \sqrt[6]{\cal V}\right)^{3/2}\nonumber\\
   & & -3 \sqrt{ |z_1+z_2|^2+\sqrt[18]{\cal V}} \left({{\bar a}_{\bar 1}} {\cal V}^{2/3} \left({a_2}-{a_1}
   \sqrt{\cal V}\right)+ {z_1} {{\bar z}_{\bar 1}}+ {{\bar z}_{\bar 1}} {z_2}+
   {z_1} {{\bar z}_{\bar 2}}+ {z_2} {{\bar z}_{\bar 2}}+{\cal V}^{2/3}+{{\bar a}_{\bar 2}} \left(\sqrt{\cal V}
   {a_1}+{a_2}\right) \sqrt[6]{\cal V}\right)\nonumber\\
   & & +\left( |z_1+z_2|^2+\sqrt[18]{\cal V}\right)^{3/2}-V\Biggr)\Biggr]\Biggr\},
   \end{eqnarray*}
   where
   \begin{eqnarray}
   & & {\cal W}_1\equiv 2 {\cal V}^{11/6}
   \sqrt{{{\bar a}_{\bar 1}} {\cal V}^{2/3} \left({a_2}-{a_1} \sqrt{\cal V}\right)+ |z_1|^2+ {{\bar z}_{\bar 1}} {z_2}+ {z_1} {{\bar z}_{\bar 2}}+
   |z_2|^2+{\cal V}^{2/3}+{{\bar a}_{\bar 2}} \left(\sqrt{\cal V} {a_1}+{a_2}\right) \sqrt[6]{\cal V}}
   \nonumber\\
& &\times   \Biggl(\left({{\bar a}_{\bar 1}} {\cal V}^{2/3} \left({a_2}-{a_1} \sqrt{\cal V}\right)+ |z_1|^2 + {{\bar z}_{\bar 1}} {z_2}+ {z_1} {{\bar z}_{\bar 2}} + |z_2|^2 +{\cal V}^{2/3}+{{\bar a}_{\bar 2}} \left(\sqrt{\cal V} {a_1}+{a_2}\right)
   \sqrt[6]{\cal V}\right)^{3/2}\nonumber\\
   & & -\left( |z_1+z_2|^2 + \sqrt[18]{\cal V}\right)^{3/2}+\sum_{\beta\in H_2^-}n^0_\beta\Biggr)^2\nonumber\\
   & & \hskip-0.8in\times \left(-\left(
   |z_1+z_2|^2+\sqrt[18]{\cal V}\right)^{3/2}+\left(-{a_1}
   {{\bar a}_{\bar 1}} {\cal V}^{7/6}+({{\bar a}_{\bar 1}} {a_2}+{a_1} {{\bar a}_{\bar 2}}) {\cal V}^{2/3}+{\cal V}^{2/3}+{a_2}
   {{\bar a}_{\bar 2}} \sqrt[6]{\cal V}+ |z_1+z_2|^2\right)^{3/2}+\sum_{\beta\in H_2^-}n^0_\beta\right).\nonumber\\
   & &
   \end{eqnarray}
      which when expanded around  around $z_i\sim{\cal V}^{\frac{1}{36}},{\bar z}_{\bar i}\sim {\cal V}^{\frac{1}{36}},
{\bar a}_{\bar I}\sim{\cal V}^{-\frac{1}{4}}, a_2\sim{\cal V}^{-\frac{1}{4}}$, yields as the coefficient of
$\left(a_1-{\cal V}^{-\frac{1}{4}}\right)$:
   \begin{eqnarray*}
 & &   \frac{1}{{\cal V}^{11/6}}\Biggl\{\left(\sqrt[4]{\cal V}+\frac{1}{\sqrt[4]{\cal V}}\right) \left(-3
   \sqrt[3]{\cal V} {\phi_0}+\left(3 {\phi_0} \sqrt[3]{\cal V}-2 \sqrt{\cal V}\right)^{2/3}+2 \sqrt{\cal V}\right)\nonumber\\
& & \times   \Biggl[\left(-\frac{{\cal V}^{5/12}-{\cal V}^{11/12}}{2 \left(4 \sqrt[18]{\cal V} +2
   \sqrt[6]{\cal V}+\frac{1}{\sqrt[3]{\cal V}}\right)^{3/2} {\cal Z}^3}-\frac{9 \left({\cal V}^{5/12}-{\cal V}^{11/12}\right)}{2{\cal Z}^4}\right)\nonumber\\
   & & \times \Biggl[ \left(2 \left(4 \sqrt[18]{\cal V}
   +2 \sqrt[6]{\cal V}+\frac{1}{\sqrt[3]{\cal V}}\right)^{3/2}+\left((4 +1)
   \sqrt[18]{\cal V}\right)^{3/2}-V+\frac{3 \left(-4 {\cal V}^{7/18} -2 \sqrt{\cal V}-1\right) \sqrt{(4
   +1) \sqrt[18]{\cal V}}}{\sqrt[3]{\cal V}}\right) \sqrt[36]{\cal V}\nonumber\\
   & & \times \left(-3 \sqrt[3]{\cal V}
   {\phi_0}+\left(3 {\phi_0} \sqrt[3]{\cal V}-2 \sqrt{\cal V}\right)^{2/3}+2 \sqrt{\cal V}\right)\nonumber\\
   & & \hskip-0.3in+\frac{6
    \left(\sqrt{(4 +1) \sqrt[18]{\cal V}}-\sqrt{4 \sqrt[18]{\cal V} +2
   \sqrt[6]{\cal V}+\frac{1}{\sqrt[3]{\cal V}}}\right) \left(-4 {\cal V}^{7/18} -2 \sqrt{\cal V}-1\right)
   \left(-3 \sqrt[3]{\cal V} {\phi_0}+\left(3 {\phi_0} \sqrt[3]{\cal V}-2 \sqrt{\cal V}\right)^{2/3}+2
   \sqrt{\cal V}\right)}{{\cal V}^{11/36}}-\nonumber\\
   & & \hskip-0.9in\frac{\left(12 \left(-3 {\phi_0}+3 \sqrt[6]{\cal V}+\frac{2
   \left({\phi_0}-\sqrt[6]{\cal V}\right)}{\sqrt[3]{3 {\phi_0} \sqrt[3]{\cal V}-2 \sqrt{\cal V}}}\right)
   {\cal V}^{11/36}-3  \left(-3 \sqrt[3]{\cal V} {\phi_0}+\left(3 {\phi_0} \sqrt[3]{\cal V}-2
   \sqrt{\cal V}\right)^{2/3}+2 \sqrt{\cal V}\right) \sqrt[36]{\cal V}\right) \left(4 {\cal V}^{7/18} +2
   \sqrt{\cal V}+1\right){\cal Z}}{\sqrt[3]{\cal V}}\Biggr]\nonumber\\
   & & +\frac{1}{\sqrt{4 \sqrt[18]{\cal V} +2
   \sqrt[6]{\cal V}+\frac{1}{\sqrt[3]{\cal V}}} {\cal Z}^3}\nonumber\\
   & &\hskip-0.3in \times\Biggl\{\frac{1}{\sqrt[3]{\cal V}}\biggl\{\left(12 \left(-3
   {\phi_0}+3 \sqrt[6]{\cal V}+\frac{2 \left({\phi_0}-\sqrt[6]{\cal V}\right)}{\sqrt[3]{3 {\phi_0}
   \sqrt[3]{\cal V}-2 \sqrt{\cal V}}}\right) {\cal V}^{11/36}-3  \left(-3 \sqrt[3]{\cal V} {\phi_0}+\left(3
   {\phi_0} \sqrt[3]{\cal V}-2 \sqrt{\cal V}\right)^{2/3}+2 \sqrt{\cal V}\right) \sqrt[36]{\cal V}\right)
   \end{eqnarray*}

   \begin{eqnarray*}
& &    \times\left(-\frac{3}{2} \sqrt{4 \sqrt[18]{\cal V} +2 \sqrt[6]{\cal V}+\frac{1}{\sqrt[3]{\cal V}}} \left(4
   {\cal V}^{7/18} +2 \sqrt{\cal V}+1\right) \left({\cal V}^{5/12}-{\cal V}^{11/12}\right)-{\cal Z} \left({\cal V}^{3/4}-{\cal V}^{5/4}\right)\right)\biggr\}\nonumber\\
   & & +\frac{6
    \left(-3 \sqrt[3]{\cal V} {\phi_0}+\left(3 {\phi_0} \sqrt[3]{\cal V}-2
   \sqrt{\cal V}\right)^{2/3}+2 \sqrt{\cal V}\right) }{{\cal V}^{11/36}}\nonumber\\
   & & \times \left(\left(\sqrt{(4 +1) \sqrt[18]{\cal V}}-\sqrt{4
   \sqrt[18]{\cal V} +2 \sqrt[6]{\cal V}+\frac{1}{\sqrt[3]{\cal V}}}\right)
   \left({\cal V}^{5/4}-{\cal V}^{3/4}\right)-\frac{\left(-4 {\cal V}^{7/18} -2 \sqrt{\cal V}-1\right)
   \left({\cal V}^{5/12}-{\cal V}^{11/12}\right)}{2 \sqrt{4 \sqrt[18]{\cal V} +2
   \sqrt[6]{\cal V}+\frac{1}{\sqrt[3]{\cal V}}}}\right)\nonumber\\
   & & + \left(-3 \sqrt[3]{\cal V}
   {\phi_0}+\left(3 {\phi_0} \sqrt[3]{\cal V}-2 \sqrt{\cal V}\right)^{2/3}+2 \sqrt{\cal V}\right) \sqrt[36]{\cal V}\nonumber\\
& & \times   \left(3 \sqrt{4 \sqrt[18]{\cal V} +2 \sqrt[6]{\cal V}+\frac{1}{\sqrt[3]{\cal V}}}
   \left({\cal V}^{5/12}-{\cal V}^{11/12}\right)+\frac{3 \sqrt{ \sqrt[18]{\cal V}}
   \left({\cal V}^{5/4}-{\cal V}^{3/4}\right)}{\sqrt[3]{\cal V}}\right)\Biggr\}\Biggr]\Biggr\}
   \end{eqnarray*}
   This yields $-{\cal V}^{-\frac{31}{18}} $ as in (\ref{eq:Higgsino_quark_sq_2}).

   Similarly,
   \begin{eqnarray*}
   \label{eq:exphalfKDDWa_1_exc_aff}
& & e^{\frac{K}{2}}\left(\left(\partial_i\partial_{a_1}K\right)W+\partial_iKD_{a_1}W +
\partial_{a_1}K\partial_iW - \left(\partial_iK\partial_{a_1}K\right)W\right)\nonumber\\
& & \sim\frac{1}{\cal W}\Biggl\{
   \left({z_1}^{18}-3 {\phi_0} {z_2}^6 {z_1}^6+{z_2}^{18}+\left(-{z_1}^{18}+3
   {\phi_0} {z_2}^6 {z_1}^6-{z_2}^{18}\right)^{2/3}\right)\nonumber\\
   & & \times \Biggl[6
   \left({{\bar a}_{\bar 2}} {\cal V}^{2/3}-{{\bar a}_{\bar 1}} {\cal V}^{7/6}\right) \left({z_1}^{18}-3 {\phi_0}
   {z_2}^6 {z_1}^6+{z_2}^{18}+\left(-{z_1}^{18}+3 {\phi_0} {z_2}^6
   {z_1}^6-{z_2}^{18}\right)^{2/3}\right) \nonumber\\
   & & \times({{\bar z}_{\bar 1}}+{{\bar z}_{\bar 2}}) \left({{\bar a}_{\bar 1}}
   {\cal V}^{2/3} \left({a_2}-{a_1} \sqrt{\cal V}\right)+ {z_1} {{\bar z}_{\bar 1}}+
   {{\bar z}_{\bar 1}} {z_2}+ {z_1} {{\bar z}_{\bar 2}}+ {z_2}
   {{\bar z}_{\bar 2}}+{\cal V}^{2/3}+{{\bar a}_{\bar 2}} \left(\sqrt{\cal V} {a_1}+{a_2}\right) \sqrt[6]{\cal V}\right)
   \nonumber\\
& & \hskip-0.4in\times   \left(\sqrt{{{\bar a}_{\bar 1}} {\cal V}^{2/3} \left({a_2}-{a_1} \sqrt{\cal V}\right)+ {z_1}
   {{\bar z}_{\bar 1}}+ {{\bar z}_{\bar 1}} {z_2}+ {z_1} {{\bar z}_{\bar 2}}+
   {z_2} {{\bar z}_{\bar 2}}+{\cal V}^{2/3}+{{\bar a}_{\bar 2}} \left(\sqrt{\cal V} {a_1}+{a_2}\right)
   \sqrt[6]{\cal V}}-\sqrt{ ({z_1}+{z_2})
   ({{\bar z}_{\bar 1}}+{{\bar z}_{\bar 2}})+\sqrt[18]{\cal V}}\right)\nonumber\\
   & & -2 \left({{\bar a}_{\bar 2}} {\cal V}^{2/3}-{{\bar a}_{\bar 1}}
   {\cal V}^{7/6}\right) \Biggl(12 \left(3 {z_1}^{17}-3 {\phi_0} {z_2}^6 {z_1}^5-\frac{2
   \left({z_1}^{17}-{\phi_0} {z_1}^5 {z_2}^6\right)}{\sqrt[3]{-{z_1}^{18}+3
   {\phi_0} {z_2}^6 {z_1}^6-{z_2}^{18}}}\right)\nonumber\\
   & & - (2 {z_1}+{z_2})
   \left({z_1}^{18}-3 {\phi_0} {z_2}^6 {z_1}^6+{z_2}^{18}+\left(-{z_1}^{18}+3
   {\phi_0} {z_2}^6 {z_1}^6-{z_2}^{18}\right)^{2/3}\right)\Biggr) \nonumber\\
   & & \times\left({{\bar a}_{\bar 1}}
   {\cal V}^{2/3} \left({a_2}-{a_1} \sqrt{\cal V}\right)+ {z_1} {{\bar z}_{\bar 1}}+
   {{\bar z}_{\bar 1}} {z_2}+ {z_1} {{\bar z}_{\bar 2}}+ {z_2}
   {{\bar z}_{\bar 2}}+{\cal V}^{2/3}+{{\bar a}_{\bar 2}} \left(\sqrt{\cal V} {a_1}+{a_2}\right) \sqrt[6]{\cal V}\right)\nonumber\\
& & \times   \Biggl[\left({{\bar a}_{\bar 1}} {\cal V}^{2/3} \left({a_2}-{a_1} \sqrt{\cal V}\right)+ {z_1}
   {{\bar z}_{\bar 1}}+ {{\bar z}_{\bar 1}} {z_2}+ {z_1} {{\bar z}_{\bar 2}}+
   {z_2} {{\bar z}_{\bar 2}}+{\cal V}^{2/3}+{{\bar a}_{\bar 2}} \left(\sqrt{\cal V} {a_1}+{a_2}\right)
   \sqrt[6]{\cal V}\right)^{3/2}\nonumber\\
   & & -\left( ({z_1}+{z_2})
   ({{\bar z}_{\bar 1}}+{{\bar z}_{\bar 2}})+\sqrt[18]{\cal V}\right)^{3/2}+{\sum_{\beta\in H_2^-(CY_3)}n^0_\beta}\Biggr]
   \end{eqnarray*}

   \begin{eqnarray*}
   & & -
   \left({{\bar a}_{\bar 1}} \sqrt{\cal V}-{{\bar a}_{\bar 2}}\right) {\cal V}^{2/3} \left({z_1}^{18}-3 {\phi_0}
   {z_2}^6 {z_1}^6+{z_2}^{18}+\left(-{z_1}^{18}+3 {\phi_0} {z_2}^6
   {z_1}^6-{z_2}^{18}\right)^{2/3}\right) ({{\bar z}_{\bar 1}}+{{\bar z}_{\bar 2}})\nonumber\\
   & & \times \Biggl(2
   \left({{\bar a}_{\bar 1}} {\cal V}^{2/3} \left({a_2}-{a_1} \sqrt{\cal V}\right)+ {z_1}
   {{\bar z}_{\bar 1}}+ {{\bar z}_{\bar 1}} {z_2}+ {z_1} {{\bar z}_{\bar 2}}+
   {z_2} {{\bar z}_{\bar 2}}+{\cal V}^{2/3}+{{\bar a}_{\bar 2}} \left(\sqrt{\cal V} {a_1}+{a_2}\right)
   \sqrt[6]{\cal V}\right)^{3/2}\nonumber\\
   & & \hskip-0.3in-3 \sqrt{ ({z_1}+{z_2})
   ({{\bar z}_{\bar 1}}+{{\bar z}_{\bar 2}})+\sqrt[18]{\cal V}} \left({{\bar a}_{\bar 1}} {\cal V}^{2/3} \left({a_2}-{a_1}
   \sqrt{\cal V}\right)+ {z_1} {{\bar z}_{\bar 1}}+ {{\bar z}_{\bar 1}} {z_2}+
   {z_1} {{\bar z}_{\bar 2}}+ {z_2} {{\bar z}_{\bar 2}}+{\cal V}^{2/3}+{{\bar a}_{\bar 2}} \left(\sqrt{\cal V}
   {a_1}+{a_2}\right) \sqrt[6]{\cal V}\right)\nonumber\\
   & & +\left( ({z_1}+{z_2})
   ({{\bar z}_{\bar 1}}+{{\bar z}_{\bar 2}})+\sqrt[18]{\cal V}\right)^{3/2}-{\sum_{\beta\in H_2^-(CY_3)}n^0_\beta}\Biggr)\Biggr]\Biggr\},
   \end{eqnarray*}
   where
   \begin{eqnarray}
   & & {\cal W}\equiv 2 {\cal V}^2
   \sqrt{{{\bar a}_{\bar 1}} {\cal V}^{2/3} \left({a_2}-{a_1} \sqrt{\cal V}\right)+ {z_1}
   {{\bar z}_{\bar 1}}+ {{\bar z}_{\bar 1}} {z_2}+ {z_1} {{\bar z}_{\bar 2}}+
   {z_2} {{\bar z}_{\bar 2}}+{\cal V}^{2/3}+{{\bar a}_{\bar 2}} \left(\sqrt{\cal V} {a_1}+{a_2}\right) \sqrt[6]{\cal V}}\times\nonumber\\
   & &
   \Biggl(\left({{\bar a}_{\bar 1}} {\cal V}^{2/3} \left({a_2}-{a_1} \sqrt{\cal V}\right)+ |z_1|^2+ {{\bar z}_{\bar 1}} {z_2}+ {z_1} {{\bar z}_{\bar 2}}+  |z_2|^2 +{\cal V}^{2/3}+{{\bar a}_{\bar 2}} \left(\sqrt{\cal V} {a_1}+{a_2}\right)
   \sqrt[6]{\cal V}\right)^{3/2}\nonumber\\
   & & -\left( |z_1+z_2|^2+\sqrt[18]{\cal V}\right)^{3/2}+{\sum_{\beta\in H_2^-(CY_3)}n^0_\beta}\Biggr)^2
   \Biggl(-\left(
   ({z_1}+{z_2}) ({{\bar z}_{\bar 1}}+{{\bar z}_{\bar 2}})+\sqrt[18]{\cal V}\right)^{3/2}\nonumber\\
   & & +\left(-|a_1|^2 {\cal V}^{7/6}+({{\bar a}_{\bar 1}} {a_2}+{a_1} {{\bar a}_{\bar 2}}) {\cal V}^{2/3}+{\cal V}^{2/3}+|a_2|^2 \sqrt[6]{\cal V}+ |{z_1}+{z_2}|^2\right)^{3/2}+{\sum_{\beta\in H_2^-(CY_3)}n^0_\beta}\Biggr).\nonumber\\
   & &
   \end{eqnarray}
   which when expanded around  around $z_i\sim{\cal V}^{\frac{1}{36}},{\bar z}_{\bar i}\sim {\cal V}^{\frac{1}{36}},
{\bar a}_{\bar I}\sim{\cal V}^{-\frac{1}{4}}, a_2\sim{\cal V}^{-\frac{1}{4}}$, yields as the coefficient of
$\left(a_1-{\cal V}^{-\frac{1}{4}}\right)$:
\begin{eqnarray*}
& & \hskip-0.3in\frac{1}{{\cal V}^2}\Biggl\{\left(-3 \sqrt[3]{\cal V} {\phi_0}+\left(3 {\phi_0}
   \sqrt[3]{\cal V}-2 \sqrt{\cal V}\right)^{2/3}+2 \sqrt{\cal V}\right) \Biggl[\left(-\frac{{\cal V}^{5/12}-{\cal V}^{11/12}}{2
   {\cal Z}^3 \left(4 \sqrt[18]{\cal V} +2
   \sqrt[6]{\cal V}+\frac{1}{\sqrt[3]{\cal V}}\right)^{3/2}}-\frac{9 \left({\cal V}^{5/12}-{\cal V}^{11/12}\right)}{2
  {\cal Z}^4}\right)\nonumber\\
  & & \hskip-0.8in\times \Biggl( \left({\cal Z}-\frac{3 \left(-4 {\cal V}^{7/18} -2
   \sqrt{\cal V}-1\right) \sqrt{ \sqrt[18]{\cal V}}}{\sqrt[3]{\cal V}}\right)
   \left(\sqrt{\cal V}-1\right) {\cal V}^{4/9} \left(-3 \sqrt[3]{\cal V} {\phi_0}+\left(3 {\phi_0} \sqrt[3]{\cal V}-2
   \sqrt{\cal V}\right)^{2/3}+2 \sqrt{\cal V}\right)\nonumber\\
   & &\hskip-0.93in -6  \left(\sqrt{
   \sqrt[18]{\cal V}}-\sqrt{4 \sqrt[18]{\cal V} +2 \sqrt[6]{\cal V}+\frac{1}{\sqrt[3]{\cal V}}}\right) \left(-4
   {\cal V}^{7/18} -2 \sqrt{\cal V}-1\right) \left(\sqrt{\cal V}-1\right) \sqrt[9]{\cal V} \left(-3 \sqrt[3]{\cal V}
   {\phi_0}+\left(3 {\phi_0} \sqrt[3]{\cal V}-2 \sqrt{\cal V}\right)^{2/3}+2
   \sqrt{\cal V}\right)\nonumber\\
   & & +\hskip-0.3in {\cal Z} \left(12 \left(-3 {\phi_0}+3 \sqrt[6]{\cal V}+\frac{2
   \left({\phi_0}-\sqrt[6]{\cal V}\right)}{\sqrt[3]{3 {\phi_0} \sqrt[3]{\cal V}-2 \sqrt{\cal V}}}\right)
   {\cal V}^{11/36}-3  \left(-3 \sqrt[3]{\cal V} {\phi_0}+\left(3 {\phi_0} \sqrt[3]{\cal V}-2
   \sqrt{\cal V}\right)^{2/3}+2 \sqrt{\cal V}\right) \sqrt[36]{\cal V}\right)
   \end{eqnarray*}

   \begin{eqnarray*}
   & & \times\left(\sqrt{\cal V}-1\right) \left(4
   {\cal V}^{7/18} +2 \sqrt{\cal V}+1\right) \sqrt[12]{\cal V}\Biggr)+\nonumber\\
    & & \frac{1}{{\cal Z}^3 \sqrt{4 \sqrt[18]{\cal V} +2
   \sqrt[6]{\cal V}+\frac{1}{\sqrt[3]{\cal V}}}}\Biggl\{\Biggl(12 \left(-3 {\phi_0}+3
   \sqrt[6]{\cal V}+\frac{2 \left({\phi_0}-\sqrt[6]{\cal V}\right)}{\sqrt[3]{3 {\phi_0} \sqrt[3]{\cal V}-2
   \sqrt{\cal V}}}\right) {\cal V}^{11/36}\nonumber\\
   & & -3  \left(-3 \sqrt[3]{\cal V} {\phi_0}+\left(3 {\phi_0}
   \sqrt[3]{\cal V}-2 \sqrt{\cal V}\right)^{2/3}+2 \sqrt{\cal V}\right) \sqrt[36]{\cal V}\Biggr)\nonumber\\
    & & \times\left(\sqrt{\cal V}-1\right)
   \sqrt[12]{\cal V} \left(\frac{3}{2} \sqrt{4 \sqrt[18]{\cal V} +2
   \sqrt[6]{\cal V}+\frac{1}{\sqrt[3]{\cal V}}} \left(4 {\cal V}^{7/18} +2 \sqrt{\cal V}+1\right)
   \left({\cal V}^{5/12}-{\cal V}^{11/12}\right)+{\cal Z} \left({\cal V}^{3/4}-{\cal V}^{5/4}\right)\right)
   \end{eqnarray*}

   \begin{eqnarray*}
   & & -6
   \left(\sqrt{\cal V}-1\right) \left(-3 \sqrt[3]{\cal V} {\phi_0}+\left(3 {\phi_0} \sqrt[3]{\cal V}-2
   \sqrt{\cal V}\right)^{2/3}+2 \sqrt{\cal V}\right) \sqrt[9]{\cal V}\nonumber\\
   & & \times \left(\left(\sqrt{
   \sqrt[18]{\cal V}}-\sqrt{4 \sqrt[18]{\cal V} +2 \sqrt[6]{\cal V}+\frac{1}{\sqrt[3]{\cal V}}}\right)
   \left({\cal V}^{5/4}-{\cal V}^{3/4}\right)-\frac{\left(-4 {\cal V}^{7/18} -2 \sqrt{\cal V}-1\right)
   \left({\cal V}^{5/12}-{\cal V}^{11/12}\right)}{2 \sqrt{4 \sqrt[18]{\cal V} +2
   \sqrt[6]{\cal V}+\frac{1}{\sqrt[3]{\cal V}}}}\right)+\nonumber\\
   & &  \left(\sqrt{\cal V}-1\right) \left(-3 \sqrt[3]{\cal V}
   {\phi_0}+\left(3 {\phi_0} \sqrt[3]{\cal V}-2 \sqrt{\cal V}\right)^{2/3}+2 \sqrt{\cal V}\right) {\cal V}^{4/9}\nonumber\\
   & & \times
   \left(-3 \sqrt{4 \sqrt[18]{\cal V} +2 \sqrt[6]{\cal V}+\frac{1}{\sqrt[3]{\cal V}}}
   \left({\cal V}^{5/12}-{\cal V}^{11/12}\right)-\frac{3 \sqrt{ \sqrt[18]{\cal V}}
   \left({\cal V}^{5/4}-{\cal V}^{3/4}\right)}{\sqrt[3]{\cal V}}\right)\Biggr\}\Biggr]\Biggr\}
   \nonumber\\
   & &
\end{eqnarray*}
This yields ${\cal V}^{-\frac{11}{9}}$ as in (\ref{eq:Higgsino_quark_sq_2}).

\section{Large Volume Ricci-Flat Metric}
\setcounter{equation}{0} \seceqdd

To work out the Ricci tensor in section {\bf 5}, in this appendix, we list the values of the independent components of the metric:
\begin{itemize}
\item
\begin{eqnarray*}
& & g_{1{\bar 1}}=\frac{X_{11}}{Y_{11}},\ {\rm where}\nonumber\\
& & X_{11}=h^{z_1^2{\bar z}_{\bar 1}^2} {r_1} \Biggl(\left(h^{z_1^2{\bar z}_{\bar 1}^2}\sqrt[18]{\cal V} z_1 {\bar z}_1 (z_1+2 z_2) ({\bar z}_1+2
   {\bar z}_2)+{r2} \left(\sqrt[18]{\cal V}-(2 z_1+z_2) (2 {\bar z}_1+{\bar z}_2)\right)\right) \epsilon
   ^2+\nonumber\\
   & & \sqrt[36]{\cal V} \Biggl[h^{z_1^2{\bar z}_{\bar 1}^2}\sqrt[18]{\cal V} \left((2 {\bar z}_1+{\bar z}_2) {\bar z}_4 z_1^2+2 \left(z_4
   {\bar z}_1^2+2 {\bar z}_2 z_4 {\bar z}_1+2 z_2 {\bar z}_4 {\bar z}_1+z_2 {\bar z}_2
   {\bar z}_4\right) z_1+{\bar z}_1 z_2 ({\bar z}_1+2 {\bar z}_2) z_4\right)\nonumber\\
   & & \hskip -0.6in-{r_2} (2 {\bar z}_1
   z_4+{\bar z}_2 z_4+2 z_1 {\bar z}_4+z_2 {\bar z}_4)\Biggr] \epsilon +h^{z_1^2{\bar z}_{\bar 1}^2}\sqrt[18]{\cal V}
   \left(z_4 {\bar z}_1^2+2 {\bar z}_2 z_4 {\bar z}_1+\sqrt[18]{\cal V} {\bar z}_4\right) \left(\sqrt[18]{\cal V}
   z_4+z_1 (z_1+2 z_2) {\bar z}_4\right)\Biggr)\nonumber\\
   \end{eqnarray*}

   \begin{eqnarray*}
   & & Y_{11}=\Biggl(\epsilon  \left({r2}+h^{z_1^2{\bar z}_{\bar 1}^2}
   \left(\left(z_1^2+z_2 z_1+z_2^2\right) \left({\bar z}_1^2+{\bar z}_2
   {\bar z}_1+{\bar z}_2^2\right)-\sqrt[18]{\cal V} (z_1+z_2) ({\bar z}_1+{\bar z}_2)\right)\right)\nonumber\\
   & & +h^{z_1^2{\bar z}_{\bar 1}^2}\sqrt[36]{\cal V}
   \Biggl[({\bar z}_1+{\bar z}_2) {\bar z}_4 z_1^2+\left(z_4 {\bar z}_1^2+{\bar z}_2 z_4
   {\bar z}_1+z_2 {\bar z}_4 {\bar z}_1+{\bar z}_2^2 z_4+z_2 {\bar z}_2 {\bar z}_4\right)
   z_1\nonumber\\
   & & -\sqrt[18]{\cal V} ({\bar z}_1 z_4+{\bar z}_2 z_4+(z_1+z_2) {\bar z}_4)+z_2
   \left(z_4 {\bar z}_1^2+{\bar z}_2 z_4 {\bar z}_1+z_2 {\bar z}_4 {\bar z}_1+{\bar z}_2^2
   z_4+z_2 {\bar z}_2 {\bar z}_4\right)\Biggr]\Biggr)^2
   \end{eqnarray*}

\item
\begin{eqnarray*}
& & g_{1{\bar 2}}=\frac{X_{12}}{Y_{12}}\ {\rm where}\nonumber\\
& & X_{12}=h^{z_1^2{\bar z}_{\bar 1}^2}r_1 \Bigg(\left(h^{z_1^2{\bar z}_{\bar 1}^2}\sqrt[18]{\cal V} z_1 (z_1+2 z_2) {\bar z}_2 (2
   {\bar z}_1+{\bar z}_2)+{r2} \left(\sqrt[18]{\cal V}-(2 z_1+z_2) ({\bar z}_1+2 {\bar z}_2)\right)\right)
   \epsilon ^2\nonumber\\
   & & \hskip-0.5in+\sqrt[36]{\cal V} \Biggl[h^{z_1^2{\bar z}_{\bar 1}^2}\sqrt[18]{\cal V} \left(({\bar z}_1+2 {\bar z}_2) {\bar z}_4 z_1^2+2 \left(z_4
   {\bar z}_2^2+2 {\bar z}_1 z_4 {\bar z}_2+2 z_2 {\bar z}_4 {\bar z}_2+{\bar z}_1 z_2
   {\bar z}_4\right) z_1+z_2 {\bar z}_2 (2 {\bar z}_1+{\bar z}_2) z_4\right)\nonumber\\
   & &\hskip-0.6in -{r_2} ({\bar z}_1
   z_4+2 {\bar z}_2 z_4+2 z_1 {\bar z}_4+z_2 {\bar z}_4)\Biggr] \epsilon +h^{z_1^2{\bar z}_{\bar 1}^2}\sqrt[18]{\cal V}
   \left(z_4 {\bar z}_2^2+2 {\bar z}_1 z_4 {\bar z}_2+\sqrt[18]{\cal V} {\bar z}_4\right) \left(\sqrt[18]{\cal V}
   z_4+z_1 (z_1+2 z_2) {\bar z}_4\right)\Biggr)\nonumber\\
   & & Y_{12}=\Biggl(\epsilon  \left({r2}+h
   \left(\left(z_1^2+z_2 z_1+z_2^2\right) \left({\bar z}_1^2+{\bar z}_2
   {\bar z}_1+{\bar z}_2^2\right)-\sqrt[18]{\cal V} (z_1+z_2) ({\bar z}_1+{\bar z}_2)\right)\right)\nonumber\\
   & & +h^{z_1^2{\bar z}_{\bar 1}^2}\sqrt[36]{\cal V}
   \Biggl[({\bar z}_1+{\bar z}_2) {\bar z}_4 z_1^2+\left(z_4 {\bar z}_1^2+{\bar z}_2 z_4
   {\bar z}_1+z_2 {\bar z}_4 {\bar z}_1+{\bar z}_2^2 z_4+z_2 {\bar z}_2 {\bar z}_4\right)
   z_1\nonumber\\
   & & -\sqrt[18]{\cal V} ({\bar z}_1 z_4+{\bar z}_2 z_4+(z_1+z_2) {\bar z}_4)+z_2
   \left(z_4 {\bar z}_1^2+{\bar z}_2 z_4 {\bar z}_1+z_2 {\bar z}_4 {\bar z}_1+{\bar z}_2^2
   z_4+z_2 {\bar z}_2 {\bar z}_4\right)\Biggr]\Biggr)^2
   \end{eqnarray*}

\item
\begin{eqnarray*}
& & g_{1{\bar 4}}=\frac{X_{14}}{Y_{14}},\ {\rm where}\nonumber\\
   & & X_{14}=-h^{z_1^2{\bar z}_{\bar 1}^2}r_1 \sqrt[36]{\cal V} \Biggl(\epsilon  \left({r2} \left((2 z_1+z_2)
   ({\bar z}_1+{\bar z}_2)-\sqrt[18]{\cal V}\right)-h^{z_1^2{\bar z}_{\bar 1}^2}\sqrt[18]{\cal V} z_1 {\bar z}_1 (z_1+2 z_2)
   {\bar z}_2\right)\nonumber\\
   & & +h^{z_1^2{\bar z}_{\bar 1}^2}\sqrt[36]{\cal V} ({\bar z}_1+{\bar z}_2) \left(-\sqrt[18]{\cal V} (2 z_1+z_2)
   ({\bar z}_1+{\bar z}_2)+z_1 (z_1+2 z_2) \left({\bar z}_1^2+{\bar z}_2
   {\bar z}_1+{\bar z}_2^2\right)+\sqrt[9]{\cal V}\right) z_4\Biggr)\nonumber\\
   & & Y_{14}=\Biggl(\epsilon  \left({r_2}+h^{z_1^2{\bar z}_{\bar 1}^2}
   \left(\left(z_1^2+z_2 z_1+z_2^2\right) \left({\bar z}_1^2+{\bar z}_2
   {\bar z}_1+{\bar z}_2^2\right)-\sqrt[18]{\cal V} (z_1+z_2) ({\bar z}_1+{\bar z}_2)\right)\right)\nonumber\\
   & & +h^{z_1^2{\bar z}_{\bar 1}^2}\sqrt[36]{\cal V}
   \Biggl[({\bar z}_1+{\bar z}_2) {\bar z}_4 z_1^2+\left(z_4 {\bar z}_1^2+{\bar z}_2 z_4
   {\bar z}_1+z_2 {\bar z}_4 {\bar z}_1+{\bar z}_2^2 z_4+z_2 {\bar z}_2 {\bar z}_4\right)
   z_1\nonumber\\
   & & -\sqrt[18]{\cal V} ({\bar z}_1 z_4+{\bar z}_2 z_4+(z_1+z_2) {\bar z}_4)+z_2
   \left(z_4 {\bar z}_1^2+{\bar z}_2 z_4 {\bar z}_1+z_2 {\bar z}_4 {\bar z}_1+{\bar z}_2^2
   z_4+z_2 {\bar z}_2 {\bar z}_4\right)\Biggr]\Biggr)^2
   \end{eqnarray*}

\item
\begin{eqnarray*}
& & g_{2{\bar 2}}=\frac{X_{22}}{Y_{22}},\ {\rm where}\nonumber\\
& & X_{22}=h^{z_1^2{\bar z}_{\bar 1}^2}r_1 \Biggl(\left(h^{z_1^2{\bar z}_{\bar 1}^2}\sqrt[18]{\cal V} z_2 (2 z_1+z_2) {\bar z}_2 (2
   {\bar z}_1+{\bar z}_2)+{r2} \left(\sqrt[18]{\cal V}-(z_1+2 z_2) ({\bar z}_1+2 {\bar z}_2)\right)\right)
   \epsilon ^2\nonumber\\
   & & +\sqrt[36]{\cal V} \Biggl[h^{z_1^2{\bar z}_{\bar 1}^2}\sqrt[18]{\cal V} \Biggl(z_2 \left(2 z_4 {\bar z}_2^2+4 {\bar z}_1 z_4
   {\bar z}_2+2 z_2 {\bar z}_4 {\bar z}_2+{\bar z}_1 z_2 {\bar z}_4\right)\nonumber\\
   & & +z_1 \left(z_4
   {\bar z}_2^2+2 {\bar z}_1 z_4 {\bar z}_2+4 z_2 {\bar z}_4 {\bar z}_2+2 {\bar z}_1 z_2
   {\bar z}_4\right)\Biggr)-{r_2} ({\bar z}_1 z_4+2 {\bar z}_2 z_4+z_1 {\bar z}_4+2 z_2
   {\bar z}_4)\Biggr] \epsilon\nonumber\\
    & & +h^{z_1^2{\bar z}_{\bar 1}^2}\sqrt[18]{\cal V} \left(z_4 {\bar z}_2^2+2 {\bar z}_1 z_4
   {\bar z}_2 +\sqrt[18]{\cal V} {\bar z}_4\right) \left(\sqrt[18]{\cal V} z_4+z_2 (2 z_1+z_2)
   {\bar z}_4\right)\Biggr)\nonumber\\
   & & Y_{22}=\Biggl(\epsilon  \left({r2}+h^{z_1^2{\bar z}_{\bar 1}^2}\left(\left(z_1^2+z_2 z_1+z_2^2\right)
   \left({\bar z}_1^2+{\bar z}_2 {\bar z}_1+{\bar z}_2^2\right)-\sqrt[18]{\cal V} (z_1+z_2)
   ({\bar z}_1+{\bar z}_2)\right)\right)\nonumber\\
   & & +h^{z_1^2{\bar z}_{\bar 1}^2}\sqrt[36]{\cal V} \Biggl(({\bar z}_1+{\bar z}_2) {\bar z}_4
   z_1^2+\left(z_4 {\bar z}_1^2+{\bar z}_2 z_4 {\bar z}_1+z_2 {\bar z}_4
   {\bar z}_1+{\bar z}_2^2 z_4+z_2 {\bar z}_2 {\bar z}_4\right) z_1\nonumber\\
   & & -\sqrt[18]{\cal V} ({\bar z}_1
   z_4+{\bar z}_2 z_4+(z_1+z_2) {\bar z}_4)+z_2 \left(z_4 {\bar z}_1^2+{\bar z}_2
   z_4 {\bar z}_1+z_2 {\bar z}_4 {\bar z}_1+{\bar z}_2^2 z_4+z_2 {\bar z}_2
   {\bar z}_4\right)\Biggr)\Biggr)^2
   \end{eqnarray*}

\item
     \begin{eqnarray*}
& & g_{2{\bar 4}}=\frac{X_{24}}{Y_{24}},\ {\rm where}\nonumber\\
& & X_{24}=-h^{z_1^2{\bar z}_{\bar 1}^2}r_1 \sqrt[36]{\cal V} \Biggl(\epsilon  \left({r2} \left((z_1+2 z_2)
   ({\bar z}_1+{\bar z}_2)-\sqrt[18]{\cal V}\right)-h^{z_1^2{\bar z}_{\bar 1}^2}\sqrt[18]{\cal V} {\bar z}_1 z_2 (2 z_1+z_2)
   {\bar z}_2\right)\nonumber\\
   & & + h^{z_1^2{\bar z}_{\bar 1}^2}\sqrt[36]{\cal V} ({\bar z}_1+{\bar z}_2) \left(-\sqrt[18]{\cal V} (z_1+2 z_2)
   ({\bar z}_1+{\bar z}_2)+z_2 (2 z_1+z_2) \left({\bar z}_1^2+{\bar z}_2
   {\bar z}_1+{\bar z}_2^2\right)+\sqrt[9]{\cal V}\right) z_4\Biggr)\nonumber\\
   & & Y_{24}=\Biggl(\epsilon  \left({r2}+h
   \left(\left(z_1^2+z_2 z_1+z_2^2\right) \left({\bar z}_1^2+{\bar z}_2
   {\bar z}_1+{\bar z}_2^2\right)-\sqrt[18]{\cal V} (z_1+z_2) ({\bar z}_1+{\bar z}_2)\right)\right)\nonumber\\
   & & +h^{z_1^2{\bar z}_{\bar 1}^2}\sqrt[36]{\cal V}
   \Biggl[({\bar z}_1+{\bar z}_2) {\bar z}_4 z_1^2+\left(z_4 {\bar z}_1^2+{\bar z}_2 z_4
   {\bar z}_1+z_2 {\bar z}_4 {\bar z}_1+{\bar z}_2^2 z_4+z_2 {\bar z}_2 {\bar z}_4\right)
   z_1\nonumber\\
   & & -\sqrt[18]{\cal V} ({\bar z}_1 z_4+{\bar z}_2 z_4+(z_1+z_2) {\bar z}_4)+z_2
   \left(z_4 {\bar z}_1^2+{\bar z}_2 z_4 {\bar z}_1+z_2 {\bar z}_4 {\bar z}_1+{\bar z}_2^2
   z_4+z_2 {\bar z}_2 {\bar z}_4\right)\Biggr]\Biggr)^2
\end{eqnarray*}

\item

   \begin{eqnarray*}
& & g_{4{\bar 4}}=\frac{X_{44}}{Y_{44}},\ {\rm where}\nonumber\\
& & X_{44}=\left(h^{z_1^2{\bar z}_{\bar 1}^2}\right)^2 r_1\sqrt[18]{V} \left(\sqrt[18]{V} (r_1z_1+r_1z_2)-\left(r_1z_1^2+r_1z_2
   r_1z_1+r_1z_2^2\right) (r_1{\bar z}_1+r_1{\bar z}_2)\right)\nonumber\\
   & & \times \left(\sqrt[18]{V}
   (r_1{\bar z}_1+r_1{\bar z}_2)-(r_1z_1+r_1z_2) \left(r_1{\bar z}_1^2+r_1{\bar z}_2
   r_1{\bar z}_1+r_1{\bar z}_2^2\right)\right)\nonumber\\
   & & \hskip-0.8in Y_{44}=\Biggl(\epsilon  \left(r_1{r2}+h \left(\left(r_1z_1^2+r_1z_2
   r_1z_1+r_1z_2^2\right) \left(r_1{\bar z}_1^2+r_1^2{\bar z}_2{\bar z}_1+r_1{\bar z}_2^2\right)-\sqrt[18]{V}
   |r_1z_1+r_1z_2|^2\right)\right)\nonumber\\
   & & \hskip-0.6in+h^{z_1^2{\bar z}_{\bar 1}^2} \sqrt[36]{V} \Biggl[(r_1{\bar z}_1+r_1{\bar z}_2) r_1{\bar z}_4
   r_1z_1^2+\left(r_1^2z_4 {\bar z}_1^2+r_1^3{\bar z}_2z_4 {\bar z}_1+r_1^3z_2 {\bar z}_4
   {\bar z}_1+r_1^2{\bar z}_2^2z_4+r_1^3z_2 {\bar z}_2 {\bar z}_4\right) r_1z_1\nonumber\\
   & &\hskip-0.93in -\sqrt[18]{V} (r_1^2{\bar z}_1
   z_4+r_1^2{\bar z}_2 z_4+(r_1z_1+r_1z_2) r_1{\bar z}_4)+r_1z_2 \left(r_1^2z_4 {\bar z}_1^2+r_1^3{\bar z}_2z_4{\bar z}_1+r_1^3z_2 {\bar z}_4 {\bar z}_1+r_1^2{\bar z}_2^2 z_4+r_1^3z_2 {\bar z}_2
   {\bar z}_4\right)\Biggr]\Biggr)^2\nonumber\\
   & &
   \end{eqnarray*}
\end{itemize}

The affine connection components are:
\begin{eqnarray*}
& & \Gamma^i_{1i}=\frac{\Gamma_1}{\Gamma_2},\ {\rm where}\nonumber\\
& & \Gamma_1=-\frac{1}{z_1-z_2}\Biggl[\sqrt[36]{\cal V} \Biggl[{r2} \sqrt[36]{\cal V} z_1 \epsilon ^2+\left({r2} \left(\sqrt[18]{\cal V} z_4+\left(2
   z_1^2+3 z_2 z_1+z_2^2\right) {\bar z}_4\right)-h^{z_1^2{\bar z}_{\bar 1}^2}\sqrt[9]{\cal V} z_1 (z_1+2 z_2)
   {\bar z}_4\right) \epsilon\nonumber\\
    & &\hskip-0.8in -h^{z_1^2{\bar z}_{\bar 1}^2}\sqrt[12]{\cal V} (z_1+z_2) {\bar z}_4 \left(\sqrt[18]{\cal V} z_4+z_1
   (z_1+2 z_2) {\bar z}_4\right)\Biggr] \Biggl(2 h^{z_1^2{\bar z}_{\bar 1}^2}\Biggl[\epsilon  \left({r2} \left((z_1+z_2)
   ({\bar z}_1+2 {\bar z}_2)-\sqrt[18]{\cal V}\right)-h^{z_1^2{\bar z}_{\bar 1}^2}\sqrt[18]{\cal V} z_1 z_2 {\bar z}_2 (2
   {\bar z}_1+{\bar z}_2)\right)\nonumber\\
   & & +h^{z_1^2{\bar z}_{\bar 1}^2}\sqrt[36]{\cal V} (z_1+z_2) \left(\left(z_1^2+z_2
   z_1+z_2^2\right) {\bar z}_2 (2 {\bar z}_1+{\bar z}_2)-\sqrt[18]{\cal V} (z_1+z_2) ({\bar z}_1+2
   {\bar z}_2)+\sqrt[9]{\cal V}\right) {\bar z}_4\Biggr]\nonumber\\
   & & \hskip-0.9in\times \left(\epsilon  \left((2 z_1+z_2)
   \left({\bar z}_1^2+{\bar z}_2 {\bar z}_1+{\bar z}_2^2\right)-\sqrt[18]{\cal V}
   ({\bar z}_1+{\bar z}_2)\right)+\sqrt[36]{\cal V} \left(z_4 {\bar z}_1^2+({\bar z}_2 z_4+(2 z_1+z_2)
   {\bar z}_4) {\bar z}_1+{\bar z}_2^2 z_4+(2 z_1+z_2) {\bar z}_2 {\bar z}_4-\sqrt[18]{\cal V}
   {\bar z}_4\right)\right)\nonumber\\
   & & \hskip-0.8in+\Biggl(\epsilon  \left({r2} ({\bar z}_1+2 {\bar z}_2)-h^{z_1^2{\bar z}_{\bar 1}^2}\sqrt[18]{\cal V} z_2 {\bar z}_2
   (2 {\bar z}_1+{\bar z}_2)\right)+h^{z_1^2{\bar z}_{\bar 1}^2}\sqrt[36]{\cal V} \Biggl(\left(3 z_1^2+4 z_2 z_1+2 z_2^2\right)
   {\bar z}_2 (2 {\bar z}_1+{\bar z}_2)-2 \sqrt[18]{\cal V} (z_1+z_2) ({\bar z}_1+2
   {\bar z}_2)\nonumber\\
   & & +\sqrt[9]{\cal V}\Biggr) {\bar z}_4\Biggr)\Biggl(h^{z_1^2{\bar z}_{\bar 1}^2}\sqrt[36]{\cal V} \Biggl[-({\bar z}_1+{\bar z}_2) {\bar z}_4
   z_1^2-\left(z_4 {\bar z}_1^2+{\bar z}_2 z_4 {\bar z}_1+z_2 {\bar z}_4
   {\bar z}_1+{\bar z}_2^2 z_4+z_2 {\bar z}_2 {\bar z}_4\right) z_1\nonumber\\
   & & +\sqrt[18]{\cal V} ({\bar z}_1
   z_4+{\bar z}_2 z_4+(z_1+z_2) {\bar z}_4)-z_2 \left(z_4 {\bar z}_1^2+{\bar z}_2
   z_4 {\bar z}_1+z_2 {\bar z}_4 {\bar z}_1+{\bar z}_2^2 z_4+z_2 {\bar z}_2
   {\bar z}_4\right)\Biggr]\nonumber\\
   & & -\epsilon  \left({r_2}+h^{z_1^2{\bar z}_{\bar 1}^2}\left(\left(z_1^2+z_2 z_1+z_2^2\right)
   \left({\bar z}_1^2+{\bar z}_2 {\bar z}_1+{\bar z}_2^2\right)-\sqrt[18]{\cal V} (z_1+z_2)
   ({\bar z}_1+{\bar z}_2)\right)\right)\Biggr)\Biggr)\Biggr]
   \nonumber\\
& &  +\frac{1}{z_1-z_2}\Biggl[\sqrt[36]{\cal V} \Biggl({r2} \sqrt[36]{\cal V}
   z_2 \epsilon ^2+\left({r_2} \left(\sqrt[18]{\cal V} z_4+\left(z_1^2+3 z_2 z_1+2
   z_2^2\right) {\bar z}_4\right)-h^{z_1^2{\bar z}_{\bar 1}^2}\sqrt[9]{\cal V} z_2 (2 z_1+z_2) {\bar z}_4\right) \epsilon\nonumber\\
   \end{eqnarray*}

   \begin{eqnarray*}
   & & \hskip-0.6in -h^{z_1^2{\bar z}_{\bar 1}^2}
   \sqrt[12]{\cal V} (z_1+z_2) {\bar z}_4 \left(\sqrt[18]{\cal V} z_4+z_2 (2 z_1+z_2)
   {\bar z}_4\right)\Biggr) \Biggl(2 h^{z_1^2{\bar z}_{\bar 1}^2}\Biggl[h^{z_1^2{\bar z}_{\bar 1}^2}\sqrt[36]{\cal V} \left(\left(z_1^3+2 z_2 z_1^2+2 z_2^2
   z_1+z_2^3\right) {\bar z}_4-\epsilon  \sqrt[36]{\cal V} z_1 z_2\right) {\bar z}_1^2\nonumber\\
    & & \hskip-0.8in+2 \left(\epsilon
   \left({r2} (z_1+z_2)-h^{z_1^2{\bar z}_{\bar 1}^2}\sqrt[18]{\cal V} z_1 z_2 {\bar z}_2\right)+h^{z_1^2{\bar z}_{\bar 1}^2}\sqrt[36]{\cal V}
   (z_1+z_2) \left(\left(z_1^2+z_2 z_1+z_2^2\right) {\bar z}_2-\sqrt[18]{\cal V}
   (z_1+z_2)\right) {\bar z}_4\right) {\bar z}_1\nonumber\\
   & & +\left(\sqrt[18]{\cal V}-(z_1+z_2) {\bar z}_2\right)
   \left(h^{z_1^2{\bar z}_{\bar 1}^2}\sqrt[12]{\cal V} (z_1+z_2) {\bar z}_4-\epsilon  {r2}\right)\Biggr]\nonumber\\
   & & \hskip-0.9in\times \left(\epsilon  \left((2
   z_1+z_2) \left({\bar z}_1^2+{\bar z}_2 {\bar z}_1+{\bar z}_2^2\right)-\sqrt[18]{\cal V}
   ({\bar z}_1+{\bar z}_2)\right)+\sqrt[36]{\cal V} \left(z_4 {\bar z}_1^2+({\bar z}_2 z_4+(2 z_1+z_2)
   {\bar z}_4) {\bar z}_1+{\bar z}_2^2 z_4+(2 z_1+z_2) {\bar z}_2 {\bar z}_4-\sqrt[18]{\cal V}
   {\bar z}_4\right)\right)\nonumber\\
   & & +\Biggl(\epsilon  \left({r2} (2 {\bar z}_1+{\bar z}_2)-h^{z_1^2{\bar z}_{\bar 1}^2}\sqrt[18]{\cal V} {\bar z}_1 z_2
   ({\bar z}_1+2 {\bar z}_2)\right)+h^{z_1^2{\bar z}_{\bar 1}^2}\sqrt[36]{\cal V} \Biggl[-2 \sqrt[18]{\cal V} (z_1+z_2) (2
   {\bar z}_1+{\bar z}_2)\nonumber\\
   & &\hskip-0.6in +{\bar z}_1 \left(3 z_1^2+4 z_2 z_1+2 z_2^2\right) ({\bar z}_1+2
   {\bar z}_2)+\sqrt[9]{\cal V}\Biggr] {\bar z}_4\Biggr)\Biggl(h^{z_1^2{\bar z}_{\bar 1}^2}\sqrt[36]{\cal V} \Biggl[-({\bar z}_1+{\bar z}_2) {\bar z}_4
   z_1^2-\left(z_4 {\bar z}_1^2+{\bar z}_2 z_4 {\bar z}_1+z_2 {\bar z}_4
   {\bar z}_1+{\bar z}_2^2 z_4+z_2 {\bar z}_2 {\bar z}_4\right) z_1\nonumber\\
   & & +\sqrt[18]{\cal V} ({\bar z}_1
   z_4+{\bar z}_2 z_4+(z_1+z_2) {\bar z}_4)-z_2 \left(z_4 {\bar z}_1^2+{\bar z}_2
   z_4 {\bar z}_1+z_2 {\bar z}_4 {\bar z}_1+{\bar z}_2^2 z_4+z_2 {\bar z}_2
   {\bar z}_4\right)\Biggr]\nonumber\\
   & & -\epsilon  \left({r2}+h^{z_1^2{\bar z}_{\bar 1}^2}\left(\left(z_1^2+z_2 z_1+z_2^2\right)
   \left({\bar z}_1^2+{\bar z}_2 {\bar z}_1+{\bar z}_2^2\right)-\sqrt[18]{\cal V} (z_1+z_2)
   ({\bar z}_1+{\bar z}_2)\right)\right)\Biggr)\Biggr)\Biggr]\nonumber\\
   & & +\epsilon  h^{z_1^2{\bar z}_{\bar 1}^2}\sqrt[18]{\cal V} \left({r2}
   \epsilon ^2+\left(\sqrt[9]{\cal V} h+{r2}\right) z_4 {\bar z}_4\right) \Biggl(-2 h^{z_1^2{\bar z}_{\bar 1}^2}\left(\sqrt[18]{\cal V}
   (z_1+z_2)-\left(z_1^2+z_2 z_1+z_2^2\right) ({\bar z}_1+{\bar z}_2)\right)\nonumber\\
& & \times   \left(\sqrt[18]{\cal V} ({\bar z}_1+{\bar z}_2)-(z_1+z_2) \left({\bar z}_1^2+{\bar z}_2
   {\bar z}_1+{\bar z}_2^2\right)\right)\nonumber\\
    & & \hskip-0.9in\times\left(\epsilon  \left((2 z_1+z_2) \left({\bar z}_1^2+{\bar z}_2
   {\bar z}_1+{\bar z}_2^2\right)-\sqrt[18]{\cal V} ({\bar z}_1+{\bar z}_2)\right)+\sqrt[36]{\cal V} \left(z_4
   {\bar z}_1^2+({\bar z}_2 z_4+(2 z_1+z_2) {\bar z}_4) {\bar z}_1+{\bar z}_2^2 z_4+(2
   z_1+z_2) {\bar z}_2 {\bar z}_4-\sqrt[18]{\cal V} {\bar z}_4\right)\right)\nonumber\\
   & & +\left(-{\bar z}_1^2-{\bar z}_2
   {\bar z}_1-{\bar z}_2^2\right) \left(\sqrt[18]{\cal V} (z_1+z_2)-\left(z_1^2+z_2
   z_1+z_2^2\right) ({\bar z}_1+{\bar z}_2)\right)\nonumber\\
   & & \times \Biggl[\epsilon  \left({r2}+h^{z_1^2{\bar z}_{\bar 1}^2}
   \left(\left(z_1^2+z_2 z_1+z_2^2\right) \left({\bar z}_1^2+{\bar z}_2
   {\bar z}_1+{\bar z}_2^2\right)-\sqrt[18]{\cal V} (z_1+z_2) ({\bar z}_1+{\bar z}_2)\right)\right)+h^{z_1^2{\bar z}_{\bar 1}^2}\sqrt[36]{\cal V}\nonumber\\
& &\times   \Biggl(({\bar z}_1+{\bar z}_2) {\bar z}_4 z_1^2+\left(z_4 {\bar z}_1^2+{\bar z}_2 z_4
   {\bar z}_1+z_2 {\bar z}_4 {\bar z}_1+{\bar z}_2^2 z_4+z_2 {\bar z}_2 {\bar z}_4\right)
   z_1-\sqrt[18]{\cal V} ({\bar z}_1 z_4+{\bar z}_2 z_4+(z_1+z_2) {\bar z}_4)\nonumber\\
   & & +z_2
   \left(z_4 {\bar z}_1^2+{\bar z}_2 z_4 {\bar z}_1+z_2 {\bar z}_4 {\bar z}_1+{\bar z}_2^2
   z_4+z_2 {\bar z}_2 {\bar z}_4\right)\Biggr)\Biggr]\nonumber\\
   & & +\left(\sqrt[18]{\cal V}-(2 z_1+z_2)
   ({\bar z}_1+{\bar z}_2)\right) \left(\sqrt[18]{\cal V} ({\bar z}_1+{\bar z}_2)-(z_1+z_2)
   \left({\bar z}_1^2+{\bar z}_2 {\bar z}_1+{\bar z}_2^2\right)\right) \nonumber\\
   & & \times\Biggl[\epsilon  \left({r_2}+h
   \left(\left(z_1^2+z_2 z_1+z_2^2\right) \left({\bar z}_1^2+{\bar z}_2
   {\bar z}_1+{\bar z}_2^2\right)-\sqrt[18]{\cal V} (z_1+z_2) ({\bar z}_1+{\bar z}_2)\right)\right)\nonumber\\
   & & \hskip-0.3in+h^{z_1^2{\bar z}_{\bar 1}^2}\sqrt[36]{\cal V}
   \Biggl(({\bar z}_1+{\bar z}_2) {\bar z}_4 z_1^2+\left(z_4 {\bar z}_1^2+{\bar z}_2 z_4
   {\bar z}_1+z_2 {\bar z}_4 {\bar z}_1+{\bar z}_2^2 z_4+z_2 {\bar z}_2 {\bar z}_4\right)
   z_1\nonumber\\
   & & -\sqrt[18]{\cal V} ({\bar z}_1 z_4+{\bar z}_2 z_4+(z_1+z_2) {\bar z}_4)+z_2
   \left(z_4 {\bar z}_1^2+{\bar z}_2 z_4 {\bar z}_1+z_2 {\bar z}_4 {\bar z}_1+{\bar z}_2^2
   z_4+z_2 {\bar z}_2 {\bar z}_4\right)\Biggr)\Biggr]\Biggr)\nonumber\\
   \end{eqnarray*}

   \begin{eqnarray*}
   & & -\frac{1}{{\bar z}_1-{\bar z}_2}\Biggl[\sqrt[36]{\cal V} \Biggl({r2} \sqrt[36]{\cal V}
   {\bar z}_1 \epsilon ^2+\Biggl({r2} \left(2 z_4 {\bar z}_1^2+3 {\bar z}_2 z_4 {\bar z}_1+{\bar z}_2^2
   z_4+\sqrt[18]{\cal V} {\bar z}_4\right)\nonumber\\
   & & -h^{z_1^2{\bar z}_{\bar 1}^2}\sqrt[9]{\cal V} {\bar z}_1 ({\bar z}_1+2 {\bar z}_2) z_4\Biggr) \epsilon
   -h^{z_1^2{\bar z}_{\bar 1}^2}\sqrt[12]{\cal V} ({\bar z}_1+{\bar z}_2) z_4 \left(z_4 {\bar z}_1^2+2 {\bar z}_2 z_4
   {\bar z}_1+\sqrt[18]{\cal V} {\bar z}_4\right)\Biggr)\nonumber\\
    & & \times\Biggl(2 h^{z_1^2{\bar z}_{\bar 1}^2}\Biggl(\epsilon  \Biggl[{r_2} \left((z_1+2 z_2)
   ({\bar z}_1+{\bar z}_2)-\sqrt[18]{\cal V}\right) -h^{z_1^2{\bar z}_{\bar 1}^2}\sqrt[18]{\cal V} {\bar z}_1 z_2 (2 z_1+z_2)
   {\bar z}_2\Biggr]+h^{z_1^2{\bar z}_{\bar 1}^2}\sqrt[36]{\cal V}\nonumber\\
    & & \times({\bar z}_1+{\bar z}_2) \left(-\sqrt[18]{\cal V} (z_1+2 z_2)
   ({\bar z}_1+{\bar z}_2)+z_2 (2 z_1+z_2) \left({\bar z}_1^2+{\bar z}_2
   {\bar z}_1+{\bar z}_2^2\right)+\sqrt[9]{\cal V}\right) z_4\Biggr)\nonumber\\
    & & \hskip-0.91in\times\Biggl(\epsilon  \left((2 z_1+z_2)
   \left({\bar z}_1^2+{\bar z}_2 {\bar z}_1+{\bar z}_2^2\right)-\sqrt[18]{\cal V}
   ({\bar z}_1+{\bar z}_2)\right)+\sqrt[36]{\cal V}\left(z_4 {\bar z}_1^2+({\bar z}_2 z_4+(2 z_1+z_2)
   {\bar z}_4) {\bar z}_1+{\bar z}_2^2 z_4+(2 z_1+z_2) {\bar z}_2 {\bar z}_4-\sqrt[18]{\cal V}
   {\bar z}_4\right)\Biggr)\nonumber\\
   & & \hskip-0.9in-\Biggl(\epsilon  \left({r2} ({\bar z}_1+{\bar z}_2)-2 h^{z_1^2{\bar z}_{\bar 1}^2}\sqrt[18]{\cal V} {\bar z}_1 z_2
   {\bar z}_2\right)+h^{z_1^2{\bar z}_{\bar 1}^2}\sqrt[36]{\cal V} ({\bar z}_1+{\bar z}_2) \left(2 z_2 \left({\bar z}_1^2+{\bar z}_2
   {\bar z}_1+{\bar z}_2^2\right)-\sqrt[18]{\cal V} ({\bar z}_1+{\bar z}_2)\right) z_4\Biggr)\nonumber\\
   & & \times \Biggl(\epsilon
   \left({r2}+h^{z_1^2{\bar z}_{\bar 1}^2}\left(\left(z_1^2+z_2 z_1+z_2^2\right) \left({\bar z}_1^2+{\bar z}_2
   {\bar z}_1+{\bar z}_2^2\right)-\sqrt[18]{\cal V} (z_1+z_2) ({\bar z}_1+{\bar z}_2)\right)\right)\nonumber\\
   & & +h^{z_1^2{\bar z}_{\bar 1}^2}\sqrt[36]{\cal V}
   \Biggl[({\bar z}_1+{\bar z}_2) {\bar z}_4 z_1^2+\left(z_4 {\bar z}_1^2+{\bar z}_2 z_4
   {\bar z}_1+z_2 {\bar z}_4 {\bar z}_1+{\bar z}_2^2 z_4+z_2 {\bar z}_2 {\bar z}_4\right)
   z_1\nonumber\\
   & & -\sqrt[18]{\cal V} ({\bar z}_1 z_4+{\bar z}_2 z_4+(z_1+z_2) {\bar z}_4)+z_2
   \left(z_4 {\bar z}_1^2+{\bar z}_2 z_4 {\bar z}_1+z_2 {\bar z}_4 {\bar z}_1+{\bar z}_2^2
   z_4+z_2 {\bar z}_2 {\bar z}_4\right)\Biggr]\Biggr)\Biggr)\Biggr]\nonumber\\
   & & +\frac{1}{{\bar z}_1-{\bar z}_2}\Biggl[\sqrt[36]{\cal V}
   \Biggl({r2} \sqrt[36]{\cal V} {\bar z}_2 \epsilon ^2+\left({r2} \left(z_4 {\bar z}_1^2+3 {\bar z}_2 z_4
   {\bar z}_1+2 {\bar z}_2^2 z_4+\sqrt[18]{\cal V} {\bar z}_4\right)-h^{z_1^2{\bar z}_{\bar 1}^2}\sqrt[9]{\cal V} {\bar z}_2 (2
   {\bar z}_1+{\bar z}_2) z_4\right) \epsilon\nonumber\\
    & &\hskip-0.7in -h^{z_1^2{\bar z}_{\bar 1}^2}\sqrt[12]{\cal V} ({\bar z}_1+{\bar z}_2) z_4 \left(z_4
   {\bar z}_2^2+2 {\bar z}_1 z_4 {\bar z}_2+\sqrt[18]{\cal V} {\bar z}_4\right)\Biggr) \Biggl[2 h^{z_1^2{\bar z}_{\bar 1}^2}\Biggl(\epsilon
   \left({r2} \left((2 z_1+z_2) ({\bar z}_1+{\bar z}_2)-\sqrt[18]{\cal V}\right)-h^{z_1^2{\bar z}_{\bar 1}^2}\sqrt[18]{\cal V} z_1
   {\bar z}_1 (z_1+2 z_2) {\bar z}_2\right)\nonumber\\
   & & +h^{z_1^2{\bar z}_{\bar 1}^2}\sqrt[36]{\cal V} ({\bar z}_1+{\bar z}_2) \left(-\sqrt[18]{\cal V} (2
   z_1+z_2) ({\bar z}_1+{\bar z}_2)+z_1 (z_1+2 z_2) \left({\bar z}_1^2+{\bar z}_2
   {\bar z}_1+{\bar z}_2^2\right)+\sqrt[9]{\cal V}\right) z_4\Biggr)\nonumber\\
    & & \hskip-0.85in\times\left(\epsilon  \left((2 z_1+z_2)
   \left({\bar z}_1^2+{\bar z}_2 {\bar z}_1+{\bar z}_2^2\right)-\sqrt[18]{\cal V}
   ({\bar z}_1+{\bar z}_2)\right)+\sqrt[36]{\cal V} \left(z_4 {\bar z}_1^2+({\bar z}_2 z_4+(2 z_1+z_2)
   {\bar z}_4) {\bar z}_1+{\bar z}_2^2 z_4+(2 z_1+z_2) {\bar z}_2 {\bar z}_4-\sqrt[18]{\cal V}
   {\bar z}_4\right)\right)\nonumber\\
   & &\hskip-0.8in -2 \left(\epsilon  \left({r2} ({\bar z}_1+{\bar z}_2)-h^{z_1^2{\bar z}_{\bar 1}^2}\sqrt[18]{\cal V} {\bar z}_1
   (z_1+z_2) {\bar z}_2\right)+h^{z_1^2{\bar z}_{\bar 1}^2}\sqrt[36]{\cal V} ({\bar z}_1+{\bar z}_2) \left((z_1+z_2)
   \left({\bar z}_1^2+{\bar z}_2 {\bar z}_1+{\bar z}_2^2\right)-\sqrt[18]{\cal V} ({\bar z}_1+{\bar z}_2)\right)
   z_4\right)\nonumber\\
    & & \times\Biggl(\epsilon  \left({r2}+h^{z_1^2{\bar z}_{\bar 1}^2}\left(\left(z_1^2+z_2 z_1+z_2^2\right)
   \left({\bar z}_1^2+{\bar z}_2 {\bar z}_1+{\bar z}_2^2\right)-\sqrt[18]{\cal V} (z_1+z_2)
   ({\bar z}_1+{\bar z}_2)\right)\right)+h^{z_1^2{\bar z}_{\bar 1}^2}\sqrt[36]{\cal V} \nonumber\\
   & & \hskip-0.9in\times\Biggl(({\bar z}_1+{\bar z}_2) {\bar z}_4
   z_1^2+\left(z_4 {\bar z}_1^2+{\bar z}_2 z_4 {\bar z}_1+z_2 {\bar z}_4
   {\bar z}_1+{\bar z}_2^2 z_4+z_2 {\bar z}_2 {\bar z}_4\right) z_1-\sqrt[18]{\cal V} ({\bar z}_1
   z_4+{\bar z}_2 z_4+(z_1+z_2) {\bar z}_4)+z_2 \Biggl(z_4 {\bar z}_1^2+{\bar z}_2
   z_4 {\bar z}_1+z_2 {\bar z}_4 {\bar z}_1\nonumber\\
   & & +{\bar z}_2^2 z_4+z_2 {\bar z}_2
   {\bar z}_4\Biggr)\Biggr)\Biggr)\Biggr]\Biggr]\nonumber\\
   \end{eqnarray*}

   \begin{eqnarray*}
   & & +\frac{1}{|z_1-z_2|^2}\Biggl[\Biggl(\epsilon  \Biggl({r_2} \left(-\sqrt[18]{\cal V}
   (2 z_1 {\bar z}_1+3 z_2 {\bar z}_1+3 z_1 {\bar z}_2+4 z_2 {\bar z}_2)+\left(z_1^2+3
   z_2 z_1+2 z_2^2\right) \left({\bar z}_1^2+3 {\bar z}_2 {\bar z}_1+2
   {\bar z}_2^2\right)+\sqrt[9]{\cal V}\right)\nonumber\\
   & &\hskip-0.9in -h^{z_1^2{\bar z}_{\bar 1}^2}\sqrt[18]{\cal V} z_2 (2 z_1+z_2) {\bar z}_2 (2
   {\bar z}_1+{\bar z}_2) \left((z_1+z_2) ({\bar z}_1+{\bar z}_2)-\sqrt[18]{\cal V}\right)\Biggr)+h^{z_1^2{\bar z}_{\bar 1}^2}\sqrt[12]{\cal V}
   \Biggl[-\Biggl(2 ({\bar z}_2 z_4+z_2 {\bar z}_4) {\bar z}_1^2+{\bar z}_2 (3 {\bar z}_2 z_4+2
   z_2 {\bar z}_4) {\bar z}_1\nonumber\\
   & & \hskip-0.9in+{\bar z}_2^2 ({\bar z}_2 z_4+2 z_2 {\bar z}_4)\Biggr)   z_1^2-z_2 \left((2 {\bar z}_2 z_4+3 z_2 {\bar z}_4) {\bar z}_1^2+3 {\bar z}_2 ({\bar z}_2
   z_4+z_2 {\bar z}_4) {\bar z}_1+{\bar z}_2^2 ({\bar z}_2 z_4+3 z_2 {\bar z}_4)\right)
   z_1\nonumber\\
   & & -z_2^2 \left((2 {\bar z}_2 z_4+z_2 {\bar z}_4) {\bar z}_1^2+{\bar z}_2 (3 {\bar z}_2
   z_4+z_2 {\bar z}_4) {\bar z}_1+{\bar z}_2^2 ({\bar z}_2 z_4+z_2
   {\bar z}_4)\right)\nonumber\\
   & & +\sqrt[18]{\cal V} \left(z_2 \left(z_4 {\bar z}_2^2+2 {\bar z}_1 z_4
   {\bar z}_2+z_2 {\bar z}_4 {\bar z}_2+{\bar z}_1 z_2 {\bar z}_4\right)+z_1 \left(z_4
   {\bar z}_2^2+2 {\bar z}_1 z_4 {\bar z}_2+2 z_2 {\bar z}_4 {\bar z}_2+2 {\bar z}_1 z_2
   {\bar z}_4\right)\right)\Biggr]\Biggr)\nonumber\\
    & & \times\Biggl(2 \Biggl[\epsilon  \left({r2}+h^{z_1^2{\bar z}_{\bar 1}^2}\left(\left(z_1^2+z_2
   z_1+z_2^2\right) \left({\bar z}_1^2+{\bar z}_2 {\bar z}_1+{\bar z}_2^2\right)-\sqrt[18]{\cal V}
   (z_1+z_2) ({\bar z}_1+{\bar z}_2)\right)\right)\nonumber\\
   & & +h^{z_1^2{\bar z}_{\bar 1}^2}\sqrt[36]{\cal V} \Biggl(({\bar z}_1+{\bar z}_2) {\bar z}_4
   z_1^2+\left(z_4 {\bar z}_1^2+{\bar z}_2 z_4 {\bar z}_1+z_2 {\bar z}_4
   {\bar z}_1+{\bar z}_2^2 z_4+z_2 {\bar z}_2 {\bar z}_4\right) z_1-\sqrt[18]{\cal V} ({\bar z}_1
   z_4+{\bar z}_2 z_4+(z_1+z_2) {\bar z}_4)\nonumber\\
   & & +z_2 \left(z_4 {\bar z}_1^2+{\bar z}_2
   z_4 {\bar z}_1+z_2 {\bar z}_4 {\bar z}_1+{\bar z}_2^2 z_4+z_2 {\bar z}_2
   {\bar z}_4\right)\Biggr)\Biggr] \Biggl(\left(h^{z_1^2{\bar z}_{\bar 1}^2}\sqrt[18]{\cal V} {\bar z}_1 (z_1+z_2) ({\bar z}_1+2
   {\bar z}_2)-{r2} (2 {\bar z}_1+{\bar z}_2)\right) \epsilon ^2\nonumber\\
   & & +\left(h^{z_1^2{\bar z}_{\bar 1}^2}\sqrt[12]{\cal V} \left(z_4 {\bar z}_1^2+2
   ({\bar z}_2 z_4+(z_1+z_2) {\bar z}_4) {\bar z}_1+(z_1+z_2) {\bar z}_2
   {\bar z}_4\right)-{r_2} \sqrt[36]{\cal V} {\bar z}_4\right) \epsilon\nonumber\\
    & & +h^{z_1^2{\bar z}_{\bar 1}^2}\sqrt[18]{\cal V} (z_1+z_2) {\bar z}_4
   \left(z_4 {\bar z}_1^2+2 {\bar z}_2 z_4 {\bar z}_1+\sqrt[18]{\cal V} {\bar z}_4\right)\Biggr)-2 h^{z_1^2{\bar z}_{\bar 1}^2}
   \Biggl(\epsilon  \left((2 z_1+z_2) \left({\bar z}_1^2+{\bar z}_2
   {\bar z}_1+{\bar z}_2^2\right)-\sqrt[18]{\cal V} ({\bar z}_1+{\bar z}_2)\right)\nonumber\\
   & & +\sqrt[36]{\cal V} \left(z_4
   {\bar z}_1^2+({\bar z}_2 z_4+(2 z_1+z_2) {\bar z}_4) {\bar z}_1+{\bar z}_2^2 z_4+(2
   z_1+z_2) {\bar z}_2 {\bar z}_4-\sqrt[18]{\cal V} {\bar z}_4\right)\Biggr) \nonumber\\
   & & \times\Biggl[\left(h^{z_1^2{\bar z}_{\bar 1}^2}\sqrt[18]{\cal V} z_1
   {\bar z}_1 (z_1+2 z_2) ({\bar z}_1+2 {\bar z}_2)+{r2} \left(\sqrt[18]{\cal V}-(2 z_1+z_2) (2
   {\bar z}_1+{\bar z}_2)\right)\right) \epsilon ^2\nonumber\\
   & & +\sqrt[36]{\cal V} \Biggl(h^{z_1^2{\bar z}_{\bar 1}^2}\sqrt[18]{\cal V} \left((2 {\bar z}_1+{\bar z}_2)
   {\bar z}_4 z_1^2+2 \left(z_4 {\bar z}_1^2+2 {\bar z}_2 z_4 {\bar z}_1+2 z_2 {\bar z}_4
   {\bar z}_1+z_2 {\bar z}_2 {\bar z}_4\right) z_1+{\bar z}_1 z_2 ({\bar z}_1+2 {\bar z}_2)
   z_4\right)\nonumber\\
   & & -{r_2} (2 {\bar z}_1 z_4+{\bar z}_2 z_4+2 z_1 {\bar z}_4+z_2
   {\bar z}_4)\Biggr) \epsilon +h^{z_1^2{\bar z}_{\bar 1}^2}\sqrt[18]{\cal V} \left(z_4 {\bar z}_1^2+2 {\bar z}_2 z_4
   {\bar z}_1+\sqrt[18]{\cal V} {\bar z}_4\right)\nonumber\\
   & & \hskip-0.8in\times \left(\sqrt[18]{\cal V} z_4+z_1 (z_1+2 z_2)
   {\bar z}_4\right)\Biggr]\Biggr)\Biggr]-\frac{1}{|z_1-z_2|^2}\Biggl[\Biggl(\epsilon  \Biggl[{r_2}
   \Biggl(-\sqrt[18]{\cal V} (3 z_1 {\bar z}_1+2 z_2 {\bar z}_1+4 z_1 {\bar z}_2+3 z_2
   {\bar z}_2)+\left(2 z_1^2+3 z_2 z_1+z_2^2\right)\nonumber\\
   & & \hskip-0.8in\times \left({\bar z}_1^2+3 {\bar z}_2 {\bar z}_1+2
   {\bar z}_2^2\right)+\sqrt[9]{\cal V}\Biggr) -h^{z_1^2{\bar z}_{\bar 1}^2}\sqrt[18]{\cal V} z_1 (z_1+2 z_2) {\bar z}_2 (2
   {\bar z}_1+{\bar z}_2) \left((z_1+z_2) ({\bar z}_1+{\bar z}_2)-\sqrt[18]{\cal V}\right)\Biggr]\nonumber\\
   & & +h^{z_1^2{\bar z}_{\bar 1}^2}\sqrt[12]{\cal V}
   \Biggl[-\left({\bar z}_1^2+{\bar z}_2 {\bar z}_1+{\bar z}_2^2\right) {\bar z}_4 z_1^3-\left((2 {\bar z}_2
   z_4+3 z_2 {\bar z}_4) {\bar z}_1^2+3 {\bar z}_2 ({\bar z}_2 z_4+z_2 {\bar z}_4)
   {\bar z}_1+{\bar z}_2^2 ({\bar z}_2 z_4+3 z_2 {\bar z}_4)\right) z_1^2\nonumber\\
   & & -z_2 \left(2
   ({\bar z}_2 z_4+z_2 {\bar z}_4) {\bar z}_1^2+{\bar z}_2 (3 {\bar z}_2 z_4+2 z_2
   {\bar z}_4) {\bar z}_1+{\bar z}_2^2 ({\bar z}_2 z_4+2 z_2 {\bar z}_4)\right) z_1-z_2^2
   {\bar z}_2 \nonumber\\
   & & \times\left(2 {\bar z}_1^2+3 {\bar z}_2 {\bar z}_1+{\bar z}_2^2\right) z_4+\sqrt[18]{\cal V}
   \left(({\bar z}_1+{\bar z}_2) {\bar z}_4 z_1^2+\left(z_4 {\bar z}_2^2+2 {\bar z}_1 z_4
   {\bar z}_2+2 z_2 {\bar z}_4 {\bar z}_2+2 {\bar z}_1 z_2 {\bar z}_4\right) z_1+z_2
   {\bar z}_2 (2 {\bar z}_1+{\bar z}_2) z_4\right)\Biggr]\Biggr)
   \end{eqnarray*}

\begin{eqnarray*}
    & & \times\Biggl[\Biggl(\epsilon  \left({r_2}+h
   \left(\left(z_1^2+z_2 z_1+z_2^2\right) \left({\bar z}_1^2+{\bar z}_2
   {\bar z}_1+{\bar z}_2^2\right)-\sqrt[18]{\cal V} (z_1+z_2) ({\bar z}_1+{\bar z}_2)\right)\right)\nonumber\\
   & & +h^{z_1^2{\bar z}_{\bar 1}^2}\sqrt[36]{\cal V}
   \Biggl(({\bar z}_1+{\bar z}_2) {\bar z}_4 z_1^2+\left(z_4 {\bar z}_1^2+{\bar z}_2 z_4
   {\bar z}_1+z_2 {\bar z}_4 {\bar z}_1+{\bar z}_2^2 z_4+z_2 {\bar z}_2 {\bar z}_4\right)
   z_1\nonumber\\
   & & -\sqrt[18]{\cal V} ({\bar z}_1 z_4+{\bar z}_2 z_4+(z_1+z_2) {\bar z}_4)+z_2
   \left(z_4 {\bar z}_1^2+{\bar z}_2 z_4 {\bar z}_1+z_2 {\bar z}_4 {\bar z}_1+{\bar z}_2^2
   z_4+z_2 {\bar z}_2 {\bar z}_4\right)\Biggr)\Biggr)\nonumber\\
    & & \times\Biggl(\left(2 h^{z_1^2{\bar z}_{\bar 1}^2}\sqrt[18]{\cal V} (z_1+z_2)
   {\bar z}_2 (2 {\bar z}_1+{\bar z}_2)-2 {r2} ({\bar z}_1+2 {\bar z}_2)\right) \epsilon ^2\nonumber\\
   & & +2 \left(h^{z_1^2{\bar z}_{\bar 1}^2}\sqrt[12]{\cal V}
   ({\bar z}_1 (2 {\bar z}_2 z_4+(z_1+z_2) {\bar z}_4)+{\bar z}_2 ({\bar z}_2 z_4+2
   (z_1+z_2) {\bar z}_4))-{r2} \sqrt[36]{\cal V} {\bar z}_4\right) \epsilon +2 h^{z_1^2{\bar z}_{\bar 1}^2}\sqrt[18]{\cal V}
   (z_1+z_2) {\bar z}_4\nonumber\\
   & & \times \left(z_4 {\bar z}_2^2+2 {\bar z}_1 z_4 {\bar z}_2+\sqrt[18]{\cal V}
   {\bar z}_4\right)\Biggr)-2 h^{z_1^2{\bar z}_{\bar 1}^2}\Biggl(\epsilon  \left((2 z_1+z_2) \left({\bar z}_1^2+{\bar z}_2
   {\bar z}_1+{\bar z}_2^2\right)-\sqrt[18]{\cal V} ({\bar z}_1+{\bar z}_2)\right)+\sqrt[36]{\cal V}\nonumber\\
   & & \times \left(z_4
   {\bar z}_1^2+({\bar z}_2 z_4+(2 z_1+z_2) {\bar z}_4) {\bar z}_1+{\bar z}_2^2 z_4+(2
   z_1+z_2) {\bar z}_2 {\bar z}_4-\sqrt[18]{\cal V} {\bar z}_4\right)\Biggr)\nonumber\\
   & & \times \Biggl[\left(h^{z_1^2{\bar z}_{\bar 1}^2}\sqrt[18]{\cal V} z_1
   (z_1+2 z_2) {\bar z}_2 (2 {\bar z}_1+{\bar z}_2)+{r2} \left(\sqrt[18]{\cal V}-(2 z_1+z_2)
   ({\bar z}_1+2 {\bar z}_2)\right)\right) \epsilon ^2\nonumber\\
   & & +\sqrt[36]{\cal V} \Biggl(h^{z_1^2{\bar z}_{\bar 1}^2}\sqrt[18]{\cal V} \left(({\bar z}_1+2 {\bar z}_2)
   {\bar z}_4 z_1^2+2 \left(z_4 {\bar z}_2^2+2 {\bar z}_1 z_4 {\bar z}_2+2 z_2 {\bar z}_4
   {\bar z}_2+{\bar z}_1 z_2 {\bar z}_4\right) z_1+z_2 {\bar z}_2 (2 {\bar z}_1+{\bar z}_2)
   z_4\right)\nonumber\\
   & & -{r_2} ({\bar z}_1 z_4+2 {\bar z}_2 z_4+2 z_1 {\bar z}_4+z_2
   {\bar z}_4)\Biggr) \epsilon +h^{z_1^2{\bar z}_{\bar 1}^2}\sqrt[18]{\cal V} \left(z_4 {\bar z}_2^2+2 {\bar z}_1 z_4
   {\bar z}_2+\sqrt[18]{\cal V} {\bar z}_4\right) \left(\sqrt[18]{\cal V} z_4+z_1 (z_1+2 z_2)
   {\bar z}_4\right)\Biggr]\Biggr]\Biggr]\nonumber\\
   & & -\frac{1}{|z_1-z_2|^2}\Biggl[\Biggl(-h^{z_1^2{\bar z}_{\bar 1}^2}\sqrt[18]{\cal V}
   \left(\epsilon  z_2 \left(2 z_1^2+3 z_2 z_1+z_2^2\right)+\sqrt[36]{\cal V}
   \left(z_1^2+z_2 z_1+z_2^2\right) z_4\right) {\bar z}_1^3\nonumber\\
   & & +\Biggl[\epsilon  \left(2 {r_2}
   \left(z_1^2+3 z_2 z_1+2 z_2^2\right)+h^{z_1^2{\bar z}_{\bar 1}^2}\sqrt[18]{\cal V} z_2 (2 z_1+z_2)
   \left(\sqrt[18]{\cal V}-3 (z_1+z_2) {\bar z}_2\right)\right)\nonumber\\
   & & +h^{z_1^2{\bar z}_{\bar 1}^2}\sqrt[12]{\cal V} \left(-(3 {\bar z}_2 z_4+2
   z_2 {\bar z}_4) z_1^2-3 z_2 ({\bar z}_2 z_4+z_2 {\bar z}_4) z_1+\sqrt[18]{\cal V}
   (z_1+z_2) z_4-z_2^2 (3 {\bar z}_2 z_4+z_2 {\bar z}_4)\right)\Biggr]
   {\bar z}_1^2\nonumber\\
   & & -\Biggl(\epsilon  \left(2 h^{z_1^2{\bar z}_{\bar 1}^2}\sqrt[18]{\cal V} z_2 (2 z_1+z_2) {\bar z}_2
   \left((z_1+z_2) {\bar z}_2-\sqrt[18]{\cal V}\right)+{r2} \left(\sqrt[18]{\cal V} (3 z_1+4 z_2)-3
   \left(z_1^2+3 z_2 z_1+2 z_2^2\right) {\bar z}_2\right)\right)\nonumber\\
   & & \hskip-0.8in+h^{z_1^2{\bar z}_{\bar 1}^2}\sqrt[12]{\cal V} \left({\bar z}_2
   \left(2 ({\bar z}_2 z_4+z_2 {\bar z}_4) z_1^2+z_2 (2 {\bar z}_2 z_4+3 z_2
   {\bar z}_4) z_1+z_2^2 (2 {\bar z}_2 z_4+z_2 {\bar z}_4)\right)-\sqrt[18]{\cal V} (2 z_1
   ({\bar z}_2 z_4+z_2 {\bar z}_4)+z_2 (2 {\bar z}_2 z_4+z_2 {\bar z}_4))\right)\Biggr)\nonumber\\
& & {\bar z}_1+\left(\sqrt[18]{\cal V}-(z_1+z_2) {\bar z}_2\right) \left(\epsilon  {r2}
   \left(\sqrt[18]{\cal V}-(z_1+2 z_2) {\bar z}_2\right)+h^{z_1^2{\bar z}_{\bar 1}^2}\sqrt[12]{\cal V} z_2 (2 z_1+z_2) {\bar z}_2
   {\bar z}_4\right)\Biggr)\nonumber\\
    & & \times\Biggl[\Biggl(\epsilon  \left({r_2}+h^{z_1^2{\bar z}_{\bar 1}^2}\left(\left(z_1^2+z_2
   z_1+z_2^2\right) \left({\bar z}_1^2+{\bar z}_2 {\bar z}_1+{\bar z}_2^2\right)-\sqrt[18]{\cal V}
   (z_1+z_2) ({\bar z}_1+{\bar z}_2)\right)\right)
   \end{eqnarray*}

   \begin{eqnarray*}
   & & +h^{z_1^2{\bar z}_{\bar 1}^2}\sqrt[36]{\cal V} \Biggl(({\bar z}_1+{\bar z}_2) {\bar z}_4
   z_1^2+\left(z_4 {\bar z}_1^2+{\bar z}_2 z_4 {\bar z}_1+z_2 {\bar z}_4
   {\bar z}_1+{\bar z}_2^2 z_4+z_2 {\bar z}_2 {\bar z}_4\right) z_1\nonumber\\
   & & -\sqrt[18]{\cal V} ({\bar z}_1
   z_4+{\bar z}_2 z_4+(z_1+z_2) {\bar z}_4)+z_2 \left(z_4 {\bar z}_1^2+{\bar z}_2
   z_4 {\bar z}_1+z_2 {\bar z}_4 {\bar z}_1+{\bar z}_2^2 z_4+z_2 {\bar z}_2
   {\bar z}_4\right)\Biggr)\Biggr)\nonumber\\
    & & \hskip-0.3in\times\Biggl(\left(2 h^{z_1^2{\bar z}_{\bar 1}^2}\sqrt[18]{\cal V} {\bar z}_1 z_2 ({\bar z}_1+2 {\bar z}_2)-{r_2}
   (2 {\bar z}_1+{\bar z}_2)\right) \epsilon ^2+\left(h^{z_1^2{\bar z}_{\bar 1}^2}\sqrt[12]{\cal V} \left(z_4 {\bar z}_1^2+2 {\bar z}_2 z_4
   {\bar z}_1+4 z_2 {\bar z}_4 {\bar z}_1+2 z_2 {\bar z}_2 {\bar z}_4\right)-{r_2} \sqrt[36]{\cal V}
   {\bar z}_4\right) \epsilon\nonumber\\
    & & +2 h^{z_1^2{\bar z}_{\bar 1}^2}\sqrt[18]{\cal V} z_2 {\bar z}_4 \left(z_4 {\bar z}_1^2+2 {\bar z}_2 z_4
   {\bar z}_1+\sqrt[18]{\cal V} {\bar z}_4\right)\Biggr)-2 h^{z_1^2{\bar z}_{\bar 1}^2}\Biggl[\epsilon  \left((2 z_1+z_2)
   \left({\bar z}_1^2+{\bar z}_2 {\bar z}_1+{\bar z}_2^2\right)-\sqrt[18]{\cal V}
   ({\bar z}_1+{\bar z}_2)\right)\nonumber\\
   & & +\sqrt[36]{\cal V} \left(z_4 {\bar z}_1^2+({\bar z}_2 z_4+(2 z_1+z_2)
   {\bar z}_4) {\bar z}_1+{\bar z}_2^2 z_4+(2 z_1+z_2) {\bar z}_2 {\bar z}_4-\sqrt[18]{\cal V}
   {\bar z}_4\right)\Biggr]\nonumber\\
    & & \times\Biggl[\left(h^{z_1^2{\bar z}_{\bar 1}^2}\sqrt[18]{\cal V} {\bar z}_1 z_2 (2 z_1+z_2) ({\bar z}_1+2
   {\bar z}_2)+{r2} \left(\sqrt[18]{\cal V}-(z_1+2 z_2) (2 {\bar z}_1+{\bar z}_2)\right)\right) \epsilon
   ^2\nonumber\\
   & & \hskip-0.9in+\sqrt[36]{\cal V} \Biggl[h^{z_1^2{\bar z}_{\bar 1}^2}\sqrt[18]{\cal V} \left(z_2 \left(2 z_4 {\bar z}_1^2+4 {\bar z}_2 z_4 {\bar z}_1+2
   z_2 {\bar z}_4 {\bar z}_1+z_2 {\bar z}_2 {\bar z}_4\right)+z_1 \left(z_4 {\bar z}_1^2+2
   {\bar z}_2 z_4 {\bar z}_1+4 z_2 {\bar z}_4 {\bar z}_1+2 z_2 {\bar z}_2
   {\bar z}_4\right)\right)\nonumber\\
   & & -{r_2} (2 {\bar z}_1 z_4+{\bar z}_2 z_4+z_1 {\bar z}_4+2 z_2
   {\bar z}_4)\Biggr] \epsilon +h^{z_1^2{\bar z}_{\bar 1}^2}\sqrt[18]{\cal V} \left(z_4 {\bar z}_1^2+2 {\bar z}_2 z_4
   {\bar z}_1+\sqrt[18]{\cal V} {\bar z}_4\right) \left(\sqrt[18]{\cal V} z_4+z_2 (2 z_1+z_2)
   {\bar z}_4\right)\Biggr]\Biggr]\Biggr]\nonumber\\
& &    +\frac{1}{|z_1-z_2|^2}\Biggl[\Biggl(-h^{z_1^2{\bar z}_{\bar 1}^2}\sqrt[18]{\cal V}
   \left(\epsilon  z_1 \left(z_1^2+3 z_2 z_1+2 z_2^2\right)+\sqrt[36]{\cal V}
   \left(z_1^2+z_2 z_1+z_2^2\right) z_4\right) {\bar z}_1^3+\Biggl[\epsilon  \Biggl(2 r_2
   \left(2 z_1^2+3 z_2 z_1+z_2^2\right)\nonumber\\
   & & +h^{z_1^2{\bar z}_{\bar 1}^2}\sqrt[18]{\cal V} z_1 (z_1+2 z_2)
   \left(\sqrt[18]{\cal V}-3 (z_1+z_2) {\bar z}_2\right)\Biggr)\nonumber\\
   & & +h^{z_1^2{\bar z}_{\bar 1}^2}\sqrt[12]{\cal V} \left(-{\bar z}_4 z_1^3-3
   ({\bar z}_2 z_4+z_2 {\bar z}_4) z_1^2-z_2 (3 {\bar z}_2 z_4+2 z_2 {\bar z}_4)
   z_1+\sqrt[18]{\cal V} (z_1+z_2) z_4-3 z_2^2 {\bar z}_2 z_4\right)\Biggr]
   {\bar z}_1^2\nonumber\\
   & & -\Biggl[\epsilon  \left(2 h^{z_1^2{\bar z}_{\bar 1}^2}\sqrt[18]{\cal V} z_1 (z_1+2 z_2) {\bar z}_2
   \left((z_1+z_2) {\bar z}_2-\sqrt[18]{\cal V}\right)+r_2 \left(\sqrt[18]{\cal V} (4 z_1+3 z_2)-3 \left(2
   z_1^2+3 z_2 z_1+z_2^2\right) {\bar z}_2\right)\right)\nonumber\\
   & & \hskip-0.8in+h^{z_1^2{\bar z}_{\bar 1}^2}\sqrt[12]{\cal V} \left({\bar z}_2
   \left({\bar z}_4 z_1^3+(2 {\bar z}_2 z_4+3 z_2 {\bar z}_4) z_1^2+2 z_2 ({\bar z}_2
   z_4+z_2 {\bar z}_4) z_1+2 z_2^2 {\bar z}_2 z_4\right)-\sqrt[18]{\cal V} \left({\bar z}_4
   z_1^2+2 ({\bar z}_2 z_4+z_2 {\bar z}_4) z_1+2 z_2 {\bar z}_2
   z_4\right)\right)\Biggr] {\bar z}_1\nonumber\\
   & & +\left(\sqrt[18]{\cal V}-(z_1+z_2) {\bar z}_2\right) \left(\epsilon
   r_2 \left(\sqrt[18]{\cal V}-(2 z_1+z_2) {\bar z}_2\right)+h^{z_1^2{\bar z}_{\bar 1}^2}\sqrt[12]{\cal V} z_1 (z_1+2 z_2)
   {\bar z}_2 {\bar z}_4\right)\Biggr)\nonumber\\
    & & \hskip-0.9in\times\Biggl(\Biggl[\epsilon  \left(r_2+h^{z_1^2{\bar z}_{\bar 1}^2}\left(\left(z_1^2+z_2
   z_1+z_2^2\right) \left({\bar z}_1^2+{\bar z}_2 {\bar z}_1+{\bar z}_2^2\right)-\sqrt[18]{\cal V}
   (z_1+z_2) ({\bar z}_1+{\bar z}_2)\right)\right)\nonumber\\
   & & +h^{z_1^2{\bar z}_{\bar 1}^2}\sqrt[36]{\cal V} \Biggl(({\bar z}_1+{\bar z}_2) {\bar z}_4
   z_1^2+\left(z_4 {\bar z}_1^2+{\bar z}_2 z_4 {\bar z}_1+z_2 {\bar z}_4
   {\bar z}_1+{\bar z}_2^2 z_4+z_2 {\bar z}_2 {\bar z}_4\right) z_1\nonumber\\
   & & -\sqrt[18]{\cal V} ({\bar z}_1
   z_4+{\bar z}_2 z_4+(z_1+z_2) {\bar z}_4)+z_2 \left(z_4 {\bar z}_1^2+{\bar z}_2
   z_4 {\bar z}_1+z_2 {\bar z}_4 {\bar z}_1+{\bar z}_2^2 z_4+z_2 {\bar z}_2
   {\bar z}_4\right)\Biggr)\Biggr]\nonumber\\
   \end{eqnarray*}

   \begin{eqnarray*}
   & & \hskip-0.4in\times\Biggl(\left(2 h^{z_1^2{\bar z}_{\bar 1}^2}\sqrt[18]{\cal V} z_2 {\bar z}_2 (2 {\bar z}_1+{\bar z}_2)-r_2
   ({\bar z}_1+2 {\bar z}_2)\right) \epsilon ^2+\left(h^{z_1^2{\bar z}_{\bar 1}^2}\sqrt[12]{\cal V} \left(z_4 {\bar z}_2^2+2 {\bar z}_1 z_4
   {\bar z}_2+4 z_2 {\bar z}_4 {\bar z}_2+2 {\bar z}_1 z_2 {\bar z}_4\right)-r_2 \sqrt[36]{\cal V}
   {\bar z}_4\right) \epsilon \nonumber\\
   & & +2 h^{z_1^2{\bar z}_{\bar 1}^2}\sqrt[18]{\cal V} z_2 {\bar z}_4 \left(z_4 {\bar z}_2^2+2 {\bar z}_1 z_4
   {\bar z}_2+\sqrt[18]{\cal V} {\bar z}_4\right)\Biggr)-2 h^{z_1^2{\bar z}_{\bar 1}^2}\Biggl(\epsilon  \left((2 z_1+z_2)
   \left({\bar z}_1^2+{\bar z}_2 {\bar z}_1+{\bar z}_2^2\right)-\sqrt[18]{\cal V}
   ({\bar z}_1+{\bar z}_2)\right)\nonumber\\
   & & +\sqrt[36]{\cal V} \left(z_4 {\bar z}_1^2+({\bar z}_2 z_4+(2 z_1+z_2)
   {\bar z}_4) {\bar z}_1+{\bar z}_2^2 z_4+(2 z_1+z_2) {\bar z}_2 {\bar z}_4-\sqrt[18]{\cal V}
   {\bar z}_4\right)\Biggr)\nonumber\\
    & & \times\Biggl(\left(h^{z_1^2{\bar z}_{\bar 1}^2}\sqrt[18]{\cal V} z_2 (2 z_1+z_2) {\bar z}_2 (2
   {\bar z}_1+{\bar z}_2)+r_2 \left(\sqrt[18]{\cal V}-(z_1+2 z_2) ({\bar z}_1+2 {\bar z}_2)\right)\right)
   \epsilon ^2+\sqrt[36]{\cal V}\nonumber\\
    & & \times\Biggl(h^{z_1^2{\bar z}_{\bar 1}^2}\sqrt[18]{\cal V} \left(z_2 \left(2 z_4 {\bar z}_2^2+4 {\bar z}_1 z_4
   {\bar z}_2+2 z_2 {\bar z}_4 {\bar z}_2+{\bar z}_1 z_2 {\bar z}_4\right)+z_1 \left(z_4
   {\bar z}_2^2+2 {\bar z}_1 z_4 {\bar z}_2+4 z_2 {\bar z}_4 {\bar z}_2+2 {\bar z}_1 z_2
   {\bar z}_4\right)\right)\nonumber\\
   & & -r_2 ({\bar z}_1 z_4+2 {\bar z}_2 z_4+z_1 {\bar z}_4+2 z_2
   {\bar z}_4)\Biggr) \epsilon +h^{z_1^2{\bar z}_{\bar 1}^2}\sqrt[18]{\cal V} \left(z_4 {\bar z}_2^2+2 {\bar z}_1 z_4
   {\bar z}_2+\sqrt[18]{\cal V} {\bar z}_4\right) \left(\sqrt[18]{\cal V} z_4+z_2 (2 z_1+z_2)
   {\bar z}_4\right)\Biggr)\Biggr)\Biggr]\nonumber\\
& & \Gamma_2=   \sqrt[18]{\cal V} \Biggl(r_2
   \left((z_1+z_2) ({\bar z}_1+{\bar z}_2)-\sqrt[18]{\cal V}\right) \epsilon ^2-r_2 \sqrt[36]{\cal V} ({\bar z}_1
   z_4+{\bar z}_2 z_4+(z_1+z_2) {\bar z}_4) \epsilon \nonumber\\
   & & +h^{z_1^2{\bar z}_{\bar 1}^2}\sqrt[9]{\cal V} (z_1+z_2)
   ({\bar z}_1+{\bar z}_2) z_4 {\bar z}_4\Biggr)\nonumber\\
    & & \times\Biggl(\epsilon  \left(r_2+h^{z_1^2{\bar z}_{\bar 1}^2}\left(\left(z_1^2+z_2
   z_1+z_2^2\right) \left({\bar z}_1^2+{\bar z}_2 {\bar z}_1+{\bar z}_2^2\right)-\sqrt[18]{\cal V}
   (z_1+z_2) ({\bar z}_1+{\bar z}_2)\right)\right)\nonumber\\
   & & +h^{z_1^2{\bar z}_{\bar 1}^2}\sqrt[36]{\cal V} \Biggl\{({\bar z}_1+{\bar z}_2) {\bar z}_4
   z_1^2+\left(z_4 {\bar z}_1^2+{\bar z}_2 z_4 {\bar z}_1+z_2 {\bar z}_4
   {\bar z}_1+{\bar z}_2^2 z_4+z_2 {\bar z}_2 {\bar z}_4\right) z_1-\sqrt[18]{\cal V} ({\bar z}_1
   z_4+{\bar z}_2 z_4+(z_1+z_2) {\bar z}_4)\nonumber\\
   & & +z_2 \left(z_4 {\bar z}_1^2+{\bar z}_2
   z_4 {\bar z}_1+z_2 {\bar z}_4 {\bar z}_1+{\bar z}_2^2 z_4+z_2 {\bar z}_2
   {\bar z}_4\right)\Biggr\}\Biggr)^2
   \end{eqnarray*}

\end{document}